\documentclass{jpp}
\usepackage{graphicx}
\usepackage{color}
\usepackage[utf8x]{inputenc} 
\usepackage[T1]{fontenc}
\usepackage{amsmath}
\usepackage{hyperref}

\shorttitle{Particle Acceleration and transport in the Galaxy}
\shortauthor{E. Amato \& S. Casanova}

\title{On particle acceleration and transport in plasmas in the Galaxy: theory and observations}

\author{Elena Amato\aff{1,2}
  \corresp{\email{amato@arcetri.astro.it}},
 \and Sabrina Casanova\aff{3,4}}

\affiliation{\aff{1} INAF - Osservatorio Astrofisico di Arcetri, Largo E. Fermi, 5, 50125, Firenze, Italy
\aff{2} Universit\`a degli Studi di Firenze, Via Sansone 1, 50019, Sesto Fiorentino (FI), Italy 
\aff{3} Institute of Nuclear Physics PAN, Radzikowskiego 152, 31-342 Kraków, Poland
\aff{4} Max Planck Institute for Nuclear Physics, Saupfercheckweg 1, 69117, Heidelberg, Germany }

\begin{document}

\maketitle

\begin{abstract}
Accelerated particles are ubiquitous in the Cosmos and play a fundamental role in many processes governing the evolution of the Universe at all scales, from the sub-AU ones relevant for the formation and evolution of stars and planets to the Mpc ones involved in Galaxy assembly. We reveal the presence of energetic particles in many classes of astrophysical sources thanks to their production of non-thermal radiation, and we detect them directly at Earth as Cosmic Rays. In the last two decades both direct and indirect observations have provided us a wealth of new, high quality data about Cosmic Rays and their interactions both in sources and during propagation, in the Galaxy and in the solar system. Some of the new data have confirmed existing theories about particle acceleration and propagation and their interplay with the environment in which they occur. Some others have brought about interesting surprises, whose interpretation is not straightforward within the standard framework and may require a change of paradigm in terms of our ideas about the origin of cosmic rays of different species or in different energy ranges. In this article, we will focus on Cosmic Rays of galactic origin, namely with energies below a few PeV, where a steepening is observed in the spectrum of energetic particles detected at the Earth. We review the recent observational findings and the current status of the theory about the origin and propagation of Galactic Cosmic Rays.
\end{abstract}

\section{Introduction}
\label{sec:intro}
Cosmic Rays have been known and studied for more than a century now (see e.g. \cite{Amato14,Blasi13} for recent reviews). They are highly energetic charged particles, mainly protons and He nuclei, with a minor fraction of heavier nuclei (1\%), electrons (2\%) and anti-matter particles (positrons and anti-protons, 1 \textperthousand). Their origin was associated with Supernova (SN) explosions in the Galaxy already in the 1930s and the mechanism by which they would be accelerated up to very high energies in the blast waves emerging from SN explosions was proposed in the late 1970s (see \S~\ref{sec:observations}). 

This association, however, has not yet found direct proof, and, as we discuss below, some recent developments, in both theory and high energy astrophysical observations, have actually cast doubts on whether Supernova Remnants (SNRs) should be considered as the primary CR accelerators in the Galaxy throughout the entire energy range, up to PeV ($10^{15}$ eV) energies. At the same time, direct CR measurements have shown that particle transport is more complex than traditionally considered, and that additional sources are likely needed at least for CR positrons (see e.g.~\cite{amatoblasi18} for a recent review, and Sec.~\ref{sec:cranti} for more details).

 In recent years it has become increasingly clear that understanding the processes that govern the acceleration and transport of energetic particles is a necessary key to unveil a number of unsettled questions in Astrophysics. Energetic particles - which we will often refer to as Cosmic Rays, in the following - play a role in many astrophysical systems. Just to mention a few, from the smallest to the largest scales, CRs affect the behaviour of planetary magnetospheres \citep{Griessmeier15}, they are believed to be the main ionizing agents penetrating dense gas clouds and regulating star formation \citep{Padovani20}, they are suspected to be the primary drivers of galactic winds and the rulers of galactic feedback \citep{Buck20}, and might even have a role in the generation of intergalactic magnetic fields \citep{Donnert18}. In some of these situations the interaction between CRs and the ambient plasma is simply described in terms of momentum exchange, in some others the nature of CRs as charged particles, carrying an electric current, is instead fundamental. In all cases, the correct description of the interaction between CRs and the ambient medium is an extremely challenging plasma physics problem (see e.g. \cite{Zweibel13}).

All the information we can use to try to understand this problem comes from two kinds of observations: direct detection of CRs at Earth and detection of non-thermal radiation from astrophysical sources. In the last two decades both types of observations have experienced enormous developments. 

In terms of indirect detection, most of the news have come from high energy telescopes observing the sky in the X-ray and gamma-ray bands. The launch of {\it Chandra} was quickly followed by two important results in terms of CR physics. First of all, the detection of non-thermal keV photons from SNRs highlighted the presence of multi-TeV electrons in these sources: since in most of the acceleration mechanisms considered in Astrophysics - and certainly in the one proposed for SNRs - the acceleration process only depends on particle rigidity, this was also taken as indirect evidence of the presence of multi-TeV hadrons in SNRs. In addition, the excellent spatial resolution of {\it Chandra} allowed to resolve the region of non-thermal X-ray emission, showing it to be very thin, with a typical size of few \% of pc (see \cite{Vink12} for a review). Such a small size, when interpreted as due to synchrotron energy losses of the emitting electrons, indicates the presence of a magnetic field of several $\times$ 100 $\mu$G downstream of the SNR shock. Such a large value of the magnetic field cannot simply result from shock compression of the typical interstellar field ($B_{\rm ISM}\approx 3 \mu$G), but rather requires an extremely efficient amplification process to be at work. 

At the same time of these discoveries, also theory was making a major step forward with the identification of a mechanism through which CRs could provide magnetic field amplification well beyond the quasi-linear theory limit of $\delta B/B\approx 1$. The process invoked is the non-resonant streaming instability \citep{Bell04} which we will discuss more in \S~\ref{sec:cracc}: this is a current induced instability that operates in situations where the CR energy density dominates over that of the background magnetic field. This condition is easily realized in young, fast expanding SNRs, if CRs are accelerated with efficiency of order 10\%, as required by the SNR-CR connection paradigm. The X-ray observations showing evidence for amplified magnetic fields were then taken as a proof of this connection: the amplified magnetic field implied that efficient CR acceleration was ongoing, and provided, at the same time, the ideal conditions for CRs to reach very high energies.

At higher photon energies, the gamma-ray satellites {\it AGILE} and {\it Fermi}, observing the sky in the 20 MeV - 300 GeV energy range, and the Cherenkov detectors H.E.S.S., VERITAS and MAGIC operating between tens of GeV and $\sim 10$ TeV, were providing, in the meantime, interesting but somewhat surprising data. For the first time, direct evidence of relativistic protons was found in two middle-aged ($\sim 20-30$ kyr old) SNRs interacting with molecular clouds, W44 \citep{FermiW44,AgileW44} and IC443 \citep{AgileIC443,FermiIC443}. These sources were not expected to be efficient CR accelerators and indeed the proton spectrum inferred from observations is cut-off at relatively low energies, $\sim 10$ GeV in W44, $\sim 10$ TeV in IC443. Follow-up theoretical studies showed that, in fact, at least in the case of W44, we are likely not witnessing fresh acceleration of CRs, with particles directly extracted from the interstellar plasma crossing the shock, but rather re-acceleration of particles from the galactic CR pool \citep[e.g.][]{,Uchiyama+10,Cardillo+16}. As far as young sources are concerned, gamma-ray observations have been partly disappointing, failing to provide indisputable evidence of hadronic acceleration \citep{FunkRev}, and often showing steeper spectra than theoretically predicted in the case of efficient acceleration \citep{Amato14}. We will discuss gamma-ray observations of SNRs and the implications of these findings on CR acceleration in SNRs in \S~\ref{sec:observations} and \S~\ref{sec:cracc}, respectively.

In terms of direct detection, experiments such as PAMELA, AMS-02 and CREAM have provided us with very precise measurements of the CR spectrum and composition up to TeV energies. Among the many discoveries, the most important ones in terms of implications on our understanding of CR acceleration and transport are: 1) the detection of a hardening in the spectra of protons, He nuclei, and virtually all primary nuclei \citep{Adriani11pHe,AMS02p,AMS02He,CREAM}, at $R\approx 300$ GV, where $R=cp/Ze$ is the particle rigidity, with $p$ the particle momentum, $Z$ its atomic number, $e$ the electron charge and $c$ the speed of light; 2) the hardening found in the spectrum of secondary CRs at $R\approx 200$ GV, with a change of spectral slope that is $\sim$ twice as large as that of primaries \citep{AMS02Sec}; 3) the energy dependence of the ratio between secondary and primary cosmic rays, that has now been measured with excellent statistics up to TV rigidities \citep{AMS02BC} and provides invaluable constraints on the energy dependence of particle transport in the Galaxy; 4) a rise in the fraction of positrons-to-electrons at energies larger than 30 GeV \citep{PamelaPos,AMS02frac}, possibly suggesting the presence of an additional source of positrons in the Galaxy; 5) the spectrum of anti-p which is unexpectedly very close to that of protons and positrons \citep{AMS02Pbar}. We will discuss these discoveries and their implications for the origin of Cosmic Rays in \S~\ref{sec:crtransp}.

\section{Testing the SNR paradigm}
\label{sec:observations}
 While already in the 1930s \cite{PhysRev.46.76.2} suggested that CRs might originate from supernovae, the SNR paradigm was later formulated based on energetic arguments: SN explosions in the Galaxy can easily provide the power needed to sustain the CR population \citep{1957RvMP...29..235M,1964ocr..book.....G}.  Assuming that the locally measured CR energy density, $w_{cr}$ $\sim$ 1 eV/cm$^3$, is representative of the CR energy density everywhere in the Galaxy, the CR production rate is $ \dot{W}_{CR} = V w_{cr}/t_{conf}$ with $t_{conf}$ and $V$ the CR confinement time and volume respectively. Using $t_{conf}\approx 15$ Myr \citep{Yanasak2001} and describing the Galaxy as a cylinder of radius $R_d\approx 15$kpc and height $H_d\approx 5$ kpc \citep{EvoliBe,MorlinoFe20} we can estimate
 $\dot W_{CR}\approx 1-3 \times {10}^{41}$ erg/s. 
 SN events happen in the Galaxy about every 30 years and typically release 10$^{51}$ erg. Thus the power needed to sustain the galactic CR population turns out to be about 10 $\%$ of the power provided by SN events. The appeal of the SNR hypothesis was then increased by the formulation of the theory of diffusive shock acceleration \citep{Bell78,BlandOstr78}: this acceleration process, expected to be active at most shock waves, predicts the spectrum of accelerated particles to be a power-law, with an index close to $-2$ in the case of a strong shock, such as the blast wave of a SN explosion. Such a spectrum perfectly fits what inferred for the sources of Galactic CRs, lending support to the association. An equally remarkable, but much more recent finding, concerns the acceleration efficiency: kinetic simulations of diffusive shock acceleration (DSA) in a regime that is close to represent a SN blast wave have finally become available in the last decade, and they show that for parallel shock waves (magnetic field perpendicular to the shock surface) the acceleration efficiency is 10-20\% \citep{Caprioli14a}, exactly as required for the SNR-CR paradigm to work. 

Testing the SNR paradigm for the origin of CRs has long been one of the top priorities in High Energy Astrophysics. Since the early days, photons with energies in the range between GeVs and TeVs have provided a unique tracer of the CR population far from Earth: the interactions of CRs with the environment are expected, indeed, to produce radiation in this energy range. The process that most directly reveals the presence of hadrons is the decay of neutral pions produced when CR hadrons collide inelastically with ambient gas in the ISM: this process is commonly referred to as the {\it hadronic} production mechanism. As a rule of thumb, an inelastic nuclear collision will produce gamma rays with $\sim$ 10 \% of the energy of the parent cosmic ray, so, for instance, photons of several tens to hundreds of TeV for CRs close to PeVs. Typically, the spectrum of the hadronic radiation at TeV mimics the parent CR spectrum shifted to lower energy by a factor 20-30 \citep{PhysRevD.74.034018,PhysRevD.90.123014}. In the same energy range, radiation can originate from {\it leptonic} production mechanisms, mainly inverse Compton (IC) scattering of CR electrons off ambient radiation fields, and non thermal electron bremsstrahlung \citep{RevModPhys.42.237}, which plays an important role in dense gas regions at sub GeV to GeV energies. A recurring difficulty when trying to infer the CR population from the GeV and TeV $\gamma$-ray emission from SNRs is to break the hadronic-leptonic degeneracy and unveil the dominant emission process: the spectral and morphological features of the emission are crucial to this task. 

The gamma-ray spectrum is also affected by the particles' energy losses. For gamma-ray emitting particles, a number of loss mechanisms that are important at lower energies, like ionization and bremsstrahlung, have a negligible impact. The most relevant loss processes affect electrons, whose synchrotron and Inverse Compton emission must be properly taken into account when trying to disentangle the radiative contribution from leptons and hadrons. Electron synchrotron, particularly efficient in the strong magnetic fields believed to be associated with particle accelerators, has a fundamental role in cooling the CR electron population and thus shape the leptonic $\gamma$-ray spectrum. Therefore, observations of non-thermal radio and X-ray synchrotron emission can provide invaluable constraints on the electron population, and clues to solve the hadronic-leptonic degeneracy.

The SNR paradigm implies that young SNRs accelerate CRs up to the knee and that on average each SNR provides roughly 10$^{50}$ erg in accelerated particles. In order to test whether SNRs are the main contributors to the Galactic CR population, it is thus crucial to assess the energetics in electrons and protons and the spectra of accelerated particles in young SNRs.

Studies of the GeV-to-TeV spectra of the remnants, of their morphology and of the spatial correlation between TeV gamma-rays and the distribution of ambient gas and of X-ray radiation are the three most powerful instruments to extract the population of accelerated particles in the SNRs. GeV to TeV gamma-ray studies of both young SNRs \citep{1994A&A...287..959D}, and dense molecular clouds close to middle-aged SNRs  \citep{1994A&A...285..645A} were undertaken to assess the role of SNRs in the origin of Galactic CRs, by understanding and disentangling the dynamics of the coexisting and competing processes of acceleration and escape or release of particles in the interstellar medium (ISM). A useful repository of high energy observations of SNRs is the Manitoba Catalogue (\url{http://snrcat.physics.umanitoba.ca/}). 

Several young shell-type SNRs have been observed at TeV energies. Among these, the best studied ones are Cas~A \citep{2001A&A...370..112A,2007A&A...474..937A,2017MNRAS.472.2956A}, Tycho \citep{2011ApJ...730L..20A,2017ApJ...836...23A}, and RX~J1713.7-3946 \citep{2004Natur.432...75A,2018A&A...612A...6H}. We will discuss the latter as an example of how the SNR paradigm is tested with radiation from X-rays to gamma-rays. 

As an example of multi-wavelength studies concerning middle-aged SNRs showing the so-called pion bump, a clear signature of the presence of relativistic protons, we will discuss the case of the brightest GeV source among these: W44.

\subsection{Young supernova remnants: the case of RX J1713.7-3946}
\label{sec:rxj}
\begin{figure}
    \centering
    \includegraphics[width=\textwidth]{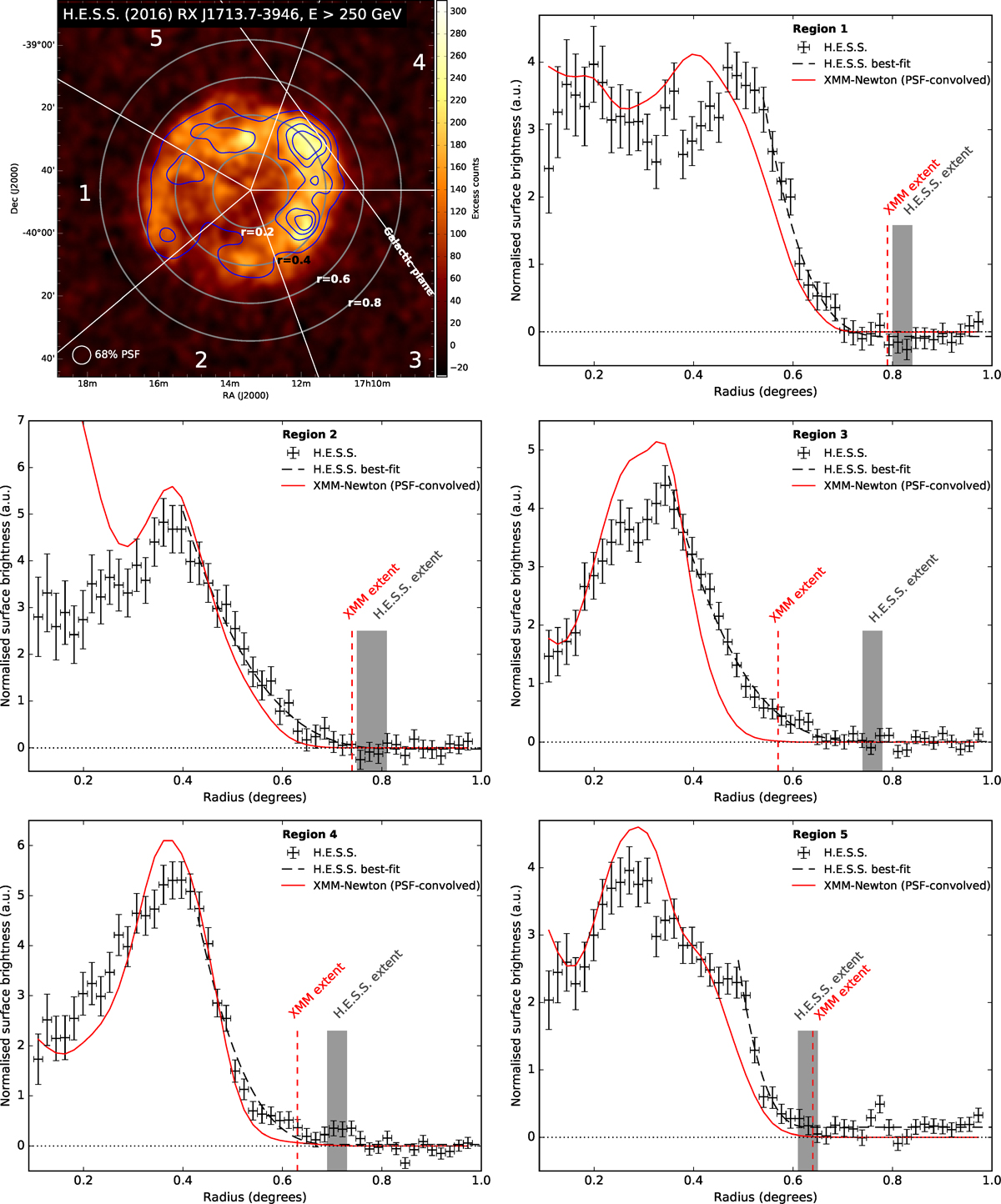}
    \caption{Top left panel: H.E.S.S. gamma-ray skymap of RX~J1713.7-3946 \citep{ 2004Natur.432...75A,2018A&A...612A...6H}. In blue the contours of the shell as observed by XMM in X-rays \citep{2004A&A...427..199C}. The 
    morphology of the TeV shell resembles closely the X-ray shell. Top right panel and following: radial profiles of the emission at selected places along the shell from H.E.S.S. and XMM-Newton. In four out of five regions of the shell, the TeV supernova remnant is more extended than the X-ray one as a result of possible escape of high energy particles from the shell. The coordinate "Radius", used for the one-dimensional profiles, refers to the mean radius "r" of the shell over which average is performed.}
    \label{RXJmap}
\end{figure}

The supernova remnant RX J1713.7-3946, possibly associated with a star explosion seen by Chinese astronomers in the year AD393 \citep{article1996,article1997}, is one of the brightest sources of gamma rays in the TeV energy range. The sky map of RX J1713.7-3946, obtained with the H.E.S.S. telescope, is shown in the top left panel of Fig.~\ref{RXJmap} \citep{2018A&A...612A...6H}. The morphology of the TeV shell resembles closely the XMM X-ray shell \citep{2004A&A...427..199C}, plotted with blue contours in the top left panel of the figure. The other five panels of Fig.~\ref{RXJmap} show the radial profiles at keV and TeV photon energies from different regions of the shell: in four cases, the TeV supernova remnant extends, in radius, beyond the X-ray emitting shell  \citep{2004Natur.432...75A,2007Natur.449..576U,2008ApJ...685..988T,2018A&A...612A...6H}. 
The main shock position and extent are visible in the X-ray data and the $\gamma$-ray emission extending further is either due to accelerated particles escaping the acceleration (shock) region or particles in the shock precursor region. 

VHE (Very High Energy) electrons, up to hundreds of TeV, are accelerated in the shell of RX J1713-3946. These particles emit hard X-rays through synchrotron processes and TeV radiation through inverse Compton scattering off CMB photons very efficiently. A crucial piece of evidence in support of efficient particle acceleration comes from hard X-ray measurements with Suzaku \citep{2007Natur.449..576U}. Suzaku detected X-ray hot-spots brightening and disappearing within one year timescale. If the X-ray variability is associated to the acceleration and immediate synchrotron cooling of the accelerated electrons, then the magnetic field in the shell must be roughly two orders of magnitude higher than the average magnetic field in the Galaxy, about 100 $\mu$Gauss. Broadband X-ray spectrometric measurements of RXJ1713.7-3946 indicate also that electron acceleration proceeds in the most effective {\it Bohm-diffusion} regime. Finally, the presence of strongly amplified magnetic fields lends further support to the idea that not only electrons, but also protons up to 100 TeV are accelerated in the shell, as these amplified magnetic fields would be the result of CR induced instabilities \citep{2007Natur.449..576U}. 

\begin{figure}
    \centering
    \includegraphics[width=\textwidth]{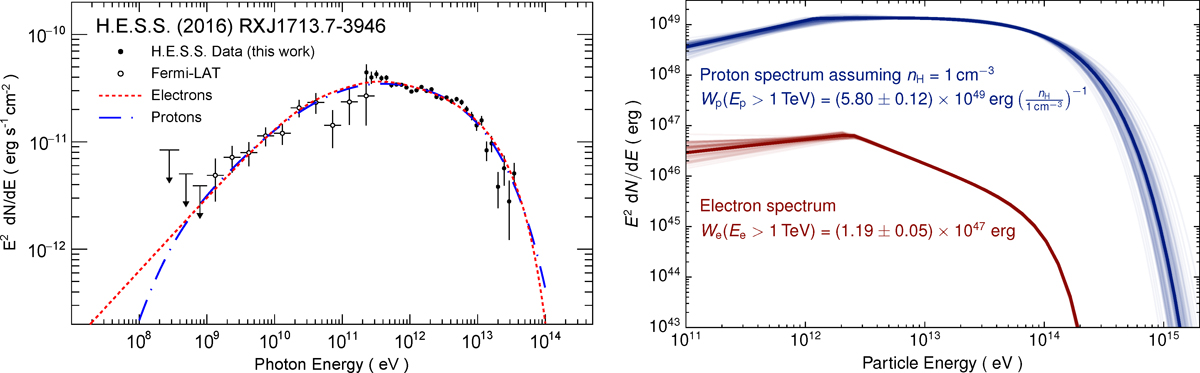}
 {\caption{{\it Left Panel:} The black data points represent the GeV (Fermi-LAT) and TeV (H.E.S.S.) spectral energy distribution of the radiation from the shell of RX J1713.7-3946. The lines represent the multi-wavelength spectra predicted by NAIMA \cite{naima}, assuming either a leptonic or hadronic origin of the emission. {\it Right Panel:} The spectral energy distributions of the electrons and protons producing the leptonic and hadronic emissions shown on the left panel.}
    \label{RXJparticle}}
\end{figure}

The high-energy part of the spectrum of RX~J1713.7-3946, from X-rays to TeV photons, can be used to infer the population of particles producing this emission. The particle spectrum of the young SNR in the {\it Left panel} of Fig.~\ref{RXJparticle} is characterised by a spectral index close to -2, as predicted by DSA, up to a few TeV (=${10}^{12}$ eV). While the hard spectrum shows that particles are efficiently accelerated in the shell of this young SNR, the cut-off at few TeV suggests that RX~J1713.7-3946 is currently accelerating electrons and protons up to 100 TeV but not up to PeV energies \citep{2018A&A...612A...6H}. 

As mentioned, protons are likely accelerated in the supernova shell through the same mechanism accelerating electrons, but it is not clear what fraction of the energy input by the supernova explosion goes into the acceleration of electrons and what fraction goes into the acceleration of protons. Both protons or electrons can be, in fact, responsible for the TeV emission from RX~J1713.7-3946 ({\it leptonic-hadronic degeneracy}, see e.g. \cite{MorlinoRXJ}). If electrons are responsible for the emission from the shell, by inverse Compton scattering of the ambient radiation fields, one speaks of {\it leptonic scenario}. On the other hand, if the emission is produced by protons colliding with the ambient gas, one speaks of {\it hadronic scenario}. In Fig.~\ref{RXJparticle} the measurements of the shell emission carried out by Fermi LAT \citep{2011ApJ...734...28A}, HESS \citep{2018A&A...612A...6H} and Suzaku \citep{2008ApJ...685..988T} are compared to leptonic ({\it Left panel}) and hadronic ({\it Right panel}) model predictions by \cite{naima}. The total energetics in accelerated electrons and protons in the relevant leptonic and hadronic models of gamma-rays can be estimated by assuming a distance to the source of about 1~kpc \citep{article1996,article1997}. The required budget in electrons is determined only by the reported gamma-ray fluxes, if the target radiation is the CMB: one finds $W_{\rm e} \simeq 1.2 \times 10^{47}$erg. On the other hand, the total energy budget of protons in hadronic models depends on the highly uncertain ambient gas density: $W_{\rm p} \simeq 5 \times {10}^{49} (n_H/1 \mathrm{ cm^{-3})^{-1}}$erg. The leptonic model in the {\it Left panel} corresponds to the electron IC scattering off CMB photons and an infrared radiation field with energy density 0.415\,eV cm$^{-3}$ and temperature T = 26.5 K.

The leptonic model requires a break in the electron spectrum at 2.5\,TeV with the index of the electron energy distribution changing from $\Gamma=1.7$ to $\Gamma=3$ beyond the break. If due to synchrotron cooling, this spectral break implies a magnetic field of about $140~\mu{\mathrm{ G}}$, which is much larger than what the X-ray flux allows within the same scenario, $B\approx 15 \mu \mathrm{G}$. Similar difficulties arise if one tries to explain the break as a result of Inverse Compton cooling, in which case a photon field of energy density of 140\,eV\,cm$^{-3}$, would be required, about 100 times larger than the average galactic value.

If the emission is hadronic in origin, a spectrum of protons with index $\Gamma=1.5$, steepening to $\Gamma=1.9$ at about 1 TeV, and with an exponential cut-off at 79\,TeV is required. The break in the proton population spectrum can be explained if the hadronic emission is produced mostly in dense clumpy regions. In fact, low energy protons can be efficiently excluded from dense gas region during the timescale of $\sim$ 1000 years since the SNR explosion. The exclusion of low energy cosmic rays would also explain the hard $\gamma$-rays detected by the Fermi LAT telescope \citep{2012ApJ...746...82F,2014MNRAS.445L..70G,2012ApJ...744...71I}, and the lack of thermal X-ray emission from the shell. The latter has traditionally been one of the strongest argument against a possible hadronic origin of the emission from the shell of RXJ1713.7-3946, since it implies an ambient gas density as low as $0.1 \rm cm^{-3}$ \citep{2010ApJ...712..287E,2012ApJ...746...82F,2014MNRAS.445L..70G}. Finally hadronic gamma-ray production in gas condensations results in a narrow angular distribution of the radiation, which is beyond the reach of the current generation of Imaging Air Cherenkov Telescopes (IACTs), but could be tested with the upcoming Cherenkov Telescope Array (CTA) which will have an angular resolution of about 1-2\,arcmin \citep{2010ApJ...708..965Z,2018APh...100...69A}. 


The conclusion one reaches, after investigating the spectrum and morphology of RX~J1713.7$-$3946 as currently known, is that neither the hadronic nor the leptonic scenario is fully satisfactory. Each emission mechanism has strengths and weaknesses when compared with observations, suggesting that the ambient conditions might differ in different parts of the remnant, making one or the other process locally dominant. While for some specific SNRs one of the two scenarios might indeed be favored (see e.g. \cite{FunkRev}), the general conclusions is that no known SNR has proved to accelerate particles beyond 100 TeV. 

\subsection{Molecular Clouds close to middle aged SNRs: the case of W44}
\label{sec:mc}
Molecular clouds are regions of the Galaxy, typically a few tens of parsecs in radius, where the density of cold molecular gas is often orders of magnitude higher than elsewhere in the diffuse 
ISM.  Stars are believed to be born in these clouds. Radio observations of the rotational $1\rightarrow 0$ line emission of carbon monoxide are mainly used to trace the distribution of molecular gas \citep{2015ARA&A..53..583H}. The cloud Galactocentric distance is usually estimated using a kinematical distance method, through each the radial velocity of the cloud is related to the rotation velocity of the Galaxy. Cross calibrations with the distance of spiral arms or known objects with precise parallax determination are also carried out \citep{2009ApJ...699.1153R,2014ApJ...783..130R}.

Giant molecular clouds, which are typically 5 to 200 parsecs in diameter and have masses of 10 thousand to 10 million solar masses \citep{Murray_2011}, are excellent laboratories for CR physics. In these clouds, the hadronic channel of gamma-ray production is enhanced by the high target density, and easily dominates over the leptonic production mechanism. Contrary to the warmer atomic gas phase, which is homogeneously distributed in the Galaxy, one or a few giant molecular clouds are essentially the dominant contributions to the gas column density along a given direction in the sky. The emission enhancement associated with the clouds makes it possible to precisely locate along a given line of sight where the CR population produces the emission. Molecular clouds are thus used to perform a sort of {\it tomography}, and obtain a three dimensional view of the CR distribution in the Galaxy \citep{1981Natur.292..430I,1996A&A...309..917A,2001SSRv...99..187A,2010PASJ...62..769C,PhysRevD.101.083018,Baghmanyan_2020}. 

The plasma conditions in molecular clouds are generally different from those in the diffuse ISM. In addition to the gas density, also the magnetic field energy density and turbulence level are enhanced (see e.g. \cite{Crutcher12} for a comprehensive review). This will lead to a suppression of the diffusion coefficient and effective exclusion of lower energy CRs from the cloud
\citep{2007ApESS.309..365G}. If CRs can penetrate clouds, the $\gamma$-ray emission from $\pi^0$-decay depends only upon the total mass of the cloud, $M_{cl}$, its distance from the Earth, $d$, and the CR flux within the cloud, $\Phi_{CR}$. The latter is thus determined as $\Phi_{CR} \propto \frac{\Phi_{\gamma} d^2} {M_{cl}}$, where ${\Phi_{\gamma}}$ is the $\gamma$-ray flux from the cloud. Lower energy cosmic rays can be effectively excluded from penetrating clouds, which results in peculiar features in the gamma-ray spectrum from clouds, such as hardenings with respect to the average interstellar spectrum \citep{2007ApESS.309..365G}.

The role of molecular clouds in testing the SNR paradigm is made particularly crucial by the time evolution of particle acceleration in SNRs. Within a DSA scenario, SNRs accelerate the highest energy CRs (in principle up to a few PeVs, but see \S~\ref{sec:cracc}) at the transition between the free expansion and the Sedov phase, which typically happens a few tens to a few hundred years after the supernova explosion, depending on the explosion type and properties of the surrounding medium. During the Sedov phase the SN shock slows down and the magnetic field intensity decreases, so that the most energetic particles cannot be confined any longer and are free to escape. In practice, a SNR can accelerate the highest energy particles only for a short time. This fact, coupled to the low rate of PeV particles accelerating events (see \S~\ref{sec:cracc}) makes the chances of observing a SNR when it still acts as a {\it PeVatron} very low. The runaway CRs can illuminate molecular clouds located close to the SNRs and this enhanced gamma-ray emission can thus provide crucial insights on the parental population of runaway CRs, which would otherwise escape the SNR without leaving a footprint  \citep{1979ICRC....1..191M,1981Natur.292..430I,2007ApESS.309..365G, 1996A&A...309..917A,2001SSRv...99..187A}. The emission produced by these runaway CRs is essential to trace back acceleration up to PeV energies.

Gamma rays in association with dense molecular clouds located close to SNRs have been detected both at GeV and TeV energies. Depending on the location of massive clouds, on the acceleration history and on the timescales of the particle escape into the interstellar medium (which depend in turn on the diffusion coefficient), a broad variety of energy distributions of gamma rays is produced, from very hard spectra (much harder than the spectrum of the SNR itself) to very steep ones \citep{1996A&A...309..917A,2009MNRAS.396.1629G,2010PASJ...62.1127C}.  

W44 is a 20,000 years old SNR located on the Galactic Plane at a distance of roughly 3 kpc from the Sun. W44, which is the brightest middle-aged SNR at GeV energies, is a perfect laboratory to test the coexisting processes of acceleration and escape of cosmic rays from SNRs. Due to the slow shock speed, high energy particles are expected to have already escaped from this source. A population of protons with spectral index close to -2.3 and a cutoff at about 80 GeV is likely responsible for the gamma-ray emission from the remnant measured by AGILE and Fermi-LAT \citep{FermiW44,AgileW44}. The presence of such a low energy cutoff might be the effect of the escape of the highest energy cosmic rays. Indeed, the remnant is thought to be currently a rather poor accelerator and most of its gamma-ray emission can be interpreted as due to re-acceleration of ambient CRs, rather than to acceleration of fresh particles \citep{Uchiyama+10,Cardillo+16}.

Because of their short lives, SNRs are often found within the giant molecular gas complexes where they were born as massive stars. W44 resides within a giant gas complex of 10$^6$ solar masses, homogeneously distributed over an extended region surrounding the remnant \citep{2001ApJ...547..792D}. North west and south east of the remnant, \citet{2012ApJ...749L..35U} discovered two bright sources, which the authors associated to runaway cosmic rays colliding with the dense gas clouds. \cite{Peron_2020} re-analysed both the Fermi-LAT data and the gas data from the clouds around W44 and noted that, despite the gas is homogeneously distributed, the GeV emissivity is enhanced only in the two regions which can be associated to regions of enhanced CR density, {\it CR clouds}, rather than gas clouds as previously thought. This phenomenon suggests the existence of a preferential path for CR escape, likely linked to the magnetic field structure in the vicinity of this source.

\begin{figure}
    \centering
    \includegraphics[width=0.49\textwidth]{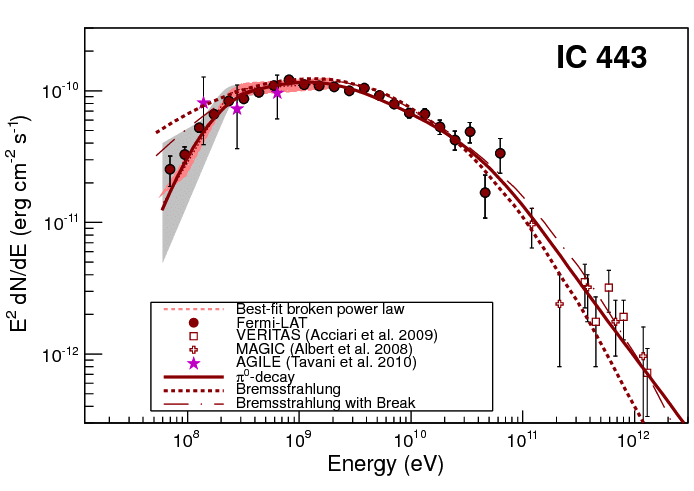}
    \includegraphics[width=0.49\textwidth]{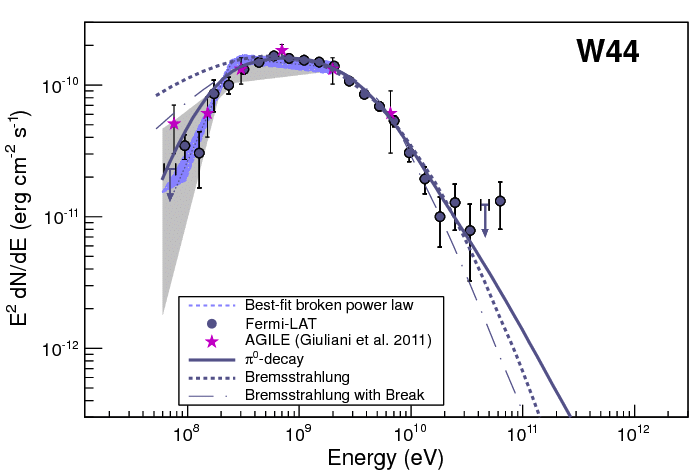}
    \caption{ Gamma-ray spectra of IC443 ({\it Left Panel}) and W44 ({\it Right Panel}) as measured with the Fermi LAT \citep{FermiW44}, in blue, and AGILE \citep{AgileW44}, in magenta. The pion bump feature at about 1 GeV is evident in the spectra of both SNRs. For IC433 the data-points of MAGIC \citep{2007ApJ...664L..87A} and VERITAS \citep{2009ApJ...698L.133A} are also shown. Color-shaded areas denote the best-fit broadband smooth broken power law (60 MeV to 2 GeV); gray-shaded bands show Fermi LAT systematic errors below 2 GeV. The data points at the highest energy (4 $ \times {10}^{10}$ - 1$ \times {10}^{11}$ eV) suggest a hardening of the spectrum, which might be produced by runaway CRs illuminating a molecular cloud (MC) located in front of the shells.}
    \label{W44}
\end{figure}

\section{Recent developments on CR acceleration in SNRs}
\label{sec:cracc}
In spite of all the recent observational developments, providing several hints of efficient CR acceleration in SNRs, a number of unsettled questions remain and challenge the association. In particular, one of the strongest arguments to look for alternative particle accelerators has to do with the difficulties SNRs have at reaching PeV energies. This aspect of the CR-SNR connection has long appeared as the most delicate. 

As already mentioned, CR acceleration in SNRs is thought to be well described within the framework of diffusive shock acceleration (DSA). Acceleration mechanisms alternative to DSA have also been proposed and studied (see e.g. \cite{Lazarian20}). In this article, however, we focus on DSA, which is still the best studied and the most promising, in this context, to reach the highest energies. The idea at the basis of the theory is that particles gain energy each time they cross the shock thanks to the discontinuity of the fluid velocity field at the shock, that leaves an unscreened electric field. The energy gain of the particle is a constant fraction ($v_{\rm sh}/c$ with $v_{\rm sh}$ the shock velocity) of its energy before the crossing. The particle moves diffusively between crossings, scattering on low frequency magnetic turbulence and continuously changing its pitch-angle. Reaching high energies requires a large number of crossings, which have to occur in a time shorter than the minimum between the duration of the system $t_{\rm life}$ and the time-scale over which energy losses become important $t_{\rm loss}$. In the case of protons losses are negligible and the energy is limited by the time for which the SNR is active as an accelerator, $t_{\rm life}<10^4$ yr. Then $t_{\rm life}$ must be compared with the acceleration time, $t_{\rm acc}$, which depends on how quickly the particle is able to go back to the shock after each crossing: we can estimate $t_{\rm acc}(E)\approx D(E)/v_{\rm sh}^2$, where $D(E)$ is the particle diffusion coefficient, typically an increasing function of the particle energy $E$. For relatively low turbulence levels one can estimate $D(E)$ in quasi-linear theory, writing it as (see e.g. \cite{Amato14}):
\begin{equation}
D(E)\approx \frac{v(E) r_L(E)}{3}\frac{1}{kW(k)}
\label{eq:D(E)}
\end{equation}
where $v(E)$ is the particle velocity, $r_L(E)$ its Larmor radius, and $W(k)$ is the spectrum of magnetic fluctuations causing the diffusion, with
\begin{equation}
\int_{1/L}^\infty W(k')dk'=\frac{\delta B^2}{B_0^2}\ ,
    \label{eq:wk}
\end{equation}
and $k$ the wave-number resonant with the particle of energy $E$, namely $k=1/r_L(E)$.

Assuming that the scattering is due to turbulence that is injected by SNRs on a typical scale $L\approx$ 100 pc with $\delta B/B_0\approx 1$, and then develops a Kolmogorov-type spectrum $W(k)\propto k^{-5/3}$, from Eqs.~\ref{eq:D(E)} and \ref{eq:wk} above\footnote{one may wonder about the appropriateness of using quasi-linear theory for turbulence levels as large as $\delta B/B_0\approx 1$: while this is not fully justified on theoretical ground, numerical simulations find it to provide a decent description of wave-particle interactions (see e.g. \cite{Holcomb19})}, one can estimate the diffusion coefficient as $D(E)=7\times 10^{27}$ (E/{GeV})$^{1/3}$ cm$^2$/s, which, as we will discuss below, is not very different from what CR observations indicate ($D(1 \mathrm{GeV})\approx 10^{28}$ cm$^2$/s). 
If this estimate of the diffusion coefficient were the appropriate one to describe particle transport in the vicinity of a SNR shock, the maximum achievable energy in these systems would only be $E_{\rm max}\sim$ few GeV, and could only improve to $\sim$ 100 TeV if particle transport were described by Bohm diffusion, namely $\delta B/B_0\approx 1$ at all relevant scales \citep{LagageCesarsky83}. It is then clear that in order to reach PeV energies, not only efficient - Bohm like - scattering is needed, but also largely amplified magnetic fields. 

As already mentioned above, this is exactly what X-ray observations of young SNRs were found to show: aside from the short variability time-scales already discussed in Sec.~\ref{sec:rxj}, X-ray synchrotron emission is seen to be confined to rims of thickness $\Delta \sim {\rm few} \times$ 0.01 pc \citep{Vink12}. The extremely small value of $\Delta$ can be interpreted either as a result of radiative losses, that kill the electron population with increasing distance from the shock, or as a result of magnetic field damping \citep{Rettig2012}. If we interpret the thickness of the rims as the distance traveled by the emitting electrons in a synchrotron loss time, we can write $\Delta\approx \sqrt{D(E)\ t_{\rm sync}(E)}=0.04\ {\rm pc}\ B_{-4}^{-3/2}$ with $B_{-4}$ the magnetic field downstream of the shock in units of $10^{-4}$G. It is clear then that a magnetic field amplified by a factor of $\sim$100 with respect to the interstellar value ($B_{\rm ISM}\approx 3 \mu$G) is implied in the rims, if they are to be interpreted as the result of radiative losses. On the other hand, interpreting $\Delta$ as a result of magnetic damping requires a value of the magnetic field which is still in the same range estimated above \citep{Pohl+05}, and hence largely amplified. 

There are several mechanisms by which a magnetic field can be amplified at a shock \citep{Bykov+12}: some involve fluid instabilities \citep{GiacaloneJokipii07,Ohira16}, some involve MHD instabilities and CR related effects \citep{Beresnyak+09}, some others CR induced instabilities, either of resonant type (\cite{AmatoBlasi06} and references therein) or of non-resonant type \citep{Bell04,Reville+08,AmatoBlasi09}. Except for purely fluid instabilities \citep{GiacaloneJokipii07}, all other classes require efficient CR acceleration, and in this sense the detection of amplified magnetic fields can be considered as indirect evidence of efficient CR acceleration in SNRs. However the different mechanisms proposed are not equivalent in terms of the spectrum of magnetic fluctuations they produce, and hence in terms of consequences on particle acceleration. Particle scattering is only efficient with resonant waves. Hence magnetic field amplification (MFA, hereafter) will increase scattering efficiency and lead to high maximum energies of the accelerated particles only if there is enough power on scales comparable with the particle Larmor radii. The most promising MFA mechanism in this sense turns out to be the non-resonant CR streaming instability (Bell's instability).

The super-Alfv\'enic streaming of energetic particles has long been known to induce the growth of magnetic fluctuations at wavelengths that are resonant with the gyro-radii of the exciting particles \citep{Skilling75}: this is the so-called resonant streaming instability. While creating fluctuations on the right scales, this instability can only lead to MFA up to a level $\delta B/B_0\lesssim 1$, and hence it is not powerful enough to explain the field strength deduced in SNRs, nor to guarantee particle acceleration up to the knee. In more recent times, however, it has been recognized that a more powerful instability arises if the CR current is large enough to twist the ambient magnetic field on the scale of the Larmor radius ($r_{L,0}$) of the particles that carry the current ($J_{\rm CR}>(c/4 \pi)(B_0/r_{L,0})$), or, equivalently, if the CR energy density, $U_{\rm CR}$, is larger than $c/v_D$ times the magnetic energy density \citep{Bell04}: 
\begin{equation}
U_{\rm CR}>\frac{c}{v_D} \frac{B_0^2}{4 \pi}\ .
\label{eq:bell2}
\end{equation}
where $v_D$ is the bulk velocity of CRs. If this condition is satisfied, the magnetic field grows very rapidly. The basic physical process can be described a follows: the CR current induces a compensating return current in the ISM plasma; the force $\vec J_{\rm ret}\wedge \vec B$ induces transverse plasma motion; the current associated to this motion acts as a source of magnetic field; the result is that the magnetic field lines associated to right-hand polarized waves are stretched and for these modes the field is amplified.

In the vicinity of a shock that is accelerating particles, Eq.~\ref{eq:bell2} can be turned into a condition for the acceleration efficiency $\xi_{\rm CR}$:
\begin{equation}
    \xi_{\rm CR}>\frac{c B_0^2}{8 \pi \rho_{\rm ISM} v_{\rm sh}^3}\ .
    \label{eq:bell3}
\end{equation}
Here $\xi_{\rm CR}$ is the fraction of incoming flow energy ($\rho_{\rm ISM} v_{\rm sh}^2/2$) that is converted into accelerated particles, $\rho_{\rm ISM}$ is the mass density of the ISM plasma and $v_{\rm sh}$ is the velocity of the blast wave, also coincident with the CR bulk velocity. Eq.~\ref{eq:bell3} makes it clear that the possibility for the non-resonant streaming instability to operate strongly depends on the shock velocity: detailed calculations show that it is indeed a viable MFA mechanism in the vicinity of the fast shocks of young SNRs \citep{AmatoBlasi09}, but then stops working early during the Sedov-Taylor phase of expansion of the blast wave, typically after few hundred to few thousand years, depending on the ambient medium density and magnetic field strength.  

The growth initially occurs on very small scales ($k\approx (4\pi/c)(J_{\rm CR}/B_0>> r_{\rm L,0}$, but quickly the power moves to larger and larger scales \cite{RiquelmeSpitkovsky09}, possibly due to some mean-field dynamo process \citep{Bykov11, Rogachevskii12}, and numerical simulations show that the final outcome of the instability, when it develops at a shock, is a spectrum of fluctuations with $W(k)\propto k^{-1}$, leading to Bohm diffusion of the particles \citep{CaprioliSpitkovsky14C}. When saturation is reached, the magnetic energy density is a substantial fraction of the CR energy density \citep{CaprioliSpitkovsky14B}. 

The level of MFA that the non-resonant streaming instability can provide is in the correct range to explain the magnetic field strength inferred in SNRs \citep{Schure+12}, and  as we just mentioned the associated scattering is efficient \citep{CaprioliSpitkovsky14C}, and yet recent studies have cast doubt on the fact that these sources might be able to accelerate particles up to PeV energies \citep{Cardillo+15}. In fact, assuming that MFA is primarily due to the streaming of particles that leave the acceleration region, it is possible to build a description of the shock as a self-regulating system: when the level of turbulence in the upstream is low, a large fraction of particles can escape from the shock; when this happens, however, the large current in the upstream causes the growth of turbulence and escape is reduced; if the fraction of escaping particles becomes too small, then the turbulence level is reduced again favouring escape. The self-regulation mechanism qualitatively illustrated above translates into a quantitative prescription for the maximum energy of shock accelerated particles as a function of the system parameters. The current of escaping particles, which is what determines the level of MFA, is in turn determined by the shock velocity and the spectrum of accelerated particles, including its total energy content (which determines the amount of energy available), its slope and the maximum particle energy (which determine the current, once the energy content is fixed). When writing the equation describing $E_{\rm max}$ as a function of time during the expansion of a SN blast wave, one finds that $E_{\rm max}$ is an ever decreasing function of time during the SNR evolution, so that the relevant value of $E_{\rm max}$ is that reached at the beginning of the Sedov-Taylor phase, namely the highest possible after a sufficient amount of mass has been processed by the blast wave. It is clear then that a higher maximum energy will be achieved in systems that enter the Sedov-Taylor phase earlier in time after the explosion. This occurs for type II explosions expanding in the dense and slow wind of a progenitor red super-giant star. In this case, assuming a $\propto E^{-2}$ spectrum, the maximum achievable energy reads:
\begin{equation}
E_{\rm max}\approx 0.5 \left(\frac{\xi_{\rm CR}}{0.1}\right) 
\left(\frac{E_{\rm SN}}{10^{51}\ {\rm erg}}\right)
\left(\frac{M_{\rm ej}}{M_\odot}\right)^{-1}
\left(\frac{\dot M}{10^{-5} M_\odot /{\rm yr}}\right)^{\frac{1}{2}}
\left(\frac{v_w}{10\ {\rm km\ s}^{-1}}\right)^{-\frac{1}{2}}
{\rm PeV}
\label{eq:emax}
\end{equation}
where $\xi_{\rm CR}$ is the CR acceleration efficiency, $E_{\rm SN}$ and $M_{\rm ej}$ are the energy released and the mass of material ejected in the SN explosion, and $\dot M$ and $v_w$ are the mass loss rate and speed of the progenitor's wind. If one takes into account that the mass of the ejecta in a type II SN explosion is more likely around $10\ M_\odot$, it is immediately clear that reaching the {\it knee} is very challenging in this framework. One is forced to invoke extremely energetic events, or extreme acceleration efficiency, or finally extreme properties of the progenitor. In all cases these must be rare events. In fact, the product of $\xi_{\rm CR}$, $E_{\rm SN}$ and the frequency in the Galaxy of the PeV producing SN explosions are constrained by the overall flux of CRs measured flux at Earth. When all of this is taken into account, one determines an event rate that cannot exceed a few in $10^4$ yr \citep{Cristofari20}. 

In reality the possibility for SNRs to reach the {\it knee} becomes even more challenging when taking into account the fact that the accelerated CR spectrum is likely $\propto E^{-p}$ with $p>2$. Such a spectrum corresponds to a smaller current for a given $E_{\rm max}$ and total energy in accelerated particles, hence requiring even more extreme parameters to reach $E_{\rm max}\approx$ PeV (e.g., \cite{Cardillo+15}). On the other hand a spectrum steeper than $E^{-2}$ is exactly what gamma-ray observations of the majority of young SNRs require and what recent progress on propagation also seems to require \citep{Evoli+19}: the CR spectrum released in the ISM will be steeper than $E^{-2}$ only if the spectrum in the source is itself steeper than $E^{-2}$ \citep{SchureBell14,Cardillo+15} and a steep injection spectrum of CRs is exactly what the most recent CR data seem to suggest, as we discuss in \S~\ref{sec:crtransp}.

\section{Cosmic ray transport in the Galaxy}
\label{sec:crtransp}
As mentioned in \S~\ref{sec:intro} the last decade has also been rich of observational progress providing interesting constraints on the properties of CR transport throughout the Galaxy. In this section we will review the main findings and try to put them in a coherent theoretical framework.

The first important discovery was that of a hardening in the spectrum of protons and He nuclei \citep{Adriani11pHe, Aguilar+15p,Aguilar+15He} and also, though with somewhat lower statistics, of heavier nuclei \citep{Ahn+10}. Within the framework of diffusive transport, the detection of a break in the spectrum of CRs can be interpreted as a signature of a change either in the injection spectrum or in the diffusion coefficient.  During propagation through the Galaxy, CRs primarily loose energy due to adiabatic expansion and ionization. In addition, particles of a given species can be also lost due to spallation or decay. However, if one focuses on stable primary CRs, with energy above a few tens of GeV, energy losses during propagation can be neglected and a very simple expression for the diffuse steady state spectrum in the Galaxy can be found. If we assume that CRs are injected in the Galaxy at a constant rate
\begin{equation}
    Q_{\rm p}(E)\propto E^{-\gamma_{\rm inj}}\ ,
    \label{eq:crinj}
\end{equation}
and that particles then propagate diffusively with an average diffusion coefficient 
\begin{equation}
    D(E)\propto E^\delta
    \label{eq:crdiff}
\end{equation}
the steady state spectrum of stable CR nuclei in the Galaxy will be given by the product between injection and confinement time, $Q(E)\times \tau_{\rm esc}$. This will read, for primary nuclei:
\begin{equation}
N_{\rm p}(E)\propto Q(E)\frac{H^2}{D(E)}\propto E^{-\gamma_{\rm inj}-\delta}\ ,
    \label{eq:nprim}
\end{equation}
where the confinement time has been taken to be $\tau_{\rm esc}=H^2/D(E)$, with $H$ the size of the magnetized halo in which CRs are confined (see e.g. \citep{amatoblasi18} for a more refined description). It is then clear then that the detected hardening implies a change in $\gamma_{\rm inj}$ or $\delta$. Several models have been proposed invoking either of the two \citep[][and references therein]{TomassettiHardening,HorandelHardening}. A possibility is that this feature is signaling the importance of non-linear effects in CR propagation. An early suggestion \citep{Blasi+12} was that the break could point to the transition between scattering in self-generated and external turbulence. It had indeed been suggested since the '70s \citep{Wentzel74,Cesarsky80} that at scales comparable with the Larmor radius of GeV particles, CRs could be an important source of turbulence in the galaxy through the resonant streaming instability. At larger scales, on the other hand, CRs become too few, given their steep spectrum, and the main source of scattering would become the large scale turbulence present in the Galaxy. The latter is usually assumed to be injected by SNRs on a scale of order tens to 100 pc and then cascade to smaller scales developing a Kolmogorov type spectrum $k^{-5/3}$. Such a spectrum translates in a diffusion coefficient with $\delta\approx 1/3$, flatter than the low energy value $\delta\approx 0.7$ that is appropriate to describe CR self-generated turbulence. A back of the envelope calculation places the transition between self-generated and external turbulence at a CR rigidity in the range 200-300 GV \citep{Blasi+12}, tantalizingly close to the value of 336 GV at which AMS-02 detects the hardening in the proton spectrum. 

This explanation of the hardening as due to a change in the properties of galactic transport entails a clear prediction for the spectrum of secondary CR nuclei. These are injected in the Galaxy as a result of spallation of primaries. If we approximate the spallation cross-section as independent of energy, their injection will be
\begin{equation}
Q_{\rm sec}(E)\propto N_p(E) \sigma_{\rm sp}c n_{\rm ISM}\propto E^{-\gamma_{\rm inj}-\delta}\ ,
    \label{eq:secinj}
\end{equation}
where $n_{\rm ISM}$ is the target gas density. Their equilibrium spectrum in the Galaxy will then read:
\begin{equation}
N_{\rm sec}(E)\propto Q_{\rm sec}(E)\tau_{\rm esc}\propto E^{-\gamma_{\rm inj}-2\delta},
    \label{eq:nsec}
\end{equation}
which implies that any hardening $\Delta$ of the spectrum of primaries due to a change in the slope of the diffusion coefficient will reflect in a hardening of the spectrum of secondaries equal to $2\Delta$: this expectation is perfectly consistent with the analysis of secondaries performed by AMS-02 \citep{Aguilar+18sec}. 

Within this interpretation of the observed hardenings, the transport of CRs through the Galaxy becomes a complex non-linear problem, where the diffusion coefficient and the spectra of all nuclei need to be determined self-consistently. Once this complex problem is solved, however, and all available AMS-02 data are reproduced, the low energy Voyager data \citep{Stone+13} are automatically reproduced \citep{Aloisio+15}. One important result that comes out of this analysis is directly related to the Boron-over-Carbon (B/C) measurements of AMS-02 at high energies \citep{Aguilar+16BC}. The B/C ratio has traditionally been considered the primary indicator of CR transport: B is the most abundant stable secondary and its mostly produced by the spallation of C, though the contribution by N and O is not negligible. In practice the B/C ratio provides a direct measurement of the grammage (mass per unit surface, or mass density integrated along the path-length) encountered by CRs during their propagation from their sources to Earth. If one compares Eqs.~\ref{eq:nsec} and \ref{eq:nprim}, for the spectrum of secondary and primary nuclei, respectively, one immediately sees that the ratio between the two is expected to scale with energy exactly as the diffusion coefficient. In fact this ratio, as measured by AMS-02 \citep{Aguilar+16BC}, well agrees with a high energy slope of the diffusion coefficient $\delta\approx 0.4$ \citep{Aloisio+15,Evoli+19}, but an additional, energy independent contribution seems to be required to well reproduce not only the highest, but also the lowest energy \citep{Cummings+16} available data. The additional grammage needed, $X_s \approx 0.15\ {\rm g\ cm}^{-2}$, is independent of energy and of the order of what particles can accumulate within a SNR - or any source with an ambient density of order $1\ {\rm cm}^{-3}$ and a duration $\approx 10^4\ {\rm yr}$ - during acceleration \citep{Aloisio+15,Bresci+19}. Within this modeling, namely if hardenings are interpreted as a result of turbulence self-generation at low energies, the contribution $X_s$ turns out to be fundamental also for explaining another recent surprise found in CR data, namely the spectrum of anti-protons, which we will discuss in \S~\ref{sec:cranti}.

Before concluding this section we would like to remark a most important consequence of a scenario in which CR propagation at high energy is described by a diffusion coefficient $\delta\approx 0.4$: CRs must be injected in the galaxy with a spectrum $E^{-\gamma_{\rm inj}}$ with $\gamma_{\rm inj}\approx 2.3$. This combination of injection and propagation parameters, and in particular the relatively weak energy dependence of the diffusion coefficient, helps to explain the low level of anisotropy detected at TeV energies, as shown by \cite{BlasiAmato12a,BlasiAmato12b,Sveshnikova+13}. On the other hand, as discussed in \S~\ref{sec:cracc}, such a steep spectrum makes it very difficult for SNRs to be the primary sources of PeV CRs.

Within a DSA scenario SNRs accelerate the highest energy CRs (up to at least a few PeV) at the transition between the free expansion and the Sedov-Taylor phase, which typically happens a few hundred years after the supernova explosion. During the Sedov-Taylor phase the SN shock slows down and the magnetic field intensity decreases, so that the shock cannot confine any longer the most energetic particles, which escape the SNR. This means that a SNR can act as a PeVatron for a relatively short time. Considering the rate of SN explosions in the entire Galaxy (about 3 per century), the chances to observe a SNR when it is still a PeVatron are thus very low, and any proof of emission from PeV CRs even from young SNRs is challenging to find. 

\section{Alternative CR sources}

While gamma-ray observations have proven that SNRs are efficient accelerators of cosmic ray electrons, and possibly protons, up to 100\,TeV, the hypothesis that acceleration of cosmic rays proceeds up to PeV energies has been rejected in all known young SNRs because of the clear cut-offs detected at several TeVs in their spectra. 

The GeV-to-TeV radiation from young shell-type SNRs, which had been long expected to provide final evidence to settle the question of the origin of cosmic rays, can be either of hadronic or leptonic origin. Typically, a 1~TeV $\gamma$-ray photon is emitted either by an electron or a proton of about 10~TeV. The cooling time for 10 TeV electrons is $\approx 5 \times 10^4$yr (for IC scattering off the CMB photons) while for protons the cooling time is $5 \times 10^7 (n/1 \rm cm^{-3})^{-1}$yr (see e.g. \cite{2004vhec.book.....A}). Thus the ratio of the luminosity in IC gamma-rays to $\pi^0$-decay gamma-rays is of the order of $10^3 \times (\frac{W_e}{W_p}) \times {(\frac{n}{\mathrm{1cm}^{-3}})}^{-1}$. The leptonic contribution to the emission is thus dominant over the hadronic one unless $\frac{W_e}{W_p} << 10^{-3}$ or alternatively if the gas density inside the shell is $n >> 1 \, \rm {cm^{-3}}$. This latter condition, however, if realized, would have the drawback of slowing down the shock wave very quickly and thus prevent efficient acceleration. On the other hand, a much larger energy density in protons than in electrons could result from rapid cooling of the electron population, especially in the presence of an amplified magnetic field, as inferred for young SNRs. If $B>10\ \mu$G, only a small fraction of their energy $w_{MBR}/w_{\rm B} \approx 0.1 (\rm {B}/10 \mu \rm G)^{-2}$, is released in IC gamma-rays, making the conditions for detection of $\pi^0$ decay more favourable.

 To recap, no known SNR has proved to accelerate particles beyond 100 TeV. Additionally, the power in accelerated protons within SNRs derived from $\gamma$-ray observations depends on the highly uncertain local gas density $n$ and on the magnetic and radiation fields within the remnant. As a result, it has so far been impossible to unequivocally prove that the population of SNRs in the Galaxy injects the necessary power (about 10$^{50}$erg per supernova event) to sustain the CR population. In fact simulations of the SNR population show how the choice of acceleration and ISM parameters lead to remarkably different SNR populations as CR sources \citep{2017MNRAS.471..201C}.
 
Some other classes of astrophysical sources, such as super-bubbles or star forming regions (see \S~\ref{sec:CR:SC}) or remnants of GRBs in our Galaxy \citep{2006ApJ...642L.153A}, and, in particular, the Centre of our Galaxy (see \S~\ref{sec:CR:GC}), have long been proposed as alternative accelerators of particles and major contributors to the population of Galactic cosmic rays up to PeV energies \citep{1980ApJ...237..236C,1982ApJ...253..188V,1983SSRv...36..173C,2004ApJ...601L..75T,2006ApJ...644.1118R,2001AstL...27..625B,2004A&A...424..747P,2005ApJ...628..738H,2006ApJ...642L.153A,2019NatAs...3..561A}.


\subsection{Galactic Centre}
\label{sec:CR:GC}

A breakthrough in the quest for the origin of the highest energy CRs in the Galaxy was the discovery of a powerful PeVatron in the Centre of the Milky Way \citep{2016Natur.531..476H}. 

The nucleus of the Milky Way is a very active region with numerous sources of non-thermal radiation and constitutes a unique laboratory for the study of very high energy astrophysical processes 
within the Galaxy and in external active galactic nuclei. The 
Galactic Centre (GC) is thought to host a super-massive 
black hole (SMBH) of $2.6 \times 10^6$ solar masses \citep{2008ApJ...689.1044G,Gillessen_2009} located very close to the dynamical centre of the Galaxy and coincident with the compact radio source Sagittarius A* (Sgr A*). Sgr A* emits radiation in X-rays and infrared through accretion of mass onto the BH. The region within 400 pc from the Centre of the Galaxy, called the Central Molecular Zone (CMZ), contains around 5\% of the total Galactic molecular gas \citep{1998ApJS..118..455O,1999ApJS..120....1T}. The molecular CO line towards the GC being somehow optically thick due to the high gas density, the gas distribution there is mapped additionally with other lines, such the CS radio line \citep{1999ApJS..120....1T}. While the presence of this dense gas and dust prevents observations of this region in the optical and ultra-violet wavelengths, the Galactic Centre region is extremely bright at radio, infrared, X-ray and gamma-ray frequencies. 

At very high energy \citep{2018A&A...612A...6H}, the GC hosts a bright point-like source, HESS J1745-290, which, within errors, coincides spatially with Sgr A* and presents an energy spectrum with a steepening below 10 TeV. While the association of HESS J1745-290 with Sgr A* is well motivated in terms of spatial coincidence and of required energetics, {the pulsar wind nebula (PWN)} candidate G359.95-0.04 is also a viable counterpart of HESS J1745-290. HESS J1745-290 is surrounded by an extended component of VHE gamma-ray emission, correlated spatially with the CMZ. The spectrum of this radiation is a pure power law, with a spectral index -2.3 and without evidence of a spectral cutoff. 

Because of their different energy distributions, the central point-like source and the extended emission cannot have the same origin. The diffuse VHE gamma-ray emission of the CMZ could be in principle produced through interactions of either relativistic protons with the ambient gas or of relativistic electrons with the radiation fields. However, a leptonic origin of the diffuse gamma-rays can be excluded. The power-law acceleration spectra of electrons should extend to about 100 TeV, which is extremely difficult because of the severe Inverse Compton and synchrotron radiative losses in the GC region. For the same reason, leptons hardly can escape the sites of their acceleration and propagate over the tens of parsec extended region of the HESS diffuse emission. 

The spatial correlation between the gamma-ray emission from the Galactic Ridge and the ambient gas supports a hadronic origin of the emission. The spectrum of the parental protons -- with a spectral index close to -2.4 -- should extend to energies close to 1\,PeV.  Assuming a cutoff in the parent proton spectrum, the corresponding secondary gamma-ray spectrum deviates from the HESS data at 68$\%$, 90$\%$ and 95$\%$ confidence levels for cutoffs at 2.9\,PeV, 0.6\,PeV and 0.4\,PeV, respectively. This makes the discovery of the diffuse emission from the Galactic Centre the first robust claim of detection of a Galactic cosmic ray PeVatron. 

The CR  energy density in the CMZ, obtained combining the gamma-ray and gas distributions, is an order of magnitude higher than the {\it sea} of cosmic rays. The radial distribution of CRs follows a 1/r dependence, where r is the distance to the GC. This means that the PeVatron should be located within 10 pc from the GC and the total energy output in protons above 10 TeV over the whole region should amount to W$_{CR}$ = 10$^{49}$ erg. Such a modest energy output could be provided by a single supernova remnant event, such as Sgr A East, a Pulsar Wind Nebula. If the super-massive black hole, Sgr~A*, is the GC PeVatron, then the particles producing the HESS extended emission are accelerated either in the vicinity of the SMBH, close to event horizon, or at the termination of a relativistic outflow, either a jet or a wind. This jet or wind should be injected close to the black hole and carry a substantial fraction of energy, extracted from the accretion disk. Indeed, the power required by the inferred CR distribution corresponds to $\gtrsim 1\%$ of the accretion power of the central black hole, and is 2-3 orders of magnitude larger than the bolometric luminosity of Sgr~A*. The source, however, could have been more active in the past \citep{2016Natur.531..476H}. 

Alternatively, the extended TeV emission from the Galactic center could be the combined result of acceleration within three powerful star clusters, the Arches, the Quintuplet and the Nuclear cluster \citep{2016Natur.531..476H,2019NatAs...3..561A}. In this case, the continuous injection over millions of years, required by the 1/r distribution of the diffuse emission, could result from the characteristic ages of massive clusters, roughly 10$^6$ years \citep{2019NatAs...3..561A}. 

All the proposals discussed above assume that particle transport in the GC vicinity can be described through the same simple model of a spatially uniform diffusion coefficient adopted for the rest of the Galaxy. If this assumption is released, however, the GC excess can be interpreted as a result of a spatially dependent diffusion coefficient, as was proposed by \cite{gaggero2017prl}, based on Fermi-LAT data showing the spatial dependence of the gamma-ray galactic diffuse emission.

\subsection{Star Clusters}
\label{sec:CR:SC}

Collective stellar winds and SNR shocks in clusters and associations of massive stars have long been suggested as possible, alternative or additional contributors to the Galactic cosmic ray flux \citep{1980ApJ...237..236C,1982ApJ...253..188V,1983SSRv...36..173C}. Core-collapse SN progenitor stars and colliding wind binaries evolve in giant molecular clouds and mostly remain close to their birthplaces in groups of loosely bound associations or dense stellar clusters. The winds of multiple massive stars in such systems can collide and form collective cluster winds which drive a giant bubble, a so called superbubble, filled with a hot (T = 10$^6$~K) and tenuous (n $<$ 0.01~cm$^{−3}$) plasma. 
At the termination shock of the stellar cluster wind, turbulence can build up, in the form of MHD fluctuations and weak shocks \citep{2004ApJ...601L..75T,2006ApJ...644.1118R,2001AstL...27..625B}. Turbulence in the superbubble interiors can accelerate particles to very high energies, not only through the 1$^{\rm st}$ order Fermi process, but also via the 2$^{\rm nd}$ order mechanism \citep{2001AstL...27..625B}. Supernova explosions of massive stars in thin and hot superbubbles can also produce efficient particle acceleration at the boundary of the superbubbles or at MHD turbulence and further amplify existing MHD turbulence \citep{2010A&A...510A.101F}. It has also been recognized that multiple shocks can result in efficient acceleration beyond PeV energies \citep{2000APh....13..161K}. The interaction of the accelerated particles with the ambient medium - often including dense molecular clouds - or with electromagnetic fields, leads to the efficient production of VHE gamma rays. 

Recent observations at TeV energies of massive star-forming regions and stellar 
clusters, such as 30~Doradus in the LMC \citep{2015Sci...347..406H}, Westerlund 1 \citep{2012A&A...537A.114A} and the Cygnus region \citep{2002A&A...393L..37A,2007ApJ...664L..91A,2007ApJ...658.1062K,2009ApJ...700L.127A,2012ApJ...745L..22B} in our own Galaxy, support the hypothesis that star forming regions are sites of high energy particle acceleration, and give new impulse to the $\gamma$-ray research in this field. 
The primary objectives of these gamma-ray observations are: 1) to constrain the fraction of mechanical energy of the stellar wind transferred to relativistic particles and hence gamma rays; 2) to unveil the physics of particle acceleration and propagation in Galactic stellar clusters and superbubbles.
Furthermore, high-energy phenomena are receiving increasing attention 
also from the point of view of their impact on
the life cycle of interstellar matter and star-formation processes.
The rate and efficiency of the star formation process depends, in fact, on the balance between the self-gravity of dense molecular cores and the countervailing forces which act to support the clouds. The most important of these are likely to be thermal pressure, turbulence, and magnetic fields. In order for magnetic support to be effective, a population of ionized particles must be present in the core. Since molecular clouds are opaque to ultraviolet radiation from stars, the main ionizing agent is thought to be low-energy CRs, and the magnetic support of the cloud is critically dependent on their abundance.

The Cygnus region hosts some of the most remarkable star-forming systems in the Milky Way, including Cygnus X, a star forming region at only 1.5 kpc from the Sun, with a total mass in molecular 
gas of a few million solar masses - at least 10 times the total mass in all other close-by star-forming regions, such as Carina or Orion - and a total mechanical stellar wind energy input of $10^{39}$ erg s$^{-1}$, which corresponds to several per cent of the kinetic energy input by SNe in the entire Galaxy. Cygnus X hosts many young star clusters and several groups of O- and B-type stars, called OB associations. One of these associations, Cygnus OB2, contains 65 O stars and nearly 500 B stars. These super stars have created cavities filled with hot, thin gas surrounded by ridges of cool, dense gas where stars are now forming, which strongly emit at GeV energies, called the Fermi Cocoon \citep{2011Sci...334.1103A}. At TeV energies the Cygnus region shows two distinctive regions. One is possibly connected to the Cygnus\,X complex, the star association Cygnus\,OB2 and the Fermi cocoon observed at GeV energies.
In the other region detected at TeV energies, the Cyg\,OB1 region, the Milagro collaboration discovered MGRO\,J2019+37 \citep{2007ApJ...664L..91A}, a very hard and extended source, possibly related to the massive star-forming region associated with the HII region Sharpless\,104 (Sh\,2-104) \citep{2004ApJ...601L..75T}. Particle acceleration in shocks driven by the winds from the Wolf-Rayet stars in the young cluster Berk 87 in the Cyg OB1 association have also been proposed as a possible origin of the VHE gamma rays \citep{2007MNRAS.382..367B,2007MNRAS.377..920B}. VERITAS resolved the Milagro source into two sources, one of which, VER\,J2019+378, is a bright, 1 degree extended source, that likely accounts for the bulk of the Milagro emission, coincident with the star formation region Sh\,2-104. Its spectrum in the range 1-30\,TeV is well fitted with a power-law model of photon index 1.75, among the hardest values measured in the VHE band. The TeV counterpart of the Fermi-Cocoon has been studied with the ARGO detector up to about 10 TeV and with the HAWC detector up to 200 TeV \citep{2014ApJ...790..152B,Hona:2019g7}. The spectral energy distribution shows a significant softening at a few TeV: this is revealed by the comparison between the ARGO and HAWC data and the Fermi-LAT data. This break in the $\gamma$-ray spectrum might hint at a cut-off in the injected CR spectrum  \citep{Hona:2019g7}, or possibly be explained as due to suppressed diffusion in the high turbulence environment of the Cygnus super-bubble, which would confine low energy particles while higher energy ones escape.

\citet{2019NatAs...3..561A} argued that three ultra-compact clusters located in the Galactic Centre power the HESS diffuse emission from the CMZ. The $1/r$ dependence of the CR density on distance from the star cluster is, in fact, a distinct signature of continuous injection of CRs over the cluster lifetime and following diffusion through ISM. The efficiency of conversion of kinetic energy of powerful stellar winds can be as high as 10 percent. This implies that the population of young massive stars can provide production of CRs at a rate of up to $10^{41}$ erg/s, which is sufficient to support the flux of Galactic CRs without invoking other source populations. This Galactic center PeVatrons together with the other massive star forming regions, such as Westerlund\,1 and Cyg\,OB2, would represent the major factories of Galactic CRs \citep{2019NatAs...3..561A}. A broader review of this subject, including a discussion of the subtleties associated with acceleration and propagation of CRs in these environments, may be found in \cite{Bykov+20}.

\section{Implications of anti-matter data and CR leptons}
\label{sec:cranti}
One final subject we want to address in this paper concerns the implications of recent observational results on the spectrum of CR electrons, positrons and anti-protons. While electrons can be directly accelerated from the ISM plasma, positrons and anti-protons are extremely few in the ordinary ISM and hence they have long been considered to be purely secondaries, namely a byproduct of CR interactions during propagation in the Galaxy. If one assumes that $e^+$ and $\bar p$ are pure secondaries, the straightforward expectation is that the ratio between their fluxes and that of protons (their primary parent particles) should monotonically decrease with energy, as is true for the B/C ratio discussed in \S~\ref{sec:crtransp}. In fact, the ratio between $e^+$ and $p$ should decrease even faster, because positrons are subject to radiation losses which further steepen the spectrum at high energies (see e.g. \cite{amatoblasi18} for extended discussion). Direct observations of the flux of positrons \citep{Adriani+09,Aguilar+13pos} and anti-protons \citep{Aguilar+16pbar} contradict this expectation, showing that the spectra of $e^+$, $\bar p$ and $p$ are all parallel to one another. 

Several different scenarios have been proposed in the literature to explain this finding (e.g. \cite{Blum+13,Cowsik+14,Cowsik+16,Lipari17,Eichler17}). An interesting possibility is that CRs accumulate sizeable grammage in the vicinity of their sources, where the diffusion coefficient is reduced with respect to the Galactic average. Here diffusion is energy dependent and this is where most of the Boron is produced. Then, in the rest of the Galaxy, diffusion must be energy independent and faster than usually assumed. Most $e^+$ and $\bar p$ would be produced by particles that accumulate most of their grammage in this second region: at a given rigidity their parent particles are $\sim 10$ times more energetic than those that produce B nuclei, and hence spend a shorter time in the source vicinity. 

While it is not fully clear that these assumptions would not violate other constraints, like those provided by anisotropy, a possible physical justification of such a scenario could stem from the effects of self-generated turbulence (see Sec.~\ref{sec:cracc}) which could considerably reduce the diffusion coefficient in the vicinity of CR sources. In terms of theory the viability of this scenario is not fully clear, especially due to the unknowns related to the abundance of neutrals, which can effectively damp the CR induced resonant streaming instability \citep{DAngelo+16,Nava+16}. However, there are observational indications that enhanced confinement indeed occurs, at least in the vicinity of some sources. A striking example are the so-called TeV haloes \citep{Abeysekara+17} surrounding evolved pulsars. Many more cases are expected to be found with next-generation gamma-ray telescopes, such as CTA.

In reality, given the present uncertainties in many of the parameters involved, among which, most notably, the $\bar p$ production cross-section \citep{Korsmeier+18}, the $\bar p/p$ ratio is not in a statistically significant disagreement with the standard description of CR propagation through the Galaxy, nor with the B/C ratio, when all available information is taken into account and the subtle effects of re-acceleration are included in the description \citep{Bresci+19}. Indeed the high precision of the currently available data prompts theory to move forward and include second-order effects: one of these is the finite probability that during propagation particles might encounter an active CR accelerator (e.g. a SNR shock) and be re-accelerated. This has no effect on CR primaries, but flattens the spectrum of CR secondaries. Taking this phenomenon into account, together with the hardening of primaries, the energy dependence of the $\bar p$ production cross section, and, finally, the source grammage mentioned in \S~\ref{sec:crtransp}, allows one to reproduce the $B/C$ ratio, the $\bar p/p$ ratio and all available data on stable nuclei \citep{Bresci+19}. 

Differently from $\bar p$, the spectrum of $e^+$ is not easy to reproduce in the standard CR propagation scenario, unless additional sources of primary positrons are invoked. 

The leptonic cosmic ray population is a small fraction of the total CR population. For instance at $\sim$1 GeV the ratio of the hadronic versus leptonic fluxes measured at Earth is roughly 100 : 1 (https://lpsc.in2p3.fr/crdb/). Differently from hadrons, leptons are heavily affected by energy losses due to synchrotron and inverse Compton processes in the interstellar magnetic and radiation fields. If positrons were purely secondary products of CR interactions, their spectrum would have to be steeper than that of protons at energies where losses are important. As recently shown by \cite{2020PhRvD.101j3030D}, in order for this not to happen, the escape time from the Galaxy would have to be so short that protons would not be able to produce the observed $\bar p$ flux.

\begin{figure}
    \centering
    \includegraphics[width=0.49\textwidth]{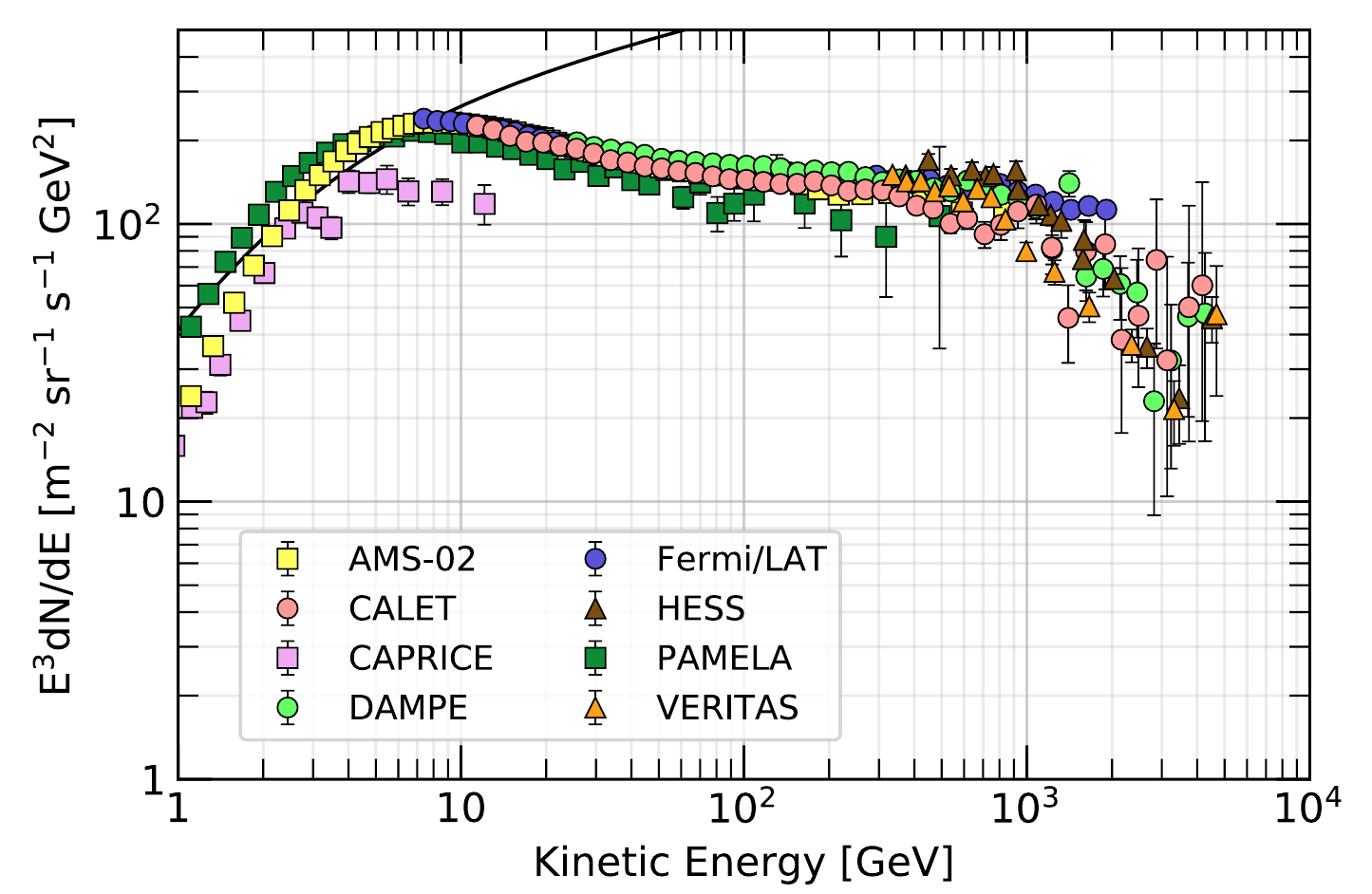}
     \includegraphics[width=0.49\textwidth]{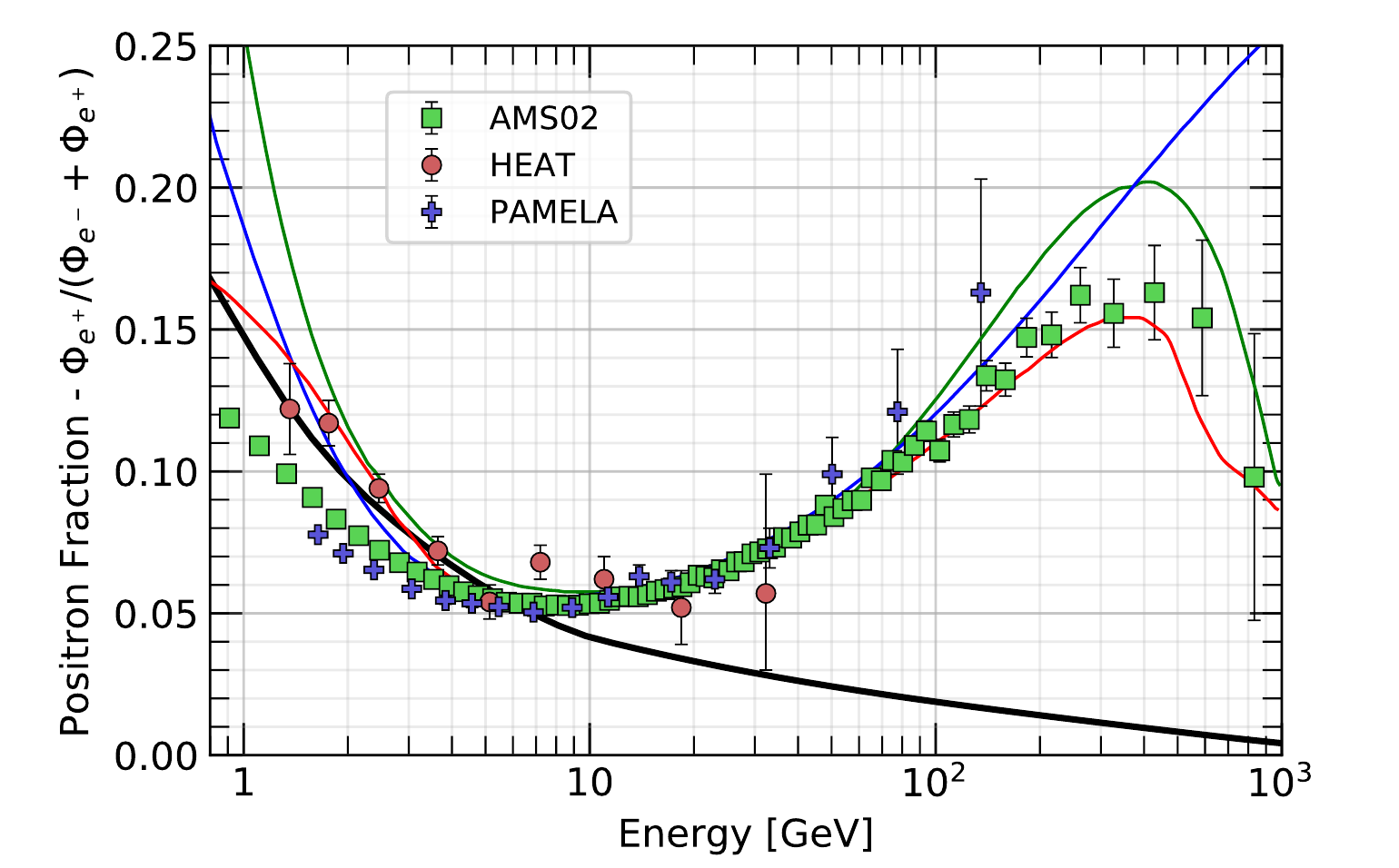}
    \caption{{\it Left Panel:} Differential spectrum of electrons and positrons multiplied by E$^3$. Cosmic-ray electrons can be measured using space-born instruments such as AMS or Fermi-LAT up to 1 $\sim$TeV and CALET or DAMPE up to $\sim$10 TeV. Ground-based Cherenkov telescopes, such as H.E.S.S., MAGIC and VERITAS, which benefit from very large effective areas, have measured the flux up to 20 TeV. The black line shows the proton spectrum multiplied by 0.01 \citep{2007APh....28..154S}. {\it Right Panel:} Positron fraction. Not all experiments are able to measure the electrons and positrons separately. Here measurements by \citet{PhysRevLett.122.101101,2009Natur.458..607A,2004PhRvL..93x1102B} are reported, together with a number of model predictions: the black curve is for pure secondary production in the standard scenario \citep{1998ApJ...493..694M}; the blue one includes a more sophisticated propagation model \citep{PhysRevLett.111.021102}; the green one invokes dark matter decay \citep{2013IJMPA..2830040I}; the red one includes a contribution from pulsars \citep{PhysRevD.88.023001}. The Figure is from \cite{2020PTEP.2020h3C01P}.}
    \label{fig:my_label}
\end{figure}

The anomalous positron spectrum has been at the origin of a huge number of papers, proposing either astrophysical or particle physics (Dark Matter annihilation related) solutions \citep{Aguilar+19pos}. Limiting the discussion to astrophysical scenarios, important constraints on the plausibility of the different proposals come from the maximum age and distance of the highest energy lepton sources. As shown in Fig.~\ref{fig:my_label}, electrons and positrons are now detected up 20 TeV. The electron spectrum shows roughly a constant slope up to an energy of about 1 TeV, above which the spectrum becomes steeper. One possibility to interpret the break at 1 TeV is to relate it to the transition between a regime where a large number of lepton accelerators contribute to the spectrum, to a regime where only a few or even only one single electron source very close to the Sun contributes \citep{Recchia19}.

The number of contributing sources depends, in turn, on the average intensity of the magnetic and radiation fields and on the diffusion coefficient in the Galaxy. The latter, in particular, seems to need some revision in light of the recent measurements of secondary nuclei, both stable and unstable ones, provided by AMS-02 \citep{Aguilar+18sec}. 

For the standard average values of the magnetic and radiation fields in the Galaxy, the cooling time of leptons with energies above few tens of GeV can be approximately estimated as: 
\begin{equation}
 t_{\rm cool} \approx 10^6\, \left[\left(\frac{U_{\rm rad}}{0.3\ {\rm eV}{\rm cm}^{-3}}\right)+\left(\frac{B}{3\mu G}\right)^2\right]^{-1}\left(\frac{E}{1 \mathrm{TeV}}\right)^{-1}\ .   
\end{equation} Losses limit the distance $\lambda$ from which leptons of a given energy can reach us to $\lambda \approx 2\sqrt{D(E)t_{\rm cool}}(E)$, with $D(E)$ the galactic diffusion coefficient. 

Using for $D(E)$ the estimate provided by \cite{EvoliBe}, which allows to reproduce all the available AMS-02 data on both primary and secondary nuclei (both stable and unstable), one can estimate that the highest energy leptons detected at the Earth cannot come from further than $\lambda({\rm 1 TeV})\approx 3.5$ kpc. This is a much larger distance than typically estimated in the past (see e.g. \cite{PhysRevD.52.3265}) and ensures that many sources contribute to the lepton flux even at the highest energies \citep{EvoliPosPRL}.

The most promising sources of primary positrons are likely Pulsar Wind Nebulae (PWNe). These are the nebulae formed by the highly relativistic magnetized wind produced by a fast spinning, highly magnetized neutron star (see e.g. \cite{Amato19} for a recent review). The magnetosphere of such a star is a very efficient anti-matter factory and the pulsar wind contains electrons and positrons in about equal amounts. At the wind termination shock, these particles are accelerated up to PeV energies with a spectrum in the form of a broken power-law, harder than $E^{-2}$ at energies below $\approx 500$ GeV and softer than $E^{-2}$ at higher energies. As discussed by \cite{Bykov+17}, the accelerated particles are released in the ISM after the pulsar leaves its associated Supernova Remnant to become a Bow Shock Pulsar Wind Nebula. Once taken into account, the contribution of CR leptons from these sources allows to very well explain both the overall flux of leptons and the positron fraction with very reasonable values of the few free parameters \citep{EvoliPosPRL}.

\section{Conclusions and Outlook}
In this article we have tried to review some recent theoretical and observational developments in the quest for the origin of galactic Cosmic Rays. High energy observations, in the X-rays, gamma-rays and VHE gamma-rays have opened new possibilities to test the long standing paradigm that wants these particles mainly produced in the blast waves of supernova explosions. 

X-ray observations have provided undisputable proof that SNRs can accelerate electrons up to tens of TeV, and at the same time have shown the presence of amplified magnetic fields in their environment. These magnetic fields might be taken as a hint of efficient particle acceleration, although alternatives are possible. For sure their strengths are in the range needed to make SNRs act as PeVatrons, accelerating particles up to the {\it knee} energy, where the galactic cosmic ray proton spectrum is expected to end. In spite of this, theory points to a rare occurrence and an extremely short duration of the PeVatron phase for SNRs (Sec.~\ref{sec:cracc}).

Multi-TeV photons (10-100 TeV photons) are the most direct messengers of the presence of protons in the critical knee region. Secondary electrons, produced as decay products of charged pions, may provide complimentary information through their X-ray synchrotron emission, but this channel has a lower radiative efficiency and secondary leptons are often subdominant with respect to primaries. In the end, the spectrum of the highest energy gamma-rays qualifies as the best source of information about the extension of the parent proton spectrum, and hence is the only direct tool to identify the PeVatrons and obtain essential information about the physics of particle acceleration and the formation of the {\it knee}. 

Detailed gamma-ray studies of young objects in this class have failed to show clear evidence of acceleration above $\sim$ 100 TeV (Sec.~\ref{sec:rxj}). Clear evidence of accelerated hadrons has only been found in middle-aged SNRs interacting with molecular clouds. The wealth of target for nuclear interactions makes these objects ideal candidates to be bright in gamma-rays of hadronic origin, without contamination from leptonic Inverse Compton emission. However, these sources are not very efficient as accelerators and the maximum energy they achieve is limited in all cases to $\lesssim$ 10 TeV (Sec.~\ref{sec:mc}). 

At the same time, observations in the TeV band have highlighted the existence of other kinds of sources showing relatively hard gamma-ray spectra extending above 100 TeV, in association with the galactic center region (Sec.~\ref{sec:CR:GC}) and with young stellar clusters (Sec.~\ref{sec:CR:SC}). How propagation in these complex environments, very likely more turbulent than the average ISM in the Galaxy, affects the spectrum finally released is yet to be thoroughly investigated.

On the other hand, direct measurements of CRs, and in particular high statistics spectra of primary and secondary nuclei, have provided new important insights on the properties of CR transport in the Galaxy, highlighting the presence of a change in propagation properties in the energy range 200-300 GeV. A possible interpretation of this finding is that below such energy CRs are mostly scattered by self-generated turbulence, another bit of information that stresses the importance of non-linearities in CR physics, and the importance of understanding how these particles affect not only the environment of their acceleration sites, but also the properties of the ISM at large (Sec.~\ref{sec:crtransp}).

Finally, a fundamentally important piece of information that we discussed, has to do with the anti-matter component of CRs (Sec.~\ref{sec:cranti}). Direct detection experiments showed that the leptonic anti-matter component of CRs can hardly be interpreted as of pure secondary origin. Additional sources of primary positrons seem to be required. The most natural candidates are pulsars and their nebulae, well known to be excellent positron factories. These will release their positrons only at late stages of their evolution, after leaving the parent SNRs. Crucial information about the particle release by these sources is likely to come from deeper studies of the gamma-ray halos that have been found to surround some of them \cite{TevHalos}. 

In a time when a combination of new and surprising observations and theoretical difficulties are pushing the CR physics community to a deep re-evaluation of the SNR-CR connection, crucial tests are expected to come from high energy astrophysical observations. All the recent surprises and issues that we have discussed through this article suggest that different astrophysical sources might contribute at different energies and at different levels to the CR population. This prompts to the need of a census of all possible CR contributors, a purpose that is well served by an unbiased survey of the Galactic Plane in the crucial multi-TeV energy range. Such survey is currently being undertaken with the high altitude water Cherenkov detector, HAWC \citep{2017APS..APR.X4008M}. A PeVatron search conducted with detailed spectroscopy, higher sensitivity above 10 TeV and higher angular resolution is a key scientific goal of the upcoming Cherenkov Telescope Array \citep{2019scta.book.....C}. In the search for PeVatrons in the Galaxy, CTA will benefit from a wider energy range and a much better angular resolution: while the latter will be essential to establish correlations with the gas maps and sources observed in other wavelengths, the former will finally give us access to the hundreds of TeV energy range, which is the realm where the hadronic mechanism is believed to be strongly dominant over the leptonic one.

\section*{Acknowledgments}
We thank the anonymous reviewers for their careful reading of the manuscript and their many useful comments, which in our opinion have substantially improved the paper. 

EA acknowledges support by INAF and ASI through Grants SKA-CTA INAF 2016, INAF-MAINSTREAM 2018 and ASI/INAF No. 2017-14-H.O and by the National Science Foundation under Grant No. NSF PHY-1748958. 

SC acknowledges the support from Polish Science Centre grant, DEC-2017/27/B/ST9/02272.

\bibliographystyle{jpp}
\cite{article-crossref} and \cite{article-full}.
\newcommand{\noopsort}[1]{} \newcommand{\printfirst}[2]{#1}
  \newcommand{\singleletter}[1]{#1} \newcommand{\switchargs}[2]{#2#1}

\bibliography{crrev.bib}

\begin{thebibliography}{185}
\expandafter\ifx\csname natexlab\endcsname\relax\def\natexlab#1{#1}\fi
\def\au#1{#1} \def\ed#1{#1} \def\yr#1{#1}\def\at#1{#1}\def\jt#1{\textit{#1}}
  \def\bt#1{#1}\def\bvol#1{\textbf{#1}} \def\vol#1{#1} \def\pg#1{#1}
  \def\publ#1{#1}\def\arxiv#1{#1}\def\org#1{#1}\def\st#1{\textit{#1}}

\bibitem[{Abdo} {\em et~al.\/}(2011){Abdo}, {Ackermann}, {Ajello}, {Allafort},
  {Baldini}, {Ballet}, {Barbiellini}, {Baring}, {Bastieri}, {Bellazzini},
  {Berenji}, {Blandford}, {Bloom}, {Bonamente}, {Borgland}, {Bouvier}, {Brand
  t}, {Bregeon}, {Brigida}, {Bruel}, {Buehler}, {Buson}, {Caliandro},
  {Cameron}, {Caraveo}, {Casandjian}, {Cecchi}, {Chaty}, {Chekhtman}, {Cheung},
  {Chiang}, {Cillis}, {Ciprini}, {Claus}, {Cohen-Tanugi}, {Conrad}, {Corbel},
  {Cutini}, {de Angelis}, {de Palma}, {Dermer}, {Digel}, {Silva}, {Drell},
  {Drlica-Wagner}, {Dubois}, {Dumora}, {Favuzzi}, {Ferrara}, {Fortin},
  {Frailis}, {Fukazawa}, {Fukui}, {Funk}, {Fusco}, {Gargano}, {Gasparrini},
  {Gehrels}, {Germani}, {Giglietto}, {Giordano}, {Giroletti}, {Glanzman},
  {Godfrey}, {Grenier}, {Grondin}, {Guiriec}, {Hadasch}, {Hanabata}, {Harding},
  {Hayashida}, {Hayashi}, {Hays}, {Horan}, {Jackson}, {J{\'o}hannesson},
  {Johnson}, {Kamae}, {Katagiri}, {Kataoka}, {Kerr}, {Kn{\"o}dlseder}, {Kuss},
  {Lande}, {Latronico}, {Lee}, {Lemoine-Goumard}, {Longo}, {Loparco},
  {Lovellette}, {Lubrano}, {Madejski}, {Makeev}, {Mazziotta}, {McEnery},
  {Michelson}, {Mignani}, {Mitthumsiri}, {Mizuno}, {Moiseev}, {Monte},
  {Monzani}, {Morselli}, {Moskalenko}, {Murgia}, {Naumann-Godo}, {Nolan},
  {Norris}, {Nuss}, {Ohsugi}, {Okumura}, {Orlando}, {Ormes}, {Paneque},
  {Parent}, {Pelassa}, {Pesce-Rollins}, {Pierbattista}, {Piron}, {Pohl},
  {Porter}, {Rain{\`o}}, {Rando}, {Razzano}, {Reimer}, {Reposeur}, {Ritz},
  {Romani}, {Roth}, {Sadrozinski}, {Saz Parkinson}, {Sgr{\`o}}, {Smith},
  {Smith}, {Spandre}, {Spinelli}, {Strickman}, {Tajima}, {Takahashi},
  {Takahashi}, {Tanaka}, {Thayer}, {Thayer}, {Thompson}, {Tibaldo}, {Tibolla},
  {Torres}, {Tosti}, {Tramacere}, {Troja}, {Uchiyama}, {Vandenbroucke},
  {Vasileiou}, {Vianello}, {Vilchez}, {Vitale}, {Waite}, {Wang}, {Winer},
  {Wood}, {Yamamoto}, {Yamazaki}, {Yang} \& {Ziegler}]{2011ApJ...734...28A}
{\sc \au{{Abdo}, A.~A.}, \au{{Ackermann}, M.}, \au{{Ajello}, M.},
  \au{{Allafort}, A.}, \au{{Baldini}, L.}, \au{{Ballet}, J.},
  \au{{Barbiellini}, G.}, \au{{Baring}, M.~G.}, \au{{Bastieri}, D.},
  \au{{Bellazzini}, R.}, \au{{Berenji}, B.}, \au{{Blandford}, R.~D.},
  \au{{Bloom}, E.~D.}, \au{{Bonamente}, E.}, \au{{Borgland}, A.~W.},
  \au{{Bouvier}, A.}, \au{{Brand t}, T.~J.}, \au{{Bregeon}, J.}, \au{{Brigida},
  M.}, \au{{Bruel}, P.}, \au{{Buehler}, R.}, \au{{Buson}, S.}, \au{{Caliandro},
  G.~A.}, \au{{Cameron}, R.~A.}, \au{{Caraveo}, P.~A.}, \au{{Casandjian},
  J.~M.}, \au{{Cecchi}, C.}, \au{{Chaty}, S.}, \au{{Chekhtman}, A.},
  \au{{Cheung}, C.~C.}, \au{{Chiang}, J.}, \au{{Cillis}, A.~N.}, \au{{Ciprini},
  S.}, \au{{Claus}, R.}, \au{{Cohen-Tanugi}, J.}, \au{{Conrad}, J.},
  \au{{Corbel}, S.}, \au{{Cutini}, S.}, \au{{de Angelis}, A.}, \au{{de Palma},
  F.}, \au{{Dermer}, C.~D.}, \au{{Digel}, S.~W.}, \au{{Silva}, E. do Couto~e.},
  \au{{Drell}, P.~S.}, \au{{Drlica-Wagner}, A.}, \au{{Dubois}, R.},
  \au{{Dumora}, D.}, \au{{Favuzzi}, C.}, \au{{Ferrara}, E.~C.}, \au{{Fortin},
  P.}, \au{{Frailis}, M.}, \au{{Fukazawa}, Y.}, \au{{Fukui}, Y.}, \au{{Funk},
  S.}, \au{{Fusco}, P.}, \au{{Gargano}, F.}, \au{{Gasparrini}, D.},
  \au{{Gehrels}, N.}, \au{{Germani}, S.}, \au{{Giglietto}, N.}, \au{{Giordano},
  F.}, \au{{Giroletti}, M.}, \au{{Glanzman}, T.}, \au{{Godfrey}, G.},
  \au{{Grenier}, I.~A.}, \au{{Grondin}, M.~H.}, \au{{Guiriec}, S.},
  \au{{Hadasch}, D.}, \au{{Hanabata}, Y.}, \au{{Harding}, A.~K.},
  \au{{Hayashida}, M.}, \au{{Hayashi}, K.}, \au{{Hays}, E.}, \au{{Horan}, D.},
  \au{{Jackson}, M.~S.}, \au{{J{\'o}hannesson}, G.}, \au{{Johnson}, A.~S.},
  \au{{Kamae}, T.}, \au{{Katagiri}, H.}, \au{{Kataoka}, J.}, \au{{Kerr}, M.},
  \au{{Kn{\"o}dlseder}, J.}, \au{{Kuss}, M.}, \au{{Lande}, J.},
  \au{{Latronico}, L.}, \au{{Lee}, S.~H.}, \au{{Lemoine-Goumard}, M.},
  \au{{Longo}, F.}, \au{{Loparco}, F.}, \au{{Lovellette}, M.~N.},
  \au{{Lubrano}, P.}, \au{{Madejski}, G.~M.}, \au{{Makeev}, A.},
  \au{{Mazziotta}, M.~N.}, \au{{McEnery}, J.~E.}, \au{{Michelson}, P.~F.},
  \au{{Mignani}, R.~P.}, \au{{Mitthumsiri}, W.}, \au{{Mizuno}, T.},
  \au{{Moiseev}, A.~A.}, \au{{Monte}, C.}, \au{{Monzani}, M.~E.},
  \au{{Morselli}, A.}, \au{{Moskalenko}, I.~V.}, \au{{Murgia}, S.},
  \au{{Naumann-Godo}, M.}, \au{{Nolan}, P.~L.}, \au{{Norris}, J.~P.},
  \au{{Nuss}, E.}, \au{{Ohsugi}, T.}, \au{{Okumura}, A.}, \au{{Orlando}, E.},
  \au{{Ormes}, J.~F.}, \au{{Paneque}, D.}, \au{{Parent}, D.}, \au{{Pelassa},
  V.}, \au{{Pesce-Rollins}, M.}, \au{{Pierbattista}, M.}, \au{{Piron}, F.},
  \au{{Pohl}, M.}, \au{{Porter}, T.~A.}, \au{{Rain{\`o}}, S.}, \au{{Rando},
  R.}, \au{{Razzano}, M.}, \au{{Reimer}, O.}, \au{{Reposeur}, T.}, \au{{Ritz},
  S.}, \au{{Romani}, R.~W.}, \au{{Roth}, M.}, \au{{Sadrozinski}, H.~F.~W.},
  \au{{Saz Parkinson}, P.~M.}, \au{{Sgr{\`o}}, C.}, \au{{Smith}, D.~A.},
  \au{{Smith}, P.~D.}, \au{{Spandre}, G.}, \au{{Spinelli}, P.},
  \au{{Strickman}, M.~S.}, \au{{Tajima}, H.}, \au{{Takahashi}, H.},
  \au{{Takahashi}, T.}, \au{{Tanaka}, T.}, \au{{Thayer}, J.~G.}, \au{{Thayer},
  J.~B.}, \au{{Thompson}, D.~J.}, \au{{Tibaldo}, L.}, \au{{Tibolla}, O.},
  \au{{Torres}, D.~F.}, \au{{Tosti}, G.}, \au{{Tramacere}, A.}, \au{{Troja},
  E.}, \au{{Uchiyama}, Y.}, \au{{Vandenbroucke}, J.}, \au{{Vasileiou}, V.},
  \au{{Vianello}, G.}, \au{{Vilchez}, N.}, \au{{Vitale}, V.}, \au{{Waite},
  A.~P.}, \au{{Wang}, P.}, \au{{Winer}, B.~L.}, \au{{Wood}, K.~S.},
  \au{{Yamamoto}, H.}, \au{{Yamazaki}, R.}, \au{{Yang}, Z.} \& \au{{Ziegler},
  M.}} \yr{2011}  \at{{Observations of the Young Supernova Remnant RX
  J1713.7-3946 with the Fermi Large Area Telescope}}.  \jt{ApJ}
  \bvol{734}~(1),  \pg{28},  \arxiv{arXiv: 1103.5727}.

\bibitem[{Abdo} {\em et~al.\/}(2010{\natexlab{{\em a\/}}}){Abdo}, {Ackermann},
  {Ajello}, {Baldini}, {Ballet}, {Barbiellini}, {Baring}, {Bastieri},
  {Baughman}, {Bechtol}, {Bellazzini}, {Berenji}, {Blandford}, {Bloom},
  {Bonamente}, {Borgland}, {Bregeon}, {Brez}, {Brigida}, {Bruel}, {Burnett},
  {Buson}, {Caliandro}, {Cameron}, {Caraveo}, {Casandjian}, {Cecchi},
  {{\c{C}}elik}, {Chekhtman}, {Cheung}, {Chiang}, {Ciprini}, {Claus},
  {Cognard}, {Cohen-Tanugi}, {Cominsky}, {Conrad}, {Cutini}, {Dermer}, {de
  Angelis}, {de Palma}, {Digel}, {do Couto e Silva}, {Drell}, {Dubois},
  {Dumora}, {Espinoza}, {Farnier}, {Favuzzi}, {Fegan}, {Focke}, {Fortin},
  {Frailis}, {Fukazawa}, {Funk}, {Fusco}, {Gargano}, {Gasparrini}, {Gehrels},
  {Germani}, {Giavitto}, {Giebels}, {Giglietto}, {Giordano}, {Glanzman},
  {Godfrey}, {Grenier}, {Grondin}, {Grove}, {Guillemot}, {Guiriec}, {Hanabata},
  {Harding}, {Hayashida}, {Hays}, {Hughes}, {Jackson}, {J{\'o}hannesson},
  {Johnson}, {Johnson}, {Johnson}, {Kamae}, {Katagiri}, {Kataoka}, {Katsuta},
  {Kawai}, {Kerr}, {Kn{\"o}dlseder}, {Kocian}, {Kramer}, {Kuss}, {Lande},
  {Latronico}, {Lemoine-Goumard}, {Longo}, {Loparco}, {Lott}, {Lovellette},
  {Lubrano}, {Lyne}, {Madejski}, {Makeev}, {Mazziotta}, {McEnery}, {Meurer},
  {Michelson}, {Mitthumsiri}, {Mizuno}, {Monte}, {Monzani}, {Morselli},
  {Moskalenko}, {Murgia}, {Nakamori}, {Nolan}, {Norris}, {Noutsos}, {Nuss},
  {Ohsugi}, {Omodei}, {Orlando}, {Ormes}, {Paneque}, {Parent}, {Pelassa},
  {Pepe}, {Pesce-Rollins}, {Piron}, {Porter}, {Rain{\`o}}, {Rando}, {Razzano},
  {Reimer}, {Reimer}, {Reposeur}, {Rochester}, {Rodriguez}, {Romani}, {Roth},
  {Ryde}, {Sadrozinski}, {Sanchez}, {Sander}, {Parkinson}, {Scargle},
  {Sgr{\`o}}, {Siskind}, {Smith}, {Smith}, {Spand re}, {Spinelli}, {Stappers},
  {Stecker}, {Strickman}, {Suson}, {Tajima}, {Takahashi}, {Takahashi},
  {Tanaka}, {Thayer}, {Thayer}, {Theureau}, {Thompson}, {Tibaldo}, {Tibolla},
  {Torres}, {Tosti}, {Tramacere}, {Uchiyama}, {Usher}, {Vasileiou}, {Venter},
  {Vilchez}, {Vitale}, {Waite}, {Wang}, {Winer}, {Wood}, {Yamazaki}, {Ylinen}
  \& {Ziegler}]{FermiW44}
{\sc \au{{Abdo}, A.~A.}, \au{{Ackermann}, M.}, \au{{Ajello}, M.},
  \au{{Baldini}, L.}, \au{{Ballet}, J.}, \au{{Barbiellini}, G.}, \au{{Baring},
  M.~G.}, \au{{Bastieri}, D.}, \au{{Baughman}, B.~M.}, \au{{Bechtol}, K.},
  \au{{Bellazzini}, R.}, \au{{Berenji}, B.}, \au{{Blandford}, R.~D.},
  \au{{Bloom}, E.~D.}, \au{{Bonamente}, E.}, \au{{Borgland}, A.~W.},
  \au{{Bregeon}, J.}, \au{{Brez}, A.}, \au{{Brigida}, M.}, \au{{Bruel}, P.},
  \au{{Burnett}, T.~H.}, \au{{Buson}, S.}, \au{{Caliandro}, G.~A.},
  \au{{Cameron}, R.~A.}, \au{{Caraveo}, P.~A.}, \au{{Casandjian}, J.~M.},
  \au{{Cecchi}, C.}, \au{{{\c{C}}elik}, {\"O}.}, \au{{Chekhtman}, A.},
  \au{{Cheung}, C.~C.}, \au{{Chiang}, J.}, \au{{Ciprini}, S.}, \au{{Claus},
  R.}, \au{{Cognard}, I.}, \au{{Cohen-Tanugi}, J.}, \au{{Cominsky}, L.~R.},
  \au{{Conrad}, J.}, \au{{Cutini}, S.}, \au{{Dermer}, C.~D.}, \au{{de Angelis},
  A.}, \au{{de Palma}, F.}, \au{{Digel}, S.~W.}, \au{{do Couto e Silva}, E.},
  \au{{Drell}, P.~S.}, \au{{Dubois}, R.}, \au{{Dumora}, D.}, \au{{Espinoza},
  C.}, \au{{Farnier}, C.}, \au{{Favuzzi}, C.}, \au{{Fegan}, S.~J.},
  \au{{Focke}, W.~B.}, \au{{Fortin}, P.}, \au{{Frailis}, M.}, \au{{Fukazawa},
  Y.}, \au{{Funk}, S.}, \au{{Fusco}, P.}, \au{{Gargano}, F.}, \au{{Gasparrini},
  D.}, \au{{Gehrels}, N.}, \au{{Germani}, S.}, \au{{Giavitto}, G.},
  \au{{Giebels}, B.}, \au{{Giglietto}, N.}, \au{{Giordano}, F.},
  \au{{Glanzman}, T.}, \au{{Godfrey}, G.}, \au{{Grenier}, I.~A.},
  \au{{Grondin}, M.~H.}, \au{{Grove}, J.~E.}, \au{{Guillemot}, L.},
  \au{{Guiriec}, S.}, \au{{Hanabata}, Y.}, \au{{Harding}, A.~K.},
  \au{{Hayashida}, M.}, \au{{Hays}, E.}, \au{{Hughes}, R.~E.}, \au{{Jackson},
  M.~S.}, \au{{J{\'o}hannesson}, G.}, \au{{Johnson}, A.~S.}, \au{{Johnson},
  T.~J.}, \au{{Johnson}, W.~N.}, \au{{Kamae}, T.}, \au{{Katagiri}, H.},
  \au{{Kataoka}, J.}, \au{{Katsuta}, J.}, \au{{Kawai}, N.}, \au{{Kerr}, M.},
  \au{{Kn{\"o}dlseder}, J.}, \au{{Kocian}, M.~L.}, \au{{Kramer}, M.},
  \au{{Kuss}, M.}, \au{{Lande}, J.}, \au{{Latronico}, L.},
  \au{{Lemoine-Goumard}, M.}, \au{{Longo}, F.}, \au{{Loparco}, F.}, \au{{Lott},
  B.}, \au{{Lovellette}, M.~N.}, \au{{Lubrano}, P.}, \au{{Lyne}, A.~G.},
  \au{{Madejski}, G.~M.}, \au{{Makeev}, A.}, \au{{Mazziotta}, M.~N.},
  \au{{McEnery}, J.~E.}, \au{{Meurer}, C.}, \au{{Michelson}, P.~F.},
  \au{{Mitthumsiri}, W.}, \au{{Mizuno}, T.}, \au{{Monte}, C.}, \au{{Monzani},
  M.~E.}, \au{{Morselli}, A.}, \au{{Moskalenko}, I.~V.}, \au{{Murgia}, S.},
  \au{{Nakamori}, T.}, \au{{Nolan}, P.~L.}, \au{{Norris}, J.~P.},
  \au{{Noutsos}, A.}, \au{{Nuss}, E.}, \au{{Ohsugi}, T.}, \au{{Omodei}, N.},
  \au{{Orlando}, E.}, \au{{Ormes}, J.~F.}, \au{{Paneque}, D.}, \au{{Parent},
  D.}, \au{{Pelassa}, V.}, \au{{Pepe}, M.}, \au{{Pesce-Rollins}, M.},
  \au{{Piron}, F.}, \au{{Porter}, T.~A.}, \au{{Rain{\`o}}, S.}, \au{{Rando},
  R.}, \au{{Razzano}, M.}, \au{{Reimer}, A.}, \au{{Reimer}, O.},
  \au{{Reposeur}, T.}, \au{{Rochester}, L.~S.}, \au{{Rodriguez}, A.~Y.},
  \au{{Romani}, R.~W.}, \au{{Roth}, M.}, \au{{Ryde}, F.}, \au{{Sadrozinski},
  H.~F.~W.}, \au{{Sanchez}, D.}, \au{{Sander}, A.}, \au{{Parkinson},
  P.~M.~Saz}, \au{{Scargle}, J.~D.}, \au{{Sgr{\`o}}, C.}, \au{{Siskind},
  E.~J.}, \au{{Smith}, D.~A.}, \au{{Smith}, P.~D.}, \au{{Spand re}, G.},
  \au{{Spinelli}, P.}, \au{{Stappers}, B.~W.}, \au{{Stecker}, F.~W.},
  \au{{Strickman}, M.~S.}, \au{{Suson}, D.~J.}, \au{{Tajima}, H.},
  \au{{Takahashi}, H.}, \au{{Takahashi}, T.}, \au{{Tanaka}, T.}, \au{{Thayer},
  J.~B.}, \au{{Thayer}, J.~G.}, \au{{Theureau}, G.}, \au{{Thompson}, D.~J.},
  \au{{Tibaldo}, L.}, \au{{Tibolla}, O.}, \au{{Torres}, D.~F.}, \au{{Tosti},
  G.}, \au{{Tramacere}, A.}, \au{{Uchiyama}, Y.}, \au{{Usher}, T.~L.},
  \au{{Vasileiou}, V.}, \au{{Venter}, C.}, \au{{Vilchez}, N.}, \au{{Vitale},
  V.}, \au{{Waite}, A.~P.}, \au{{Wang}, P.}, \au{{Winer}, B.~L.}, \au{{Wood},
  K.~S.}, \au{{Yamazaki}, R.}, \au{{Ylinen}, T.} \& \au{{Ziegler}, M.}}
  \yr{2010{\natexlab{{\em a\/}}}}  \at{{Gamma-Ray Emission from the Shell of
  Supernova Remnant W44 Revealed by the Fermi LAT}}.  \jt{Science}
  \bvol{327}~(5969),  \pg{1103}.

\bibitem[{Abdo} {\em et~al.\/}(2010{\natexlab{{\em b\/}}}){Abdo}, {Ackermann},
  {Ajello}, {Baldini}, {Ballet}, {Barbiellini}, {Bastieri}, {Baughman},
  {Bechtol}, {Bellazzini}, {Berenji}, {Blandford}, {Bloom}, {Bonamente},
  {Borgland}, {Bregeon}, {Brez}, {Brigida}, {Bruel}, {Burnett}, {Buson},
  {Caliandro}, {Cameron}, {Caraveo}, {Casand jian}, {Cecchi}, {{\c{C}}elik},
  {Chekhtman}, {Cheung}, {Chiang}, {Cillis}, {Ciprini}, {Claus},
  {Cohen-Tanugi}, {Cominsky}, {Conrad}, {Cutini}, {Dermer}, {de Angelis}, {de
  Palma}, {Silva}, {Drell}, {Drlica-Wagner}, {Dubois}, {Dumora}, {Farnier},
  {Favuzzi}, {Fegan}, {Focke}, {Fortin}, {Frailis}, {Fukazawa}, {Funk},
  {Fusco}, {Gargano}, {Gasparrini}, {Gehrels}, {Germani}, {Giavitto},
  {Giebels}, {Giglietto}, {Giordano}, {Glanzman}, {Godfrey}, {Grenier},
  {Grondin}, {Grove}, {Guillemot}, {Guiriec}, {Hanabata}, {Harding},
  {Hayashida}, {Hughes}, {Jackson}, {J{\'o}hannesson}, {Johnson}, {Johnson},
  {Johnson}, {Kamae}, {Katagiri}, {Kataoka}, {Kawai}, {Kerr}, {Kn{\"o}dlseder},
  {Kocian}, {Kuss}, {Land e}, {Latronico}, {Lee}, {Lemoine-Goumard}, {Longo},
  {Loparco}, {Lott}, {Lovellette}, {Lubrano}, {Madejski}, {Makeev},
  {Mazziotta}, {Meurer}, {Michelson}, {Mitthumsiri}, {Moiseev}, {Monte},
  {Monzani}, {Morselli}, {Moskalenko}, {Murgia}, {Nakamori}, {Nolan}, {Norris},
  {Nuss}, {Ohsugi}, {Orlando}, {Ormes}, {Ozaki}, {Paneque}, {Panetta},
  {Parent}, {Pelassa}, {Pepe}, {Pesce-Rollins}, {Piron}, {Porter}, {Rain{\`o}},
  {Rando}, {Razzano}, {Reimer}, {Reimer}, {Reposeur}, {Rochester}, {Rodriguez},
  {Romani}, {Roth}, {Ryde}, {Sadrozinski}, {Sanchez}, {Sander}, {Saz
  Parkinson}, {Scargle}, {Sgr{\`o}}, {Siskind}, {Smith}, {Smith}, {Spandre},
  {Spinelli}, {Strickman}, {Strong}, {Suson}, {Tajima}, {Takahashi},
  {Takahashi}, {Tanaka}, {Thayer}, {Thayer}, {Thompson}, {Tibaldo}, {Torres},
  {Tosti}, {Tramacere}, {Uchiyama}, {Usher}, {Van Etten}, {Vasileiou},
  {Venter}, {Vilchez}, {Vitale}, {Waite}, {Wang}, {Winer}, {Wood}, {Ylinen} \&
  {Ziegler}]{FermiIC443}
{\sc \au{{Abdo}, A.~A.}, \au{{Ackermann}, M.}, \au{{Ajello}, M.},
  \au{{Baldini}, L.}, \au{{Ballet}, J.}, \au{{Barbiellini}, G.},
  \au{{Bastieri}, D.}, \au{{Baughman}, B.~M.}, \au{{Bechtol}, K.},
  \au{{Bellazzini}, R.}, \au{{Berenji}, B.}, \au{{Blandford}, R.~D.},
  \au{{Bloom}, E.~D.}, \au{{Bonamente}, E.}, \au{{Borgland}, A.~W.},
  \au{{Bregeon}, J.}, \au{{Brez}, A.}, \au{{Brigida}, M.}, \au{{Bruel}, P.},
  \au{{Burnett}, T.~H.}, \au{{Buson}, S.}, \au{{Caliandro}, G.~A.},
  \au{{Cameron}, R.~A.}, \au{{Caraveo}, P.~A.}, \au{{Casand jian}, J.~M.},
  \au{{Cecchi}, C.}, \au{{{\c{C}}elik}, {\"O}.}, \au{{Chekhtman}, A.},
  \au{{Cheung}, C.~C.}, \au{{Chiang}, J.}, \au{{Cillis}, A.~N.}, \au{{Ciprini},
  S.}, \au{{Claus}, R.}, \au{{Cohen-Tanugi}, J.}, \au{{Cominsky}, L.~R.},
  \au{{Conrad}, J.}, \au{{Cutini}, S.}, \au{{Dermer}, C.~D.}, \au{{de Angelis},
  A.}, \au{{de Palma}, F.}, \au{{Silva}, E. do Couto~e.}, \au{{Drell}, P.~S.},
  \au{{Drlica-Wagner}, A.}, \au{{Dubois}, R.}, \au{{Dumora}, D.},
  \au{{Farnier}, C.}, \au{{Favuzzi}, C.}, \au{{Fegan}, S.~J.}, \au{{Focke},
  W.~B.}, \au{{Fortin}, P.}, \au{{Frailis}, M.}, \au{{Fukazawa}, Y.},
  \au{{Funk}, S.}, \au{{Fusco}, P.}, \au{{Gargano}, F.}, \au{{Gasparrini}, D.},
  \au{{Gehrels}, N.}, \au{{Germani}, S.}, \au{{Giavitto}, G.}, \au{{Giebels},
  B.}, \au{{Giglietto}, N.}, \au{{Giordano}, F.}, \au{{Glanzman}, T.},
  \au{{Godfrey}, G.}, \au{{Grenier}, I.~A.}, \au{{Grondin}, M.~H.},
  \au{{Grove}, J.~E.}, \au{{Guillemot}, L.}, \au{{Guiriec}, S.},
  \au{{Hanabata}, Y.}, \au{{Harding}, A.~K.}, \au{{Hayashida}, M.},
  \au{{Hughes}, R.~E.}, \au{{Jackson}, M.~S.}, \au{{J{\'o}hannesson}, G.},
  \au{{Johnson}, A.~S.}, \au{{Johnson}, T.~J.}, \au{{Johnson}, W.~N.},
  \au{{Kamae}, T.}, \au{{Katagiri}, H.}, \au{{Kataoka}, J.}, \au{{Kawai}, N.},
  \au{{Kerr}, M.}, \au{{Kn{\"o}dlseder}, J.}, \au{{Kocian}, M.~L.}, \au{{Kuss},
  M.}, \au{{Land e}, J.}, \au{{Latronico}, L.}, \au{{Lee}, S.~H.},
  \au{{Lemoine-Goumard}, M.}, \au{{Longo}, F.}, \au{{Loparco}, F.}, \au{{Lott},
  B.}, \au{{Lovellette}, M.~N.}, \au{{Lubrano}, P.}, \au{{Madejski}, G.~M.},
  \au{{Makeev}, A.}, \au{{Mazziotta}, M.~N.}, \au{{Meurer}, C.},
  \au{{Michelson}, P.~F.}, \au{{Mitthumsiri}, W.}, \au{{Moiseev}, A.~A.},
  \au{{Monte}, C.}, \au{{Monzani}, M.~E.}, \au{{Morselli}, A.},
  \au{{Moskalenko}, I.~V.}, \au{{Murgia}, S.}, \au{{Nakamori}, T.},
  \au{{Nolan}, P.~L.}, \au{{Norris}, J.~P.}, \au{{Nuss}, E.}, \au{{Ohsugi},
  T.}, \au{{Orlando}, E.}, \au{{Ormes}, J.~F.}, \au{{Ozaki}, M.},
  \au{{Paneque}, D.}, \au{{Panetta}, J.~H.}, \au{{Parent}, D.}, \au{{Pelassa},
  V.}, \au{{Pepe}, M.}, \au{{Pesce-Rollins}, M.}, \au{{Piron}, F.},
  \au{{Porter}, T.~A.}, \au{{Rain{\`o}}, S.}, \au{{Rando}, R.}, \au{{Razzano},
  M.}, \au{{Reimer}, A.}, \au{{Reimer}, O.}, \au{{Reposeur}, T.},
  \au{{Rochester}, L.~S.}, \au{{Rodriguez}, A.~Y.}, \au{{Romani}, R.~W.},
  \au{{Roth}, M.}, \au{{Ryde}, F.}, \au{{Sadrozinski}, H.~F.~W.},
  \au{{Sanchez}, D.}, \au{{Sander}, A.}, \au{{Saz Parkinson}, P.~M.},
  \au{{Scargle}, J.~D.}, \au{{Sgr{\`o}}, C.}, \au{{Siskind}, E.~J.},
  \au{{Smith}, D.~A.}, \au{{Smith}, P.~D.}, \au{{Spandre}, G.}, \au{{Spinelli},
  P.}, \au{{Strickman}, M.~S.}, \au{{Strong}, A.~W.}, \au{{Suson}, D.~J.},
  \au{{Tajima}, H.}, \au{{Takahashi}, H.}, \au{{Takahashi}, T.}, \au{{Tanaka},
  T.}, \au{{Thayer}, J.~B.}, \au{{Thayer}, J.~G.}, \au{{Thompson}, D.~J.},
  \au{{Tibaldo}, L.}, \au{{Torres}, D.~F.}, \au{{Tosti}, G.}, \au{{Tramacere},
  A.}, \au{{Uchiyama}, Y.}, \au{{Usher}, T.~L.}, \au{{Van Etten}, A.},
  \au{{Vasileiou}, V.}, \au{{Venter}, C.}, \au{{Vilchez}, N.}, \au{{Vitale},
  V.}, \au{{Waite}, A.~P.}, \au{{Wang}, P.}, \au{{Winer}, B.~L.}, \au{{Wood},
  K.~S.}, \au{{Ylinen}, T.} \& \au{{Ziegler}, M.}} \yr{2010{\natexlab{{\em
  b\/}}}}  \at{{Observation of Supernova Remnant IC 443 with the Fermi Large
  Area Telescope}}.  \jt{ApJ}  \bvol{712}~(1),  \pg{459--468},  \arxiv{arXiv:
  1002.2198}.

\bibitem[{Abdo} {\em et~al.\/}(2007){Abdo}, {Allen}, {Berley}, {Casanova},
  {Chen}, {Coyne}, {Dingus}, {Ellsworth}, {Fleysher}, {Fleysher}, {Gonzalez},
  {Goodman}, {Hays}, {Hoffman}, {Hopper}, {H{\"u}ntemeyer}, {Kolterman},
  {Lansdell}, {Linnemann}, {McEnery}, {Mincer}, {Nemethy}, {Noyes}, {Ryan},
  {Saz Parkinson}, {Shoup}, {Sinnis}, {Smith}, {Sullivan}, {Vasileiou},
  {Walker}, {Williams}, {Xu} \& {Yodh}]{2007ApJ...664L..91A}
{\sc \au{{Abdo}, A.~A.}, \au{{Allen}, B.}, \au{{Berley}, D.}, \au{{Casanova},
  S.}, \au{{Chen}, C.}, \au{{Coyne}, D.~G.}, \au{{Dingus}, B.~L.},
  \au{{Ellsworth}, R.~W.}, \au{{Fleysher}, L.}, \au{{Fleysher}, R.},
  \au{{Gonzalez}, M.~M.}, \au{{Goodman}, J.~A.}, \au{{Hays}, E.},
  \au{{Hoffman}, C.~M.}, \au{{Hopper}, B.}, \au{{H{\"u}ntemeyer}, P.~H.},
  \au{{Kolterman}, B.~E.}, \au{{Lansdell}, C.~P.}, \au{{Linnemann}, J.~T.},
  \au{{McEnery}, J.~E.}, \au{{Mincer}, A.~I.}, \au{{Nemethy}, P.}, \au{{Noyes},
  D.}, \au{{Ryan}, J.~M.}, \au{{Saz Parkinson}, P.~M.}, \au{{Shoup}, A.},
  \au{{Sinnis}, G.}, \au{{Smith}, A.~J.}, \au{{Sullivan}, G.~W.},
  \au{{Vasileiou}, V.}, \au{{Walker}, G.~P.}, \au{{Williams}, D.~A.}, \au{{Xu},
  X.~W.} \& \au{{Yodh}, G.~B.}} \yr{2007}  \at{{TeV Gamma-Ray Sources from a
  Survey of the Galactic Plane with Milagro}}.  \jt{ApJL}  \bvol{664}~(2),
  \pg{L91--L94},  \arxiv{arXiv: 0705.0707}.

\bibitem[{Abdo} {\em et~al.\/}(2009){Abdo}, {Allen}, {Aune}, {Berley}, {Chen},
  {Christopher}, {DeYoung}, {Dingus}, {Ellsworth}, {Gonzalez}, {Goodman},
  {Hays}, {Hoffman}, {H{\"u}ntemeyer}, {Kolterman}, {Linnemann}, {McEnery},
  {Morgan}, {Mincer}, {Nemethy}, {Pretz}, {Ryan}, {Saz Parkinson}, {Shoup},
  {Sinnis}, {Smith}, {Vasileiou}, {Walker}, {Williams} \&
  {Yodh}]{2009ApJ...700L.127A}
{\sc \au{{Abdo}, A.~A.}, \au{{Allen}, B.~T.}, \au{{Aune}, T.}, \au{{Berley},
  D.}, \au{{Chen}, C.}, \au{{Christopher}, G.~E.}, \au{{DeYoung}, T.},
  \au{{Dingus}, B.~L.}, \au{{Ellsworth}, R.~W.}, \au{{Gonzalez}, M.~M.},
  \au{{Goodman}, J.~A.}, \au{{Hays}, E.}, \au{{Hoffman}, C.~M.},
  \au{{H{\"u}ntemeyer}, P.~H.}, \au{{Kolterman}, B.~E.}, \au{{Linnemann},
  J.~T.}, \au{{McEnery}, J.~E.}, \au{{Morgan}, T.}, \au{{Mincer}, A.~I.},
  \au{{Nemethy}, P.}, \au{{Pretz}, J.}, \au{{Ryan}, J.~M.}, \au{{Saz
  Parkinson}, P.~M.}, \au{{Shoup}, A.}, \au{{Sinnis}, G.}, \au{{Smith}, A.~J.},
  \au{{Vasileiou}, V.}, \au{{Walker}, G.~P.}, \au{{Williams}, D.~A.} \&
  \au{{Yodh}, G.~B.}} \yr{2009}  \at{{Milagro Observations of Multi-TeV
  Emission from Galactic Sources in the Fermi Bright Source List}}.  \jt{ApJL}
  \bvol{700}~(2),  \pg{L127--L131},  \arxiv{arXiv: 0904.1018}.

\bibitem[{Abeysekara} {\em et~al.\/}(2017{\natexlab{{\em a\/}}}){Abeysekara},
  {Albert}, {Alfaro}, {Alvarez}, {{\'A}lvarez}, {Arceo},
  {Arteaga-Vel{\'a}zquez}, {Avila Rojas}, {Ayala Solares}, {Barber},
  {Bautista-Elivar}, {Becerril}, {Belmont-Moreno}, {BenZvi}, {Berley},
  {Bernal}, {Braun}, {Brisbois}, {Caballero-Mora}, {Capistr{\'a}n},
  {Carrami{\~n}ana}, {Casanova}, {Castillo}, {Cotti}, {Cotzomi}, {Couti{\~n}o
  de Le{\'o}n}, {De Le{\'o}n}, {De la Fuente}, {Dingus}, {DuVernois},
  {D{\'\i}az-V{\'e}lez}, {Ellsworth}, {Engel}, {Enr{\'\i}quez-Rivera},
  {Fiorino}, {Fraija}, {Garc{\'\i}a-Gonz{\'a}lez}, {Garfias}, {Gerhardt},
  {Gonz{\'a}lez Mu{\~n}oz}, {Gonz{\'a}lez}, {Goodman}, {Hampel-Arias},
  {Harding}, {Hern{\'a}ndez}, {Hern{\'a}ndez-Almada}, {Hinton}, {Hona}, {Hui},
  {H{\"u}ntemeyer}, {Iriarte}, {Jardin-Blicq}, {Joshi}, {Kaufmann}, {Kieda},
  {Lara}, {Lauer}, {Lee}, {Lennarz}, {Vargas}, {Linnemann}, {Longinotti}, {Luis
  Raya}, {Luna-Garc{\'\i}a}, {L{\'o}pez-Coto}, {Malone}, {Marinelli},
  {Martinez}, {Martinez-Castellanos}, {Mart{\'\i}nez-Castro},
  {Mart{\'\i}nez-Huerta}, {Matthews}, {Mirand a-Romagnoli}, {Moreno},
  {Mostaf{\'a}}, {Nellen}, {Newbold}, {Nisa}, {Noriega-Papaqui}, {Pelayo},
  {Pretz}, {P{\'e}rez-P{\'e}rez}, {Ren}, {Rho}, {Rivi{\`e}re},
  {Rosa-Gonz{\'a}lez}, {Rosenberg}, {Ruiz-Velasco}, {Salazar}, {Salesa Greus},
  {Sand oval}, {Schneider}, {Schoorlemmer}, {Sinnis}, {Smith}, {Springer},
  {Surajbali}, {Taboada}, {Tibolla}, {Tollefson}, {Torres}, {Ukwatta},
  {Vianello}, {Weisgarber}, {Westerhoff}, {Wisher}, {Wood}, {Yapici}, {Yodh},
  {Younk}, {Zepeda}, {Zhou}, {Guo}, {Hahn}, {Li} \& {Zhang}]{Abeysekara+17}
{\sc \au{{Abeysekara}, A.~U.}, \au{{Albert}, A.}, \au{{Alfaro}, R.},
  \au{{Alvarez}, C.}, \au{{{\'A}lvarez}, J.~D.}, \au{{Arceo}, R.},
  \au{{Arteaga-Vel{\'a}zquez}, J.~C.}, \au{{Avila Rojas}, D.}, \au{{Ayala
  Solares}, H.~A.}, \au{{Barber}, A.~S.}, \au{{Bautista-Elivar}, N.},
  \au{{Becerril}, A.}, \au{{Belmont-Moreno}, E.}, \au{{BenZvi}, S.~Y.},
  \au{{Berley}, D.}, \au{{Bernal}, A.}, \au{{Braun}, J.}, \au{{Brisbois}, C.},
  \au{{Caballero-Mora}, K.~S.}, \au{{Capistr{\'a}n}, T.},
  \au{{Carrami{\~n}ana}, A.}, \au{{Casanova}, S.}, \au{{Castillo}, M.},
  \au{{Cotti}, U.}, \au{{Cotzomi}, J.}, \au{{Couti{\~n}o de Le{\'o}n}, S.},
  \au{{De Le{\'o}n}, C.}, \au{{De la Fuente}, E.}, \au{{Dingus}, B.~L.},
  \au{{DuVernois}, M.~A.}, \au{{D{\'\i}az-V{\'e}lez}, J.~C.}, \au{{Ellsworth},
  R.~W.}, \au{{Engel}, K.}, \au{{Enr{\'\i}quez-Rivera}, O.}, \au{{Fiorino},
  D.~W.}, \au{{Fraija}, N.}, \au{{Garc{\'\i}a-Gonz{\'a}lez}, J.~A.},
  \au{{Garfias}, F.}, \au{{Gerhardt}, M.}, \au{{Gonz{\'a}lez Mu{\~n}oz}, A.},
  \au{{Gonz{\'a}lez}, M.~M.}, \au{{Goodman}, J.~A.}, \au{{Hampel-Arias}, Z.},
  \au{{Harding}, J.~P.}, \au{{Hern{\'a}ndez}, S.}, \au{{Hern{\'a}ndez-Almada},
  A.}, \au{{Hinton}, J.}, \au{{Hona}, B.}, \au{{Hui}, C.~M.},
  \au{{H{\"u}ntemeyer}, P.}, \au{{Iriarte}, A.}, \au{{Jardin-Blicq}, A.},
  \au{{Joshi}, V.}, \au{{Kaufmann}, S.}, \au{{Kieda}, D.}, \au{{Lara}, A.},
  \au{{Lauer}, R.~J.}, \au{{Lee}, W.~H.}, \au{{Lennarz}, D.}, \au{{Vargas},
  H.~Le{\'o}n}, \au{{Linnemann}, J.~T.}, \au{{Longinotti}, A.~L.}, \au{{Luis
  Raya}, G.}, \au{{Luna-Garc{\'\i}a}, R.}, \au{{L{\'o}pez-Coto}, R.},
  \au{{Malone}, K.}, \au{{Marinelli}, S.~S.}, \au{{Martinez}, O.},
  \au{{Martinez-Castellanos}, I.}, \au{{Mart{\'\i}nez-Castro}, J.},
  \au{{Mart{\'\i}nez-Huerta}, H.}, \au{{Matthews}, J.~A.}, \au{{Mirand
  a-Romagnoli}, P.}, \au{{Moreno}, E.}, \au{{Mostaf{\'a}}, M.}, \au{{Nellen},
  L.}, \au{{Newbold}, M.}, \au{{Nisa}, M.~U.}, \au{{Noriega-Papaqui}, R.},
  \au{{Pelayo}, R.}, \au{{Pretz}, J.}, \au{{P{\'e}rez-P{\'e}rez}, E.~G.},
  \au{{Ren}, Z.}, \au{{Rho}, C.~D.}, \au{{Rivi{\`e}re}, C.},
  \au{{Rosa-Gonz{\'a}lez}, D.}, \au{{Rosenberg}, M.}, \au{{Ruiz-Velasco}, E.},
  \au{{Salazar}, H.}, \au{{Salesa Greus}, F.}, \au{{Sand oval}, A.},
  \au{{Schneider}, M.}, \au{{Schoorlemmer}, H.}, \au{{Sinnis}, G.},
  \au{{Smith}, A.~J.}, \au{{Springer}, R.~W.}, \au{{Surajbali}, P.},
  \au{{Taboada}, I.}, \au{{Tibolla}, O.}, \au{{Tollefson}, K.}, \au{{Torres},
  I.}, \au{{Ukwatta}, T.~N.}, \au{{Vianello}, G.}, \au{{Weisgarber}, T.},
  \au{{Westerhoff}, S.}, \au{{Wisher}, I.~G.}, \au{{Wood}, J.}, \au{{Yapici},
  T.}, \au{{Yodh}, G.}, \au{{Younk}, P.~W.}, \au{{Zepeda}, A.}, \au{{Zhou},
  H.}, \au{{Guo}, F.}, \au{{Hahn}, J.}, \au{{Li}, H.} \& \au{{Zhang}, H.}}
  \yr{2017{\natexlab{{\em a\/}}}}  \at{{Extended gamma-ray sources around
  pulsars constrain the origin of the positron flux at Earth}}.  \jt{Science}
  \bvol{358}~(6365),  \pg{911--914},  \arxiv{arXiv: 1711.06223}.

\bibitem[{Abeysekara} {\em et~al.\/}(2017{\natexlab{{\em b\/}}}){Abeysekara},
  {Albert}, {Alfaro}, {Alvarez}, {{\'A}lvarez}, {Arceo},
  {Arteaga-Vel{\'a}zquez}, {Avila Rojas}, {Ayala Solares}, {Barber},
  {Bautista-Elivar}, {Becerril}, {Belmont-Moreno}, {BenZvi}, {Berley},
  {Bernal}, {Braun}, {Brisbois}, {Caballero-Mora}, {Capistr{\'a}n},
  {Carrami{\~n}ana}, {Casanova}, {Castillo}, {Cotti}, {Cotzomi}, {Couti{\~n}o
  de Le{\'o}n}, {De Le{\'o}n}, {De la Fuente}, {Dingus}, {DuVernois},
  {D{\'\i}az-V{\'e}lez}, {Ellsworth}, {Engel}, {Enr{\'\i}quez-Rivera},
  {Fiorino}, {Fraija}, {Garc{\'\i}a-Gonz{\'a}lez}, {Garfias}, {Gerhardt},
  {Gonz{\'a}lez Mu{\~n}oz}, {Gonz{\'a}lez}, {Goodman}, {Hampel-Arias},
  {Harding}, {Hern{\'a}ndez}, {Hern{\'a}ndez-Almada}, {Hinton}, {Hona}, {Hui},
  {H{\"u}ntemeyer}, {Iriarte}, {Jardin-Blicq}, {Joshi}, {Kaufmann}, {Kieda},
  {Lara}, {Lauer}, {Lee}, {Lennarz}, {Vargas}, {Linnemann}, {Longinotti}, {Luis
  Raya}, {Luna-Garc{\'\i}a}, {L{\'o}pez-Coto}, {Malone}, {Marinelli},
  {Martinez}, {Martinez-Castellanos}, {Mart{\'\i}nez-Castro},
  {Mart{\'\i}nez-Huerta}, {Matthews}, {Miranda-Romagnoli}, {Moreno},
  {Mostaf{\'a}}, {Nellen}, {Newbold}, {Nisa}, {Noriega-Papaqui}, {Pelayo},
  {Pretz}, {P{\'e}rez-P{\'e}rez}, {Ren}, {Rho}, {Rivi{\`e}re},
  {Rosa-Gonz{\'a}lez}, {Rosenberg}, {Ruiz-Velasco}, {Salazar}, {Salesa Greus},
  {Sandoval}, {Schneider}, {Schoorlemmer}, {Sinnis}, {Smith}, {Springer},
  {Surajbali}, {Taboada}, {Tibolla}, {Tollefson}, {Torres}, {Ukwatta},
  {Vianello}, {Weisgarber}, {Westerhoff}, {Wisher}, {Wood}, {Yapici}, {Yodh},
  {Younk}, {Zepeda}, {Zhou}, {Guo}, {Hahn}, {Li} \& {Zhang}]{TevHalos}
{\sc \au{{Abeysekara}, A.~U.}, \au{{Albert}, A.}, \au{{Alfaro}, R.},
  \au{{Alvarez}, C.}, \au{{{\'A}lvarez}, J.~D.}, \au{{Arceo}, R.},
  \au{{Arteaga-Vel{\'a}zquez}, J.~C.}, \au{{Avila Rojas}, D.}, \au{{Ayala
  Solares}, H.~A.}, \au{{Barber}, A.~S.}, \au{{Bautista-Elivar}, N.},
  \au{{Becerril}, A.}, \au{{Belmont-Moreno}, E.}, \au{{BenZvi}, S.~Y.},
  \au{{Berley}, D.}, \au{{Bernal}, A.}, \au{{Braun}, J.}, \au{{Brisbois}, C.},
  \au{{Caballero-Mora}, K.~S.}, \au{{Capistr{\'a}n}, T.},
  \au{{Carrami{\~n}ana}, A.}, \au{{Casanova}, S.}, \au{{Castillo}, M.},
  \au{{Cotti}, U.}, \au{{Cotzomi}, J.}, \au{{Couti{\~n}o de Le{\'o}n}, S.},
  \au{{De Le{\'o}n}, C.}, \au{{De la Fuente}, E.}, \au{{Dingus}, B.~L.},
  \au{{DuVernois}, M.~A.}, \au{{D{\'\i}az-V{\'e}lez}, J.~C.}, \au{{Ellsworth},
  R.~W.}, \au{{Engel}, K.}, \au{{Enr{\'\i}quez-Rivera}, O.}, \au{{Fiorino},
  D.~W.}, \au{{Fraija}, N.}, \au{{Garc{\'\i}a-Gonz{\'a}lez}, J.~A.},
  \au{{Garfias}, F.}, \au{{Gerhardt}, M.}, \au{{Gonz{\'a}lez Mu{\~n}oz}, A.},
  \au{{Gonz{\'a}lez}, M.~M.}, \au{{Goodman}, J.~A.}, \au{{Hampel-Arias}, Z.},
  \au{{Harding}, J.~P.}, \au{{Hern{\'a}ndez}, S.}, \au{{Hern{\'a}ndez-Almada},
  A.}, \au{{Hinton}, J.}, \au{{Hona}, B.}, \au{{Hui}, C.~M.},
  \au{{H{\"u}ntemeyer}, P.}, \au{{Iriarte}, A.}, \au{{Jardin-Blicq}, A.},
  \au{{Joshi}, V.}, \au{{Kaufmann}, S.}, \au{{Kieda}, D.}, \au{{Lara}, A.},
  \au{{Lauer}, R.~J.}, \au{{Lee}, W.~H.}, \au{{Lennarz}, D.}, \au{{Vargas},
  H.~Le{\'o}n}, \au{{Linnemann}, J.~T.}, \au{{Longinotti}, A.~L.}, \au{{Luis
  Raya}, G.}, \au{{Luna-Garc{\'\i}a}, R.}, \au{{L{\'o}pez-Coto}, R.},
  \au{{Malone}, K.}, \au{{Marinelli}, S.~S.}, \au{{Martinez}, O.},
  \au{{Martinez-Castellanos}, I.}, \au{{Mart{\'\i}nez-Castro}, J.},
  \au{{Mart{\'\i}nez-Huerta}, H.}, \au{{Matthews}, J.~A.},
  \au{{Miranda-Romagnoli}, P.}, \au{{Moreno}, E.}, \au{{Mostaf{\'a}}, M.},
  \au{{Nellen}, L.}, \au{{Newbold}, M.}, \au{{Nisa}, M.~U.},
  \au{{Noriega-Papaqui}, R.}, \au{{Pelayo}, R.}, \au{{Pretz}, J.},
  \au{{P{\'e}rez-P{\'e}rez}, E.~G.}, \au{{Ren}, Z.}, \au{{Rho}, C.~D.},
  \au{{Rivi{\`e}re}, C.}, \au{{Rosa-Gonz{\'a}lez}, D.}, \au{{Rosenberg}, M.},
  \au{{Ruiz-Velasco}, E.}, \au{{Salazar}, H.}, \au{{Salesa Greus}, F.},
  \au{{Sandoval}, A.}, \au{{Schneider}, M.}, \au{{Schoorlemmer}, H.},
  \au{{Sinnis}, G.}, \au{{Smith}, A.~J.}, \au{{Springer}, R.~W.},
  \au{{Surajbali}, P.}, \au{{Taboada}, I.}, \au{{Tibolla}, O.},
  \au{{Tollefson}, K.}, \au{{Torres}, I.}, \au{{Ukwatta}, T.~N.},
  \au{{Vianello}, G.}, \au{{Weisgarber}, T.}, \au{{Westerhoff}, S.},
  \au{{Wisher}, I.~G.}, \au{{Wood}, J.}, \au{{Yapici}, T.}, \au{{Yodh}, G.},
  \au{{Younk}, P.~W.}, \au{{Zepeda}, A.}, \au{{Zhou}, H.}, \au{{Guo}, F.},
  \au{{Hahn}, J.}, \au{{Li}, H.} \& \au{{Zhang}, H.}} \yr{2017{\natexlab{{\em
  b\/}}}}  \at{{Extended gamma-ray sources around pulsars constrain the origin
  of the positron flux at Earth}}.  \jt{Science}  \bvol{358}~(6365),
  \pg{911--914},  \arxiv{arXiv: 1711.06223}.

\bibitem[{Abramowski} {\em et~al.\/}(2012){Abramowski}, {Acero}, {Aharonian},
  {Akhperjanian}, {Anton}, {Balzer}, {Barnacka}, {Barres de Almeida},
  {Becherini}, {Becker}, {Behera}, {Bernl{\"o}hr}, {Birsin}, {Biteau},
  {Bochow}, {Boisson}, {Bolmont}, {Bordas}, {Brucker}, {Brun}, {Brun}, {Bulik},
  {B{\"u}sching}, {Carrigan}, {Casanova}, {Cerruti}, {Chadwick}, {Charbonnier},
  {Chaves}, {Cheesebrough}, {Chounet}, {Clapson}, {Coignet}, {Cologna},
  {Conrad}, {Dalton}, {Daniel}, {Davids}, {Degrange}, {Deil}, {Dickinson},
  {Djannati-Ata{\"\i}}, {Domainko}, {O'C. Drury}, {Dubois}, {Dubus}, {Dutson},
  {Dyks}, {Dyrda}, {Egberts}, {Eger}, {Espigat}, {Fallon}, {Farnier}, {Fegan},
  {Feinstein}, {Fernandes}, {Fiasson}, {Fontaine}, {F{\"o}rster},
  {F{\"u}{\ss}ling}, {Gallant}, {Gast}, {G{\'e}rard}, {Gerbig}, {Giebels},
  {Glicenstein}, {Gl{\"u}ck}, {Goret}, {G{\"o}ring}, {H{\"a}ffner}, {Hague},
  {Hampf}, {Hauser}, {Heinz}, {Heinzelmann}, {Henri}, {Hermann}, {Hinton},
  {Hoffmann}, {Hofmann}, {Hofverberg}, {Holler}, {Horns}, {Jacholkowska}, {de
  Jager}, {Jahn}, {Jamrozy}, {Jung}, {Kastendieck}, {Katarzy{\'n}ski}, {Katz},
  {Kaufmann}, {Keogh}, {Khangulyan}, {Kh{\'e}lifi}, {Klochkov}, {Klu{\.z}niak},
  {Kneiske}, {Komin}, {Kosack}, {Kossakowski}, {Laffon}, {Lamanna}, {Lennarz},
  {Lohse}, {Lopatin}, {Lu}, {Marandon}, {Marcowith}, {Masbou}, {Maurin},
  {Maxted}, {Mayer}, {McComb}, {Medina}, {M{\'e}hault}, {Moderski}, {Moulin},
  {Naumann}, {Naumann-Godo}, {de Naurois}, {Nedbal}, {Nekrassov}, {Nguyen},
  {Nicholas}, {Niemiec}, {Nolan}, {Ohm}, {de O{\~n}a Wilhelmi}, {Opitz},
  {Ostrowski}, {Oya}, {Panter}, {Paz Arribas}, {Pedaletti}, {Pelletier},
  {Petrucci}, {Pita}, {P{\"u}hlhofer}, {Punch}, {Quirrenbach}, {Raue},
  {Rayner}, {Reimer}, {Reimer}, {Renaud}, {de Los Reyes}, {Rieger}, {Ripken},
  {Rob}, {Rosier-Lees}, {Rowell}, {Rudak}, {Rulten}, {Ruppel}, {Sahakian},
  {Sanchez}, {Santangelo}, {Schlickeiser}, {Sch{\"o}ck}, {Schulz}, {Schwanke},
  {Schwarzburg}, {Schwemmer}, {Sheidaei}, {Sikora}, {Skilton}, {Sol},
  {Spengler}, {Stawarz}, {Steenkamp}, {Stegmann}, {Stinzing}, {Stycz},
  {Sushch}, {Szostek}, {Tavernet}, {Terrier}, {Tluczykont}, {Valerius}, {van
  Eldik}, {Vasileiadis}, {Venter}, {Vialle}, {Viana}, {Vincent}, {V{\"o}lk},
  {Volpe}, {Vorobiov}, {Vorster}, {Wagner}, {Ward}, {White}, {Wierzcholska},
  {Zacharias}, {Zajczyk}, {Zdziarski}, {Zech} \&
  {Zechlin}]{2012A&A...537A.114A}
{\sc \au{{Abramowski}, A.}, \au{{Acero}, F.}, \au{{Aharonian}, F.},
  \au{{Akhperjanian}, A.~G.}, \au{{Anton}, G.}, \au{{Balzer}, A.},
  \au{{Barnacka}, A.}, \au{{Barres de Almeida}, U.}, \au{{Becherini}, Y.},
  \au{{Becker}, J.}, \au{{Behera}, B.}, \au{{Bernl{\"o}hr}, K.}, \au{{Birsin},
  E.}, \au{{Biteau}, J.}, \au{{Bochow}, A.}, \au{{Boisson}, C.}, \au{{Bolmont},
  J.}, \au{{Bordas}, P.}, \au{{Brucker}, J.}, \au{{Brun}, F.}, \au{{Brun}, P.},
  \au{{Bulik}, T.}, \au{{B{\"u}sching}, I.}, \au{{Carrigan}, S.},
  \au{{Casanova}, S.}, \au{{Cerruti}, M.}, \au{{Chadwick}, P.~M.},
  \au{{Charbonnier}, A.}, \au{{Chaves}, R.~C.~G.}, \au{{Cheesebrough}, A.},
  \au{{Chounet}, L.~M.}, \au{{Clapson}, A.~C.}, \au{{Coignet}, G.},
  \au{{Cologna}, G.}, \au{{Conrad}, J.}, \au{{Dalton}, M.}, \au{{Daniel},
  M.~K.}, \au{{Davids}, I.~D.}, \au{{Degrange}, B.}, \au{{Deil}, C.},
  \au{{Dickinson}, H.~J.}, \au{{Djannati-Ata{\"\i}}, A.}, \au{{Domainko}, W.},
  \au{{O'C. Drury}, L.}, \au{{Dubois}, F.}, \au{{Dubus}, G.}, \au{{Dutson},
  K.}, \au{{Dyks}, J.}, \au{{Dyrda}, M.}, \au{{Egberts}, K.}, \au{{Eger}, P.},
  \au{{Espigat}, P.}, \au{{Fallon}, L.}, \au{{Farnier}, C.}, \au{{Fegan}, S.},
  \au{{Feinstein}, F.}, \au{{Fernandes}, M.~V.}, \au{{Fiasson}, A.},
  \au{{Fontaine}, G.}, \au{{F{\"o}rster}, A.}, \au{{F{\"u}{\ss}ling}, M.},
  \au{{Gallant}, Y.~A.}, \au{{Gast}, H.}, \au{{G{\'e}rard}, L.}, \au{{Gerbig},
  D.}, \au{{Giebels}, B.}, \au{{Glicenstein}, J.~F.}, \au{{Gl{\"u}ck}, B.},
  \au{{Goret}, P.}, \au{{G{\"o}ring}, D.}, \au{{H{\"a}ffner}, S.}, \au{{Hague},
  J.~D.}, \au{{Hampf}, D.}, \au{{Hauser}, M.}, \au{{Heinz}, S.},
  \au{{Heinzelmann}, G.}, \au{{Henri}, G.}, \au{{Hermann}, G.}, \au{{Hinton},
  J.~A.}, \au{{Hoffmann}, A.}, \au{{Hofmann}, W.}, \au{{Hofverberg}, P.},
  \au{{Holler}, M.}, \au{{Horns}, D.}, \au{{Jacholkowska}, A.}, \au{{de Jager},
  O.~C.}, \au{{Jahn}, C.}, \au{{Jamrozy}, M.}, \au{{Jung}, I.},
  \au{{Kastendieck}, M.~A.}, \au{{Katarzy{\'n}ski}, K.}, \au{{Katz}, U.},
  \au{{Kaufmann}, S.}, \au{{Keogh}, D.}, \au{{Khangulyan}, D.},
  \au{{Kh{\'e}lifi}, B.}, \au{{Klochkov}, D.}, \au{{Klu{\.z}niak}, W.},
  \au{{Kneiske}, T.}, \au{{Komin}, Nu.}, \au{{Kosack}, K.}, \au{{Kossakowski},
  R.}, \au{{Laffon}, H.}, \au{{Lamanna}, G.}, \au{{Lennarz}, D.}, \au{{Lohse},
  T.}, \au{{Lopatin}, A.}, \au{{Lu}, C.~C.}, \au{{Marandon}, V.},
  \au{{Marcowith}, A.}, \au{{Masbou}, J.}, \au{{Maurin}, D.}, \au{{Maxted},
  N.}, \au{{Mayer}, M.}, \au{{McComb}, T.~J.~L.}, \au{{Medina}, M.~C.},
  \au{{M{\'e}hault}, J.}, \au{{Moderski}, R.}, \au{{Moulin}, E.},
  \au{{Naumann}, C.~L.}, \au{{Naumann-Godo}, M.}, \au{{de Naurois}, M.},
  \au{{Nedbal}, D.}, \au{{Nekrassov}, D.}, \au{{Nguyen}, N.}, \au{{Nicholas},
  B.}, \au{{Niemiec}, J.}, \au{{Nolan}, S.~J.}, \au{{Ohm}, S.}, \au{{de O{\~n}a
  Wilhelmi}, E.}, \au{{Opitz}, B.}, \au{{Ostrowski}, M.}, \au{{Oya}, I.},
  \au{{Panter}, M.}, \au{{Paz Arribas}, M.}, \au{{Pedaletti}, G.},
  \au{{Pelletier}, G.}, \au{{Petrucci}, P.~O.}, \au{{Pita}, S.},
  \au{{P{\"u}hlhofer}, G.}, \au{{Punch}, M.}, \au{{Quirrenbach}, A.},
  \au{{Raue}, M.}, \au{{Rayner}, S.~M.}, \au{{Reimer}, A.}, \au{{Reimer}, O.},
  \au{{Renaud}, M.}, \au{{de Los Reyes}, R.}, \au{{Rieger}, F.}, \au{{Ripken},
  J.}, \au{{Rob}, L.}, \au{{Rosier-Lees}, S.}, \au{{Rowell}, G.}, \au{{Rudak},
  B.}, \au{{Rulten}, C.~B.}, \au{{Ruppel}, J.}, \au{{Sahakian}, V.},
  \au{{Sanchez}, D.}, \au{{Santangelo}, A.}, \au{{Schlickeiser}, R.},
  \au{{Sch{\"o}ck}, F.~M.}, \au{{Schulz}, A.}, \au{{Schwanke}, U.},
  \au{{Schwarzburg}, S.}, \au{{Schwemmer}, S.}, \au{{Sheidaei}, F.},
  \au{{Sikora}, M.}, \au{{Skilton}, J.~L.}, \au{{Sol}, H.}, \au{{Spengler},
  G.}, \au{{Stawarz}, {\L}.}, \au{{Steenkamp}, R.}, \au{{Stegmann}, C.},
  \au{{Stinzing}, F.}, \au{{Stycz}, K.}, \au{{Sushch}, I.}, \au{{Szostek}, A.},
  \au{{Tavernet}, J.~P.}, \au{{Terrier}, R.}, \au{{Tluczykont}, M.},
  \au{{Valerius}, K.}, \au{{van Eldik}, C.}, \au{{Vasileiadis}, G.},
  \au{{Venter}, C.}, \au{{Vialle}, J.~P.}, \au{{Viana}, A.}, \au{{Vincent},
  P.}, \au{{V{\"o}lk}, H.~J.}, \au{{Volpe}, F.}, \au{{Vorobiov}, S.},
  \au{{Vorster}, M.}, \au{{Wagner}, S.~J.}, \au{{Ward}, M.}, \au{{White}, R.},
  \au{{Wierzcholska}, A.}, \au{{Zacharias}, M.}, \au{{Zajczyk}, A.},
  \au{{Zdziarski}, A.~A.}, \au{{Zech}, A.} \& \au{{Zechlin}, H.~S.}} \yr{2012}
  \at{{Discovery of extended VHE {\ensuremath{\gamma}}-ray emission from the
  vicinity of the young massive stellar cluster Westerlund 1}}.  \jt{Astronomy
  and Astrophysics}  \bvol{537},  \pg{A114},  \arxiv{arXiv: 1111.2043}.

\bibitem[{Acciari} {\em et~al.\/}(2009){Acciari}, {Aliu}, {Arlen}, {Aune},
  {Bautista}, {Beilicke}, {Benbow}, {Bradbury}, {Buckley}, {Bugaev}, {Butt},
  {Byrum}, {Cannon}, {Celik}, {Cesarini}, {Chow}, {Ciupik}, {Cogan}, {Colin},
  {Cui}, {Daniel}, {Dickherber}, {Duke}, {Dwarkadas}, {Ergin}, {Fegan},
  {Finley}, {Finnegan}, {Fortin}, {Fortson}, {Furniss}, {Gall}, {Gibbs},
  {Gillanders}, {Godambe}, {Grube}, {Guenette}, {Gyuk}, {Hanna}, {Hays},
  {Holder}, {Horan}, {Hui}, {Humensky}, {Imran}, {Kaaret}, {Karlsson},
  {Kertzman}, {Kieda}, {Kildea}, {Konopelko}, {Krawczynski}, {Krennrich},
  {Lang}, {LeBohec}, {Maier}, {McCann}, {McCutcheon}, {Millis}, {Moriarty},
  {Ong}, {Otte}, {Pandel}, {Perkins}, {Pohl}, {Quinn}, {Ragan}, {Reyes},
  {Reynolds}, {Roache}, {Rose}, {Schroedter}, {Sembroski}, {Smith}, {Steele},
  {Swordy}, {Theiling}, {Toner}, {Valcarcel}, {Varlotta}, {Vassiliev},
  {Vincent}, {Wagner}, {Wakely}, {Ward}, {Weekes}, {Weinstein}, {Weisgarber},
  {Williams}, {Wissel}, {Wood} \& {Zitzer}]{2009ApJ...698L.133A}
{\sc \au{{Acciari}, V.~A.}, \au{{Aliu}, E.}, \au{{Arlen}, T.}, \au{{Aune}, T.},
  \au{{Bautista}, M.}, \au{{Beilicke}, M.}, \au{{Benbow}, W.}, \au{{Bradbury},
  S.~M.}, \au{{Buckley}, J.~H.}, \au{{Bugaev}, V.}, \au{{Butt}, Y.},
  \au{{Byrum}, K.}, \au{{Cannon}, A.}, \au{{Celik}, O.}, \au{{Cesarini}, A.},
  \au{{Chow}, Y.~C.}, \au{{Ciupik}, L.}, \au{{Cogan}, P.}, \au{{Colin}, P.},
  \au{{Cui}, W.}, \au{{Daniel}, M.~K.}, \au{{Dickherber}, R.}, \au{{Duke}, C.},
  \au{{Dwarkadas}, V.~V.}, \au{{Ergin}, T.}, \au{{Fegan}, S.~J.}, \au{{Finley},
  J.~P.}, \au{{Finnegan}, G.}, \au{{Fortin}, P.}, \au{{Fortson}, L.},
  \au{{Furniss}, A.}, \au{{Gall}, D.}, \au{{Gibbs}, K.}, \au{{Gillanders},
  G.~H.}, \au{{Godambe}, S.}, \au{{Grube}, J.}, \au{{Guenette}, R.},
  \au{{Gyuk}, G.}, \au{{Hanna}, D.}, \au{{Hays}, E.}, \au{{Holder}, J.},
  \au{{Horan}, D.}, \au{{Hui}, C.~M.}, \au{{Humensky}, T.~B.}, \au{{Imran},
  A.}, \au{{Kaaret}, P.}, \au{{Karlsson}, N.}, \au{{Kertzman}, M.},
  \au{{Kieda}, D.}, \au{{Kildea}, J.}, \au{{Konopelko}, A.}, \au{{Krawczynski},
  H.}, \au{{Krennrich}, F.}, \au{{Lang}, M.~J.}, \au{{LeBohec}, S.},
  \au{{Maier}, G.}, \au{{McCann}, A.}, \au{{McCutcheon}, M.}, \au{{Millis},
  J.}, \au{{Moriarty}, P.}, \au{{Ong}, R.~A.}, \au{{Otte}, A.~N.},
  \au{{Pandel}, D.}, \au{{Perkins}, J.~S.}, \au{{Pohl}, M.}, \au{{Quinn}, J.},
  \au{{Ragan}, K.}, \au{{Reyes}, L.~C.}, \au{{Reynolds}, P.~T.}, \au{{Roache},
  E.}, \au{{Rose}, H.~J.}, \au{{Schroedter}, M.}, \au{{Sembroski}, G.~H.},
  \au{{Smith}, A.~W.}, \au{{Steele}, D.}, \au{{Swordy}, S.~P.}, \au{{Theiling},
  M.}, \au{{Toner}, J.~A.}, \au{{Valcarcel}, L.}, \au{{Varlotta}, A.},
  \au{{Vassiliev}, V.~V.}, \au{{Vincent}, S.}, \au{{Wagner}, R.~G.},
  \au{{Wakely}, S.~P.}, \au{{Ward}, J.~E.}, \au{{Weekes}, T.~C.},
  \au{{Weinstein}, A.}, \au{{Weisgarber}, T.}, \au{{Williams}, D.~A.},
  \au{{Wissel}, S.}, \au{{Wood}, M.} \& \au{{Zitzer}, B.}} \yr{2009}
  \at{{Observation of Extended Very High Energy Emission from the Supernova
  Remnant IC 443 with VERITAS}}.  \jt{ApJL}  \bvol{698}~(2),  \pg{L133--L137},
  \arxiv{arXiv: 0905.3291}.

\bibitem[{Acciari} {\em et~al.\/}(2011){Acciari}, {Aliu}, {Arlen}, {Aune},
  {Beilicke}, {Benbow}, {Bradbury}, {Buckley}, {Bugaev}, {Byrum}, {Cannon},
  {Cesarini}, {Ciupik}, {Collins-Hughes}, {Cui}, {Dickherber}, {Duke},
  {Errando}, {Finley}, {Finnegan}, {Fortson}, {Furniss}, {Galante}, {Gall},
  {Gillanders}, {Godambe}, {Griffin}, {Grube}, {Guenette}, {Gyuk}, {Hanna},
  {Holder}, {Hughes}, {Hui}, {Humensky}, {Kaaret}, {Karlsson}, {Kertzman},
  {Kieda}, {Krawczynski}, {Krennrich}, {Lang}, {LeBohec}, {Madhavan}, {Maier},
  {Majumdar}, {McArthur}, {McCann}, {Moriarty}, {Mukherjee}, {Ong}, {Orr},
  {Otte}, {Pandel}, {Park}, {Perkins}, {Pohl}, {Quinn}, {Ragan}, {Reyes},
  {Reynolds}, {Roache}, {Rose}, {Saxon}, {Schroedter}, {Sembroski}, {Senturk},
  {Slane}, {Smith}, {Te{\v{s}}i{\'c}}, {Theiling}, {Thibadeau}, {Tsurusaki},
  {Varlotta}, {Vassiliev}, {Vincent}, {Vivier}, {Wakely}, {Ward}, {Weekes},
  {Weinstein}, {Weisgarber}, {Williams}, {Wood} \&
  {Zitzer}]{2011ApJ...730L..20A}
{\sc \au{{Acciari}, V.~A.}, \au{{Aliu}, E.}, \au{{Arlen}, T.}, \au{{Aune}, T.},
  \au{{Beilicke}, M.}, \au{{Benbow}, W.}, \au{{Bradbury}, S.~M.},
  \au{{Buckley}, J.~H.}, \au{{Bugaev}, V.}, \au{{Byrum}, K.}, \au{{Cannon},
  A.}, \au{{Cesarini}, A.}, \au{{Ciupik}, L.}, \au{{Collins-Hughes}, E.},
  \au{{Cui}, W.}, \au{{Dickherber}, R.}, \au{{Duke}, C.}, \au{{Errando}, M.},
  \au{{Finley}, J.~P.}, \au{{Finnegan}, G.}, \au{{Fortson}, L.}, \au{{Furniss},
  A.}, \au{{Galante}, N.}, \au{{Gall}, D.}, \au{{Gillanders}, G.~H.},
  \au{{Godambe}, S.}, \au{{Griffin}, S.}, \au{{Grube}, J.}, \au{{Guenette},
  R.}, \au{{Gyuk}, G.}, \au{{Hanna}, D.}, \au{{Holder}, J.}, \au{{Hughes},
  J.~P.}, \au{{Hui}, C.~M.}, \au{{Humensky}, T.~B.}, \au{{Kaaret}, P.},
  \au{{Karlsson}, N.}, \au{{Kertzman}, M.}, \au{{Kieda}, D.},
  \au{{Krawczynski}, H.}, \au{{Krennrich}, F.}, \au{{Lang}, M.~J.},
  \au{{LeBohec}, S.}, \au{{Madhavan}, A.~S.}, \au{{Maier}, G.}, \au{{Majumdar},
  P.}, \au{{McArthur}, S.}, \au{{McCann}, A.}, \au{{Moriarty}, P.},
  \au{{Mukherjee}, R.}, \au{{Ong}, R.~A.}, \au{{Orr}, M.}, \au{{Otte}, A.~N.},
  \au{{Pandel}, D.}, \au{{Park}, N.~H.}, \au{{Perkins}, J.~S.}, \au{{Pohl},
  M.}, \au{{Quinn}, J.}, \au{{Ragan}, K.}, \au{{Reyes}, L.~C.}, \au{{Reynolds},
  P.~T.}, \au{{Roache}, E.}, \au{{Rose}, H.~J.}, \au{{Saxon}, D.~B.},
  \au{{Schroedter}, M.}, \au{{Sembroski}, G.~H.}, \au{{Senturk}, G.~Demet},
  \au{{Slane}, P.}, \au{{Smith}, A.~W.}, \au{{Te{\v{s}}i{\'c}}, G.},
  \au{{Theiling}, M.}, \au{{Thibadeau}, S.}, \au{{Tsurusaki}, K.},
  \au{{Varlotta}, A.}, \au{{Vassiliev}, V.~V.}, \au{{Vincent}, S.},
  \au{{Vivier}, M.}, \au{{Wakely}, S.~P.}, \au{{Ward}, J.~E.}, \au{{Weekes},
  T.~C.}, \au{{Weinstein}, A.}, \au{{Weisgarber}, T.}, \au{{Williams}, D.~A.},
  \au{{Wood}, M.} \& \au{{Zitzer}, B.}} \yr{2011}  \at{{Discovery of TeV
  Gamma-ray Emission from Tycho's Supernova Remnant}}.  \jt{ApJ Letters}
  \bvol{730}~(2),  \pg{L20},  \arxiv{arXiv: 1102.3871}.

\bibitem[{Ackermann} {\em et~al.\/}(2011){Ackermann}, {Ajello}, {Allafort},
  {Baldini}, {Ballet}, {Barbiellini}, {Bastieri}, {Belfiore}, {Bellazzini},
  {Berenji}, {Bland ford}, {Bloom}, {Bonamente}, {Borgland }, {Bottacini},
  {Brigida}, {Bruel}, {Buehler}, {Buson}, {Caliandro}, {Cameron}, {Caraveo},
  {Casandjian}, {Cecchi}, {Chekhtman}, {Cheung}, {Chiang}, {Ciprini}, {Claus},
  {Cohen-Tanugi}, {de Angelis}, {de Palma}, {Dermer}, {do Couto e Silva},
  {Drell}, {Dumora}, {Favuzzi}, {Fegan}, {Focke}, {Fortin}, {Fukazawa},
  {Fusco}, {Gargano}, {Germani}, {Giglietto}, {Giordano}, {Giroletti},
  {Glanzman}, {Godfrey}, {Grenier}, {Guillemot}, {Guiriec}, {Hadasch},
  {Hanabata}, {Harding}, {Hayashida}, {Hayashi}, {Hays}, {J{\'o}hannesson},
  {Johnson}, {Kamae}, {Katagiri}, {Kataoka}, {Kerr}, {Kn{\"o}dlseder}, {Kuss},
  {Lande}, {Latronico}, {Lee}, {Longo}, {Loparco}, {Lott}, {Lovellette},
  {Lubrano}, {Martin}, {Mazziotta}, {McEnery}, {Mehault}, {Michelson},
  {Mitthumsiri}, {Mizuno}, {Monte}, {Monzani}, {Morselli}, {Moskalenko},
  {Murgia}, {Naumann-Godo}, {Nolan}, {Norris}, {Nuss}, {Ohsugi}, {Okumura},
  {Orlando}, {Ormes}, {Ozaki}, {Paneque}, {Parent}, {Pesce-Rollins},
  {Pierbattista}, {Piron}, {Pohl}, {Prokhorov}, {Rain{\`o}}, {Rando},
  {Razzano}, {Reposeur}, {Ritz}, {Parkinson}, {Sgr{\`o}}, {Siskind}, {Smith},
  {Spinelli}, {Strong}, {Takahashi}, {Tanaka}, {Thayer}, {Thayer}, {Thompson},
  {Tibaldo}, {Torres}, {Tosti}, {Tramacere}, {Troja}, {Uchiyama},
  {Vandenbroucke}, {Vasileiou}, {Vianello}, {Vitale}, {Waite}, {Wang}, {Winer},
  {Wood}, {Yang}, {Zimmer} \& {Bontemps}]{2011Sci...334.1103A}
{\sc \au{{Ackermann}, M.}, \au{{Ajello}, M.}, \au{{Allafort}, A.},
  \au{{Baldini}, L.}, \au{{Ballet}, J.}, \au{{Barbiellini}, G.},
  \au{{Bastieri}, D.}, \au{{Belfiore}, A.}, \au{{Bellazzini}, R.},
  \au{{Berenji}, B.}, \au{{Bland ford}, R.~D.}, \au{{Bloom}, E.~D.},
  \au{{Bonamente}, E.}, \au{{Borgland }, A.~W.}, \au{{Bottacini}, E.},
  \au{{Brigida}, M.}, \au{{Bruel}, P.}, \au{{Buehler}, R.}, \au{{Buson}, S.},
  \au{{Caliandro}, G.~A.}, \au{{Cameron}, R.~A.}, \au{{Caraveo}, P.~A.},
  \au{{Casandjian}, J.~M.}, \au{{Cecchi}, C.}, \au{{Chekhtman}, A.},
  \au{{Cheung}, C.~C.}, \au{{Chiang}, J.}, \au{{Ciprini}, S.}, \au{{Claus},
  R.}, \au{{Cohen-Tanugi}, J.}, \au{{de Angelis}, A.}, \au{{de Palma}, F.},
  \au{{Dermer}, C.~D.}, \au{{do Couto e Silva}, E.}, \au{{Drell}, P.~S.},
  \au{{Dumora}, D.}, \au{{Favuzzi}, C.}, \au{{Fegan}, S.~J.}, \au{{Focke},
  W.~B.}, \au{{Fortin}, P.}, \au{{Fukazawa}, Y.}, \au{{Fusco}, P.},
  \au{{Gargano}, F.}, \au{{Germani}, S.}, \au{{Giglietto}, N.}, \au{{Giordano},
  F.}, \au{{Giroletti}, M.}, \au{{Glanzman}, T.}, \au{{Godfrey}, G.},
  \au{{Grenier}, I.~A.}, \au{{Guillemot}, L.}, \au{{Guiriec}, S.},
  \au{{Hadasch}, D.}, \au{{Hanabata}, Y.}, \au{{Harding}, A.~K.},
  \au{{Hayashida}, M.}, \au{{Hayashi}, K.}, \au{{Hays}, E.},
  \au{{J{\'o}hannesson}, G.}, \au{{Johnson}, A.~S.}, \au{{Kamae}, T.},
  \au{{Katagiri}, H.}, \au{{Kataoka}, J.}, \au{{Kerr}, M.},
  \au{{Kn{\"o}dlseder}, J.}, \au{{Kuss}, M.}, \au{{Lande}, J.},
  \au{{Latronico}, L.}, \au{{Lee}, S.~H.}, \au{{Longo}, F.}, \au{{Loparco},
  F.}, \au{{Lott}, B.}, \au{{Lovellette}, M.~N.}, \au{{Lubrano}, P.},
  \au{{Martin}, P.}, \au{{Mazziotta}, M.~N.}, \au{{McEnery}, J.~E.},
  \au{{Mehault}, J.}, \au{{Michelson}, P.~F.}, \au{{Mitthumsiri}, W.},
  \au{{Mizuno}, T.}, \au{{Monte}, C.}, \au{{Monzani}, M.~E.}, \au{{Morselli},
  A.}, \au{{Moskalenko}, I.~V.}, \au{{Murgia}, S.}, \au{{Naumann-Godo}, M.},
  \au{{Nolan}, P.~L.}, \au{{Norris}, J.~P.}, \au{{Nuss}, E.}, \au{{Ohsugi},
  T.}, \au{{Okumura}, A.}, \au{{Orlando}, E.}, \au{{Ormes}, J.~F.},
  \au{{Ozaki}, M.}, \au{{Paneque}, D.}, \au{{Parent}, D.}, \au{{Pesce-Rollins},
  M.}, \au{{Pierbattista}, M.}, \au{{Piron}, F.}, \au{{Pohl}, M.},
  \au{{Prokhorov}, D.}, \au{{Rain{\`o}}, S.}, \au{{Rando}, R.}, \au{{Razzano},
  M.}, \au{{Reposeur}, T.}, \au{{Ritz}, S.}, \au{{Parkinson}, P.~M.~Saz},
  \au{{Sgr{\`o}}, C.}, \au{{Siskind}, E.~J.}, \au{{Smith}, P.~D.},
  \au{{Spinelli}, P.}, \au{{Strong}, A.~W.}, \au{{Takahashi}, H.},
  \au{{Tanaka}, T.}, \au{{Thayer}, J.~G.}, \au{{Thayer}, J.~B.},
  \au{{Thompson}, D.~J.}, \au{{Tibaldo}, L.}, \au{{Torres}, D.~F.},
  \au{{Tosti}, G.}, \au{{Tramacere}, A.}, \au{{Troja}, E.}, \au{{Uchiyama},
  Y.}, \au{{Vandenbroucke}, J.}, \au{{Vasileiou}, V.}, \au{{Vianello}, G.},
  \au{{Vitale}, V.}, \au{{Waite}, A.~P.}, \au{{Wang}, P.}, \au{{Winer}, B.~L.},
  \au{{Wood}, K.~S.}, \au{{Yang}, Z.}, \au{{Zimmer}, S.} \& \au{{Bontemps},
  S.}} \yr{2011}  \at{{A Cocoon of Freshly Accelerated Cosmic Rays Detected by
  Fermi in the Cygnus Superbubble}}.  \jt{Science}  \bvol{334}~(6059),
  \pg{1103}.

\bibitem[{Adriani} {\em et~al.\/}(2011){Adriani}, {Barbarino}, {Bazilevskaya},
  {Bellotti}, {Boezio}, {Bogomolov}, {Bonechi}, {Bongi}, {Bonvicini},
  {Borisov}, {Bottai}, {Bruno}, {Cafagna}, {Campana}, {Carbone}, {Carlson},
  {Casolino}, {Castellini}, {Consiglio}, {De Pascale}, {De Santis}, {De
  Simone}, {Di Felice}, {Galper}, {Gillard}, {Grishantseva}, {Jerse},
  {Karelin}, {Koldashov}, {Krutkov}, {Kvashnin}, {Leonov}, {Malakhov},
  {Malvezzi}, {Marcelli}, {Mayorov}, {Menn}, {Mikhailov}, {Mocchiutti},
  {Monaco}, {Mori}, {Nikonov}, {Osteria}, {Palma}, {Papini}, {Pearce},
  {Picozza}, {Pizzolotto}, {Ricci}, {Ricciarini}, {Rossetto}, {Sarkar},
  {Simon}, {Sparvoli}, {Spillantini}, {Stozhkov}, {Vacchi}, {Vannuccini},
  {Vasilyev}, {Voronov}, {Yurkin}, {Wu}, {Zampa}, {Zampa} \&
  {Zverev}]{Adriani11pHe}
{\sc \au{{Adriani}, O.}, \au{{Barbarino}, G.~C.}, \au{{Bazilevskaya}, G.~A.},
  \au{{Bellotti}, R.}, \au{{Boezio}, M.}, \au{{Bogomolov}, E.~A.},
  \au{{Bonechi}, L.}, \au{{Bongi}, M.}, \au{{Bonvicini}, V.}, \au{{Borisov},
  S.}, \au{{Bottai}, S.}, \au{{Bruno}, A.}, \au{{Cafagna}, F.}, \au{{Campana},
  D.}, \au{{Carbone}, R.}, \au{{Carlson}, P.}, \au{{Casolino}, M.},
  \au{{Castellini}, G.}, \au{{Consiglio}, L.}, \au{{De Pascale}, M.~P.},
  \au{{De Santis}, C.}, \au{{De Simone}, N.}, \au{{Di Felice}, V.},
  \au{{Galper}, A.~M.}, \au{{Gillard}, W.}, \au{{Grishantseva}, L.},
  \au{{Jerse}, G.}, \au{{Karelin}, A.~V.}, \au{{Koldashov}, S.~V.},
  \au{{Krutkov}, S.~Y.}, \au{{Kvashnin}, A.~N.}, \au{{Leonov}, A.},
  \au{{Malakhov}, V.}, \au{{Malvezzi}, V.}, \au{{Marcelli}, L.}, \au{{Mayorov},
  A.~G.}, \au{{Menn}, W.}, \au{{Mikhailov}, V.~V.}, \au{{Mocchiutti}, E.},
  \au{{Monaco}, A.}, \au{{Mori}, N.}, \au{{Nikonov}, N.}, \au{{Osteria}, G.},
  \au{{Palma}, F.}, \au{{Papini}, P.}, \au{{Pearce}, M.}, \au{{Picozza}, P.},
  \au{{Pizzolotto}, C.}, \au{{Ricci}, M.}, \au{{Ricciarini}, S.~B.},
  \au{{Rossetto}, L.}, \au{{Sarkar}, R.}, \au{{Simon}, M.}, \au{{Sparvoli},
  R.}, \au{{Spillantini}, P.}, \au{{Stozhkov}, Y.~I.}, \au{{Vacchi}, A.},
  \au{{Vannuccini}, E.}, \au{{Vasilyev}, G.}, \au{{Voronov}, S.~A.},
  \au{{Yurkin}, Y.~T.}, \au{{Wu}, J.}, \au{{Zampa}, G.}, \au{{Zampa}, N.} \&
  \au{{Zverev}, V.~G.}} \yr{2011}  \at{{PAMELA Measurements of Cosmic-Ray
  Proton and Helium Spectra}}.  \jt{Science}  \bvol{332}~(6025),  \pg{69},
  \arxiv{arXiv: 1103.4055}.

\bibitem[{Adriani} {\em et~al.\/}(2009{\natexlab{{\em a\/}}}){Adriani},
  {Barbarino}, {Bazilevskaya}, {Bellotti}, {Boezio}, {Bogomolov}, {Bonechi},
  {Bongi}, {Bonvicini}, {Bottai}, {Bruno}, {Cafagna}, {Campana}, {Carlson},
  {Casolino}, {Castellini}, {de Pascale}, {de Rosa}, {de Simone}, {di Felice},
  {Galper}, {Grishantseva}, {Hofverberg}, {Koldashov}, {Krutkov}, {Kvashnin},
  {Leonov}, {Malvezzi}, {Marcelli}, {Menn}, {Mikhailov}, {Mocchiutti}, {Orsi},
  {Osteria}, {Papini}, {Pearce}, {Picozza}, {Ricci}, {Ricciarini}, {Simon},
  {Sparvoli}, {Spillantini}, {Stozhkov}, {Vacchi}, {Vannuccini}, {Vasilyev},
  {Voronov}, {Yurkin}, {Zampa}, {Zampa} \& {Zverev}]{PamelaPos}
{\sc \au{{Adriani}, O.}, \au{{Barbarino}, G.~C.}, \au{{Bazilevskaya}, G.~A.},
  \au{{Bellotti}, R.}, \au{{Boezio}, M.}, \au{{Bogomolov}, E.~A.},
  \au{{Bonechi}, L.}, \au{{Bongi}, M.}, \au{{Bonvicini}, V.}, \au{{Bottai},
  S.}, \au{{Bruno}, A.}, \au{{Cafagna}, F.}, \au{{Campana}, D.}, \au{{Carlson},
  P.}, \au{{Casolino}, M.}, \au{{Castellini}, G.}, \au{{de Pascale}, M.~P.},
  \au{{de Rosa}, G.}, \au{{de Simone}, N.}, \au{{di Felice}, V.}, \au{{Galper},
  A.~M.}, \au{{Grishantseva}, L.}, \au{{Hofverberg}, P.}, \au{{Koldashov},
  S.~V.}, \au{{Krutkov}, S.~Y.}, \au{{Kvashnin}, A.~N.}, \au{{Leonov}, A.},
  \au{{Malvezzi}, V.}, \au{{Marcelli}, L.}, \au{{Menn}, W.}, \au{{Mikhailov},
  V.~V.}, \au{{Mocchiutti}, E.}, \au{{Orsi}, S.}, \au{{Osteria}, G.},
  \au{{Papini}, P.}, \au{{Pearce}, M.}, \au{{Picozza}, P.}, \au{{Ricci}, M.},
  \au{{Ricciarini}, S.~B.}, \au{{Simon}, M.}, \au{{Sparvoli}, R.},
  \au{{Spillantini}, P.}, \au{{Stozhkov}, Y.~I.}, \au{{Vacchi}, A.},
  \au{{Vannuccini}, E.}, \au{{Vasilyev}, G.}, \au{{Voronov}, S.~A.},
  \au{{Yurkin}, Y.~T.}, \au{{Zampa}, G.}, \au{{Zampa}, N.} \& \au{{Zverev},
  V.~G.}} \yr{2009{\natexlab{{\em a\/}}}}  \at{{An anomalous positron abundance
  in cosmic rays with energies 1.5-100GeV}}.  \jt{Nature}  \bvol{458}~(7238),
  \pg{607--609},  \arxiv{arXiv: 0810.4995}.

\bibitem[{Adriani} {\em et~al.\/}(2009{\natexlab{{\em b\/}}}){Adriani},
  {Barbarino}, {Bazilevskaya}, {Bellotti}, {Boezio}, {Bogomolov}, {Bonechi},
  {Bongi}, {Bonvicini}, {Bottai}, {Bruno}, {Cafagna}, {Campana}, {Carlson},
  {Casolino}, {Castellini}, {de Pascale}, {de Rosa}, {de Simone}, {di Felice},
  {Galper}, {Grishantseva}, {Hofverberg}, {Koldashov}, {Krutkov}, {Kvashnin},
  {Leonov}, {Malvezzi}, {Marcelli}, {Menn}, {Mikhailov}, {Mocchiutti}, {Orsi},
  {Osteria}, {Papini}, {Pearce}, {Picozza}, {Ricci}, {Ricciarini}, {Simon},
  {Sparvoli}, {Spillantini}, {Stozhkov}, {Vacchi}, {Vannuccini}, {Vasilyev},
  {Voronov}, {Yurkin}, {Zampa}, {Zampa} \& {Zverev}]{Adriani+09}
{\sc \au{{Adriani}, O.}, \au{{Barbarino}, G.~C.}, \au{{Bazilevskaya}, G.~A.},
  \au{{Bellotti}, R.}, \au{{Boezio}, M.}, \au{{Bogomolov}, E.~A.},
  \au{{Bonechi}, L.}, \au{{Bongi}, M.}, \au{{Bonvicini}, V.}, \au{{Bottai},
  S.}, \au{{Bruno}, A.}, \au{{Cafagna}, F.}, \au{{Campana}, D.}, \au{{Carlson},
  P.}, \au{{Casolino}, M.}, \au{{Castellini}, G.}, \au{{de Pascale}, M.~P.},
  \au{{de Rosa}, G.}, \au{{de Simone}, N.}, \au{{di Felice}, V.}, \au{{Galper},
  A.~M.}, \au{{Grishantseva}, L.}, \au{{Hofverberg}, P.}, \au{{Koldashov},
  S.~V.}, \au{{Krutkov}, S.~Y.}, \au{{Kvashnin}, A.~N.}, \au{{Leonov}, A.},
  \au{{Malvezzi}, V.}, \au{{Marcelli}, L.}, \au{{Menn}, W.}, \au{{Mikhailov},
  V.~V.}, \au{{Mocchiutti}, E.}, \au{{Orsi}, S.}, \au{{Osteria}, G.},
  \au{{Papini}, P.}, \au{{Pearce}, M.}, \au{{Picozza}, P.}, \au{{Ricci}, M.},
  \au{{Ricciarini}, S.~B.}, \au{{Simon}, M.}, \au{{Sparvoli}, R.},
  \au{{Spillantini}, P.}, \au{{Stozhkov}, Y.~I.}, \au{{Vacchi}, A.},
  \au{{Vannuccini}, E.}, \au{{Vasilyev}, G.}, \au{{Voronov}, S.~A.},
  \au{{Yurkin}, Y.~T.}, \au{{Zampa}, G.}, \au{{Zampa}, N.} \& \au{{Zverev},
  V.~G.}} \yr{2009{\natexlab{{\em b\/}}}}  \at{{An anomalous positron abundance
  in cosmic rays with energies 1.5-100GeV}}.  \jt{Nature}  \bvol{458}~(7238),
  \pg{607--609},  \arxiv{arXiv: 0810.4995}.

\bibitem[{Adriani} {\em et~al.\/}(2009{\natexlab{{\em c\/}}}){Adriani},
  {Barbarino}, {Bazilevskaya}, {Bellotti}, {Boezio}, {Bogomolov}, {Bonechi},
  {Bongi}, {Bonvicini}, {Bottai}, {Bruno}, {Cafagna}, {Campana}, {Carlson},
  {Casolino}, {Castellini}, {de Pascale}, {de Rosa}, {de Simone}, {di Felice},
  {Galper}, {Grishantseva}, {Hofverberg}, {Koldashov}, {Krutkov}, {Kvashnin},
  {Leonov}, {Malvezzi}, {Marcelli}, {Menn}, {Mikhailov}, {Mocchiutti}, {Orsi},
  {Osteria}, {Papini}, {Pearce}, {Picozza}, {Ricci}, {Ricciarini}, {Simon},
  {Sparvoli}, {Spillantini}, {Stozhkov}, {Vacchi}, {Vannuccini}, {Vasilyev},
  {Voronov}, {Yurkin}, {Zampa}, {Zampa} \& {Zverev}]{2009Natur.458..607A}
{\sc \au{{Adriani}, O.}, \au{{Barbarino}, G.~C.}, \au{{Bazilevskaya}, G.~A.},
  \au{{Bellotti}, R.}, \au{{Boezio}, M.}, \au{{Bogomolov}, E.~A.},
  \au{{Bonechi}, L.}, \au{{Bongi}, M.}, \au{{Bonvicini}, V.}, \au{{Bottai},
  S.}, \au{{Bruno}, A.}, \au{{Cafagna}, F.}, \au{{Campana}, D.}, \au{{Carlson},
  P.}, \au{{Casolino}, M.}, \au{{Castellini}, G.}, \au{{de Pascale}, M.~P.},
  \au{{de Rosa}, G.}, \au{{de Simone}, N.}, \au{{di Felice}, V.}, \au{{Galper},
  A.~M.}, \au{{Grishantseva}, L.}, \au{{Hofverberg}, P.}, \au{{Koldashov},
  S.~V.}, \au{{Krutkov}, S.~Y.}, \au{{Kvashnin}, A.~N.}, \au{{Leonov}, A.},
  \au{{Malvezzi}, V.}, \au{{Marcelli}, L.}, \au{{Menn}, W.}, \au{{Mikhailov},
  V.~V.}, \au{{Mocchiutti}, E.}, \au{{Orsi}, S.}, \au{{Osteria}, G.},
  \au{{Papini}, P.}, \au{{Pearce}, M.}, \au{{Picozza}, P.}, \au{{Ricci}, M.},
  \au{{Ricciarini}, S.~B.}, \au{{Simon}, M.}, \au{{Sparvoli}, R.},
  \au{{Spillantini}, P.}, \au{{Stozhkov}, Y.~I.}, \au{{Vacchi}, A.},
  \au{{Vannuccini}, E.}, \au{{Vasilyev}, G.}, \au{{Voronov}, S.~A.},
  \au{{Yurkin}, Y.~T.}, \au{{Zampa}, G.}, \au{{Zampa}, N.} \& \au{{Zverev},
  V.~G.}} \yr{2009{\natexlab{{\em c\/}}}}  \at{{An anomalous positron abundance
  in cosmic rays with energies 1.5-100GeV}}.  \jt{Nature}  \bvol{458}~(7238),
  \pg{607--609},  \arxiv{arXiv: 0810.4995}.

\bibitem[{Aguilar} {\em et~al.\/}(2015{\natexlab{{\em a\/}}}){Aguilar}, {Aisa},
  {Alpat}, {Alvino}, {Ambrosi}, {Andeen}, {Arruda}, {Attig}, {Azzarello},
  {Bachlechner}, {Barao}, {Barrau}, {Barrin}, {Bartoloni}, {Basara},
  {Battarbee}, {Battiston}, {Bazo}, {Becker}, {Behlmann}, {Beischer},
  {Berdugo}, {Bertucci}, {Bigongiari}, {Bindi}, {Bizzaglia}, {Bizzarri},
  {Boella}, {de Boer}, {Bollweg}, {Bonnivard}, {Borgia}, {Borsini}, {Boschini},
  {Bourquin}, {Burger}, {Cadoux}, {Cai}, {Capell}, {Caroff}, {Casaus},
  {Cascioli}, {Castellini}, {Cernuda}, {Cerreta}, {Cervelli}, {Chae}, {Chang},
  {Chen}, {Chen}, {Cheng}, {Chen}, {Cheng}, {Chou}, {Choumilov}, {Choutko},
  {Chung}, {Clark}, {Clavero}, {Coignet}, {Consolandi}, {Contin}, {Corti},
  {Gil}, {Coste}, {Creus}, {Crispoltoni}, {Cui}, {Dai}, {Delgado}, {Della
  Torre}, {Demirk{\"o}z}, {Derome}, {Di Falco}, {Di Masso}, {Dimiccoli},
  {D{\'\i}az}, {von Doetinchem}, {Donnini}, {Du}, {Duranti}, {D'Urso}, {Eline},
  {Eppling}, {Eronen}, {Fan}, {Farnesini}, {Feng}, {Fiandrini}, {Fiasson},
  {Finch}, {Fisher}, {Galaktionov}, {Gallucci}, {Garc{\'\i}a},
  {Garc{\'\i}a-L{\'o}pez}, {Gargiulo}, {Gast}, {Gebauer}, {Gervasi}, {Ghelfi},
  {Gillard}, {Giovacchini}, {Goglov}, {Gong}, {Goy}, {Grabski}, {Grandi},
  {Graziani}, {Guand alini}, {Guerri}, {Guo}, {Haas}, {Habiby}, {Haino}, {Han},
  {He}, {Heil}, {Hoffman}, {Hsieh}, {Huang}, {Huh}, {Incagli}, {Ionica},
  {Jang}, {Jinchi}, {Kanishev}, {Kim}, {Kim}, {Kirn}, {Kossakowski}, {Kounina},
  {Kounine}, {Koutsenko}, {Krafczyk}, {La Vacca}, {Laudi}, {Laurenti},
  {Lazzizzera}, {Lebedev}, {Lee}, {Lee}, {Leluc}, {Levi}, {Li}, {Li}, {Li},
  {Li}, {Li}, {Li}, {Li}, {Li}, {Li}, {Lim}, {Lin}, {Lipari}, {Lippert}, {Liu},
  {Liu}, {Lolli}, {Lomtadze}, {Lu}, {Lu}, {Lu}, {Luebelsmeyer}, {Luo}, {Lv},
  {Majka}, {Ma{\~n}{\'a}}, {Mar{\'\i}n}, {Martin}, {Mart{\'\i}nez}, {Masi},
  {Maurin}, {Menchaca-Rocha}, {Meng}, {Mo}, {Morescalchi}, {Mott},
  {M{\"u}ller}, {Ni}, {Nikonov}, {Nozzoli}, {Nunes}, {Obermeier}, {Oliva},
  {Orcinha}, {Palmonari}, {Palomares}, {Paniccia}, {Papi}, {Pauluzzi},
  {Pedreschi}, {Pensotti}, {Pereira}, {Picot-Clemente}, {Pilo}, {Piluso},
  {Pizzolotto}, {Plyaskin}, {Pohl}, {Poireau}, {Postaci}, {Putze}, {Quadrani},
  {Qi}, {Qin}, {Qu}, {R{\"a}ih{\"a}}, {Rancoita}, {Rapin}, {Ricol},
  {Rodr{\'\i}guez}, {Rosier-Lees}, {Rozhkov}, {Rozza}, {Sagdeev}, {Sandweiss},
  {Saouter}, {Sbarra}, {Schael}, {Schmidt}, {von Dratzig}, {Schwering},
  {Scolieri}, {Seo}, {Shan}, {Shan}, {Shi}, {Shi}, {Shi}, {Siedenburg}, {Son},
  {Spada}, {Spinella}, {Sun}, {Sun}, {Tacconi}, {Tang}, {Tang}, {Tang}, {Tao},
  {Tescaro}, {Ting}, {Ting}, {Tomassetti}, {Torsti}, {T{\"u}rko{\v{g}}lu},
  {Urban}, {Vagelli}, {Valente}, {Vannini}, {Valtonen}, {Vaurynovich},
  {Vecchi}, {Velasco}, {Vialle}, {Vitale}, {Vitillo}, {Wang}, {Wang}, {Wang},
  {Wang}, {Wang}, {Wang}, {Weng}, {Whitman}, {Wienkenh{\"o}ver}, {Wu}, {Wu},
  {Xia}, {Xie}, {Xie}, {Xiong}, {Xin}, {Xu}, {Xu}, {Yan}, {Yang}, {Yang}, {Ye},
  {Yi}, {Yu}, {Yu}, {Zeissler}, {Zhang}, {Zhang}, {Zhang}, {Zhang}, {Zheng},
  {Zhuang}, {Zhukov}, {Zichichi}, {Zimmermann}, {Zuccon}, {Zurbach} \& {AMS
  Collaboration}]{AMS02p}
{\sc \au{{Aguilar}, M.}, \au{{Aisa}, D.}, \au{{Alpat}, B.}, \au{{Alvino}, A.},
  \au{{Ambrosi}, G.}, \au{{Andeen}, K.}, \au{{Arruda}, L.}, \au{{Attig}, N.},
  \au{{Azzarello}, P.}, \au{{Bachlechner}, A.}, \au{{Barao}, F.}, \au{{Barrau},
  A.}, \au{{Barrin}, L.}, \au{{Bartoloni}, A.}, \au{{Basara}, L.},
  \au{{Battarbee}, M.}, \au{{Battiston}, R.}, \au{{Bazo}, J.}, \au{{Becker},
  U.}, \au{{Behlmann}, M.}, \au{{Beischer}, B.}, \au{{Berdugo}, J.},
  \au{{Bertucci}, B.}, \au{{Bigongiari}, G.}, \au{{Bindi}, V.},
  \au{{Bizzaglia}, S.}, \au{{Bizzarri}, M.}, \au{{Boella}, G.}, \au{{de Boer},
  W.}, \au{{Bollweg}, K.}, \au{{Bonnivard}, V.}, \au{{Borgia}, B.},
  \au{{Borsini}, S.}, \au{{Boschini}, M.~J.}, \au{{Bourquin}, M.},
  \au{{Burger}, J.}, \au{{Cadoux}, F.}, \au{{Cai}, X.~D.}, \au{{Capell}, M.},
  \au{{Caroff}, S.}, \au{{Casaus}, J.}, \au{{Cascioli}, V.}, \au{{Castellini},
  G.}, \au{{Cernuda}, I.}, \au{{Cerreta}, D.}, \au{{Cervelli}, F.}, \au{{Chae},
  M.~J.}, \au{{Chang}, Y.~H.}, \au{{Chen}, A.~I.}, \au{{Chen}, H.},
  \au{{Cheng}, G.~M.}, \au{{Chen}, H.~S.}, \au{{Cheng}, L.}, \au{{Chou},
  H.~Y.}, \au{{Choumilov}, E.}, \au{{Choutko}, V.}, \au{{Chung}, C.~H.},
  \au{{Clark}, C.}, \au{{Clavero}, R.}, \au{{Coignet}, G.}, \au{{Consolandi},
  C.}, \au{{Contin}, A.}, \au{{Corti}, C.}, \au{{Gil}, E.~Cortina},
  \au{{Coste}, B.}, \au{{Creus}, W.}, \au{{Crispoltoni}, M.}, \au{{Cui}, Z.},
  \au{{Dai}, Y.~M.}, \au{{Delgado}, C.}, \au{{Della Torre}, S.},
  \au{{Demirk{\"o}z}, M.~B.}, \au{{Derome}, L.}, \au{{Di Falco}, S.}, \au{{Di
  Masso}, L.}, \au{{Dimiccoli}, F.}, \au{{D{\'\i}az}, C.}, \au{{von
  Doetinchem}, P.}, \au{{Donnini}, F.}, \au{{Du}, W.~J.}, \au{{Duranti}, M.},
  \au{{D'Urso}, D.}, \au{{Eline}, A.}, \au{{Eppling}, F.~J.}, \au{{Eronen},
  T.}, \au{{Fan}, Y.~Y.}, \au{{Farnesini}, L.}, \au{{Feng}, J.},
  \au{{Fiandrini}, E.}, \au{{Fiasson}, A.}, \au{{Finch}, E.}, \au{{Fisher},
  P.}, \au{{Galaktionov}, Y.}, \au{{Gallucci}, G.}, \au{{Garc{\'\i}a}, B.},
  \au{{Garc{\'\i}a-L{\'o}pez}, R.}, \au{{Gargiulo}, C.}, \au{{Gast}, H.},
  \au{{Gebauer}, I.}, \au{{Gervasi}, M.}, \au{{Ghelfi}, A.}, \au{{Gillard},
  W.}, \au{{Giovacchini}, F.}, \au{{Goglov}, P.}, \au{{Gong}, J.}, \au{{Goy},
  C.}, \au{{Grabski}, V.}, \au{{Grandi}, D.}, \au{{Graziani}, M.}, \au{{Guand
  alini}, C.}, \au{{Guerri}, I.}, \au{{Guo}, K.~H.}, \au{{Haas}, D.},
  \au{{Habiby}, M.}, \au{{Haino}, S.}, \au{{Han}, K.~C.}, \au{{He}, Z.~H.},
  \au{{Heil}, M.}, \au{{Hoffman}, J.}, \au{{Hsieh}, T.~H.}, \au{{Huang},
  Z.~C.}, \au{{Huh}, C.}, \au{{Incagli}, M.}, \au{{Ionica}, M.}, \au{{Jang},
  W.~Y.}, \au{{Jinchi}, H.}, \au{{Kanishev}, K.}, \au{{Kim}, G.~N.}, \au{{Kim},
  K.~S.}, \au{{Kirn}, Th.}, \au{{Kossakowski}, R.}, \au{{Kounina}, O.},
  \au{{Kounine}, A.}, \au{{Koutsenko}, V.}, \au{{Krafczyk}, M.~S.}, \au{{La
  Vacca}, G.}, \au{{Laudi}, E.}, \au{{Laurenti}, G.}, \au{{Lazzizzera}, I.},
  \au{{Lebedev}, A.}, \au{{Lee}, H.~T.}, \au{{Lee}, S.~C.}, \au{{Leluc}, C.},
  \au{{Levi}, G.}, \au{{Li}, H.~L.}, \au{{Li}, J.~Q.}, \au{{Li}, Q.}, \au{{Li},
  Q.}, \au{{Li}, T.~X.}, \au{{Li}, W.}, \au{{Li}, Y.}, \au{{Li}, Z.~H.},
  \au{{Li}, Z.~Y.}, \au{{Lim}, S.}, \au{{Lin}, C.~H.}, \au{{Lipari}, P.},
  \au{{Lippert}, T.}, \au{{Liu}, D.}, \au{{Liu}, H.}, \au{{Lolli}, M.},
  \au{{Lomtadze}, T.}, \au{{Lu}, M.~J.}, \au{{Lu}, S.~Q.}, \au{{Lu}, Y.~S.},
  \au{{Luebelsmeyer}, K.}, \au{{Luo}, J.~Z.}, \au{{Lv}, S.~S.}, \au{{Majka},
  R.}, \au{{Ma{\~n}{\'a}}, C.}, \au{{Mar{\'\i}n}, J.}, \au{{Martin}, T.},
  \au{{Mart{\'\i}nez}, G.}, \au{{Masi}, N.}, \au{{Maurin}, D.},
  \au{{Menchaca-Rocha}, A.}, \au{{Meng}, Q.}, \au{{Mo}, D.~C.},
  \au{{Morescalchi}, L.}, \au{{Mott}, P.}, \au{{M{\"u}ller}, M.}, \au{{Ni},
  J.~Q.}, \au{{Nikonov}, N.}, \au{{Nozzoli}, F.}, \au{{Nunes}, P.},
  \au{{Obermeier}, A.}, \au{{Oliva}, A.}, \au{{Orcinha}, M.}, \au{{Palmonari},
  F.}, \au{{Palomares}, C.}, \au{{Paniccia}, M.}, \au{{Papi}, A.},
  \au{{Pauluzzi}, M.}, \au{{Pedreschi}, E.}, \au{{Pensotti}, S.},
  \au{{Pereira}, R.}, \au{{Picot-Clemente}, N.}, \au{{Pilo}, F.}, \au{{Piluso},
  A.}, \au{{Pizzolotto}, C.}, \au{{Plyaskin}, V.}, \au{{Pohl}, M.},
  \au{{Poireau}, V.}, \au{{Postaci}, E.}, \au{{Putze}, A.}, \au{{Quadrani},
  L.}, \au{{Qi}, X.~M.}, \au{{Qin}, X.}, \au{{Qu}, Z.~Y.}, \au{{R{\"a}ih{\"a}},
  T.}, \au{{Rancoita}, P.~G.}, \au{{Rapin}, D.}, \au{{Ricol}, J.~S.},
  \au{{Rodr{\'\i}guez}, I.}, \au{{Rosier-Lees}, S.}, \au{{Rozhkov}, A.},
  \au{{Rozza}, D.}, \au{{Sagdeev}, R.}, \au{{Sandweiss}, J.}, \au{{Saouter},
  P.}, \au{{Sbarra}, C.}, \au{{Schael}, S.}, \au{{Schmidt}, S.~M.}, \au{{von
  Dratzig}, A.~Schulz}, \au{{Schwering}, G.}, \au{{Scolieri}, G.}, \au{{Seo},
  E.~S.}, \au{{Shan}, B.~S.}, \au{{Shan}, Y.~H.}, \au{{Shi}, J.~Y.}, \au{{Shi},
  X.~Y.}, \au{{Shi}, Y.~M.}, \au{{Siedenburg}, T.}, \au{{Son}, D.},
  \au{{Spada}, F.}, \au{{Spinella}, F.}, \au{{Sun}, W.}, \au{{Sun}, W.~H.},
  \au{{Tacconi}, M.}, \au{{Tang}, C.~P.}, \au{{Tang}, X.~W.}, \au{{Tang},
  Z.~C.}, \au{{Tao}, L.}, \au{{Tescaro}, D.}, \au{{Ting}, Samuel C.~C.},
  \au{{Ting}, S.~M.}, \au{{Tomassetti}, N.}, \au{{Torsti}, J.},
  \au{{T{\"u}rko{\v{g}}lu}, C.}, \au{{Urban}, T.}, \au{{Vagelli}, V.},
  \au{{Valente}, E.}, \au{{Vannini}, C.}, \au{{Valtonen}, E.},
  \au{{Vaurynovich}, S.}, \au{{Vecchi}, M.}, \au{{Velasco}, M.}, \au{{Vialle},
  J.~P.}, \au{{Vitale}, V.}, \au{{Vitillo}, S.}, \au{{Wang}, L.~Q.},
  \au{{Wang}, N.~H.}, \au{{Wang}, Q.~L.}, \au{{Wang}, R.~S.}, \au{{Wang}, X.},
  \au{{Wang}, Z.~X.}, \au{{Weng}, Z.~L.}, \au{{Whitman}, K.},
  \au{{Wienkenh{\"o}ver}, J.}, \au{{Wu}, H.}, \au{{Wu}, X.}, \au{{Xia}, X.},
  \au{{Xie}, M.}, \au{{Xie}, S.}, \au{{Xiong}, R.~Q.}, \au{{Xin}, G.~M.},
  \au{{Xu}, N.~S.}, \au{{Xu}, W.}, \au{{Yan}, Q.}, \au{{Yang}, J.}, \au{{Yang},
  M.}, \au{{Ye}, Q.~H.}, \au{{Yi}, H.}, \au{{Yu}, Y.~J.}, \au{{Yu}, Z.~Q.},
  \au{{Zeissler}, S.}, \au{{Zhang}, J.~H.}, \au{{Zhang}, M.~T.}, \au{{Zhang},
  X.~B.}, \au{{Zhang}, Z.}, \au{{Zheng}, Z.~M.}, \au{{Zhuang}, H.~L.},
  \au{{Zhukov}, V.}, \au{{Zichichi}, A.}, \au{{Zimmermann}, N.}, \au{{Zuccon},
  P.}, \au{{Zurbach}, C.} \& \au{{AMS Collaboration}}} \yr{2015{\natexlab{{\em
  a\/}}}}  \at{{Precision Measurement of the Proton Flux in Primary Cosmic Rays
  from Rigidity 1 GV to 1.8 TV with the Alpha Magnetic Spectrometer on the
  International Space Station}}.  \jt{Physical Review Letters}
  \bvol{114}~(17),  \pg{171103}.

\bibitem[{Aguilar} {\em et~al.\/}(2015{\natexlab{{\em b\/}}}){Aguilar}, {Aisa},
  {Alpat}, {Alvino}, {Ambrosi}, {Andeen}, {Arruda}, {Attig}, {Azzarello},
  {Bachlechner}, {Barao}, {Barrau}, {Barrin}, {Bartoloni}, {Basara},
  {Battarbee}, {Battiston}, {Bazo}, {Becker}, {Behlmann}, {Beischer},
  {Berdugo}, {Bertucci}, {Bindi}, {Bizzaglia}, {Bizzarri}, {Boella}, {de Boer},
  {Bollweg}, {Bonnivard}, {Borgia}, {Borsini}, {Boschini}, {Bourquin},
  {Burger}, {Cadoux}, {Cai}, {Capell}, {Caroff}, {Casaus}, {Castellini},
  {Cernuda}, {Cerreta}, {Cervelli}, {Chae}, {Chang}, {Chen}, {Chen}, {Chen},
  {Chen}, {Cheng}, {Chou}, {Choumilov}, {Choutko}, {Chung}, {Clark}, {Clavero},
  {Coignet}, {Consolandi}, {Contin}, {Corti}, {Gil}, {Coste}, {Creus},
  {Crispoltoni}, {Cui}, {Dai}, {Delgado}, {Della Torre}, {Demirk{\"o}z},
  {Derome}, {Di Falco}, {Di Masso}, {Dimiccoli}, {D{\'\i}az}, {von Doetinchem},
  {Donnini}, {Duranti}, {D'Urso}, {Egorov}, {Eline}, {Eppling}, {Eronen},
  {Fan}, {Farnesini}, {Feng}, {Fiandrini}, {Fiasson}, {Finch}, {Fisher},
  {Formato}, {Galaktionov}, {Gallucci}, {Garc{\'\i}a}, {Garc{\'\i}a-L{\'o}pez},
  {Gargiulo}, {Gast}, {Gebauer}, {Gervasi}, {Ghelfi}, {Giovacchini}, {Goglov},
  {Gong}, {Goy}, {Grabski}, {Grandi}, {Graziani}, {Guand alini}, {Guerri},
  {Guo}, {Haas}, {Habiby}, {Haino}, {Han}, {He}, {Heil}, {Hoffman}, {Hsieh},
  {Huang}, {Huh}, {Incagli}, {Ionica}, {Jang}, {Jinchi}, {Kanishev}, {Kim},
  {Kim}, {Kirn}, {Korkmaz}, {Kossakowski}, {Kounina}, {Kounine}, {Koutsenko},
  {Krafczyk}, {La Vacca}, {Laudi}, {Laurenti}, {Lazzizzera}, {Lebedev}, {Lee},
  {Lee}, {Leluc}, {Li}, {Li}, {Li}, {Li}, {Li}, {Li}, {Li}, {Li}, {Li}, {Li},
  {Lim}, {Lin}, {Lipari}, {Lippert}, {Liu}, {Liu}, {Liu}, {Lolli}, {Lomtadze},
  {Lu}, {Lu}, {Lu}, {Luebelsmeyer}, {Luo}, {Luo}, {Lv}, {Majka},
  {Ma{\~n}{\'a}}, {Mar{\'\i}n}, {Martin}, {Mart{\'\i}nez}, {Masi}, {Maurin},
  {Menchaca-Rocha}, {Meng}, {Mo}, {Morescalchi}, {Mott}, {M{\"u}ller},
  {Nelson}, {Ni}, {Nikonov}, {Nozzoli}, {Nunes}, {Obermeier}, {Oliva},
  {Orcinha}, {Palmonari}, {Palomares}, {Paniccia}, {Papi}, {Pauluzzi},
  {Pedreschi}, {Pensotti}, {Pereira}, {Picot-Clemente}, {Pilo}, {Piluso},
  {Pizzolotto}, {Plyaskin}, {Pohl}, {Poireau}, {Putze}, {Quadrani}, {Qi},
  {Qin}, {Qu}, {R{\"a}ih{\"a}}, {Rancoita}, {Rapin}, {Ricol}, {Rodr{\'\i}guez},
  {Rosier-Lees}, {Rozhkov}, {Rozza}, {Sagdeev}, {Sandweiss}, {Saouter},
  {Schael}, {Schmidt}, {von Dratzig}, {Schwering}, {Scolieri}, {Seo}, {Shan},
  {Shan}, {Shi}, {Shi}, {Shi}, {Siedenburg}, {Son}, {Song}, {Spada},
  {Spinella}, {Sun}, {Sun}, {Tacconi}, {Tang}, {Tang}, {Tang}, {Tao},
  {Tescaro}, {Ting}, {Ting}, {Tomassetti}, {Torsti}, {T{\"u}rko{\v{g}}lu},
  {Urban}, {Vagelli}, {Valente}, {Vannini}, {Valtonen}, {Vaurynovich},
  {Vecchi}, {Velasco}, {Vialle}, {Vitale}, {Vitillo}, {Wang}, {Wang}, {Wang},
  {Wang}, {Wang}, {Wang}, {Weng}, {Whitman}, {Wienkenh{\"o}ver}, {Willenbrock},
  {Wu}, {Wu}, {Xia}, {Xie}, {Xie}, {Xiong}, {Xu}, {Xu}, {Yan}, {Yang}, {Yang},
  {Yang}, {Ye}, {Yi}, {Yu}, {Yu}, {Zeissler}, {Zhang}, {Zhang}, {Zhang},
  {Zhang}, {Zhang}, {Zhang}, {Zhang}, {Zheng}, {Zhuang}, {Zhukov}, {Zichichi},
  {Zimmermann}, {Zuccon} \& {AMS Collaboration}]{AMS02He}
{\sc \au{{Aguilar}, M.}, \au{{Aisa}, D.}, \au{{Alpat}, B.}, \au{{Alvino}, A.},
  \au{{Ambrosi}, G.}, \au{{Andeen}, K.}, \au{{Arruda}, L.}, \au{{Attig}, N.},
  \au{{Azzarello}, P.}, \au{{Bachlechner}, A.}, \au{{Barao}, F.}, \au{{Barrau},
  A.}, \au{{Barrin}, L.}, \au{{Bartoloni}, A.}, \au{{Basara}, L.},
  \au{{Battarbee}, M.}, \au{{Battiston}, R.}, \au{{Bazo}, J.}, \au{{Becker},
  U.}, \au{{Behlmann}, M.}, \au{{Beischer}, B.}, \au{{Berdugo}, J.},
  \au{{Bertucci}, B.}, \au{{Bindi}, V.}, \au{{Bizzaglia}, S.}, \au{{Bizzarri},
  M.}, \au{{Boella}, G.}, \au{{de Boer}, W.}, \au{{Bollweg}, K.},
  \au{{Bonnivard}, V.}, \au{{Borgia}, B.}, \au{{Borsini}, S.}, \au{{Boschini},
  M.~J.}, \au{{Bourquin}, M.}, \au{{Burger}, J.}, \au{{Cadoux}, F.}, \au{{Cai},
  X.~D.}, \au{{Capell}, M.}, \au{{Caroff}, S.}, \au{{Casaus}, J.},
  \au{{Castellini}, G.}, \au{{Cernuda}, I.}, \au{{Cerreta}, D.},
  \au{{Cervelli}, F.}, \au{{Chae}, M.~J.}, \au{{Chang}, Y.~H.}, \au{{Chen},
  A.~I.}, \au{{Chen}, G.~M.}, \au{{Chen}, H.}, \au{{Chen}, H.~S.}, \au{{Cheng},
  L.}, \au{{Chou}, H.~Y.}, \au{{Choumilov}, E.}, \au{{Choutko}, V.},
  \au{{Chung}, C.~H.}, \au{{Clark}, C.}, \au{{Clavero}, R.}, \au{{Coignet},
  G.}, \au{{Consolandi}, C.}, \au{{Contin}, A.}, \au{{Corti}, C.}, \au{{Gil},
  E.~Cortina}, \au{{Coste}, B.}, \au{{Creus}, W.}, \au{{Crispoltoni}, M.},
  \au{{Cui}, Z.}, \au{{Dai}, Y.~M.}, \au{{Delgado}, C.}, \au{{Della Torre},
  S.}, \au{{Demirk{\"o}z}, M.~B.}, \au{{Derome}, L.}, \au{{Di Falco}, S.},
  \au{{Di Masso}, L.}, \au{{Dimiccoli}, F.}, \au{{D{\'\i}az}, C.}, \au{{von
  Doetinchem}, P.}, \au{{Donnini}, F.}, \au{{Duranti}, M.}, \au{{D'Urso}, D.},
  \au{{Egorov}, A.}, \au{{Eline}, A.}, \au{{Eppling}, F.~J.}, \au{{Eronen},
  T.}, \au{{Fan}, Y.~Y.}, \au{{Farnesini}, L.}, \au{{Feng}, J.},
  \au{{Fiandrini}, E.}, \au{{Fiasson}, A.}, \au{{Finch}, E.}, \au{{Fisher},
  P.}, \au{{Formato}, V.}, \au{{Galaktionov}, Y.}, \au{{Gallucci}, G.},
  \au{{Garc{\'\i}a}, B.}, \au{{Garc{\'\i}a-L{\'o}pez}, R.}, \au{{Gargiulo},
  C.}, \au{{Gast}, H.}, \au{{Gebauer}, I.}, \au{{Gervasi}, M.}, \au{{Ghelfi},
  A.}, \au{{Giovacchini}, F.}, \au{{Goglov}, P.}, \au{{Gong}, J.}, \au{{Goy},
  C.}, \au{{Grabski}, V.}, \au{{Grandi}, D.}, \au{{Graziani}, M.}, \au{{Guand
  alini}, C.}, \au{{Guerri}, I.}, \au{{Guo}, K.~H.}, \au{{Haas}, D.},
  \au{{Habiby}, M.}, \au{{Haino}, S.}, \au{{Han}, K.~C.}, \au{{He}, Z.~H.},
  \au{{Heil}, M.}, \au{{Hoffman}, J.}, \au{{Hsieh}, T.~H.}, \au{{Huang},
  Z.~C.}, \au{{Huh}, C.}, \au{{Incagli}, M.}, \au{{Ionica}, M.}, \au{{Jang},
  W.~Y.}, \au{{Jinchi}, H.}, \au{{Kanishev}, K.}, \au{{Kim}, G.~N.}, \au{{Kim},
  K.~S.}, \au{{Kirn}, Th.}, \au{{Korkmaz}, M.~A.}, \au{{Kossakowski}, R.},
  \au{{Kounina}, O.}, \au{{Kounine}, A.}, \au{{Koutsenko}, V.}, \au{{Krafczyk},
  M.~S.}, \au{{La Vacca}, G.}, \au{{Laudi}, E.}, \au{{Laurenti}, G.},
  \au{{Lazzizzera}, I.}, \au{{Lebedev}, A.}, \au{{Lee}, H.~T.}, \au{{Lee},
  S.~C.}, \au{{Leluc}, C.}, \au{{Li}, H.~L.}, \au{{Li}, J.~Q.}, \au{{Li},
  J.~Q.}, \au{{Li}, Q.}, \au{{Li}, Q.}, \au{{Li}, T.~X.}, \au{{Li}, W.},
  \au{{Li}, Y.}, \au{{Li}, Z.~H.}, \au{{Li}, Z.~Y.}, \au{{Lim}, S.}, \au{{Lin},
  C.~H.}, \au{{Lipari}, P.}, \au{{Lippert}, T.}, \au{{Liu}, D.}, \au{{Liu},
  H.}, \au{{Liu}, Hu}, \au{{Lolli}, M.}, \au{{Lomtadze}, T.}, \au{{Lu}, M.~J.},
  \au{{Lu}, S.~Q.}, \au{{Lu}, Y.~S.}, \au{{Luebelsmeyer}, K.}, \au{{Luo}, F.},
  \au{{Luo}, J.~Z.}, \au{{Lv}, S.~S.}, \au{{Majka}, R.}, \au{{Ma{\~n}{\'a}},
  C.}, \au{{Mar{\'\i}n}, J.}, \au{{Martin}, T.}, \au{{Mart{\'\i}nez}, G.},
  \au{{Masi}, N.}, \au{{Maurin}, D.}, \au{{Menchaca-Rocha}, A.}, \au{{Meng},
  Q.}, \au{{Mo}, D.~C.}, \au{{Morescalchi}, L.}, \au{{Mott}, P.},
  \au{{M{\"u}ller}, M.}, \au{{Nelson}, T.}, \au{{Ni}, J.~Q.}, \au{{Nikonov},
  N.}, \au{{Nozzoli}, F.}, \au{{Nunes}, P.}, \au{{Obermeier}, A.}, \au{{Oliva},
  A.}, \au{{Orcinha}, M.}, \au{{Palmonari}, F.}, \au{{Palomares}, C.},
  \au{{Paniccia}, M.}, \au{{Papi}, A.}, \au{{Pauluzzi}, M.}, \au{{Pedreschi},
  E.}, \au{{Pensotti}, S.}, \au{{Pereira}, R.}, \au{{Picot-Clemente}, N.},
  \au{{Pilo}, F.}, \au{{Piluso}, A.}, \au{{Pizzolotto}, C.}, \au{{Plyaskin},
  V.}, \au{{Pohl}, M.}, \au{{Poireau}, V.}, \au{{Putze}, A.}, \au{{Quadrani},
  L.}, \au{{Qi}, X.~M.}, \au{{Qin}, X.}, \au{{Qu}, Z.~Y.}, \au{{R{\"a}ih{\"a}},
  T.}, \au{{Rancoita}, P.~G.}, \au{{Rapin}, D.}, \au{{Ricol}, J.~S.},
  \au{{Rodr{\'\i}guez}, I.}, \au{{Rosier-Lees}, S.}, \au{{Rozhkov}, A.},
  \au{{Rozza}, D.}, \au{{Sagdeev}, R.}, \au{{Sandweiss}, J.}, \au{{Saouter},
  P.}, \au{{Schael}, S.}, \au{{Schmidt}, S.~M.}, \au{{von Dratzig}, A.~Schulz},
  \au{{Schwering}, G.}, \au{{Scolieri}, G.}, \au{{Seo}, E.~S.}, \au{{Shan},
  B.~S.}, \au{{Shan}, Y.~H.}, \au{{Shi}, J.~Y.}, \au{{Shi}, X.~Y.}, \au{{Shi},
  Y.~M.}, \au{{Siedenburg}, T.}, \au{{Son}, D.}, \au{{Song}, J.~W.},
  \au{{Spada}, F.}, \au{{Spinella}, F.}, \au{{Sun}, W.}, \au{{Sun}, W.~H.},
  \au{{Tacconi}, M.}, \au{{Tang}, C.~P.}, \au{{Tang}, X.~W.}, \au{{Tang},
  Z.~C.}, \au{{Tao}, L.}, \au{{Tescaro}, D.}, \au{{Ting}, Samuel C.~C.},
  \au{{Ting}, S.~M.}, \au{{Tomassetti}, N.}, \au{{Torsti}, J.},
  \au{{T{\"u}rko{\v{g}}lu}, C.}, \au{{Urban}, T.}, \au{{Vagelli}, V.},
  \au{{Valente}, E.}, \au{{Vannini}, C.}, \au{{Valtonen}, E.},
  \au{{Vaurynovich}, S.}, \au{{Vecchi}, M.}, \au{{Velasco}, M.}, \au{{Vialle},
  J.~P.}, \au{{Vitale}, V.}, \au{{Vitillo}, S.}, \au{{Wang}, L.~Q.},
  \au{{Wang}, N.~H.}, \au{{Wang}, Q.~L.}, \au{{Wang}, R.~S.}, \au{{Wang}, X.},
  \au{{Wang}, Z.~X.}, \au{{Weng}, Z.~L.}, \au{{Whitman}, K.},
  \au{{Wienkenh{\"o}ver}, J.}, \au{{Willenbrock}, M.}, \au{{Wu}, H.}, \au{{Wu},
  X.}, \au{{Xia}, X.}, \au{{Xie}, M.}, \au{{Xie}, S.}, \au{{Xiong}, R.~Q.},
  \au{{Xu}, N.~S.}, \au{{Xu}, W.}, \au{{Yan}, Q.}, \au{{Yang}, J.}, \au{{Yang},
  M.}, \au{{Yang}, Y.}, \au{{Ye}, Q.~H.}, \au{{Yi}, H.}, \au{{Yu}, Y.~J.},
  \au{{Yu}, Z.~Q.}, \au{{Zeissler}, S.}, \au{{Zhang}, C.}, \au{{Zhang}, J.~H.},
  \au{{Zhang}, M.~T.}, \au{{Zhang}, S.~D.}, \au{{Zhang}, S.~W.}, \au{{Zhang},
  X.~B.}, \au{{Zhang}, Z.}, \au{{Zheng}, Z.~M.}, \au{{Zhuang}, H.~L.},
  \au{{Zhukov}, V.}, \au{{Zichichi}, A.}, \au{{Zimmermann}, N.}, \au{{Zuccon},
  P.} \& \au{{AMS Collaboration}}} \yr{2015{\natexlab{{\em b\/}}}}
  \at{{Precision Measurement of the Helium Flux in Primary Cosmic Rays of
  Rigidities 1.9 GV to 3 TV with the Alpha Magnetic Spectrometer on the
  International Space Station}}.  \jt{Physical Review Letters}
  \bvol{115}~(21),  \pg{211101}.

\bibitem[{Aguilar} {\em et~al.\/}(2015{\natexlab{{\em c\/}}}){Aguilar}, {Aisa},
  {Alpat}, {Alvino}, {Ambrosi}, {Andeen}, {Arruda}, {Attig}, {Azzarello},
  {Bachlechner}, {Barao}, {Barrau}, {Barrin}, {Bartoloni}, {Basara},
  {Battarbee}, {Battiston}, {Bazo}, {Becker}, {Behlmann}, {Beischer},
  {Berdugo}, {Bertucci}, {Bigongiari}, {Bindi}, {Bizzaglia}, {Bizzarri},
  {Boella}, {de Boer}, {Bollweg}, {Bonnivard}, {Borgia}, {Borsini}, {Boschini},
  {Bourquin}, {Burger}, {Cadoux}, {Cai}, {Capell}, {Caroff}, {Casaus},
  {Cascioli}, {Castellini}, {Cernuda}, {Cerreta}, {Cervelli}, {Chae}, {Chang},
  {Chen}, {Chen}, {Cheng}, {Chen}, {Cheng}, {Chou}, {Choumilov}, {Choutko},
  {Chung}, {Clark}, {Clavero}, {Coignet}, {Consolandi}, {Contin}, {Corti},
  {Gil}, {Coste}, {Creus}, {Crispoltoni}, {Cui}, {Dai}, {Delgado}, {Della
  Torre}, {Demirk{\"o}z}, {Derome}, {Di Falco}, {Di Masso}, {Dimiccoli},
  {D{\'\i}az}, {von Doetinchem}, {Donnini}, {Du}, {Duranti}, {D'Urso}, {Eline},
  {Eppling}, {Eronen}, {Fan}, {Farnesini}, {Feng}, {Fiandrini}, {Fiasson},
  {Finch}, {Fisher}, {Galaktionov}, {Gallucci}, {Garc{\'\i}a},
  {Garc{\'\i}a-L{\'o}pez}, {Gargiulo}, {Gast}, {Gebauer}, {Gervasi}, {Ghelfi},
  {Gillard}, {Giovacchini}, {Goglov}, {Gong}, {Goy}, {Grabski}, {Grandi},
  {Graziani}, {Guand alini}, {Guerri}, {Guo}, {Haas}, {Habiby}, {Haino}, {Han},
  {He}, {Heil}, {Hoffman}, {Hsieh}, {Huang}, {Huh}, {Incagli}, {Ionica},
  {Jang}, {Jinchi}, {Kanishev}, {Kim}, {Kim}, {Kirn}, {Kossakowski}, {Kounina},
  {Kounine}, {Koutsenko}, {Krafczyk}, {La Vacca}, {Laudi}, {Laurenti},
  {Lazzizzera}, {Lebedev}, {Lee}, {Lee}, {Leluc}, {Levi}, {Li}, {Li}, {Li},
  {Li}, {Li}, {Li}, {Li}, {Li}, {Li}, {Lim}, {Lin}, {Lipari}, {Lippert}, {Liu},
  {Liu}, {Lolli}, {Lomtadze}, {Lu}, {Lu}, {Lu}, {Luebelsmeyer}, {Luo}, {Lv},
  {Majka}, {Ma{\~n}{\'a}}, {Mar{\'\i}n}, {Martin}, {Mart{\'\i}nez}, {Masi},
  {Maurin}, {Menchaca-Rocha}, {Meng}, {Mo}, {Morescalchi}, {Mott},
  {M{\"u}ller}, {Ni}, {Nikonov}, {Nozzoli}, {Nunes}, {Obermeier}, {Oliva},
  {Orcinha}, {Palmonari}, {Palomares}, {Paniccia}, {Papi}, {Pauluzzi},
  {Pedreschi}, {Pensotti}, {Pereira}, {Picot-Clemente}, {Pilo}, {Piluso},
  {Pizzolotto}, {Plyaskin}, {Pohl}, {Poireau}, {Postaci}, {Putze}, {Quadrani},
  {Qi}, {Qin}, {Qu}, {R{\"a}ih{\"a}}, {Rancoita}, {Rapin}, {Ricol},
  {Rodr{\'\i}guez}, {Rosier-Lees}, {Rozhkov}, {Rozza}, {Sagdeev}, {Sandweiss},
  {Saouter}, {Sbarra}, {Schael}, {Schmidt}, {von Dratzig}, {Schwering},
  {Scolieri}, {Seo}, {Shan}, {Shan}, {Shi}, {Shi}, {Shi}, {Siedenburg}, {Son},
  {Spada}, {Spinella}, {Sun}, {Sun}, {Tacconi}, {Tang}, {Tang}, {Tang}, {Tao},
  {Tescaro}, {Ting}, {Ting}, {Tomassetti}, {Torsti}, {T{\"u}rko{\v{g}}lu},
  {Urban}, {Vagelli}, {Valente}, {Vannini}, {Valtonen}, {Vaurynovich},
  {Vecchi}, {Velasco}, {Vialle}, {Vitale}, {Vitillo}, {Wang}, {Wang}, {Wang},
  {Wang}, {Wang}, {Wang}, {Weng}, {Whitman}, {Wienkenh{\"o}ver}, {Wu}, {Wu},
  {Xia}, {Xie}, {Xie}, {Xiong}, {Xin}, {Xu}, {Xu}, {Yan}, {Yang}, {Yang}, {Ye},
  {Yi}, {Yu}, {Yu}, {Zeissler}, {Zhang}, {Zhang}, {Zhang}, {Zhang}, {Zheng},
  {Zhuang}, {Zhukov}, {Zichichi}, {Zimmermann}, {Zuccon}, {Zurbach} \& {AMS
  Collaboration}]{Aguilar+15p}
{\sc \au{{Aguilar}, M.}, \au{{Aisa}, D.}, \au{{Alpat}, B.}, \au{{Alvino}, A.},
  \au{{Ambrosi}, G.}, \au{{Andeen}, K.}, \au{{Arruda}, L.}, \au{{Attig}, N.},
  \au{{Azzarello}, P.}, \au{{Bachlechner}, A.}, \au{{Barao}, F.}, \au{{Barrau},
  A.}, \au{{Barrin}, L.}, \au{{Bartoloni}, A.}, \au{{Basara}, L.},
  \au{{Battarbee}, M.}, \au{{Battiston}, R.}, \au{{Bazo}, J.}, \au{{Becker},
  U.}, \au{{Behlmann}, M.}, \au{{Beischer}, B.}, \au{{Berdugo}, J.},
  \au{{Bertucci}, B.}, \au{{Bigongiari}, G.}, \au{{Bindi}, V.},
  \au{{Bizzaglia}, S.}, \au{{Bizzarri}, M.}, \au{{Boella}, G.}, \au{{de Boer},
  W.}, \au{{Bollweg}, K.}, \au{{Bonnivard}, V.}, \au{{Borgia}, B.},
  \au{{Borsini}, S.}, \au{{Boschini}, M.~J.}, \au{{Bourquin}, M.},
  \au{{Burger}, J.}, \au{{Cadoux}, F.}, \au{{Cai}, X.~D.}, \au{{Capell}, M.},
  \au{{Caroff}, S.}, \au{{Casaus}, J.}, \au{{Cascioli}, V.}, \au{{Castellini},
  G.}, \au{{Cernuda}, I.}, \au{{Cerreta}, D.}, \au{{Cervelli}, F.}, \au{{Chae},
  M.~J.}, \au{{Chang}, Y.~H.}, \au{{Chen}, A.~I.}, \au{{Chen}, H.},
  \au{{Cheng}, G.~M.}, \au{{Chen}, H.~S.}, \au{{Cheng}, L.}, \au{{Chou},
  H.~Y.}, \au{{Choumilov}, E.}, \au{{Choutko}, V.}, \au{{Chung}, C.~H.},
  \au{{Clark}, C.}, \au{{Clavero}, R.}, \au{{Coignet}, G.}, \au{{Consolandi},
  C.}, \au{{Contin}, A.}, \au{{Corti}, C.}, \au{{Gil}, E.~Cortina},
  \au{{Coste}, B.}, \au{{Creus}, W.}, \au{{Crispoltoni}, M.}, \au{{Cui}, Z.},
  \au{{Dai}, Y.~M.}, \au{{Delgado}, C.}, \au{{Della Torre}, S.},
  \au{{Demirk{\"o}z}, M.~B.}, \au{{Derome}, L.}, \au{{Di Falco}, S.}, \au{{Di
  Masso}, L.}, \au{{Dimiccoli}, F.}, \au{{D{\'\i}az}, C.}, \au{{von
  Doetinchem}, P.}, \au{{Donnini}, F.}, \au{{Du}, W.~J.}, \au{{Duranti}, M.},
  \au{{D'Urso}, D.}, \au{{Eline}, A.}, \au{{Eppling}, F.~J.}, \au{{Eronen},
  T.}, \au{{Fan}, Y.~Y.}, \au{{Farnesini}, L.}, \au{{Feng}, J.},
  \au{{Fiandrini}, E.}, \au{{Fiasson}, A.}, \au{{Finch}, E.}, \au{{Fisher},
  P.}, \au{{Galaktionov}, Y.}, \au{{Gallucci}, G.}, \au{{Garc{\'\i}a}, B.},
  \au{{Garc{\'\i}a-L{\'o}pez}, R.}, \au{{Gargiulo}, C.}, \au{{Gast}, H.},
  \au{{Gebauer}, I.}, \au{{Gervasi}, M.}, \au{{Ghelfi}, A.}, \au{{Gillard},
  W.}, \au{{Giovacchini}, F.}, \au{{Goglov}, P.}, \au{{Gong}, J.}, \au{{Goy},
  C.}, \au{{Grabski}, V.}, \au{{Grandi}, D.}, \au{{Graziani}, M.}, \au{{Guand
  alini}, C.}, \au{{Guerri}, I.}, \au{{Guo}, K.~H.}, \au{{Haas}, D.},
  \au{{Habiby}, M.}, \au{{Haino}, S.}, \au{{Han}, K.~C.}, \au{{He}, Z.~H.},
  \au{{Heil}, M.}, \au{{Hoffman}, J.}, \au{{Hsieh}, T.~H.}, \au{{Huang},
  Z.~C.}, \au{{Huh}, C.}, \au{{Incagli}, M.}, \au{{Ionica}, M.}, \au{{Jang},
  W.~Y.}, \au{{Jinchi}, H.}, \au{{Kanishev}, K.}, \au{{Kim}, G.~N.}, \au{{Kim},
  K.~S.}, \au{{Kirn}, Th.}, \au{{Kossakowski}, R.}, \au{{Kounina}, O.},
  \au{{Kounine}, A.}, \au{{Koutsenko}, V.}, \au{{Krafczyk}, M.~S.}, \au{{La
  Vacca}, G.}, \au{{Laudi}, E.}, \au{{Laurenti}, G.}, \au{{Lazzizzera}, I.},
  \au{{Lebedev}, A.}, \au{{Lee}, H.~T.}, \au{{Lee}, S.~C.}, \au{{Leluc}, C.},
  \au{{Levi}, G.}, \au{{Li}, H.~L.}, \au{{Li}, J.~Q.}, \au{{Li}, Q.}, \au{{Li},
  Q.}, \au{{Li}, T.~X.}, \au{{Li}, W.}, \au{{Li}, Y.}, \au{{Li}, Z.~H.},
  \au{{Li}, Z.~Y.}, \au{{Lim}, S.}, \au{{Lin}, C.~H.}, \au{{Lipari}, P.},
  \au{{Lippert}, T.}, \au{{Liu}, D.}, \au{{Liu}, H.}, \au{{Lolli}, M.},
  \au{{Lomtadze}, T.}, \au{{Lu}, M.~J.}, \au{{Lu}, S.~Q.}, \au{{Lu}, Y.~S.},
  \au{{Luebelsmeyer}, K.}, \au{{Luo}, J.~Z.}, \au{{Lv}, S.~S.}, \au{{Majka},
  R.}, \au{{Ma{\~n}{\'a}}, C.}, \au{{Mar{\'\i}n}, J.}, \au{{Martin}, T.},
  \au{{Mart{\'\i}nez}, G.}, \au{{Masi}, N.}, \au{{Maurin}, D.},
  \au{{Menchaca-Rocha}, A.}, \au{{Meng}, Q.}, \au{{Mo}, D.~C.},
  \au{{Morescalchi}, L.}, \au{{Mott}, P.}, \au{{M{\"u}ller}, M.}, \au{{Ni},
  J.~Q.}, \au{{Nikonov}, N.}, \au{{Nozzoli}, F.}, \au{{Nunes}, P.},
  \au{{Obermeier}, A.}, \au{{Oliva}, A.}, \au{{Orcinha}, M.}, \au{{Palmonari},
  F.}, \au{{Palomares}, C.}, \au{{Paniccia}, M.}, \au{{Papi}, A.},
  \au{{Pauluzzi}, M.}, \au{{Pedreschi}, E.}, \au{{Pensotti}, S.},
  \au{{Pereira}, R.}, \au{{Picot-Clemente}, N.}, \au{{Pilo}, F.}, \au{{Piluso},
  A.}, \au{{Pizzolotto}, C.}, \au{{Plyaskin}, V.}, \au{{Pohl}, M.},
  \au{{Poireau}, V.}, \au{{Postaci}, E.}, \au{{Putze}, A.}, \au{{Quadrani},
  L.}, \au{{Qi}, X.~M.}, \au{{Qin}, X.}, \au{{Qu}, Z.~Y.}, \au{{R{\"a}ih{\"a}},
  T.}, \au{{Rancoita}, P.~G.}, \au{{Rapin}, D.}, \au{{Ricol}, J.~S.},
  \au{{Rodr{\'\i}guez}, I.}, \au{{Rosier-Lees}, S.}, \au{{Rozhkov}, A.},
  \au{{Rozza}, D.}, \au{{Sagdeev}, R.}, \au{{Sandweiss}, J.}, \au{{Saouter},
  P.}, \au{{Sbarra}, C.}, \au{{Schael}, S.}, \au{{Schmidt}, S.~M.}, \au{{von
  Dratzig}, A.~Schulz}, \au{{Schwering}, G.}, \au{{Scolieri}, G.}, \au{{Seo},
  E.~S.}, \au{{Shan}, B.~S.}, \au{{Shan}, Y.~H.}, \au{{Shi}, J.~Y.}, \au{{Shi},
  X.~Y.}, \au{{Shi}, Y.~M.}, \au{{Siedenburg}, T.}, \au{{Son}, D.},
  \au{{Spada}, F.}, \au{{Spinella}, F.}, \au{{Sun}, W.}, \au{{Sun}, W.~H.},
  \au{{Tacconi}, M.}, \au{{Tang}, C.~P.}, \au{{Tang}, X.~W.}, \au{{Tang},
  Z.~C.}, \au{{Tao}, L.}, \au{{Tescaro}, D.}, \au{{Ting}, Samuel C.~C.},
  \au{{Ting}, S.~M.}, \au{{Tomassetti}, N.}, \au{{Torsti}, J.},
  \au{{T{\"u}rko{\v{g}}lu}, C.}, \au{{Urban}, T.}, \au{{Vagelli}, V.},
  \au{{Valente}, E.}, \au{{Vannini}, C.}, \au{{Valtonen}, E.},
  \au{{Vaurynovich}, S.}, \au{{Vecchi}, M.}, \au{{Velasco}, M.}, \au{{Vialle},
  J.~P.}, \au{{Vitale}, V.}, \au{{Vitillo}, S.}, \au{{Wang}, L.~Q.},
  \au{{Wang}, N.~H.}, \au{{Wang}, Q.~L.}, \au{{Wang}, R.~S.}, \au{{Wang}, X.},
  \au{{Wang}, Z.~X.}, \au{{Weng}, Z.~L.}, \au{{Whitman}, K.},
  \au{{Wienkenh{\"o}ver}, J.}, \au{{Wu}, H.}, \au{{Wu}, X.}, \au{{Xia}, X.},
  \au{{Xie}, M.}, \au{{Xie}, S.}, \au{{Xiong}, R.~Q.}, \au{{Xin}, G.~M.},
  \au{{Xu}, N.~S.}, \au{{Xu}, W.}, \au{{Yan}, Q.}, \au{{Yang}, J.}, \au{{Yang},
  M.}, \au{{Ye}, Q.~H.}, \au{{Yi}, H.}, \au{{Yu}, Y.~J.}, \au{{Yu}, Z.~Q.},
  \au{{Zeissler}, S.}, \au{{Zhang}, J.~H.}, \au{{Zhang}, M.~T.}, \au{{Zhang},
  X.~B.}, \au{{Zhang}, Z.}, \au{{Zheng}, Z.~M.}, \au{{Zhuang}, H.~L.},
  \au{{Zhukov}, V.}, \au{{Zichichi}, A.}, \au{{Zimmermann}, N.}, \au{{Zuccon},
  P.}, \au{{Zurbach}, C.} \& \au{{AMS Collaboration}}} \yr{2015{\natexlab{{\em
  c\/}}}}  \at{{Precision Measurement of the Proton Flux in Primary Cosmic Rays
  from Rigidity 1 GV to 1.8 TV with the Alpha Magnetic Spectrometer on the
  International Space Station}}.  \jt{PRL}  \bvol{114}~(17),  \pg{171103}.

\bibitem[{Aguilar} {\em et~al.\/}(2015{\natexlab{{\em d\/}}}){Aguilar}, {Aisa},
  {Alpat}, {Alvino}, {Ambrosi}, {Andeen}, {Arruda}, {Attig}, {Azzarello},
  {Bachlechner}, {Barao}, {Barrau}, {Barrin}, {Bartoloni}, {Basara},
  {Battarbee}, {Battiston}, {Bazo}, {Becker}, {Behlmann}, {Beischer},
  {Berdugo}, {Bertucci}, {Bindi}, {Bizzaglia}, {Bizzarri}, {Boella}, {de Boer},
  {Bollweg}, {Bonnivard}, {Borgia}, {Borsini}, {Boschini}, {Bourquin},
  {Burger}, {Cadoux}, {Cai}, {Capell}, {Caroff}, {Casaus}, {Castellini},
  {Cernuda}, {Cerreta}, {Cervelli}, {Chae}, {Chang}, {Chen}, {Chen}, {Chen},
  {Chen}, {Cheng}, {Chou}, {Choumilov}, {Choutko}, {Chung}, {Clark}, {Clavero},
  {Coignet}, {Consolandi}, {Contin}, {Corti}, {Gil}, {Coste}, {Creus},
  {Crispoltoni}, {Cui}, {Dai}, {Delgado}, {Della Torre}, {Demirk{\"o}z},
  {Derome}, {Di Falco}, {Di Masso}, {Dimiccoli}, {D{\'\i}az}, {von Doetinchem},
  {Donnini}, {Duranti}, {D'Urso}, {Egorov}, {Eline}, {Eppling}, {Eronen},
  {Fan}, {Farnesini}, {Feng}, {Fiandrini}, {Fiasson}, {Finch}, {Fisher},
  {Formato}, {Galaktionov}, {Gallucci}, {Garc{\'\i}a}, {Garc{\'\i}a-L{\'o}pez},
  {Gargiulo}, {Gast}, {Gebauer}, {Gervasi}, {Ghelfi}, {Giovacchini}, {Goglov},
  {Gong}, {Goy}, {Grabski}, {Grandi}, {Graziani}, {Guand alini}, {Guerri},
  {Guo}, {Haas}, {Habiby}, {Haino}, {Han}, {He}, {Heil}, {Hoffman}, {Hsieh},
  {Huang}, {Huh}, {Incagli}, {Ionica}, {Jang}, {Jinchi}, {Kanishev}, {Kim},
  {Kim}, {Kirn}, {Korkmaz}, {Kossakowski}, {Kounina}, {Kounine}, {Koutsenko},
  {Krafczyk}, {La Vacca}, {Laudi}, {Laurenti}, {Lazzizzera}, {Lebedev}, {Lee},
  {Lee}, {Leluc}, {Li}, {Li}, {Li}, {Li}, {Li}, {Li}, {Li}, {Li}, {Li}, {Li},
  {Lim}, {Lin}, {Lipari}, {Lippert}, {Liu}, {Liu}, {Liu}, {Lolli}, {Lomtadze},
  {Lu}, {Lu}, {Lu}, {Luebelsmeyer}, {Luo}, {Luo}, {Lv}, {Majka},
  {Ma{\~n}{\'a}}, {Mar{\'\i}n}, {Martin}, {Mart{\'\i}nez}, {Masi}, {Maurin},
  {Menchaca-Rocha}, {Meng}, {Mo}, {Morescalchi}, {Mott}, {M{\"u}ller},
  {Nelson}, {Ni}, {Nikonov}, {Nozzoli}, {Nunes}, {Obermeier}, {Oliva},
  {Orcinha}, {Palmonari}, {Palomares}, {Paniccia}, {Papi}, {Pauluzzi},
  {Pedreschi}, {Pensotti}, {Pereira}, {Picot-Clemente}, {Pilo}, {Piluso},
  {Pizzolotto}, {Plyaskin}, {Pohl}, {Poireau}, {Putze}, {Quadrani}, {Qi},
  {Qin}, {Qu}, {R{\"a}ih{\"a}}, {Rancoita}, {Rapin}, {Ricol}, {Rodr{\'\i}guez},
  {Rosier-Lees}, {Rozhkov}, {Rozza}, {Sagdeev}, {Sandweiss}, {Saouter},
  {Schael}, {Schmidt}, {von Dratzig}, {Schwering}, {Scolieri}, {Seo}, {Shan},
  {Shan}, {Shi}, {Shi}, {Shi}, {Siedenburg}, {Son}, {Song}, {Spada},
  {Spinella}, {Sun}, {Sun}, {Tacconi}, {Tang}, {Tang}, {Tang}, {Tao},
  {Tescaro}, {Ting}, {Ting}, {Tomassetti}, {Torsti}, {T{\"u}rko{\v{g}}lu},
  {Urban}, {Vagelli}, {Valente}, {Vannini}, {Valtonen}, {Vaurynovich},
  {Vecchi}, {Velasco}, {Vialle}, {Vitale}, {Vitillo}, {Wang}, {Wang}, {Wang},
  {Wang}, {Wang}, {Wang}, {Weng}, {Whitman}, {Wienkenh{\"o}ver}, {Willenbrock},
  {Wu}, {Wu}, {Xia}, {Xie}, {Xie}, {Xiong}, {Xu}, {Xu}, {Yan}, {Yang}, {Yang},
  {Yang}, {Ye}, {Yi}, {Yu}, {Yu}, {Zeissler}, {Zhang}, {Zhang}, {Zhang},
  {Zhang}, {Zhang}, {Zhang}, {Zhang}, {Zheng}, {Zhuang}, {Zhukov}, {Zichichi},
  {Zimmermann}, {Zuccon} \& {AMS Collaboration}]{Aguilar+15He}
{\sc \au{{Aguilar}, M.}, \au{{Aisa}, D.}, \au{{Alpat}, B.}, \au{{Alvino}, A.},
  \au{{Ambrosi}, G.}, \au{{Andeen}, K.}, \au{{Arruda}, L.}, \au{{Attig}, N.},
  \au{{Azzarello}, P.}, \au{{Bachlechner}, A.}, \au{{Barao}, F.}, \au{{Barrau},
  A.}, \au{{Barrin}, L.}, \au{{Bartoloni}, A.}, \au{{Basara}, L.},
  \au{{Battarbee}, M.}, \au{{Battiston}, R.}, \au{{Bazo}, J.}, \au{{Becker},
  U.}, \au{{Behlmann}, M.}, \au{{Beischer}, B.}, \au{{Berdugo}, J.},
  \au{{Bertucci}, B.}, \au{{Bindi}, V.}, \au{{Bizzaglia}, S.}, \au{{Bizzarri},
  M.}, \au{{Boella}, G.}, \au{{de Boer}, W.}, \au{{Bollweg}, K.},
  \au{{Bonnivard}, V.}, \au{{Borgia}, B.}, \au{{Borsini}, S.}, \au{{Boschini},
  M.~J.}, \au{{Bourquin}, M.}, \au{{Burger}, J.}, \au{{Cadoux}, F.}, \au{{Cai},
  X.~D.}, \au{{Capell}, M.}, \au{{Caroff}, S.}, \au{{Casaus}, J.},
  \au{{Castellini}, G.}, \au{{Cernuda}, I.}, \au{{Cerreta}, D.},
  \au{{Cervelli}, F.}, \au{{Chae}, M.~J.}, \au{{Chang}, Y.~H.}, \au{{Chen},
  A.~I.}, \au{{Chen}, G.~M.}, \au{{Chen}, H.}, \au{{Chen}, H.~S.}, \au{{Cheng},
  L.}, \au{{Chou}, H.~Y.}, \au{{Choumilov}, E.}, \au{{Choutko}, V.},
  \au{{Chung}, C.~H.}, \au{{Clark}, C.}, \au{{Clavero}, R.}, \au{{Coignet},
  G.}, \au{{Consolandi}, C.}, \au{{Contin}, A.}, \au{{Corti}, C.}, \au{{Gil},
  E.~Cortina}, \au{{Coste}, B.}, \au{{Creus}, W.}, \au{{Crispoltoni}, M.},
  \au{{Cui}, Z.}, \au{{Dai}, Y.~M.}, \au{{Delgado}, C.}, \au{{Della Torre},
  S.}, \au{{Demirk{\"o}z}, M.~B.}, \au{{Derome}, L.}, \au{{Di Falco}, S.},
  \au{{Di Masso}, L.}, \au{{Dimiccoli}, F.}, \au{{D{\'\i}az}, C.}, \au{{von
  Doetinchem}, P.}, \au{{Donnini}, F.}, \au{{Duranti}, M.}, \au{{D'Urso}, D.},
  \au{{Egorov}, A.}, \au{{Eline}, A.}, \au{{Eppling}, F.~J.}, \au{{Eronen},
  T.}, \au{{Fan}, Y.~Y.}, \au{{Farnesini}, L.}, \au{{Feng}, J.},
  \au{{Fiandrini}, E.}, \au{{Fiasson}, A.}, \au{{Finch}, E.}, \au{{Fisher},
  P.}, \au{{Formato}, V.}, \au{{Galaktionov}, Y.}, \au{{Gallucci}, G.},
  \au{{Garc{\'\i}a}, B.}, \au{{Garc{\'\i}a-L{\'o}pez}, R.}, \au{{Gargiulo},
  C.}, \au{{Gast}, H.}, \au{{Gebauer}, I.}, \au{{Gervasi}, M.}, \au{{Ghelfi},
  A.}, \au{{Giovacchini}, F.}, \au{{Goglov}, P.}, \au{{Gong}, J.}, \au{{Goy},
  C.}, \au{{Grabski}, V.}, \au{{Grandi}, D.}, \au{{Graziani}, M.}, \au{{Guand
  alini}, C.}, \au{{Guerri}, I.}, \au{{Guo}, K.~H.}, \au{{Haas}, D.},
  \au{{Habiby}, M.}, \au{{Haino}, S.}, \au{{Han}, K.~C.}, \au{{He}, Z.~H.},
  \au{{Heil}, M.}, \au{{Hoffman}, J.}, \au{{Hsieh}, T.~H.}, \au{{Huang},
  Z.~C.}, \au{{Huh}, C.}, \au{{Incagli}, M.}, \au{{Ionica}, M.}, \au{{Jang},
  W.~Y.}, \au{{Jinchi}, H.}, \au{{Kanishev}, K.}, \au{{Kim}, G.~N.}, \au{{Kim},
  K.~S.}, \au{{Kirn}, Th.}, \au{{Korkmaz}, M.~A.}, \au{{Kossakowski}, R.},
  \au{{Kounina}, O.}, \au{{Kounine}, A.}, \au{{Koutsenko}, V.}, \au{{Krafczyk},
  M.~S.}, \au{{La Vacca}, G.}, \au{{Laudi}, E.}, \au{{Laurenti}, G.},
  \au{{Lazzizzera}, I.}, \au{{Lebedev}, A.}, \au{{Lee}, H.~T.}, \au{{Lee},
  S.~C.}, \au{{Leluc}, C.}, \au{{Li}, H.~L.}, \au{{Li}, J.~Q.}, \au{{Li},
  J.~Q.}, \au{{Li}, Q.}, \au{{Li}, Q.}, \au{{Li}, T.~X.}, \au{{Li}, W.},
  \au{{Li}, Y.}, \au{{Li}, Z.~H.}, \au{{Li}, Z.~Y.}, \au{{Lim}, S.}, \au{{Lin},
  C.~H.}, \au{{Lipari}, P.}, \au{{Lippert}, T.}, \au{{Liu}, D.}, \au{{Liu},
  H.}, \au{{Liu}, Hu}, \au{{Lolli}, M.}, \au{{Lomtadze}, T.}, \au{{Lu}, M.~J.},
  \au{{Lu}, S.~Q.}, \au{{Lu}, Y.~S.}, \au{{Luebelsmeyer}, K.}, \au{{Luo}, F.},
  \au{{Luo}, J.~Z.}, \au{{Lv}, S.~S.}, \au{{Majka}, R.}, \au{{Ma{\~n}{\'a}},
  C.}, \au{{Mar{\'\i}n}, J.}, \au{{Martin}, T.}, \au{{Mart{\'\i}nez}, G.},
  \au{{Masi}, N.}, \au{{Maurin}, D.}, \au{{Menchaca-Rocha}, A.}, \au{{Meng},
  Q.}, \au{{Mo}, D.~C.}, \au{{Morescalchi}, L.}, \au{{Mott}, P.},
  \au{{M{\"u}ller}, M.}, \au{{Nelson}, T.}, \au{{Ni}, J.~Q.}, \au{{Nikonov},
  N.}, \au{{Nozzoli}, F.}, \au{{Nunes}, P.}, \au{{Obermeier}, A.}, \au{{Oliva},
  A.}, \au{{Orcinha}, M.}, \au{{Palmonari}, F.}, \au{{Palomares}, C.},
  \au{{Paniccia}, M.}, \au{{Papi}, A.}, \au{{Pauluzzi}, M.}, \au{{Pedreschi},
  E.}, \au{{Pensotti}, S.}, \au{{Pereira}, R.}, \au{{Picot-Clemente}, N.},
  \au{{Pilo}, F.}, \au{{Piluso}, A.}, \au{{Pizzolotto}, C.}, \au{{Plyaskin},
  V.}, \au{{Pohl}, M.}, \au{{Poireau}, V.}, \au{{Putze}, A.}, \au{{Quadrani},
  L.}, \au{{Qi}, X.~M.}, \au{{Qin}, X.}, \au{{Qu}, Z.~Y.}, \au{{R{\"a}ih{\"a}},
  T.}, \au{{Rancoita}, P.~G.}, \au{{Rapin}, D.}, \au{{Ricol}, J.~S.},
  \au{{Rodr{\'\i}guez}, I.}, \au{{Rosier-Lees}, S.}, \au{{Rozhkov}, A.},
  \au{{Rozza}, D.}, \au{{Sagdeev}, R.}, \au{{Sandweiss}, J.}, \au{{Saouter},
  P.}, \au{{Schael}, S.}, \au{{Schmidt}, S.~M.}, \au{{von Dratzig}, A.~Schulz},
  \au{{Schwering}, G.}, \au{{Scolieri}, G.}, \au{{Seo}, E.~S.}, \au{{Shan},
  B.~S.}, \au{{Shan}, Y.~H.}, \au{{Shi}, J.~Y.}, \au{{Shi}, X.~Y.}, \au{{Shi},
  Y.~M.}, \au{{Siedenburg}, T.}, \au{{Son}, D.}, \au{{Song}, J.~W.},
  \au{{Spada}, F.}, \au{{Spinella}, F.}, \au{{Sun}, W.}, \au{{Sun}, W.~H.},
  \au{{Tacconi}, M.}, \au{{Tang}, C.~P.}, \au{{Tang}, X.~W.}, \au{{Tang},
  Z.~C.}, \au{{Tao}, L.}, \au{{Tescaro}, D.}, \au{{Ting}, Samuel C.~C.},
  \au{{Ting}, S.~M.}, \au{{Tomassetti}, N.}, \au{{Torsti}, J.},
  \au{{T{\"u}rko{\v{g}}lu}, C.}, \au{{Urban}, T.}, \au{{Vagelli}, V.},
  \au{{Valente}, E.}, \au{{Vannini}, C.}, \au{{Valtonen}, E.},
  \au{{Vaurynovich}, S.}, \au{{Vecchi}, M.}, \au{{Velasco}, M.}, \au{{Vialle},
  J.~P.}, \au{{Vitale}, V.}, \au{{Vitillo}, S.}, \au{{Wang}, L.~Q.},
  \au{{Wang}, N.~H.}, \au{{Wang}, Q.~L.}, \au{{Wang}, R.~S.}, \au{{Wang}, X.},
  \au{{Wang}, Z.~X.}, \au{{Weng}, Z.~L.}, \au{{Whitman}, K.},
  \au{{Wienkenh{\"o}ver}, J.}, \au{{Willenbrock}, M.}, \au{{Wu}, H.}, \au{{Wu},
  X.}, \au{{Xia}, X.}, \au{{Xie}, M.}, \au{{Xie}, S.}, \au{{Xiong}, R.~Q.},
  \au{{Xu}, N.~S.}, \au{{Xu}, W.}, \au{{Yan}, Q.}, \au{{Yang}, J.}, \au{{Yang},
  M.}, \au{{Yang}, Y.}, \au{{Ye}, Q.~H.}, \au{{Yi}, H.}, \au{{Yu}, Y.~J.},
  \au{{Yu}, Z.~Q.}, \au{{Zeissler}, S.}, \au{{Zhang}, C.}, \au{{Zhang}, J.~H.},
  \au{{Zhang}, M.~T.}, \au{{Zhang}, S.~D.}, \au{{Zhang}, S.~W.}, \au{{Zhang},
  X.~B.}, \au{{Zhang}, Z.}, \au{{Zheng}, Z.~M.}, \au{{Zhuang}, H.~L.},
  \au{{Zhukov}, V.}, \au{{Zichichi}, A.}, \au{{Zimmermann}, N.}, \au{{Zuccon},
  P.} \& \au{{AMS Collaboration}}} \yr{2015{\natexlab{{\em d\/}}}}
  \at{{Precision Measurement of the Helium Flux in Primary Cosmic Rays of
  Rigidities 1.9 GV to 3 TV with the Alpha Magnetic Spectrometer on the
  International Space Station}}.  \jt{PRL}  \bvol{115}~(21),  \pg{211101}.

\bibitem[{Aguilar} {\em et~al.\/}(2013{\natexlab{{\em a\/}}}){Aguilar},
  {Alberti}, {Alpat}, {Alvino}, {Ambrosi}, {Andeen}, {Anderhub}, {Arruda},
  {Azzarello}, {Bachlechner}, {Barao}, {Baret}, {Barrau}, {Barrin},
  {Bartoloni}, {Basara}, {Basili}, {Batalha}, {Bates}, {Battiston}, {Bazo},
  {Becker}, {Becker}, {Behlmann}, {Beischer}, {Berdugo}, {Berges}, {Bertucci},
  {Bigongiari}, {Biland}, {Bindi}, {Bizzaglia}, {Boella}, {de Boer}, {Bollweg},
  {Bolmont}, {Borgia}, {Borsini}, {Boschini}, {Boudoul}, {Bourquin}, {Brun},
  {Bu{\'e}nerd}, {Burger}, {Burger}, {Cadoux}, {Cai}, {Capell}, {Casadei},
  {Casaus}, {Cascioli}, {Castellini}, {Cernuda}, {Cervelli}, {Chae}, {Chang},
  {Chen}, {Chen}, {Chen}, {Cheng}, {Chen}, {Cheng}, {Chernoplyiokov},
  {Chikanian}, {Choumilov}, {Choutko}, {Chung}, {Clark}, {Clavero}, {Coignet},
  {Commichau}, {Consolandi}, {Contin}, {Corti}, {Costado Dios}, {Coste},
  {Crespo}, {Cui}, {Dai}, {Delgado}, {Della Torre}, {Demirkoz}, {Dennett},
  {Derome}, {Di Falco}, {Diao}, {Diago}, {Djambazov}, {D{\'\i}az}, {von
  Doetinchem}, {Du}, {Dubois}, {Duperay}, {Duranti}, {D'Urso}, {Egorov},
  {Eline}, {Eppling}, {Eronen}, {van Es}, {Esser}, {Falvard}, {Fiandrini},
  {Fiasson}, {Finch}, {Fisher}, {Flood}, {Foglio}, {Fohey}, {Fopp}, {Fouque},
  {Galaktionov}, {Gallilee}, {Gallin-Martel}, {Gallucci}, {Garc{\'\i}a},
  {Garc{\'\i}a}, {Garc{\'\i}a-L{\'o}pez}, {Garc{\'\i}a-Tabares}, {Gargiulo},
  {Gast}, {Gebauer}, {Gentile}, {Gervasi}, {Gillard}, {Giovacchini}, {Girard},
  {Goglov}, {Gong}, {Goy-Henningsen}, {Grandi}, {Graziani}, {Grechko}, {Gross},
  {Guerri}, {de la Gu{\'\i}a}, {Guo}, {Habiby}, {Haino}, {Hauler}, {He},
  {Heil}, {Heilig}, {Hermel}, {Hofer}, {Huang}, {Hungerford}, {Incagli},
  {Ionica}, {Jacholkowska}, {Jang}, {Jinchi}, {Jongmanns}, {Journet},
  {Jungermann}, {Karpinski}, {Kim}, {Kim}, {Kirn}, {Kossakowski}, {Koulemzine},
  {Kounina}, {Kounine}, {Koutsenko}, {Krafczyk}, {Laudi}, {Laurenti},
  {Lauritzen}, {Lebedev}, {Lee}, {Lee}, {Leluc}, {Le{\'o}n Vargas}, {Lepareur},
  {Li}, {Li}, {Li}, {Li}, {Li}, {Lipari}, {Lin}, {Liu}, {Liu}, {Lomtadze},
  {Lu}, {Lucidi}, {L{\"u}belsmeyer}, {Luo}, {Lustermann}, {Lv}, {Madsen},
  {Majka}, {Malinin}, {Ma{\~n}{\'a}}, {Mar{\'\i}n}, {Martin}, {Mart{\'\i}nez},
  {Masciocchi}, {Masi}, {Maurin}, {McInturff}, {McIntyre}, {Menchaca-Rocha},
  {Meng}, {Menichelli}, {Mereu}, {Millinger}, {Mo}, {Molina}, {Mott},
  {Mujunen}, {Natale}, {Nemeth}, {Ni}, {Nikonov}, {Nozzoli}, {Nunes},
  {Obermeier}, {Oh}, {Oliva}, {Palmonari}, {Palomares}, {Paniccia}, {Papi},
  {Park}, {Pauluzzi}, {Pauss}, {Pauw}, {Pedreschi}, {Pensotti}, {Pereira},
  {Perrin}, {Pessina}, {Pierschel}, {Pilo}, {Piluso}, {Pizzolotto}, {Plyaskin},
  {Pochon}, {Pohl}, {Poireau}, {Porter}, {Pouxe}, {Putze}, {Quadrani}, {Qi},
  {Rancoita}, {Rapin}, {Ren}, {Ricol}, {Riihonen}, {Rodr{\'\i}guez}, {Roeser},
  {Rosier-Lees}, {Rossi}, {Rozhkov}, {Rozza}, {Sabellek}, {Sagdeev},
  {Sandweiss}, {Santos}, {Saouter}, {Sarchioni}, {Schael}, {Schinzel},
  {Schmanau}, {Schwering}, {Schulz von Dratzig}, {Scolieri}, {Seo}, {Shan},
  {Shi}, {Shi}, {Siedenburg}, {Siedling}, {Son}, {Spada}, {Spinella}, {Steuer},
  {Stiff}, {Sun}, {Sun}, {Sun}, {Tacconi}, {Tang}, {Tang}, {Tang}, {Tao},
  {Tassan-Viol}, {Ting}, {Ting}, {Titus}, {Tomassetti}, {Toral}, {Torsti},
  {Tsai}, {Tutt}, {Ulbricht}, {Urban}, {Vagelli}, {Valente}, {Vannini},
  {Valtonen}, {Vargas Trevino}, {Vaurynovich}, {Vecchi}, {Vergain}, {Verlaat},
  {Vescovi}, {Vialle}, {Viertel}, {Volpini}, {Wang}, {Wang}, {Wang}, {Wang},
  {Wang}, {Wang}, {Wallraff}, {Weng}, {Willenbrock}, {Wlochal}, {Wu}, {Wu},
  {Wu}, {Xiao}, {Xie}, {Xiong}, {Xin}, {Xu}, {Xu}, {Yan}, {Yang}, {Yang}, {Ye},
  {Yi}, {Yu}, {Yu}, {Zeissler}, {Zhang}, {Zhang}, {Zhang}, {Zheng}, {Zhuang},
  {Zhukov}, {Zichichi}, {Zuccon} \& {Zurbach}]{AMS02frac}
{\sc \au{{Aguilar}, M.}, \au{{Alberti}, G.}, \au{{Alpat}, B.}, \au{{Alvino},
  A.}, \au{{Ambrosi}, G.}, \au{{Andeen}, K.}, \au{{Anderhub}, H.},
  \au{{Arruda}, L.}, \au{{Azzarello}, P.}, \au{{Bachlechner}, A.}, \au{{Barao},
  F.}, \au{{Baret}, B.}, \au{{Barrau}, A.}, \au{{Barrin}, L.}, \au{{Bartoloni},
  A.}, \au{{Basara}, L.}, \au{{Basili}, A.}, \au{{Batalha}, L.}, \au{{Bates},
  J.}, \au{{Battiston}, R.}, \au{{Bazo}, J.}, \au{{Becker}, R.}, \au{{Becker},
  U.}, \au{{Behlmann}, M.}, \au{{Beischer}, B.}, \au{{Berdugo}, J.},
  \au{{Berges}, P.}, \au{{Bertucci}, B.}, \au{{Bigongiari}, G.}, \au{{Biland},
  A.}, \au{{Bindi}, V.}, \au{{Bizzaglia}, S.}, \au{{Boella}, G.}, \au{{de
  Boer}, W.}, \au{{Bollweg}, K.}, \au{{Bolmont}, J.}, \au{{Borgia}, B.},
  \au{{Borsini}, S.}, \au{{Boschini}, M.~J.}, \au{{Boudoul}, G.},
  \au{{Bourquin}, M.}, \au{{Brun}, P.}, \au{{Bu{\'e}nerd}, M.}, \au{{Burger},
  J.}, \au{{Burger}, W.}, \au{{Cadoux}, F.}, \au{{Cai}, X.~D.}, \au{{Capell},
  M.}, \au{{Casadei}, D.}, \au{{Casaus}, J.}, \au{{Cascioli}, V.},
  \au{{Castellini}, G.}, \au{{Cernuda}, I.}, \au{{Cervelli}, F.}, \au{{Chae},
  M.~J.}, \au{{Chang}, Y.~H.}, \au{{Chen}, A.~I.}, \au{{Chen}, C.~R.},
  \au{{Chen}, H.}, \au{{Cheng}, G.~M.}, \au{{Chen}, H.~S.}, \au{{Cheng}, L.},
  \au{{Chernoplyiokov}, N.}, \au{{Chikanian}, A.}, \au{{Choumilov}, E.},
  \au{{Choutko}, V.}, \au{{Chung}, C.~H.}, \au{{Clark}, C.}, \au{{Clavero},
  R.}, \au{{Coignet}, G.}, \au{{Commichau}, V.}, \au{{Consolandi}, C.},
  \au{{Contin}, A.}, \au{{Corti}, C.}, \au{{Costado Dios}, M.~T.}, \au{{Coste},
  B.}, \au{{Crespo}, D.}, \au{{Cui}, Z.}, \au{{Dai}, M.}, \au{{Delgado}, C.},
  \au{{Della Torre}, S.}, \au{{Demirkoz}, B.}, \au{{Dennett}, P.},
  \au{{Derome}, L.}, \au{{Di Falco}, S.}, \au{{Diao}, X.~H.}, \au{{Diago}, A.},
  \au{{Djambazov}, L.}, \au{{D{\'\i}az}, C.}, \au{{von Doetinchem}, P.},
  \au{{Du}, W.~J.}, \au{{Dubois}, J.~M.}, \au{{Duperay}, R.}, \au{{Duranti},
  M.}, \au{{D'Urso}, D.}, \au{{Egorov}, A.}, \au{{Eline}, A.}, \au{{Eppling},
  F.~J.}, \au{{Eronen}, T.}, \au{{van Es}, J.}, \au{{Esser}, H.},
  \au{{Falvard}, A.}, \au{{Fiandrini}, E.}, \au{{Fiasson}, A.}, \au{{Finch},
  E.}, \au{{Fisher}, P.}, \au{{Flood}, K.}, \au{{Foglio}, R.}, \au{{Fohey},
  M.}, \au{{Fopp}, S.}, \au{{Fouque}, N.}, \au{{Galaktionov}, Y.},
  \au{{Gallilee}, M.}, \au{{Gallin-Martel}, L.}, \au{{Gallucci}, G.},
  \au{{Garc{\'\i}a}, B.}, \au{{Garc{\'\i}a}, J.}, \au{{Garc{\'\i}a-L{\'o}pez},
  R.}, \au{{Garc{\'\i}a-Tabares}, L.}, \au{{Gargiulo}, C.}, \au{{Gast}, H.},
  \au{{Gebauer}, I.}, \au{{Gentile}, S.}, \au{{Gervasi}, M.}, \au{{Gillard},
  W.}, \au{{Giovacchini}, F.}, \au{{Girard}, L.}, \au{{Goglov}, P.},
  \au{{Gong}, J.}, \au{{Goy-Henningsen}, C.}, \au{{Grandi}, D.},
  \au{{Graziani}, M.}, \au{{Grechko}, A.}, \au{{Gross}, A.}, \au{{Guerri}, I.},
  \au{{de la Gu{\'\i}a}, C.}, \au{{Guo}, K.~H.}, \au{{Habiby}, M.},
  \au{{Haino}, S.}, \au{{Hauler}, F.}, \au{{He}, Z.~H.}, \au{{Heil}, M.},
  \au{{Heilig}, J.}, \au{{Hermel}, R.}, \au{{Hofer}, H.}, \au{{Huang}, Z.~C.},
  \au{{Hungerford}, W.}, \au{{Incagli}, M.}, \au{{Ionica}, M.},
  \au{{Jacholkowska}, A.}, \au{{Jang}, W.~Y.}, \au{{Jinchi}, H.},
  \au{{Jongmanns}, M.}, \au{{Journet}, L.}, \au{{Jungermann}, L.},
  \au{{Karpinski}, W.}, \au{{Kim}, G.~N.}, \au{{Kim}, K.~S.}, \au{{Kirn}, Th.},
  \au{{Kossakowski}, R.}, \au{{Koulemzine}, A.}, \au{{Kounina}, O.},
  \au{{Kounine}, A.}, \au{{Koutsenko}, V.}, \au{{Krafczyk}, M.~S.},
  \au{{Laudi}, E.}, \au{{Laurenti}, G.}, \au{{Lauritzen}, C.}, \au{{Lebedev},
  A.}, \au{{Lee}, M.~W.}, \au{{Lee}, S.~C.}, \au{{Leluc}, C.}, \au{{Le{\'o}n
  Vargas}, H.}, \au{{Lepareur}, V.}, \au{{Li}, J.~Q.}, \au{{Li}, Q.}, \au{{Li},
  T.~X.}, \au{{Li}, W.}, \au{{Li}, Z.~H.}, \au{{Lipari}, P.}, \au{{Lin},
  C.~H.}, \au{{Liu}, D.}, \au{{Liu}, H.}, \au{{Lomtadze}, T.}, \au{{Lu},
  Y.~S.}, \au{{Lucidi}, S.}, \au{{L{\"u}belsmeyer}, K.}, \au{{Luo}, J.~Z.},
  \au{{Lustermann}, W.}, \au{{Lv}, S.}, \au{{Madsen}, J.}, \au{{Majka}, R.},
  \au{{Malinin}, A.}, \au{{Ma{\~n}{\'a}}, C.}, \au{{Mar{\'\i}n}, J.},
  \au{{Martin}, T.}, \au{{Mart{\'\i}nez}, G.}, \au{{Masciocchi}, F.},
  \au{{Masi}, N.}, \au{{Maurin}, D.}, \au{{McInturff}, A.}, \au{{McIntyre},
  P.}, \au{{Menchaca-Rocha}, A.}, \au{{Meng}, Q.}, \au{{Menichelli}, M.},
  \au{{Mereu}, I.}, \au{{Millinger}, M.}, \au{{Mo}, D.~C.}, \au{{Molina}, M.},
  \au{{Mott}, P.}, \au{{Mujunen}, A.}, \au{{Natale}, S.}, \au{{Nemeth}, P.},
  \au{{Ni}, J.~Q.}, \au{{Nikonov}, N.}, \au{{Nozzoli}, F.}, \au{{Nunes}, P.},
  \au{{Obermeier}, A.}, \au{{Oh}, S.}, \au{{Oliva}, A.}, \au{{Palmonari}, F.},
  \au{{Palomares}, C.}, \au{{Paniccia}, M.}, \au{{Papi}, A.}, \au{{Park},
  W.~H.}, \au{{Pauluzzi}, M.}, \au{{Pauss}, F.}, \au{{Pauw}, A.},
  \au{{Pedreschi}, E.}, \au{{Pensotti}, S.}, \au{{Pereira}, R.}, \au{{Perrin},
  E.}, \au{{Pessina}, G.}, \au{{Pierschel}, G.}, \au{{Pilo}, F.}, \au{{Piluso},
  A.}, \au{{Pizzolotto}, C.}, \au{{Plyaskin}, V.}, \au{{Pochon}, J.},
  \au{{Pohl}, M.}, \au{{Poireau}, V.}, \au{{Porter}, S.}, \au{{Pouxe}, J.},
  \au{{Putze}, A.}, \au{{Quadrani}, L.}, \au{{Qi}, X.~N.}, \au{{Rancoita},
  P.~G.}, \au{{Rapin}, D.}, \au{{Ren}, Z.~L.}, \au{{Ricol}, J.~S.},
  \au{{Riihonen}, E.}, \au{{Rodr{\'\i}guez}, I.}, \au{{Roeser}, U.},
  \au{{Rosier-Lees}, S.}, \au{{Rossi}, L.}, \au{{Rozhkov}, A.}, \au{{Rozza},
  D.}, \au{{Sabellek}, A.}, \au{{Sagdeev}, R.}, \au{{Sandweiss}, J.},
  \au{{Santos}, B.}, \au{{Saouter}, P.}, \au{{Sarchioni}, M.}, \au{{Schael},
  S.}, \au{{Schinzel}, D.}, \au{{Schmanau}, M.}, \au{{Schwering}, G.},
  \au{{Schulz von Dratzig}, A.}, \au{{Scolieri}, G.}, \au{{Seo}, E.~S.},
  \au{{Shan}, B.~S.}, \au{{Shi}, J.~Y.}, \au{{Shi}, Y.~M.}, \au{{Siedenburg},
  T.}, \au{{Siedling}, R.}, \au{{Son}, D.}, \au{{Spada}, F.}, \au{{Spinella},
  F.}, \au{{Steuer}, M.}, \au{{Stiff}, K.}, \au{{Sun}, W.}, \au{{Sun}, W.~H.},
  \au{{Sun}, X.~H.}, \au{{Tacconi}, M.}, \au{{Tang}, C.~P.}, \au{{Tang},
  X.~W.}, \au{{Tang}, Z.~C.}, \au{{Tao}, L.}, \au{{Tassan-Viol}, J.},
  \au{{Ting}, Samuel C.~C.}, \au{{Ting}, S.~M.}, \au{{Titus}, C.},
  \au{{Tomassetti}, N.}, \au{{Toral}, F.}, \au{{Torsti}, J.}, \au{{Tsai},
  J.~R.}, \au{{Tutt}, J.~C.}, \au{{Ulbricht}, J.}, \au{{Urban}, T.},
  \au{{Vagelli}, V.}, \au{{Valente}, E.}, \au{{Vannini}, C.}, \au{{Valtonen},
  E.}, \au{{Vargas Trevino}, M.}, \au{{Vaurynovich}, S.}, \au{{Vecchi}, M.},
  \au{{Vergain}, M.}, \au{{Verlaat}, B.}, \au{{Vescovi}, C.}, \au{{Vialle},
  J.~P.}, \au{{Viertel}, G.}, \au{{Volpini}, G.}, \au{{Wang}, D.}, \au{{Wang},
  N.~H.}, \au{{Wang}, Q.~L.}, \au{{Wang}, R.~S.}, \au{{Wang}, X.}, \au{{Wang},
  Z.~X.}, \au{{Wallraff}, W.}, \au{{Weng}, Z.~L.}, \au{{Willenbrock}, M.},
  \au{{Wlochal}, M.}, \au{{Wu}, H.}, \au{{Wu}, K.~Y.}, \au{{Wu}, Z.~S.},
  \au{{Xiao}, W.~J.}, \au{{Xie}, S.}, \au{{Xiong}, R.~Q.}, \au{{Xin}, G.~M.},
  \au{{Xu}, N.~S.}, \au{{Xu}, W.}, \au{{Yan}, Q.}, \au{{Yang}, J.}, \au{{Yang},
  M.}, \au{{Ye}, Q.~H.}, \au{{Yi}, H.}, \au{{Yu}, Y.~J.}, \au{{Yu}, Z.~Q.},
  \au{{Zeissler}, S.}, \au{{Zhang}, J.~G.}, \au{{Zhang}, Z.}, \au{{Zhang},
  M.~M.}, \au{{Zheng}, Z.~M.}, \au{{Zhuang}, H.~L.}, \au{{Zhukov}, V.},
  \au{{Zichichi}, A.}, \au{{Zuccon}, P.} \& \au{{Zurbach}, C.}}
  \yr{2013{\natexlab{{\em a\/}}}}  \at{{First Result from the Alpha Magnetic
  Spectrometer on the International Space Station: Precision Measurement of the
  Positron Fraction in Primary Cosmic Rays of 0.5-350 GeV}}.  \jt{Physical
  Review Letters}  \bvol{110}~(14),  \pg{141102}.

\bibitem[{Aguilar} {\em et~al.\/}(2013{\natexlab{{\em b\/}}}){Aguilar},
  {Alberti}, {Alpat}, {Alvino}, {Ambrosi}, {Andeen}, {Anderhub}, {Arruda},
  {Azzarello}, {Bachlechner}, {Barao}, {Baret}, {Barrau}, {Barrin},
  {Bartoloni}, {Basara}, {Basili}, {Batalha}, {Bates}, {Battiston}, {Bazo},
  {Becker}, {Becker}, {Behlmann}, {Beischer}, {Berdugo}, {Berges}, {Bertucci},
  {Bigongiari}, {Biland}, {Bindi}, {Bizzaglia}, {Boella}, {de Boer}, {Bollweg},
  {Bolmont}, {Borgia}, {Borsini}, {Boschini}, {Boudoul}, {Bourquin}, {Brun},
  {Bu{\'e}nerd}, {Burger}, {Burger}, {Cadoux}, {Cai}, {Capell}, {Casadei},
  {Casaus}, {Cascioli}, {Castellini}, {Cernuda}, {Cervelli}, {Chae}, {Chang},
  {Chen}, {Chen}, {Chen}, {Cheng}, {Chen}, {Cheng}, {Chernoplyiokov},
  {Chikanian}, {Choumilov}, {Choutko}, {Chung}, {Clark}, {Clavero}, {Coignet},
  {Commichau}, {Consolandi}, {Contin}, {Corti}, {Costado Dios}, {Coste},
  {Crespo}, {Cui}, {Dai}, {Delgado}, {Della Torre}, {Demirkoz}, {Dennett},
  {Derome}, {Di Falco}, {Diao}, {Diago}, {Djambazov}, {D{\'\i}az}, {von
  Doetinchem}, {Du}, {Dubois}, {Duperay}, {Duranti}, {D'Urso}, {Egorov},
  {Eline}, {Eppling}, {Eronen}, {van Es}, {Esser}, {Falvard}, {Fiandrini},
  {Fiasson}, {Finch}, {Fisher}, {Flood}, {Foglio}, {Fohey}, {Fopp}, {Fouque},
  {Galaktionov}, {Gallilee}, {Gallin-Martel}, {Gallucci}, {Garc{\'\i}a},
  {Garc{\'\i}a}, {Garc{\'\i}a-L{\'o}pez}, {Garc{\'\i}a-Tabares}, {Gargiulo},
  {Gast}, {Gebauer}, {Gentile}, {Gervasi}, {Gillard}, {Giovacchini}, {Girard},
  {Goglov}, {Gong}, {Goy-Henningsen}, {Grandi}, {Graziani}, {Grechko}, {Gross},
  {Guerri}, {de la Gu{\'\i}a}, {Guo}, {Habiby}, {Haino}, {Hauler}, {He},
  {Heil}, {Heilig}, {Hermel}, {Hofer}, {Huang}, {Hungerford}, {Incagli},
  {Ionica}, {Jacholkowska}, {Jang}, {Jinchi}, {Jongmanns}, {Journet},
  {Jungermann}, {Karpinski}, {Kim}, {Kim}, {Kirn}, {Kossakowski}, {Koulemzine},
  {Kounina}, {Kounine}, {Koutsenko}, {Krafczyk}, {Laudi}, {Laurenti},
  {Lauritzen}, {Lebedev}, {Lee}, {Lee}, {Leluc}, {Le{\'o}n Vargas}, {Lepareur},
  {Li}, {Li}, {Li}, {Li}, {Li}, {Lipari}, {Lin}, {Liu}, {Liu}, {Lomtadze},
  {Lu}, {Lucidi}, {L{\"u}belsmeyer}, {Luo}, {Lustermann}, {Lv}, {Madsen},
  {Majka}, {Malinin}, {Ma{\~n}{\'a}}, {Mar{\'\i}n}, {Martin}, {Mart{\'\i}nez},
  {Masciocchi}, {Masi}, {Maurin}, {McInturff}, {McIntyre}, {Menchaca-Rocha},
  {Meng}, {Menichelli}, {Mereu}, {Millinger}, {Mo}, {Molina}, {Mott},
  {Mujunen}, {Natale}, {Nemeth}, {Ni}, {Nikonov}, {Nozzoli}, {Nunes},
  {Obermeier}, {Oh}, {Oliva}, {Palmonari}, {Palomares}, {Paniccia}, {Papi},
  {Park}, {Pauluzzi}, {Pauss}, {Pauw}, {Pedreschi}, {Pensotti}, {Pereira},
  {Perrin}, {Pessina}, {Pierschel}, {Pilo}, {Piluso}, {Pizzolotto}, {Plyaskin},
  {Pochon}, {Pohl}, {Poireau}, {Porter}, {Pouxe}, {Putze}, {Quadrani}, {Qi},
  {Rancoita}, {Rapin}, {Ren}, {Ricol}, {Riihonen}, {Rodr{\'\i}guez}, {Roeser},
  {Rosier-Lees}, {Rossi}, {Rozhkov}, {Rozza}, {Sabellek}, {Sagdeev},
  {Sandweiss}, {Santos}, {Saouter}, {Sarchioni}, {Schael}, {Schinzel},
  {Schmanau}, {Schwering}, {Schulz von Dratzig}, {Scolieri}, {Seo}, {Shan},
  {Shi}, {Shi}, {Siedenburg}, {Siedling}, {Son}, {Spada}, {Spinella}, {Steuer},
  {Stiff}, {Sun}, {Sun}, {Sun}, {Tacconi}, {Tang}, {Tang}, {Tang}, {Tao},
  {Tassan-Viol}, {Ting}, {Ting}, {Titus}, {Tomassetti}, {Toral}, {Torsti},
  {Tsai}, {Tutt}, {Ulbricht}, {Urban}, {Vagelli}, {Valente}, {Vannini},
  {Valtonen}, {Vargas Trevino}, {Vaurynovich}, {Vecchi}, {Vergain}, {Verlaat},
  {Vescovi}, {Vialle}, {Viertel}, {Volpini}, {Wang}, {Wang}, {Wang}, {Wang},
  {Wang}, {Wang}, {Wallraff}, {Weng}, {Willenbrock}, {Wlochal}, {Wu}, {Wu},
  {Wu}, {Xiao}, {Xie}, {Xiong}, {Xin}, {Xu}, {Xu}, {Yan}, {Yang}, {Yang}, {Ye},
  {Yi}, {Yu}, {Yu}, {Zeissler}, {Zhang}, {Zhang}, {Zhang}, {Zheng}, {Zhuang},
  {Zhukov}, {Zichichi}, {Zuccon} \& {Zurbach}]{Aguilar+13pos}
{\sc \au{{Aguilar}, M.}, \au{{Alberti}, G.}, \au{{Alpat}, B.}, \au{{Alvino},
  A.}, \au{{Ambrosi}, G.}, \au{{Andeen}, K.}, \au{{Anderhub}, H.},
  \au{{Arruda}, L.}, \au{{Azzarello}, P.}, \au{{Bachlechner}, A.}, \au{{Barao},
  F.}, \au{{Baret}, B.}, \au{{Barrau}, A.}, \au{{Barrin}, L.}, \au{{Bartoloni},
  A.}, \au{{Basara}, L.}, \au{{Basili}, A.}, \au{{Batalha}, L.}, \au{{Bates},
  J.}, \au{{Battiston}, R.}, \au{{Bazo}, J.}, \au{{Becker}, R.}, \au{{Becker},
  U.}, \au{{Behlmann}, M.}, \au{{Beischer}, B.}, \au{{Berdugo}, J.},
  \au{{Berges}, P.}, \au{{Bertucci}, B.}, \au{{Bigongiari}, G.}, \au{{Biland},
  A.}, \au{{Bindi}, V.}, \au{{Bizzaglia}, S.}, \au{{Boella}, G.}, \au{{de
  Boer}, W.}, \au{{Bollweg}, K.}, \au{{Bolmont}, J.}, \au{{Borgia}, B.},
  \au{{Borsini}, S.}, \au{{Boschini}, M.~J.}, \au{{Boudoul}, G.},
  \au{{Bourquin}, M.}, \au{{Brun}, P.}, \au{{Bu{\'e}nerd}, M.}, \au{{Burger},
  J.}, \au{{Burger}, W.}, \au{{Cadoux}, F.}, \au{{Cai}, X.~D.}, \au{{Capell},
  M.}, \au{{Casadei}, D.}, \au{{Casaus}, J.}, \au{{Cascioli}, V.},
  \au{{Castellini}, G.}, \au{{Cernuda}, I.}, \au{{Cervelli}, F.}, \au{{Chae},
  M.~J.}, \au{{Chang}, Y.~H.}, \au{{Chen}, A.~I.}, \au{{Chen}, C.~R.},
  \au{{Chen}, H.}, \au{{Cheng}, G.~M.}, \au{{Chen}, H.~S.}, \au{{Cheng}, L.},
  \au{{Chernoplyiokov}, N.}, \au{{Chikanian}, A.}, \au{{Choumilov}, E.},
  \au{{Choutko}, V.}, \au{{Chung}, C.~H.}, \au{{Clark}, C.}, \au{{Clavero},
  R.}, \au{{Coignet}, G.}, \au{{Commichau}, V.}, \au{{Consolandi}, C.},
  \au{{Contin}, A.}, \au{{Corti}, C.}, \au{{Costado Dios}, M.~T.}, \au{{Coste},
  B.}, \au{{Crespo}, D.}, \au{{Cui}, Z.}, \au{{Dai}, M.}, \au{{Delgado}, C.},
  \au{{Della Torre}, S.}, \au{{Demirkoz}, B.}, \au{{Dennett}, P.},
  \au{{Derome}, L.}, \au{{Di Falco}, S.}, \au{{Diao}, X.~H.}, \au{{Diago}, A.},
  \au{{Djambazov}, L.}, \au{{D{\'\i}az}, C.}, \au{{von Doetinchem}, P.},
  \au{{Du}, W.~J.}, \au{{Dubois}, J.~M.}, \au{{Duperay}, R.}, \au{{Duranti},
  M.}, \au{{D'Urso}, D.}, \au{{Egorov}, A.}, \au{{Eline}, A.}, \au{{Eppling},
  F.~J.}, \au{{Eronen}, T.}, \au{{van Es}, J.}, \au{{Esser}, H.},
  \au{{Falvard}, A.}, \au{{Fiandrini}, E.}, \au{{Fiasson}, A.}, \au{{Finch},
  E.}, \au{{Fisher}, P.}, \au{{Flood}, K.}, \au{{Foglio}, R.}, \au{{Fohey},
  M.}, \au{{Fopp}, S.}, \au{{Fouque}, N.}, \au{{Galaktionov}, Y.},
  \au{{Gallilee}, M.}, \au{{Gallin-Martel}, L.}, \au{{Gallucci}, G.},
  \au{{Garc{\'\i}a}, B.}, \au{{Garc{\'\i}a}, J.}, \au{{Garc{\'\i}a-L{\'o}pez},
  R.}, \au{{Garc{\'\i}a-Tabares}, L.}, \au{{Gargiulo}, C.}, \au{{Gast}, H.},
  \au{{Gebauer}, I.}, \au{{Gentile}, S.}, \au{{Gervasi}, M.}, \au{{Gillard},
  W.}, \au{{Giovacchini}, F.}, \au{{Girard}, L.}, \au{{Goglov}, P.},
  \au{{Gong}, J.}, \au{{Goy-Henningsen}, C.}, \au{{Grandi}, D.},
  \au{{Graziani}, M.}, \au{{Grechko}, A.}, \au{{Gross}, A.}, \au{{Guerri}, I.},
  \au{{de la Gu{\'\i}a}, C.}, \au{{Guo}, K.~H.}, \au{{Habiby}, M.},
  \au{{Haino}, S.}, \au{{Hauler}, F.}, \au{{He}, Z.~H.}, \au{{Heil}, M.},
  \au{{Heilig}, J.}, \au{{Hermel}, R.}, \au{{Hofer}, H.}, \au{{Huang}, Z.~C.},
  \au{{Hungerford}, W.}, \au{{Incagli}, M.}, \au{{Ionica}, M.},
  \au{{Jacholkowska}, A.}, \au{{Jang}, W.~Y.}, \au{{Jinchi}, H.},
  \au{{Jongmanns}, M.}, \au{{Journet}, L.}, \au{{Jungermann}, L.},
  \au{{Karpinski}, W.}, \au{{Kim}, G.~N.}, \au{{Kim}, K.~S.}, \au{{Kirn}, Th.},
  \au{{Kossakowski}, R.}, \au{{Koulemzine}, A.}, \au{{Kounina}, O.},
  \au{{Kounine}, A.}, \au{{Koutsenko}, V.}, \au{{Krafczyk}, M.~S.},
  \au{{Laudi}, E.}, \au{{Laurenti}, G.}, \au{{Lauritzen}, C.}, \au{{Lebedev},
  A.}, \au{{Lee}, M.~W.}, \au{{Lee}, S.~C.}, \au{{Leluc}, C.}, \au{{Le{\'o}n
  Vargas}, H.}, \au{{Lepareur}, V.}, \au{{Li}, J.~Q.}, \au{{Li}, Q.}, \au{{Li},
  T.~X.}, \au{{Li}, W.}, \au{{Li}, Z.~H.}, \au{{Lipari}, P.}, \au{{Lin},
  C.~H.}, \au{{Liu}, D.}, \au{{Liu}, H.}, \au{{Lomtadze}, T.}, \au{{Lu},
  Y.~S.}, \au{{Lucidi}, S.}, \au{{L{\"u}belsmeyer}, K.}, \au{{Luo}, J.~Z.},
  \au{{Lustermann}, W.}, \au{{Lv}, S.}, \au{{Madsen}, J.}, \au{{Majka}, R.},
  \au{{Malinin}, A.}, \au{{Ma{\~n}{\'a}}, C.}, \au{{Mar{\'\i}n}, J.},
  \au{{Martin}, T.}, \au{{Mart{\'\i}nez}, G.}, \au{{Masciocchi}, F.},
  \au{{Masi}, N.}, \au{{Maurin}, D.}, \au{{McInturff}, A.}, \au{{McIntyre},
  P.}, \au{{Menchaca-Rocha}, A.}, \au{{Meng}, Q.}, \au{{Menichelli}, M.},
  \au{{Mereu}, I.}, \au{{Millinger}, M.}, \au{{Mo}, D.~C.}, \au{{Molina}, M.},
  \au{{Mott}, P.}, \au{{Mujunen}, A.}, \au{{Natale}, S.}, \au{{Nemeth}, P.},
  \au{{Ni}, J.~Q.}, \au{{Nikonov}, N.}, \au{{Nozzoli}, F.}, \au{{Nunes}, P.},
  \au{{Obermeier}, A.}, \au{{Oh}, S.}, \au{{Oliva}, A.}, \au{{Palmonari}, F.},
  \au{{Palomares}, C.}, \au{{Paniccia}, M.}, \au{{Papi}, A.}, \au{{Park},
  W.~H.}, \au{{Pauluzzi}, M.}, \au{{Pauss}, F.}, \au{{Pauw}, A.},
  \au{{Pedreschi}, E.}, \au{{Pensotti}, S.}, \au{{Pereira}, R.}, \au{{Perrin},
  E.}, \au{{Pessina}, G.}, \au{{Pierschel}, G.}, \au{{Pilo}, F.}, \au{{Piluso},
  A.}, \au{{Pizzolotto}, C.}, \au{{Plyaskin}, V.}, \au{{Pochon}, J.},
  \au{{Pohl}, M.}, \au{{Poireau}, V.}, \au{{Porter}, S.}, \au{{Pouxe}, J.},
  \au{{Putze}, A.}, \au{{Quadrani}, L.}, \au{{Qi}, X.~N.}, \au{{Rancoita},
  P.~G.}, \au{{Rapin}, D.}, \au{{Ren}, Z.~L.}, \au{{Ricol}, J.~S.},
  \au{{Riihonen}, E.}, \au{{Rodr{\'\i}guez}, I.}, \au{{Roeser}, U.},
  \au{{Rosier-Lees}, S.}, \au{{Rossi}, L.}, \au{{Rozhkov}, A.}, \au{{Rozza},
  D.}, \au{{Sabellek}, A.}, \au{{Sagdeev}, R.}, \au{{Sandweiss}, J.},
  \au{{Santos}, B.}, \au{{Saouter}, P.}, \au{{Sarchioni}, M.}, \au{{Schael},
  S.}, \au{{Schinzel}, D.}, \au{{Schmanau}, M.}, \au{{Schwering}, G.},
  \au{{Schulz von Dratzig}, A.}, \au{{Scolieri}, G.}, \au{{Seo}, E.~S.},
  \au{{Shan}, B.~S.}, \au{{Shi}, J.~Y.}, \au{{Shi}, Y.~M.}, \au{{Siedenburg},
  T.}, \au{{Siedling}, R.}, \au{{Son}, D.}, \au{{Spada}, F.}, \au{{Spinella},
  F.}, \au{{Steuer}, M.}, \au{{Stiff}, K.}, \au{{Sun}, W.}, \au{{Sun}, W.~H.},
  \au{{Sun}, X.~H.}, \au{{Tacconi}, M.}, \au{{Tang}, C.~P.}, \au{{Tang},
  X.~W.}, \au{{Tang}, Z.~C.}, \au{{Tao}, L.}, \au{{Tassan-Viol}, J.},
  \au{{Ting}, Samuel C.~C.}, \au{{Ting}, S.~M.}, \au{{Titus}, C.},
  \au{{Tomassetti}, N.}, \au{{Toral}, F.}, \au{{Torsti}, J.}, \au{{Tsai},
  J.~R.}, \au{{Tutt}, J.~C.}, \au{{Ulbricht}, J.}, \au{{Urban}, T.},
  \au{{Vagelli}, V.}, \au{{Valente}, E.}, \au{{Vannini}, C.}, \au{{Valtonen},
  E.}, \au{{Vargas Trevino}, M.}, \au{{Vaurynovich}, S.}, \au{{Vecchi}, M.},
  \au{{Vergain}, M.}, \au{{Verlaat}, B.}, \au{{Vescovi}, C.}, \au{{Vialle},
  J.~P.}, \au{{Viertel}, G.}, \au{{Volpini}, G.}, \au{{Wang}, D.}, \au{{Wang},
  N.~H.}, \au{{Wang}, Q.~L.}, \au{{Wang}, R.~S.}, \au{{Wang}, X.}, \au{{Wang},
  Z.~X.}, \au{{Wallraff}, W.}, \au{{Weng}, Z.~L.}, \au{{Willenbrock}, M.},
  \au{{Wlochal}, M.}, \au{{Wu}, H.}, \au{{Wu}, K.~Y.}, \au{{Wu}, Z.~S.},
  \au{{Xiao}, W.~J.}, \au{{Xie}, S.}, \au{{Xiong}, R.~Q.}, \au{{Xin}, G.~M.},
  \au{{Xu}, N.~S.}, \au{{Xu}, W.}, \au{{Yan}, Q.}, \au{{Yang}, J.}, \au{{Yang},
  M.}, \au{{Ye}, Q.~H.}, \au{{Yi}, H.}, \au{{Yu}, Y.~J.}, \au{{Yu}, Z.~Q.},
  \au{{Zeissler}, S.}, \au{{Zhang}, J.~G.}, \au{{Zhang}, Z.}, \au{{Zhang},
  M.~M.}, \au{{Zheng}, Z.~M.}, \au{{Zhuang}, H.~L.}, \au{{Zhukov}, V.},
  \au{{Zichichi}, A.}, \au{{Zuccon}, P.} \& \au{{Zurbach}, C.}}
  \yr{2013{\natexlab{{\em b\/}}}}  \at{{First Result from the Alpha Magnetic
  Spectrometer on the International Space Station: Precision Measurement of the
  Positron Fraction in Primary Cosmic Rays of 0.5-350 GeV}}.  \jt{PRL}
  \bvol{110}~(14),  \pg{141102}.

\bibitem[{Aguilar} {\em et~al.\/}(2016{\natexlab{{\em a\/}}}){Aguilar}, {Ali
  Cavasonza}, {Alpat}, {Ambrosi}, {Arruda}, {Attig}, {Aupetit}, {Azzarello},
  {Bachlechner}, {Barao}, {Barrau}, {Barrin}, {Bartoloni}, {Basara},
  {Ba{\textcommabelow s}e{\c{C}}{\textsection}mez-du Pree}, {Battarbee},
  {Battiston}, {Bazo}, {Becker}, {Behlmann}, {Beischer}, {Berdugo}, {Bertucci},
  {Bindi}, {Boella}, {de Boer}, {Bollweg}, {Bonnivard}, {Borgia}, {Boschini},
  {Bourquin}, {Bueno}, {Burger}, {Cadoux}, {Cai}, {Capell}, {Caroff}, {Casaus},
  {Castellini}, {Cernuda}, {Cervelli}, {Chae}, {Chang}, {Chen}, {Chen}, {Chen},
  {Cheng}, {Chou}, {Choumilov}, {Choutko}, {Chung}, {Clark}, {Clavero},
  {Coignet}, {Consolandi}, {Contin}, {Corti}, {Coste}, {Creus}, {Crispoltoni},
  {Cui}, {Dai}, {Delgado}, {Della Torre}, {Demirk{\"o}z}, {Derome}, {Di Falco},
  {Dimiccoli}, {D{\'\i}az}, {von Doetinchem}, {Dong}, {Donnini}, {Duranti},
  {D'Urso}, {Egorov}, {Eline}, {Eronen}, {Feng}, {Fiand rini}, {Finch},
  {Fisher}, {Formato}, {Galaktionov}, {Gallucci}, {Garc{\'\i}a},
  {Garc{\'\i}a-L{\'o}pez}, {Gargiulo}, {Gast}, {Gebauer}, {Gervasi}, {Ghelfi},
  {Giovacchini}, {Goglov}, {G{\'o}mez-Coral}, {Gong}, {Goy}, {Grabski},
  {Grandi}, {Graziani}, {Guerri}, {Guo}, {Habiby}, {Haino}, {Han}, {He},
  {Heil}, {Hoffman}, {Hsieh}, {Huang}, {Huang}, {Huh}, {Incagli}, {Ionica},
  {Jang}, {Jinchi}, {Kang}, {Kanishev}, {Kim}, {Kim}, {Kirn}, {Konak},
  {Kounina}, {Kounine}, {Koutsenko}, {Krafczyk}, {La Vacca}, {Laudi},
  {Laurenti}, {Lazzizzera}, {Lebedev}, {Lee}, {Lee}, {Leluc}, {Li}, {Li}, {Li},
  {Li}, {Li}, {Li}, {Li}, {Li}, {Lim}, {Lin}, {Lipari}, {Lippert}, {Liu},
  {Liu}, {Lu}, {Lu}, {Luebelsmeyer}, {Luo}, {Luo}, {Lv}, {Majka},
  {Ma{\~n}{\'a}}, {Mar{\'\i}n}, {Martin}, {Mart{\'\i}nez}, {Masi}, {Maurin},
  {Menchaca-Rocha}, {Meng}, {Mo}, {Morescalchi}, {Mott}, {Nelson}, {Ni},
  {Nikonov}, {Nozzoli}, {Nunes}, {Oliva}, {Orcinha}, {Palmonari}, {Palomares},
  {Paniccia}, {Pauluzzi}, {Pensotti}, {Pereira}, {Picot-Clemente}, {Pilo},
  {Pizzolotto}, {Plyaskin}, {Pohl}, {Poireau}, {Putze}, {Quadrani}, {Qi},
  {Qin}, {Qu}, {R{\"a}ih{\"a}}, {Rancoita}, {Rapin}, {Ricol}, {Rodr{\'\i}guez},
  {Rosier-Lees}, {Rozhkov}, {Rozza}, {Sagdeev}, {Sandweiss}, {Saouter},
  {Schael}, {Schmidt}, {Schulz von Dratzig}, {Schwering}, {Seo}, {Shan}, {Shi},
  {Siedenburg}, {Son}, {Song}, {Sun}, {Tacconi}, {Tang}, {Tang}, {Tao},
  {Tescaro}, {Ting}, {Ting}, {Tomassetti}, {Torsti}, {T{\"u}rko{\v{g}}lu},
  {Urban}, {Vagelli}, {Valente}, {Vannini}, {Valtonen}, {V{\'a}zquez Acosta},
  {Vecchi}, {Velasco}, {Vialle}, {Vitale}, {Vitillo}, {Wang}, {Wang}, {Wang},
  {Wang}, {Wang}, {Wang}, {Wei}, {Weng}, {Whitman}, {Wienkenh{\"o}ver},
  {Willenbrock}, {Wu}, {Wu}, {Xia}, {Xiong}, {Xu}, {Yan}, {Yang}, {Yang},
  {Yang}, {Yi}, {Yu}, {Yu}, {Zeissler}, {Zhang}, {Zhang}, {Zhang}, {Zhang},
  {Zhang}, {Zhang}, {Zheng}, {Zhu}, {Zhuang}, {Zhukov}, {Zichichi},
  {Zimmermann}, {Zuccon} \& {AMS Collaboration}]{AMS02Pbar}
{\sc \au{{Aguilar}, M.}, \au{{Ali Cavasonza}, L.}, \au{{Alpat}, B.},
  \au{{Ambrosi}, G.}, \au{{Arruda}, L.}, \au{{Attig}, N.}, \au{{Aupetit}, S.},
  \au{{Azzarello}, P.}, \au{{Bachlechner}, A.}, \au{{Barao}, F.}, \au{{Barrau},
  A.}, \au{{Barrin}, L.}, \au{{Bartoloni}, A.}, \au{{Basara}, L.},
  \au{{Ba{\textcommabelow s}e{\c{C}}{\textsection}mez-du Pree}, S.},
  \au{{Battarbee}, M.}, \au{{Battiston}, R.}, \au{{Bazo}, J.}, \au{{Becker},
  U.}, \au{{Behlmann}, M.}, \au{{Beischer}, B.}, \au{{Berdugo}, J.},
  \au{{Bertucci}, B.}, \au{{Bindi}, V.}, \au{{Boella}, G.}, \au{{de Boer}, W.},
  \au{{Bollweg}, K.}, \au{{Bonnivard}, V.}, \au{{Borgia}, B.}, \au{{Boschini},
  M.~J.}, \au{{Bourquin}, M.}, \au{{Bueno}, E.~F.}, \au{{Burger}, J.},
  \au{{Cadoux}, F.}, \au{{Cai}, X.~D.}, \au{{Capell}, M.}, \au{{Caroff}, S.},
  \au{{Casaus}, J.}, \au{{Castellini}, G.}, \au{{Cernuda}, I.}, \au{{Cervelli},
  F.}, \au{{Chae}, M.~J.}, \au{{Chang}, Y.~H.}, \au{{Chen}, A.~I.}, \au{{Chen},
  G.~M.}, \au{{Chen}, H.~S.}, \au{{Cheng}, L.}, \au{{Chou}, H.~Y.},
  \au{{Choumilov}, E.}, \au{{Choutko}, V.}, \au{{Chung}, C.~H.}, \au{{Clark},
  C.}, \au{{Clavero}, R.}, \au{{Coignet}, G.}, \au{{Consolandi}, C.},
  \au{{Contin}, A.}, \au{{Corti}, C.}, \au{{Coste}, B.}, \au{{Creus}, W.},
  \au{{Crispoltoni}, M.}, \au{{Cui}, Z.}, \au{{Dai}, Y.~M.}, \au{{Delgado},
  C.}, \au{{Della Torre}, S.}, \au{{Demirk{\"o}z}, M.~B.}, \au{{Derome}, L.},
  \au{{Di Falco}, S.}, \au{{Dimiccoli}, F.}, \au{{D{\'\i}az}, C.}, \au{{von
  Doetinchem}, P.}, \au{{Dong}, F.}, \au{{Donnini}, F.}, \au{{Duranti}, M.},
  \au{{D'Urso}, D.}, \au{{Egorov}, A.}, \au{{Eline}, A.}, \au{{Eronen}, T.},
  \au{{Feng}, J.}, \au{{Fiand rini}, E.}, \au{{Finch}, E.}, \au{{Fisher}, P.},
  \au{{Formato}, V.}, \au{{Galaktionov}, Y.}, \au{{Gallucci}, G.},
  \au{{Garc{\'\i}a}, B.}, \au{{Garc{\'\i}a-L{\'o}pez}, R.~J.}, \au{{Gargiulo},
  C.}, \au{{Gast}, H.}, \au{{Gebauer}, I.}, \au{{Gervasi}, M.}, \au{{Ghelfi},
  A.}, \au{{Giovacchini}, F.}, \au{{Goglov}, P.}, \au{{G{\'o}mez-Coral},
  D.~M.}, \au{{Gong}, J.}, \au{{Goy}, C.}, \au{{Grabski}, V.}, \au{{Grandi},
  D.}, \au{{Graziani}, M.}, \au{{Guerri}, I.}, \au{{Guo}, K.~H.}, \au{{Habiby},
  M.}, \au{{Haino}, S.}, \au{{Han}, K.~C.}, \au{{He}, Z.~H.}, \au{{Heil}, M.},
  \au{{Hoffman}, J.}, \au{{Hsieh}, T.~H.}, \au{{Huang}, H.}, \au{{Huang},
  Z.~C.}, \au{{Huh}, C.}, \au{{Incagli}, M.}, \au{{Ionica}, M.}, \au{{Jang},
  W.~Y.}, \au{{Jinchi}, H.}, \au{{Kang}, S.~C.}, \au{{Kanishev}, K.},
  \au{{Kim}, G.~N.}, \au{{Kim}, K.~S.}, \au{{Kirn}, Th.}, \au{{Konak}, C.},
  \au{{Kounina}, O.}, \au{{Kounine}, A.}, \au{{Koutsenko}, V.}, \au{{Krafczyk},
  M.~S.}, \au{{La Vacca}, G.}, \au{{Laudi}, E.}, \au{{Laurenti}, G.},
  \au{{Lazzizzera}, I.}, \au{{Lebedev}, A.}, \au{{Lee}, H.~T.}, \au{{Lee},
  S.~C.}, \au{{Leluc}, C.}, \au{{Li}, H.~S.}, \au{{Li}, J.~Q.}, \au{{Li},
  J.~Q.}, \au{{Li}, Q.}, \au{{Li}, T.~X.}, \au{{Li}, W.}, \au{{Li}, Z.~H.},
  \au{{Li}, Z.~Y.}, \au{{Lim}, S.}, \au{{Lin}, C.~H.}, \au{{Lipari}, P.},
  \au{{Lippert}, T.}, \au{{Liu}, D.}, \au{{Liu}, Hu}, \au{{Lu}, S.~Q.},
  \au{{Lu}, Y.~S.}, \au{{Luebelsmeyer}, K.}, \au{{Luo}, F.}, \au{{Luo}, J.~Z.},
  \au{{Lv}, S.~S.}, \au{{Majka}, R.}, \au{{Ma{\~n}{\'a}}, C.},
  \au{{Mar{\'\i}n}, J.}, \au{{Martin}, T.}, \au{{Mart{\'\i}nez}, G.},
  \au{{Masi}, N.}, \au{{Maurin}, D.}, \au{{Menchaca-Rocha}, A.}, \au{{Meng},
  Q.}, \au{{Mo}, D.~C.}, \au{{Morescalchi}, L.}, \au{{Mott}, P.}, \au{{Nelson},
  T.}, \au{{Ni}, J.~Q.}, \au{{Nikonov}, N.}, \au{{Nozzoli}, F.}, \au{{Nunes},
  P.}, \au{{Oliva}, A.}, \au{{Orcinha}, M.}, \au{{Palmonari}, F.},
  \au{{Palomares}, C.}, \au{{Paniccia}, M.}, \au{{Pauluzzi}, M.},
  \au{{Pensotti}, S.}, \au{{Pereira}, R.}, \au{{Picot-Clemente}, N.},
  \au{{Pilo}, F.}, \au{{Pizzolotto}, C.}, \au{{Plyaskin}, V.}, \au{{Pohl}, M.},
  \au{{Poireau}, V.}, \au{{Putze}, A.}, \au{{Quadrani}, L.}, \au{{Qi}, X.~M.},
  \au{{Qin}, X.}, \au{{Qu}, Z.~Y.}, \au{{R{\"a}ih{\"a}}, T.}, \au{{Rancoita},
  P.~G.}, \au{{Rapin}, D.}, \au{{Ricol}, J.~S.}, \au{{Rodr{\'\i}guez}, I.},
  \au{{Rosier-Lees}, S.}, \au{{Rozhkov}, A.}, \au{{Rozza}, D.}, \au{{Sagdeev},
  R.}, \au{{Sandweiss}, J.}, \au{{Saouter}, P.}, \au{{Schael}, S.},
  \au{{Schmidt}, S.~M.}, \au{{Schulz von Dratzig}, A.}, \au{{Schwering}, G.},
  \au{{Seo}, E.~S.}, \au{{Shan}, B.~S.}, \au{{Shi}, J.~Y.}, \au{{Siedenburg},
  T.}, \au{{Son}, D.}, \au{{Song}, J.~W.}, \au{{Sun}, W.~H.}, \au{{Tacconi},
  M.}, \au{{Tang}, X.~W.}, \au{{Tang}, Z.~C.}, \au{{Tao}, L.}, \au{{Tescaro},
  D.}, \au{{Ting}, Samuel C.~C.}, \au{{Ting}, S.~M.}, \au{{Tomassetti}, N.},
  \au{{Torsti}, J.}, \au{{T{\"u}rko{\v{g}}lu}, C.}, \au{{Urban}, T.},
  \au{{Vagelli}, V.}, \au{{Valente}, E.}, \au{{Vannini}, C.}, \au{{Valtonen},
  E.}, \au{{V{\'a}zquez Acosta}, M.}, \au{{Vecchi}, M.}, \au{{Velasco}, M.},
  \au{{Vialle}, J.~P.}, \au{{Vitale}, V.}, \au{{Vitillo}, S.}, \au{{Wang},
  L.~Q.}, \au{{Wang}, N.~H.}, \au{{Wang}, Q.~L.}, \au{{Wang}, X.}, \au{{Wang},
  X.~Q.}, \au{{Wang}, Z.~X.}, \au{{Wei}, C.~C.}, \au{{Weng}, Z.~L.},
  \au{{Whitman}, K.}, \au{{Wienkenh{\"o}ver}, J.}, \au{{Willenbrock}, M.},
  \au{{Wu}, H.}, \au{{Wu}, X.}, \au{{Xia}, X.}, \au{{Xiong}, R.~Q.}, \au{{Xu},
  W.}, \au{{Yan}, Q.}, \au{{Yang}, J.}, \au{{Yang}, M.}, \au{{Yang}, Y.},
  \au{{Yi}, H.}, \au{{Yu}, Y.~J.}, \au{{Yu}, Z.~Q.}, \au{{Zeissler}, S.},
  \au{{Zhang}, C.}, \au{{Zhang}, J.}, \au{{Zhang}, J.~H.}, \au{{Zhang}, S.~D.},
  \au{{Zhang}, S.~W.}, \au{{Zhang}, Z.}, \au{{Zheng}, Z.~M.}, \au{{Zhu},
  Z.~Q.}, \au{{Zhuang}, H.~L.}, \au{{Zhukov}, V.}, \au{{Zichichi}, A.},
  \au{{Zimmermann}, N.}, \au{{Zuccon}, P.} \& \au{{AMS Collaboration}}}
  \yr{2016{\natexlab{{\em a\/}}}}  \at{{Antiproton Flux, Antiproton-to-Proton
  Flux Ratio, and Properties of Elementary Particle Fluxes in Primary Cosmic
  Rays Measured with the Alpha Magnetic Spectrometer on the International Space
  Station}}.  \jt{Physical Review Letters}  \bvol{117}~(9),  \pg{091103}.

\bibitem[{Aguilar} {\em et~al.\/}(2016{\natexlab{{\em b\/}}}){Aguilar}, {Ali
  Cavasonza}, {Alpat}, {Ambrosi}, {Arruda}, {Attig}, {Aupetit}, {Azzarello},
  {Bachlechner}, {Barao}, {Barrau}, {Barrin}, {Bartoloni}, {Basara},
  {Ba{\textcommabelow s}e{\c{C}}{\textsection}mez-du Pree}, {Battarbee},
  {Battiston}, {Bazo}, {Becker}, {Behlmann}, {Beischer}, {Berdugo}, {Bertucci},
  {Bindi}, {Boella}, {de Boer}, {Bollweg}, {Bonnivard}, {Borgia}, {Boschini},
  {Bourquin}, {Bueno}, {Burger}, {Cadoux}, {Cai}, {Capell}, {Caroff}, {Casaus},
  {Castellini}, {Cernuda}, {Cervelli}, {Chae}, {Chang}, {Chen}, {Chen}, {Chen},
  {Cheng}, {Chou}, {Choumilov}, {Choutko}, {Chung}, {Clark}, {Clavero},
  {Coignet}, {Consolandi}, {Contin}, {Corti}, {Coste}, {Creus}, {Crispoltoni},
  {Cui}, {Dai}, {Delgado}, {Della Torre}, {Demirk{\"o}z}, {Derome}, {Di Falco},
  {Dimiccoli}, {D{\'\i}az}, {von Doetinchem}, {Dong}, {Donnini}, {Duranti},
  {D'Urso}, {Egorov}, {Eline}, {Eronen}, {Feng}, {Fiand rini}, {Finch},
  {Fisher}, {Formato}, {Galaktionov}, {Gallucci}, {Garc{\'\i}a},
  {Garc{\'\i}a-L{\'o}pez}, {Gargiulo}, {Gast}, {Gebauer}, {Gervasi}, {Ghelfi},
  {Giovacchini}, {Goglov}, {G{\'o}mez-Coral}, {Gong}, {Goy}, {Grabski},
  {Grandi}, {Graziani}, {Guerri}, {Guo}, {Habiby}, {Haino}, {Han}, {He},
  {Heil}, {Hoffman}, {Hsieh}, {Huang}, {Huang}, {Huh}, {Incagli}, {Ionica},
  {Jang}, {Jinchi}, {Kang}, {Kanishev}, {Kim}, {Kim}, {Kirn}, {Konak},
  {Kounina}, {Kounine}, {Koutsenko}, {Krafczyk}, {La Vacca}, {Laudi},
  {Laurenti}, {Lazzizzera}, {Lebedev}, {Lee}, {Lee}, {Leluc}, {Li}, {Li}, {Li},
  {Li}, {Li}, {Li}, {Li}, {Li}, {Lim}, {Lin}, {Lipari}, {Lippert}, {Liu},
  {Liu}, {Lu}, {Lu}, {Luebelsmeyer}, {Luo}, {Luo}, {Lv}, {Majka},
  {Ma{\~n}{\'a}}, {Mar{\'\i}n}, {Martin}, {Mart{\'\i}nez}, {Masi}, {Maurin},
  {Menchaca-Rocha}, {Meng}, {Mo}, {Morescalchi}, {Mott}, {Nelson}, {Ni},
  {Nikonov}, {Nozzoli}, {Nunes}, {Oliva}, {Orcinha}, {Palmonari}, {Palomares},
  {Paniccia}, {Pauluzzi}, {Pensotti}, {Pereira}, {Picot-Clemente}, {Pilo},
  {Pizzolotto}, {Plyaskin}, {Pohl}, {Poireau}, {Putze}, {Quadrani}, {Qi},
  {Qin}, {Qu}, {R{\"a}ih{\"a}}, {Rancoita}, {Rapin}, {Ricol}, {Rodr{\'\i}guez},
  {Rosier-Lees}, {Rozhkov}, {Rozza}, {Sagdeev}, {Sandweiss}, {Saouter},
  {Schael}, {Schmidt}, {Schulz von Dratzig}, {Schwering}, {Seo}, {Shan}, {Shi},
  {Siedenburg}, {Son}, {Song}, {Sun}, {Tacconi}, {Tang}, {Tang}, {Tao},
  {Tescaro}, {Ting}, {Ting}, {Tomassetti}, {Torsti}, {T{\"u}rko{\v{g}}lu},
  {Urban}, {Vagelli}, {Valente}, {Vannini}, {Valtonen}, {V{\'a}zquez Acosta},
  {Vecchi}, {Velasco}, {Vialle}, {Vitale}, {Vitillo}, {Wang}, {Wang}, {Wang},
  {Wang}, {Wang}, {Wang}, {Wei}, {Weng}, {Whitman}, {Wienkenh{\"o}ver},
  {Willenbrock}, {Wu}, {Wu}, {Xia}, {Xiong}, {Xu}, {Yan}, {Yang}, {Yang},
  {Yang}, {Yi}, {Yu}, {Yu}, {Zeissler}, {Zhang}, {Zhang}, {Zhang}, {Zhang},
  {Zhang}, {Zhang}, {Zheng}, {Zhu}, {Zhuang}, {Zhukov}, {Zichichi},
  {Zimmermann}, {Zuccon} \& {AMS Collaboration}]{Aguilar+16pbar}
{\sc \au{{Aguilar}, M.}, \au{{Ali Cavasonza}, L.}, \au{{Alpat}, B.},
  \au{{Ambrosi}, G.}, \au{{Arruda}, L.}, \au{{Attig}, N.}, \au{{Aupetit}, S.},
  \au{{Azzarello}, P.}, \au{{Bachlechner}, A.}, \au{{Barao}, F.}, \au{{Barrau},
  A.}, \au{{Barrin}, L.}, \au{{Bartoloni}, A.}, \au{{Basara}, L.},
  \au{{Ba{\textcommabelow s}e{\c{C}}{\textsection}mez-du Pree}, S.},
  \au{{Battarbee}, M.}, \au{{Battiston}, R.}, \au{{Bazo}, J.}, \au{{Becker},
  U.}, \au{{Behlmann}, M.}, \au{{Beischer}, B.}, \au{{Berdugo}, J.},
  \au{{Bertucci}, B.}, \au{{Bindi}, V.}, \au{{Boella}, G.}, \au{{de Boer}, W.},
  \au{{Bollweg}, K.}, \au{{Bonnivard}, V.}, \au{{Borgia}, B.}, \au{{Boschini},
  M.~J.}, \au{{Bourquin}, M.}, \au{{Bueno}, E.~F.}, \au{{Burger}, J.},
  \au{{Cadoux}, F.}, \au{{Cai}, X.~D.}, \au{{Capell}, M.}, \au{{Caroff}, S.},
  \au{{Casaus}, J.}, \au{{Castellini}, G.}, \au{{Cernuda}, I.}, \au{{Cervelli},
  F.}, \au{{Chae}, M.~J.}, \au{{Chang}, Y.~H.}, \au{{Chen}, A.~I.}, \au{{Chen},
  G.~M.}, \au{{Chen}, H.~S.}, \au{{Cheng}, L.}, \au{{Chou}, H.~Y.},
  \au{{Choumilov}, E.}, \au{{Choutko}, V.}, \au{{Chung}, C.~H.}, \au{{Clark},
  C.}, \au{{Clavero}, R.}, \au{{Coignet}, G.}, \au{{Consolandi}, C.},
  \au{{Contin}, A.}, \au{{Corti}, C.}, \au{{Coste}, B.}, \au{{Creus}, W.},
  \au{{Crispoltoni}, M.}, \au{{Cui}, Z.}, \au{{Dai}, Y.~M.}, \au{{Delgado},
  C.}, \au{{Della Torre}, S.}, \au{{Demirk{\"o}z}, M.~B.}, \au{{Derome}, L.},
  \au{{Di Falco}, S.}, \au{{Dimiccoli}, F.}, \au{{D{\'\i}az}, C.}, \au{{von
  Doetinchem}, P.}, \au{{Dong}, F.}, \au{{Donnini}, F.}, \au{{Duranti}, M.},
  \au{{D'Urso}, D.}, \au{{Egorov}, A.}, \au{{Eline}, A.}, \au{{Eronen}, T.},
  \au{{Feng}, J.}, \au{{Fiand rini}, E.}, \au{{Finch}, E.}, \au{{Fisher}, P.},
  \au{{Formato}, V.}, \au{{Galaktionov}, Y.}, \au{{Gallucci}, G.},
  \au{{Garc{\'\i}a}, B.}, \au{{Garc{\'\i}a-L{\'o}pez}, R.~J.}, \au{{Gargiulo},
  C.}, \au{{Gast}, H.}, \au{{Gebauer}, I.}, \au{{Gervasi}, M.}, \au{{Ghelfi},
  A.}, \au{{Giovacchini}, F.}, \au{{Goglov}, P.}, \au{{G{\'o}mez-Coral},
  D.~M.}, \au{{Gong}, J.}, \au{{Goy}, C.}, \au{{Grabski}, V.}, \au{{Grandi},
  D.}, \au{{Graziani}, M.}, \au{{Guerri}, I.}, \au{{Guo}, K.~H.}, \au{{Habiby},
  M.}, \au{{Haino}, S.}, \au{{Han}, K.~C.}, \au{{He}, Z.~H.}, \au{{Heil}, M.},
  \au{{Hoffman}, J.}, \au{{Hsieh}, T.~H.}, \au{{Huang}, H.}, \au{{Huang},
  Z.~C.}, \au{{Huh}, C.}, \au{{Incagli}, M.}, \au{{Ionica}, M.}, \au{{Jang},
  W.~Y.}, \au{{Jinchi}, H.}, \au{{Kang}, S.~C.}, \au{{Kanishev}, K.},
  \au{{Kim}, G.~N.}, \au{{Kim}, K.~S.}, \au{{Kirn}, Th.}, \au{{Konak}, C.},
  \au{{Kounina}, O.}, \au{{Kounine}, A.}, \au{{Koutsenko}, V.}, \au{{Krafczyk},
  M.~S.}, \au{{La Vacca}, G.}, \au{{Laudi}, E.}, \au{{Laurenti}, G.},
  \au{{Lazzizzera}, I.}, \au{{Lebedev}, A.}, \au{{Lee}, H.~T.}, \au{{Lee},
  S.~C.}, \au{{Leluc}, C.}, \au{{Li}, H.~S.}, \au{{Li}, J.~Q.}, \au{{Li},
  J.~Q.}, \au{{Li}, Q.}, \au{{Li}, T.~X.}, \au{{Li}, W.}, \au{{Li}, Z.~H.},
  \au{{Li}, Z.~Y.}, \au{{Lim}, S.}, \au{{Lin}, C.~H.}, \au{{Lipari}, P.},
  \au{{Lippert}, T.}, \au{{Liu}, D.}, \au{{Liu}, Hu}, \au{{Lu}, S.~Q.},
  \au{{Lu}, Y.~S.}, \au{{Luebelsmeyer}, K.}, \au{{Luo}, F.}, \au{{Luo}, J.~Z.},
  \au{{Lv}, S.~S.}, \au{{Majka}, R.}, \au{{Ma{\~n}{\'a}}, C.},
  \au{{Mar{\'\i}n}, J.}, \au{{Martin}, T.}, \au{{Mart{\'\i}nez}, G.},
  \au{{Masi}, N.}, \au{{Maurin}, D.}, \au{{Menchaca-Rocha}, A.}, \au{{Meng},
  Q.}, \au{{Mo}, D.~C.}, \au{{Morescalchi}, L.}, \au{{Mott}, P.}, \au{{Nelson},
  T.}, \au{{Ni}, J.~Q.}, \au{{Nikonov}, N.}, \au{{Nozzoli}, F.}, \au{{Nunes},
  P.}, \au{{Oliva}, A.}, \au{{Orcinha}, M.}, \au{{Palmonari}, F.},
  \au{{Palomares}, C.}, \au{{Paniccia}, M.}, \au{{Pauluzzi}, M.},
  \au{{Pensotti}, S.}, \au{{Pereira}, R.}, \au{{Picot-Clemente}, N.},
  \au{{Pilo}, F.}, \au{{Pizzolotto}, C.}, \au{{Plyaskin}, V.}, \au{{Pohl}, M.},
  \au{{Poireau}, V.}, \au{{Putze}, A.}, \au{{Quadrani}, L.}, \au{{Qi}, X.~M.},
  \au{{Qin}, X.}, \au{{Qu}, Z.~Y.}, \au{{R{\"a}ih{\"a}}, T.}, \au{{Rancoita},
  P.~G.}, \au{{Rapin}, D.}, \au{{Ricol}, J.~S.}, \au{{Rodr{\'\i}guez}, I.},
  \au{{Rosier-Lees}, S.}, \au{{Rozhkov}, A.}, \au{{Rozza}, D.}, \au{{Sagdeev},
  R.}, \au{{Sandweiss}, J.}, \au{{Saouter}, P.}, \au{{Schael}, S.},
  \au{{Schmidt}, S.~M.}, \au{{Schulz von Dratzig}, A.}, \au{{Schwering}, G.},
  \au{{Seo}, E.~S.}, \au{{Shan}, B.~S.}, \au{{Shi}, J.~Y.}, \au{{Siedenburg},
  T.}, \au{{Son}, D.}, \au{{Song}, J.~W.}, \au{{Sun}, W.~H.}, \au{{Tacconi},
  M.}, \au{{Tang}, X.~W.}, \au{{Tang}, Z.~C.}, \au{{Tao}, L.}, \au{{Tescaro},
  D.}, \au{{Ting}, Samuel C.~C.}, \au{{Ting}, S.~M.}, \au{{Tomassetti}, N.},
  \au{{Torsti}, J.}, \au{{T{\"u}rko{\v{g}}lu}, C.}, \au{{Urban}, T.},
  \au{{Vagelli}, V.}, \au{{Valente}, E.}, \au{{Vannini}, C.}, \au{{Valtonen},
  E.}, \au{{V{\'a}zquez Acosta}, M.}, \au{{Vecchi}, M.}, \au{{Velasco}, M.},
  \au{{Vialle}, J.~P.}, \au{{Vitale}, V.}, \au{{Vitillo}, S.}, \au{{Wang},
  L.~Q.}, \au{{Wang}, N.~H.}, \au{{Wang}, Q.~L.}, \au{{Wang}, X.}, \au{{Wang},
  X.~Q.}, \au{{Wang}, Z.~X.}, \au{{Wei}, C.~C.}, \au{{Weng}, Z.~L.},
  \au{{Whitman}, K.}, \au{{Wienkenh{\"o}ver}, J.}, \au{{Willenbrock}, M.},
  \au{{Wu}, H.}, \au{{Wu}, X.}, \au{{Xia}, X.}, \au{{Xiong}, R.~Q.}, \au{{Xu},
  W.}, \au{{Yan}, Q.}, \au{{Yang}, J.}, \au{{Yang}, M.}, \au{{Yang}, Y.},
  \au{{Yi}, H.}, \au{{Yu}, Y.~J.}, \au{{Yu}, Z.~Q.}, \au{{Zeissler}, S.},
  \au{{Zhang}, C.}, \au{{Zhang}, J.}, \au{{Zhang}, J.~H.}, \au{{Zhang}, S.~D.},
  \au{{Zhang}, S.~W.}, \au{{Zhang}, Z.}, \au{{Zheng}, Z.~M.}, \au{{Zhu},
  Z.~Q.}, \au{{Zhuang}, H.~L.}, \au{{Zhukov}, V.}, \au{{Zichichi}, A.},
  \au{{Zimmermann}, N.}, \au{{Zuccon}, P.} \& \au{{AMS Collaboration}}}
  \yr{2016{\natexlab{{\em b\/}}}}  \at{{Antiproton Flux, Antiproton-to-Proton
  Flux Ratio, and Properties of Elementary Particle Fluxes in Primary Cosmic
  Rays Measured with the Alpha Magnetic Spectrometer on the International Space
  Station}}.  \jt{PRL}  \bvol{117}~(9),  \pg{091103}.

\bibitem[Aguilar {\em et~al.\/}(2019{\natexlab{{\em a\/}}})Aguilar,
  Ali~Cavasonza, Alpat, Ambrosi, Arruda, Attig, Azzarello, Bachlechner, Barao,
  Barrau, Barrin, Bartoloni, Basara, Ba\ifmmode \mbox{\c{s}}\else
  \c{s}\fi{}e\ifmmode \breve{g}\else \u{g}\fi{}mez-du Pree, Battiston, Becker,
  Behlmann, Beischer, Berdugo, Bertucci, Bindi, de~Boer, Bollweg, Borgia,
  Boschini, Bourquin, Bueno, Burger, Burger, Cai, Capell, Caroff, Casaus,
  Castellini, Cervelli, Chang, Chen, Chen, Chen, Cheng, Chou, Choutko, Chung,
  Clark, Coignet, Consolandi, Contin, Corti, Crispoltoni, Cui, Dadzie, Dai,
  Datta, Delgado, Della~Torre, Demirk\"oz, Derome, Di~Falco, Di~Felice,
  Dimiccoli, D\'{\i}az, von Doetinchem, Dong, Donnini, Duranti, Egorov, Eline,
  Eronen, Feng, Fiandrini, Fisher, Formato, Galaktionov, Garc\'{\i}a-L\'opez,
  Gargiulo, Gast, Gebauer, Gervasi, Giovacchini, G\'omez-Coral, Gong, Goy,
  Grabski, Grandi, Graziani, Guo, Haino, Han, He, Heil, Hsieh, Huang, Huang,
  Incagli, Jia, Jinchi, Kanishev, Khiali, Kirn, Konak, Kounina, Kounine,
  Koutsenko, Kulemzin, La~Vacca, Laudi, Laurenti, Lazzizzera, Lebedev, Lee,
  Lee, Leluc, Li, Li, Li, Li, Light, Lin, Lippert, Liu, Liu, Liu, Lu, Lu,
  Luebelsmeyer, Luo, Luo, Luo, Lyu, Machate, Ma\~n\'a, Mar\'{\i}n, Martin,
  Mart\'{\i}nez, Masi, Maurin, Menchaca-Rocha, Meng, Mo, Molero, Mott,
  Mussolin, Nelson, Ni, Nikonov, Nozzoli, Oliva, Orcinha, Palermo, Palmonari,
  Paniccia, Pashnin, Pauluzzi, Pensotti, Perrina, Phan, Picot-Clemente,
  Plyaskin, Pohl, Poireau, Popkow, Quadrani, Qi, Qin, Qu, Rancoita, Rapin,
  Conde, Rosier-Lees, Rozhkov, Rozza, Sagdeev, Solano, Schael, Schmidt, von
  Dratzig, Schwering, Seo, Shan, Shi, Siedenburg, Song, Sun, Tacconi, Tang,
  Tang, Tian, Ting, Ting, Tomassetti, Torsti, Urban, Vagelli, Valente,
  Valtonen, Acosta, Vecchi, Velasco, Vialle, Viz\'an, Wang, Wang, Wang, Wang,
  Wang, Wang, Wei, Weng, Wu, Xiong, Xu, Yan, Yang, Yi, Yu, Yu, Zannoni,
  Zeissler, Zhang, Zhang, Zhang, Zhang, Zhao, Zheng, Zhuang, Zhukov, Zichichi,
  Zimmermann \& Zuccon]{PhysRevLett.122.101101}
{\sc \au{Aguilar, M.}, \au{Ali~Cavasonza, L.}, \au{Alpat, B.}, \au{Ambrosi,
  G.}, \au{Arruda, L.}, \au{Attig, N.}, \au{Azzarello, P.}, \au{Bachlechner,
  A.}, \au{Barao, F.}, \au{Barrau, A.}, \au{Barrin, L.}, \au{Bartoloni, A.},
  \au{Basara, L.}, \au{Ba\ifmmode \mbox{\c{s}}\else \c{s}\fi{}e\ifmmode
  \breve{g}\else \u{g}\fi{}mez-du Pree, S.}, \au{Battiston, R.}, \au{Becker,
  U.}, \au{Behlmann, M.}, \au{Beischer, B.}, \au{Berdugo, J.}, \au{Bertucci,
  B.}, \au{Bindi, V.}, \au{de~Boer, W.}, \au{Bollweg, K.}, \au{Borgia, B.},
  \au{Boschini, M.~J.}, \au{Bourquin, M.}, \au{Bueno, E.~F.}, \au{Burger, J.},
  \au{Burger, W.~J.}, \au{Cai, X.~D.}, \au{Capell, M.}, \au{Caroff, S.},
  \au{Casaus, J.}, \au{Castellini, G.}, \au{Cervelli, F.}, \au{Chang, Y.~H.},
  \au{Chen, G.~M.}, \au{Chen, H.~S.}, \au{Chen, Y.}, \au{Cheng, L.}, \au{Chou,
  H.~Y.}, \au{Choutko, V.}, \au{Chung, C.~H.}, \au{Clark, C.}, \au{Coignet,
  G.}, \au{Consolandi, C.}, \au{Contin, A.}, \au{Corti, C.}, \au{Crispoltoni,
  M.}, \au{Cui, Z.}, \au{Dadzie, K.}, \au{Dai, Y.~M.}, \au{Datta, A.},
  \au{Delgado, C.}, \au{Della~Torre, S.}, \au{Demirk\"oz, M.~B.}, \au{Derome,
  L.}, \au{Di~Falco, S.}, \au{Di~Felice, V.}, \au{Dimiccoli, F.},
  \au{D\'{\i}az, C.}, \au{von Doetinchem, P.}, \au{Dong, F.}, \au{Donnini, F.},
  \au{Duranti, M.}, \au{Egorov, A.}, \au{Eline, A.}, \au{Eronen, T.}, \au{Feng,
  J.}, \au{Fiandrini, E.}, \au{Fisher, P.}, \au{Formato, V.}, \au{Galaktionov,
  Y.}, \au{Garc\'{\i}a-L\'opez, R.~J.}, \au{Gargiulo, C.}, \au{Gast, H.},
  \au{Gebauer, I.}, \au{Gervasi, M.}, \au{Giovacchini, F.}, \au{G\'omez-Coral,
  D.~M.}, \au{Gong, J.}, \au{Goy, C.}, \au{Grabski, V.}, \au{Grandi, D.},
  \au{Graziani, M.}, \au{Guo, K.~H.}, \au{Haino, S.}, \au{Han, K.~C.}, \au{He,
  Z.~H.}, \au{Heil, M.}, \au{Hsieh, T.~H.}, \au{Huang, H.}, \au{Huang, Z.~C.},
  \au{Incagli, M.}, \au{Jia, Yi}, \au{Jinchi, H.}, \au{Kanishev, K.},
  \au{Khiali, B.}, \au{Kirn, Th.}, \au{Konak, C.}, \au{Kounina, O.},
  \au{Kounine, A.}, \au{Koutsenko, V.}, \au{Kulemzin, A.}, \au{La~Vacca, G.},
  \au{Laudi, E.}, \au{Laurenti, G.}, \au{Lazzizzera, I.}, \au{Lebedev, A.},
  \au{Lee, H.~T.}, \au{Lee, S.~C.}, \au{Leluc, C.}, \au{Li, J.~Q.}, \au{Li,
  Q.}, \au{Li, T.~X.}, \au{Li, Z.~H.}, \au{Light, C.}, \au{Lin, C.~H.},
  \au{Lippert, T.}, \au{Liu, F.~Z.}, \au{Liu, Hu}, \au{Liu, Z.}, \au{Lu,
  S.~Q.}, \au{Lu, Y.~S.}, \au{Luebelsmeyer, K.}, \au{Luo, F.}, \au{Luo, J.~Z.},
  \au{Luo, Xi}, \au{Lyu, S.~S.}, \au{Machate, F.}, \au{Ma\~n\'a, C.},
  \au{Mar\'{\i}n, J.}, \au{Martin, T.}, \au{Mart\'{\i}nez, G.}, \au{Masi, N.},
  \au{Maurin, D.}, \au{Menchaca-Rocha, A.}, \au{Meng, Q.}, \au{Mo, D.~C.},
  \au{Molero, M.}, \au{Mott, P.}, \au{Mussolin, L.}, \au{Nelson, T.}, \au{Ni,
  J.~Q.}, \au{Nikonov, N.}, \au{Nozzoli, F.}, \au{Oliva, A.}, \au{Orcinha, M.},
  \au{Palermo, M.}, \au{Palmonari, F.}, \au{Paniccia, M.}, \au{Pashnin, A.},
  \au{Pauluzzi, M.}, \au{Pensotti, S.}, \au{Perrina, C.}, \au{Phan, H.~D.},
  \au{Picot-Clemente, N.}, \au{Plyaskin, V.}, \au{Pohl, M.}, \au{Poireau, V.},
  \au{Popkow, A.}, \au{Quadrani, L.}, \au{Qi, X.~M.}, \au{Qin, X.}, \au{Qu,
  Z.~Y.}, \au{Rancoita, P.~G.}, \au{Rapin, D.}, \au{Conde, A.~Reina},
  \au{Rosier-Lees, S.}, \au{Rozhkov, A.}, \au{Rozza, D.}, \au{Sagdeev, R.},
  \au{Solano, C.}, \au{Schael, S.}, \au{Schmidt, S.~M.}, \au{von Dratzig,
  A.~Schulz}, \au{Schwering, G.}, \au{Seo, E.~S.}, \au{Shan, B.~S.}, \au{Shi,
  J.~Y.}, \au{Siedenburg, T.}, \au{Song, J.~W.}, \au{Sun, Z.~T.}, \au{Tacconi,
  M.}, \au{Tang, X.~W.}, \au{Tang, Z.~C.}, \au{Tian, J.}, \au{Ting, Samuel
  C.~C.}, \au{Ting, S.~M.}, \au{Tomassetti, N.}, \au{Torsti, J.}, \au{Urban,
  T.}, \au{Vagelli, V.}, \au{Valente, E.}, \au{Valtonen, E.}, \au{Acosta,
  M.~V\'azquez}, \au{Vecchi, M.}, \au{Velasco, M.}, \au{Vialle, J.~P.},
  \au{Viz\'an, J.}, \au{Wang, L.~Q.}, \au{Wang, N.~H.}, \au{Wang, Q.~L.},
  \au{Wang, X.}, \au{Wang, X.~Q.}, \au{Wang, Z.~X.}, \au{Wei, J.}, \au{Weng,
  Z.~L.}, \au{Wu, H.}, \au{Xiong, R.~Q.}, \au{Xu, W.}, \au{Yan, Q.}, \au{Yang,
  Y.}, \au{Yi, H.}, \au{Yu, Y.~J.}, \au{Yu, Z.~Q.}, \au{Zannoni, M.},
  \au{Zeissler, S.}, \au{Zhang, C.}, \au{Zhang, F.}, \au{Zhang, J.~H.},
  \au{Zhang, Z.}, \au{Zhao, F.}, \au{Zheng, Z.~M.}, \au{Zhuang, H.~L.},
  \au{Zhukov, V.}, \au{Zichichi, A.}, \au{Zimmermann, N.} \& \au{Zuccon, P.}}
  \yr{2019{\natexlab{{\em a\/}}}}  \at{Towards understanding the origin of
  cosmic-ray electrons}.  \jt{Phys. Rev. Lett.}  \bvol{122},  \pg{101101}.

\bibitem[{Aguilar} {\em et~al.\/}(2018{\natexlab{{\em a\/}}}){Aguilar}, {Ali
  Cavasonza}, {Ambrosi}, {Arruda}, {Attig}, {Aupetit}, {Azzarello},
  {Bachlechner}, {Barao}, {Barrau}, {Barrin}, {Bartoloni}, {Basara},
  {Ba{\textcommabelow s}e{\v{g}}mez-du Pree}, {Battarbee}, {Battiston},
  {Becker}, {Behlmann}, {Beischer}, {Berdugo}, {Bertucci}, {Bindel}, {Bindi},
  {de Boer}, {Bollweg}, {Bonnivard}, {Borgia}, {Boschini}, {Bourquin}, {Bueno},
  {Burger}, {Burger}, {Cadoux}, {Cai}, {Capell}, {Caroff}, {Casaus},
  {Castellini}, {Cervelli}, {Chae}, {Chang}, {Chen}, {Chen}, {Chen}, {Cheng},
  {Chou}, {Choumilov}, {Choutko}, {Chung}, {Clark}, {Clavero}, {Coignet},
  {Consolandi}, {Contin}, {Corti}, {Creus}, {Crispoltoni}, {Cui}, {Dadzie},
  {Dai}, {Datta}, {Delgado}, {Della Torre}, {Demirk{\"o}z}, {Derome}, {Di
  Falco}, {Dimiccoli}, {D{\'\i}az}, {von Doetinchem}, {Dong}, {Donnini},
  {Duranti}, {D'Urso}, {Egorov}, {Eline}, {Eronen}, {Feng}, {Fiandrini},
  {Fisher}, {Formato}, {Galaktionov}, {Gallucci}, {Garc{\'\i}a-L{\'o}pez},
  {Gargiulo}, {Gast}, {Gebauer}, {Gervasi}, {Ghelfi}, {Giovacchini},
  {G{\'o}mez-Coral}, {Gong}, {Goy}, {Grabski}, {Grandi}, {Graziani}, {Guo},
  {Haino}, {Han}, {He}, {Heil}, {Hsieh}, {Huang}, {Huang}, {Huh}, {Incagli},
  {Ionica}, {Jang}, {Jia}, {Jinchi}, {Kang}, {Kanishev}, {Khiali}, {Kim},
  {Kim}, {Kirn}, {Konak}, {Kounina}, {Kounine}, {Koutsenko}, {Kulemzin}, {La
  Vacca}, {Laudi}, {Laurenti}, {Lazzizzera}, {Lebedev}, {Lee}, {Lee}, {Leluc},
  {Li}, {Li}, {Li}, {Li}, {Li}, {Li}, {Li}, {Lim}, {Lin}, {Lipari}, {Lippert},
  {Liu}, {Liu}, {Lordello}, {Lu}, {Lu}, {Luebelsmeyer}, {Luo}, {Luo}, {Lyu},
  {Machate}, {Ma{\~n}{\'a}}, {Mar{\'\i}n}, {Martin}, {Mart{\'\i}nez}, {Masi},
  {Maurin}, {Menchaca-Rocha}, {Meng}, {Mikuni}, {Mo}, {Mott}, {Nelson}, {Ni},
  {Nikonov}, {Nozzoli}, {Oliva}, {Orcinha}, {Palermo}, {Palmonari},
  {Palomares}, {Paniccia}, {Pauluzzi}, {Pensotti}, {Perrina}, {Phan},
  {Picot-Clemente}, {Pilo}, {Pizzolotto}, {Plyaskin}, {Pohl}, {Poireau},
  {Quadrani}, {Qi}, {Qin}, {Qu}, {R{\"a}ih{\"a}}, {Rancoita}, {Rapin}, {Ricol},
  {Rosier-Lees}, {Rozhkov}, {Rozza}, {Sagdeev}, {Schael}, {Schmidt}, {Schulz
  von Dratzig}, {Schwering}, {Seo}, {Shan}, {Shi}, {Siedenburg}, {Son}, {Song},
  {Tacconi}, {Tang}, {Tang}, {Tescaro}, {Ting}, {Ting}, {Tomassetti}, {Torsti},
  {T{\"u}rko{\v{g}}lu}, {Urban}, {Vagelli}, {Valente}, {Valtonen}, {V{\'a}zquez
  Acosta}, {Vecchi}, {Velasco}, {Vialle}, {Vitale}, {Wang}, {Wang}, {Wang},
  {Wang}, {Wang}, {Wang}, {Wei}, {Weng}, {Whitman}, {Wu}, {Wu}, {Xiong}, {Xu},
  {Yan}, {Yang}, {Yang}, {Yang}, {Yi}, {Yu}, {Yu}, {Zannoni}, {Zeissler},
  {Zhang}, {Zhang}, {Zhang}, {Zhang}, {Zhang}, {Zhang}, {Zheng}, {Zhuang},
  {Zhukov}, {Zichichi}, {Zimmermann}, {Zuccon} \& {AMS
  Collaboration}]{AMS02Sec}
{\sc \au{{Aguilar}, M.}, \au{{Ali Cavasonza}, L.}, \au{{Ambrosi}, G.},
  \au{{Arruda}, L.}, \au{{Attig}, N.}, \au{{Aupetit}, S.}, \au{{Azzarello},
  P.}, \au{{Bachlechner}, A.}, \au{{Barao}, F.}, \au{{Barrau}, A.},
  \au{{Barrin}, L.}, \au{{Bartoloni}, A.}, \au{{Basara}, L.},
  \au{{Ba{\textcommabelow s}e{\v{g}}mez-du Pree}, S.}, \au{{Battarbee}, M.},
  \au{{Battiston}, R.}, \au{{Becker}, U.}, \au{{Behlmann}, M.}, \au{{Beischer},
  B.}, \au{{Berdugo}, J.}, \au{{Bertucci}, B.}, \au{{Bindel}, K.~F.},
  \au{{Bindi}, V.}, \au{{de Boer}, W.}, \au{{Bollweg}, K.}, \au{{Bonnivard},
  V.}, \au{{Borgia}, B.}, \au{{Boschini}, M.~J.}, \au{{Bourquin}, M.},
  \au{{Bueno}, E.~F.}, \au{{Burger}, J.}, \au{{Burger}, W.~J.}, \au{{Cadoux},
  F.}, \au{{Cai}, X.~D.}, \au{{Capell}, M.}, \au{{Caroff}, S.}, \au{{Casaus},
  J.}, \au{{Castellini}, G.}, \au{{Cervelli}, F.}, \au{{Chae}, M.~J.},
  \au{{Chang}, Y.~H.}, \au{{Chen}, A.~I.}, \au{{Chen}, G.~M.}, \au{{Chen},
  H.~S.}, \au{{Cheng}, L.}, \au{{Chou}, H.~Y.}, \au{{Choumilov}, E.},
  \au{{Choutko}, V.}, \au{{Chung}, C.~H.}, \au{{Clark}, C.}, \au{{Clavero},
  R.}, \au{{Coignet}, G.}, \au{{Consolandi}, C.}, \au{{Contin}, A.},
  \au{{Corti}, C.}, \au{{Creus}, W.}, \au{{Crispoltoni}, M.}, \au{{Cui}, Z.},
  \au{{Dadzie}, K.}, \au{{Dai}, Y.~M.}, \au{{Datta}, A.}, \au{{Delgado}, C.},
  \au{{Della Torre}, S.}, \au{{Demirk{\"o}z}, M.~B.}, \au{{Derome}, L.},
  \au{{Di Falco}, S.}, \au{{Dimiccoli}, F.}, \au{{D{\'\i}az}, C.}, \au{{von
  Doetinchem}, P.}, \au{{Dong}, F.}, \au{{Donnini}, F.}, \au{{Duranti}, M.},
  \au{{D'Urso}, D.}, \au{{Egorov}, A.}, \au{{Eline}, A.}, \au{{Eronen}, T.},
  \au{{Feng}, J.}, \au{{Fiandrini}, E.}, \au{{Fisher}, P.}, \au{{Formato}, V.},
  \au{{Galaktionov}, Y.}, \au{{Gallucci}, G.}, \au{{Garc{\'\i}a-L{\'o}pez},
  R.~J.}, \au{{Gargiulo}, C.}, \au{{Gast}, H.}, \au{{Gebauer}, I.},
  \au{{Gervasi}, M.}, \au{{Ghelfi}, A.}, \au{{Giovacchini}, F.},
  \au{{G{\'o}mez-Coral}, D.~M.}, \au{{Gong}, J.}, \au{{Goy}, C.},
  \au{{Grabski}, V.}, \au{{Grandi}, D.}, \au{{Graziani}, M.}, \au{{Guo},
  K.~H.}, \au{{Haino}, S.}, \au{{Han}, K.~C.}, \au{{He}, Z.~H.}, \au{{Heil},
  M.}, \au{{Hsieh}, T.~H.}, \au{{Huang}, H.}, \au{{Huang}, Z.~C.}, \au{{Huh},
  C.}, \au{{Incagli}, M.}, \au{{Ionica}, M.}, \au{{Jang}, W.~Y.}, \au{{Jia},
  Yi}, \au{{Jinchi}, H.}, \au{{Kang}, S.~C.}, \au{{Kanishev}, K.},
  \au{{Khiali}, B.}, \au{{Kim}, G.~N.}, \au{{Kim}, K.~S.}, \au{{Kirn}, Th.},
  \au{{Konak}, C.}, \au{{Kounina}, O.}, \au{{Kounine}, A.}, \au{{Koutsenko},
  V.}, \au{{Kulemzin}, A.}, \au{{La Vacca}, G.}, \au{{Laudi}, E.},
  \au{{Laurenti}, G.}, \au{{Lazzizzera}, I.}, \au{{Lebedev}, A.}, \au{{Lee},
  H.~T.}, \au{{Lee}, S.~C.}, \au{{Leluc}, C.}, \au{{Li}, H.~S.}, \au{{Li},
  J.~Q.}, \au{{Li}, Q.}, \au{{Li}, T.~X.}, \au{{Li}, Y.}, \au{{Li}, Z.~H.},
  \au{{Li}, Z.~Y.}, \au{{Lim}, S.}, \au{{Lin}, C.~H.}, \au{{Lipari}, P.},
  \au{{Lippert}, T.}, \au{{Liu}, D.}, \au{{Liu}, Hu}, \au{{Lordello}, V.~D.},
  \au{{Lu}, S.~Q.}, \au{{Lu}, Y.~S.}, \au{{Luebelsmeyer}, K.}, \au{{Luo}, F.},
  \au{{Luo}, J.~Z.}, \au{{Lyu}, S.~S.}, \au{{Machate}, F.}, \au{{Ma{\~n}{\'a}},
  C.}, \au{{Mar{\'\i}n}, J.}, \au{{Martin}, T.}, \au{{Mart{\'\i}nez}, G.},
  \au{{Masi}, N.}, \au{{Maurin}, D.}, \au{{Menchaca-Rocha}, A.}, \au{{Meng},
  Q.}, \au{{Mikuni}, V.~M.}, \au{{Mo}, D.~C.}, \au{{Mott}, P.}, \au{{Nelson},
  T.}, \au{{Ni}, J.~Q.}, \au{{Nikonov}, N.}, \au{{Nozzoli}, F.}, \au{{Oliva},
  A.}, \au{{Orcinha}, M.}, \au{{Palermo}, M.}, \au{{Palmonari}, F.},
  \au{{Palomares}, C.}, \au{{Paniccia}, M.}, \au{{Pauluzzi}, M.},
  \au{{Pensotti}, S.}, \au{{Perrina}, C.}, \au{{Phan}, H.~D.},
  \au{{Picot-Clemente}, N.}, \au{{Pilo}, F.}, \au{{Pizzolotto}, C.},
  \au{{Plyaskin}, V.}, \au{{Pohl}, M.}, \au{{Poireau}, V.}, \au{{Quadrani},
  L.}, \au{{Qi}, X.~M.}, \au{{Qin}, X.}, \au{{Qu}, Z.~Y.}, \au{{R{\"a}ih{\"a}},
  T.}, \au{{Rancoita}, P.~G.}, \au{{Rapin}, D.}, \au{{Ricol}, J.~S.},
  \au{{Rosier-Lees}, S.}, \au{{Rozhkov}, A.}, \au{{Rozza}, D.}, \au{{Sagdeev},
  R.}, \au{{Schael}, S.}, \au{{Schmidt}, S.~M.}, \au{{Schulz von Dratzig}, A.},
  \au{{Schwering}, G.}, \au{{Seo}, E.~S.}, \au{{Shan}, B.~S.}, \au{{Shi},
  J.~Y.}, \au{{Siedenburg}, T.}, \au{{Son}, D.}, \au{{Song}, J.~W.},
  \au{{Tacconi}, M.}, \au{{Tang}, X.~W.}, \au{{Tang}, Z.~C.}, \au{{Tescaro},
  D.}, \au{{Ting}, Samuel C.~C.}, \au{{Ting}, S.~M.}, \au{{Tomassetti}, N.},
  \au{{Torsti}, J.}, \au{{T{\"u}rko{\v{g}}lu}, C.}, \au{{Urban}, T.},
  \au{{Vagelli}, V.}, \au{{Valente}, E.}, \au{{Valtonen}, E.}, \au{{V{\'a}zquez
  Acosta}, M.}, \au{{Vecchi}, M.}, \au{{Velasco}, M.}, \au{{Vialle}, J.~P.},
  \au{{Vitale}, V.}, \au{{Wang}, L.~Q.}, \au{{Wang}, N.~H.}, \au{{Wang},
  Q.~L.}, \au{{Wang}, X.}, \au{{Wang}, X.~Q.}, \au{{Wang}, Z.~X.}, \au{{Wei},
  C.~C.}, \au{{Weng}, Z.~L.}, \au{{Whitman}, K.}, \au{{Wu}, H.}, \au{{Wu}, X.},
  \au{{Xiong}, R.~Q.}, \au{{Xu}, W.}, \au{{Yan}, Q.}, \au{{Yang}, J.},
  \au{{Yang}, M.}, \au{{Yang}, Y.}, \au{{Yi}, H.}, \au{{Yu}, Y.~J.}, \au{{Yu},
  Z.~Q.}, \au{{Zannoni}, M.}, \au{{Zeissler}, S.}, \au{{Zhang}, C.},
  \au{{Zhang}, F.}, \au{{Zhang}, J.}, \au{{Zhang}, J.~H.}, \au{{Zhang}, S.~W.},
  \au{{Zhang}, Z.}, \au{{Zheng}, Z.~M.}, \au{{Zhuang}, H.~L.}, \au{{Zhukov},
  V.}, \au{{Zichichi}, A.}, \au{{Zimmermann}, N.}, \au{{Zuccon}, P.} \&
  \au{{AMS Collaboration}}} \yr{2018{\natexlab{{\em a\/}}}}  \at{{Observation
  of New Properties of Secondary Cosmic Rays Lithium, Beryllium, and Boron by
  the Alpha Magnetic Spectrometer on the International Space Station}}.
  \jt{Physical Review Letters}  \bvol{120}~(2),  \pg{021101}.

\bibitem[{Aguilar} {\em et~al.\/}(2016{\natexlab{{\em c\/}}}){Aguilar}, {Ali
  Cavasonza}, {Ambrosi}, {Arruda}, {Attig}, {Aupetit}, {Azzarello},
  {Bachlechner}, {Barao}, {Barrau}, {Barrin}, {Bartoloni}, {Basara},
  {Ba{\textcommabelow s}e{\v{g}}mez-du Pree}, {Battarbee}, {Battiston},
  {Becker}, {Behlmann}, {Beischer}, {Berdugo}, {Bertucci}, {Bindel}, {Bindi},
  {Boella}, {de Boer}, {Bollweg}, {Bonnivard}, {Borgia}, {Boschini},
  {Bourquin}, {Bueno}, {Burger}, {Cadoux}, {Cai}, {Capell}, {Caroff}, {Casaus},
  {Castellini}, {Cervelli}, {Chae}, {Chang}, {Chen}, {Chen}, {Chen}, {Cheng},
  {Chou}, {Choumilov}, {Choutko}, {Chung}, {Clark}, {Clavero}, {Coignet},
  {Consolandi}, {Contin}, {Corti}, {Creus}, {Crispoltoni}, {Cui}, {Dai},
  {Delgado}, {Della Torre}, {Demakov}, {Demirk{\"o}z}, {Derome}, {Di Falco},
  {Dimiccoli}, {D{\'\i}az}, {von Doetinchem}, {Dong}, {Donnini}, {Duranti},
  {D'Urso}, {Egorov}, {Eline}, {Eronen}, {Feng}, {Fiand rini}, {Finch},
  {Fisher}, {Formato}, {Galaktionov}, {Gallucci}, {Garc{\'\i}a},
  {Garc{\'\i}a-L{\'o}pez}, {Gargiulo}, {Gast}, {Gebauer}, {Gervasi}, {Ghelfi},
  {Giovacchini}, {Goglov}, {G{\'o}mez-Coral}, {Gong}, {Goy}, {Grabski},
  {Grandi}, {Graziani}, {Guo}, {Haino}, {Han}, {He}, {Heil}, {Hoffman},
  {Hsieh}, {Huang}, {Huang}, {Huh}, {Incagli}, {Ionica}, {Jang}, {Jinchi},
  {Kang}, {Kanishev}, {Kim}, {Kim}, {Kirn}, {Konak}, {Kounina}, {Kounine},
  {Koutsenko}, {Krafczyk}, {La Vacca}, {Laudi}, {Laurenti}, {Lazzizzera},
  {Lebedev}, {Lee}, {Lee}, {Leluc}, {Li}, {Li}, {Li}, {Li}, {Li}, {Li}, {Li},
  {Li}, {Li}, {Lim}, {Lin}, {Lipari}, {Lippert}, {Liu}, {Liu}, {Lordello},
  {Lu}, {Lu}, {Luebelsmeyer}, {Luo}, {Luo}, {Lv}, {Machate}, {Majka},
  {Ma{\~n}{\'a}}, {Mar{\'\i}n}, {Martin}, {Mart{\'\i}nez}, {Masi}, {Maurin},
  {Menchaca-Rocha}, {Meng}, {Mikuni}, {Mo}, {Morescalchi}, {Mott}, {Nelson},
  {Ni}, {Nikonov}, {Nozzoli}, {Oliva}, {Orcinha}, {Palmonari}, {Palomares},
  {Paniccia}, {Pauluzzi}, {Pensotti}, {Pereira}, {Picot-Clemente}, {Pilo},
  {Pizzolotto}, {Plyaskin}, {Pohl}, {Poireau}, {Putze}, {Quadrani}, {Qi},
  {Qin}, {Qu}, {R{\"a}ih{\"a}}, {Rancoita}, {Rapin}, {Ricol}, {Rosier-Lees},
  {Rozhkov}, {Rozza}, {Sagdeev}, {Sandweiss}, {Saouter}, {Schael}, {Schmidt},
  {Schulz von Dratzig}, {Schwering}, {Seo}, {Shan}, {Shi}, {Siedenburg}, {Son},
  {Song}, {Sun}, {Tacconi}, {Tang}, {Tang}, {Tao}, {Tescaro}, {Ting}, {Ting},
  {Tomassetti}, {Torsti}, {T{\"u}rko{\v{g}}lu}, {Urban}, {Vagelli}, {Valente},
  {Vannini}, {Valtonen}, {V{\'a}zquez Acosta}, {Vecchi}, {Velasco}, {Vialle},
  {Vitale}, {Vitillo}, {Wang}, {Wang}, {Wang}, {Wang}, {Wang}, {Wang}, {Wei},
  {Weng}, {Whitman}, {Wienkenh{\"o}ver}, {Wu}, {Wu}, {Xia}, {Xiong}, {Xu},
  {Yan}, {Yang}, {Yang}, {Yang}, {Yi}, {Yu}, {Yu}, {Zeissler}, {Zhang},
  {Zhang}, {Zhang}, {Zhang}, {Zhang}, {Zhang}, {Zheng}, {Zhu}, {Zhuang},
  {Zhukov}, {Zichichi}, {Zimmermann}, {Zuccon} \& {AMS Collaboration}]{AMS02BC}
{\sc \au{{Aguilar}, M.}, \au{{Ali Cavasonza}, L.}, \au{{Ambrosi}, G.},
  \au{{Arruda}, L.}, \au{{Attig}, N.}, \au{{Aupetit}, S.}, \au{{Azzarello},
  P.}, \au{{Bachlechner}, A.}, \au{{Barao}, F.}, \au{{Barrau}, A.},
  \au{{Barrin}, L.}, \au{{Bartoloni}, A.}, \au{{Basara}, L.},
  \au{{Ba{\textcommabelow s}e{\v{g}}mez-du Pree}, S.}, \au{{Battarbee}, M.},
  \au{{Battiston}, R.}, \au{{Becker}, U.}, \au{{Behlmann}, M.}, \au{{Beischer},
  B.}, \au{{Berdugo}, J.}, \au{{Bertucci}, B.}, \au{{Bindel}, K.~F.},
  \au{{Bindi}, V.}, \au{{Boella}, G.}, \au{{de Boer}, W.}, \au{{Bollweg}, K.},
  \au{{Bonnivard}, V.}, \au{{Borgia}, B.}, \au{{Boschini}, M.~J.},
  \au{{Bourquin}, M.}, \au{{Bueno}, E.~F.}, \au{{Burger}, J.}, \au{{Cadoux},
  F.}, \au{{Cai}, X.~D.}, \au{{Capell}, M.}, \au{{Caroff}, S.}, \au{{Casaus},
  J.}, \au{{Castellini}, G.}, \au{{Cervelli}, F.}, \au{{Chae}, M.~J.},
  \au{{Chang}, Y.~H.}, \au{{Chen}, A.~I.}, \au{{Chen}, G.~M.}, \au{{Chen},
  H.~S.}, \au{{Cheng}, L.}, \au{{Chou}, H.~Y.}, \au{{Choumilov}, E.},
  \au{{Choutko}, V.}, \au{{Chung}, C.~H.}, \au{{Clark}, C.}, \au{{Clavero},
  R.}, \au{{Coignet}, G.}, \au{{Consolandi}, C.}, \au{{Contin}, A.},
  \au{{Corti}, C.}, \au{{Creus}, W.}, \au{{Crispoltoni}, M.}, \au{{Cui}, Z.},
  \au{{Dai}, Y.~M.}, \au{{Delgado}, C.}, \au{{Della Torre}, S.}, \au{{Demakov},
  O.}, \au{{Demirk{\"o}z}, M.~B.}, \au{{Derome}, L.}, \au{{Di Falco}, S.},
  \au{{Dimiccoli}, F.}, \au{{D{\'\i}az}, C.}, \au{{von Doetinchem}, P.},
  \au{{Dong}, F.}, \au{{Donnini}, F.}, \au{{Duranti}, M.}, \au{{D'Urso}, D.},
  \au{{Egorov}, A.}, \au{{Eline}, A.}, \au{{Eronen}, T.}, \au{{Feng}, J.},
  \au{{Fiand rini}, E.}, \au{{Finch}, E.}, \au{{Fisher}, P.}, \au{{Formato},
  V.}, \au{{Galaktionov}, Y.}, \au{{Gallucci}, G.}, \au{{Garc{\'\i}a}, B.},
  \au{{Garc{\'\i}a-L{\'o}pez}, R.~J.}, \au{{Gargiulo}, C.}, \au{{Gast}, H.},
  \au{{Gebauer}, I.}, \au{{Gervasi}, M.}, \au{{Ghelfi}, A.}, \au{{Giovacchini},
  F.}, \au{{Goglov}, P.}, \au{{G{\'o}mez-Coral}, D.~M.}, \au{{Gong}, J.},
  \au{{Goy}, C.}, \au{{Grabski}, V.}, \au{{Grandi}, D.}, \au{{Graziani}, M.},
  \au{{Guo}, K.~H.}, \au{{Haino}, S.}, \au{{Han}, K.~C.}, \au{{He}, Z.~H.},
  \au{{Heil}, M.}, \au{{Hoffman}, J.}, \au{{Hsieh}, T.~H.}, \au{{Huang}, H.},
  \au{{Huang}, Z.~C.}, \au{{Huh}, C.}, \au{{Incagli}, M.}, \au{{Ionica}, M.},
  \au{{Jang}, W.~Y.}, \au{{Jinchi}, H.}, \au{{Kang}, S.~C.}, \au{{Kanishev},
  K.}, \au{{Kim}, G.~N.}, \au{{Kim}, K.~S.}, \au{{Kirn}, Th.}, \au{{Konak},
  C.}, \au{{Kounina}, O.}, \au{{Kounine}, A.}, \au{{Koutsenko}, V.},
  \au{{Krafczyk}, M.~S.}, \au{{La Vacca}, G.}, \au{{Laudi}, E.},
  \au{{Laurenti}, G.}, \au{{Lazzizzera}, I.}, \au{{Lebedev}, A.}, \au{{Lee},
  H.~T.}, \au{{Lee}, S.~C.}, \au{{Leluc}, C.}, \au{{Li}, H.~S.}, \au{{Li},
  J.~Q.}, \au{{Li}, J.~Q.}, \au{{Li}, Q.}, \au{{Li}, T.~X.}, \au{{Li}, W.},
  \au{{Li}, Y.}, \au{{Li}, Z.~H.}, \au{{Li}, Z.~Y.}, \au{{Lim}, S.}, \au{{Lin},
  C.~H.}, \au{{Lipari}, P.}, \au{{Lippert}, T.}, \au{{Liu}, D.}, \au{{Liu},
  Hu}, \au{{Lordello}, V.~D.}, \au{{Lu}, S.~Q.}, \au{{Lu}, Y.~S.},
  \au{{Luebelsmeyer}, K.}, \au{{Luo}, F.}, \au{{Luo}, J.~Z.}, \au{{Lv}, S.~S.},
  \au{{Machate}, F.}, \au{{Majka}, R.}, \au{{Ma{\~n}{\'a}}, C.},
  \au{{Mar{\'\i}n}, J.}, \au{{Martin}, T.}, \au{{Mart{\'\i}nez}, G.},
  \au{{Masi}, N.}, \au{{Maurin}, D.}, \au{{Menchaca-Rocha}, A.}, \au{{Meng},
  Q.}, \au{{Mikuni}, V.~M.}, \au{{Mo}, D.~C.}, \au{{Morescalchi}, L.},
  \au{{Mott}, P.}, \au{{Nelson}, T.}, \au{{Ni}, J.~Q.}, \au{{Nikonov}, N.},
  \au{{Nozzoli}, F.}, \au{{Oliva}, A.}, \au{{Orcinha}, M.}, \au{{Palmonari},
  F.}, \au{{Palomares}, C.}, \au{{Paniccia}, M.}, \au{{Pauluzzi}, M.},
  \au{{Pensotti}, S.}, \au{{Pereira}, R.}, \au{{Picot-Clemente}, N.},
  \au{{Pilo}, F.}, \au{{Pizzolotto}, C.}, \au{{Plyaskin}, V.}, \au{{Pohl}, M.},
  \au{{Poireau}, V.}, \au{{Putze}, A.}, \au{{Quadrani}, L.}, \au{{Qi}, X.~M.},
  \au{{Qin}, X.}, \au{{Qu}, Z.~Y.}, \au{{R{\"a}ih{\"a}}, T.}, \au{{Rancoita},
  P.~G.}, \au{{Rapin}, D.}, \au{{Ricol}, J.~S.}, \au{{Rosier-Lees}, S.},
  \au{{Rozhkov}, A.}, \au{{Rozza}, D.}, \au{{Sagdeev}, R.}, \au{{Sandweiss},
  J.}, \au{{Saouter}, P.}, \au{{Schael}, S.}, \au{{Schmidt}, S.~M.},
  \au{{Schulz von Dratzig}, A.}, \au{{Schwering}, G.}, \au{{Seo}, E.~S.},
  \au{{Shan}, B.~S.}, \au{{Shi}, J.~Y.}, \au{{Siedenburg}, T.}, \au{{Son}, D.},
  \au{{Song}, J.~W.}, \au{{Sun}, W.~H.}, \au{{Tacconi}, M.}, \au{{Tang},
  X.~W.}, \au{{Tang}, Z.~C.}, \au{{Tao}, L.}, \au{{Tescaro}, D.}, \au{{Ting},
  Samuel C.~C.}, \au{{Ting}, S.~M.}, \au{{Tomassetti}, N.}, \au{{Torsti}, J.},
  \au{{T{\"u}rko{\v{g}}lu}, C.}, \au{{Urban}, T.}, \au{{Vagelli}, V.},
  \au{{Valente}, E.}, \au{{Vannini}, C.}, \au{{Valtonen}, E.}, \au{{V{\'a}zquez
  Acosta}, M.}, \au{{Vecchi}, M.}, \au{{Velasco}, M.}, \au{{Vialle}, J.~P.},
  \au{{Vitale}, V.}, \au{{Vitillo}, S.}, \au{{Wang}, L.~Q.}, \au{{Wang},
  N.~H.}, \au{{Wang}, Q.~L.}, \au{{Wang}, X.}, \au{{Wang}, X.~Q.}, \au{{Wang},
  Z.~X.}, \au{{Wei}, C.~C.}, \au{{Weng}, Z.~L.}, \au{{Whitman}, K.},
  \au{{Wienkenh{\"o}ver}, J.}, \au{{Wu}, H.}, \au{{Wu}, X.}, \au{{Xia}, X.},
  \au{{Xiong}, R.~Q.}, \au{{Xu}, W.}, \au{{Yan}, Q.}, \au{{Yang}, J.},
  \au{{Yang}, M.}, \au{{Yang}, Y.}, \au{{Yi}, H.}, \au{{Yu}, Y.~J.}, \au{{Yu},
  Z.~Q.}, \au{{Zeissler}, S.}, \au{{Zhang}, C.}, \au{{Zhang}, J.}, \au{{Zhang},
  J.~H.}, \au{{Zhang}, S.~D.}, \au{{Zhang}, S.~W.}, \au{{Zhang}, Z.},
  \au{{Zheng}, Z.~M.}, \au{{Zhu}, Z.~Q.}, \au{{Zhuang}, H.~L.}, \au{{Zhukov},
  V.}, \au{{Zichichi}, A.}, \au{{Zimmermann}, N.}, \au{{Zuccon}, P.} \&
  \au{{AMS Collaboration}}} \yr{2016{\natexlab{{\em c\/}}}}  \at{{Precision
  Measurement of the Boron to Carbon Flux Ratio in Cosmic Rays from 1.9 GV to
  2.6 TV with the Alpha Magnetic Spectrometer on the International Space
  Station}}.  \jt{Physical Review Letters}  \bvol{117}~(23),  \pg{231102}.

\bibitem[{Aguilar} {\em et~al.\/}(2018{\natexlab{{\em b\/}}}){Aguilar}, {Ali
  Cavasonza}, {Ambrosi}, {Arruda}, {Attig}, {Aupetit}, {Azzarello},
  {Bachlechner}, {Barao}, {Barrau}, {Barrin}, {Bartoloni}, {Basara},
  {Ba{\textcommabelow s}e{\v{g}}mez-du Pree}, {Battarbee}, {Battiston},
  {Becker}, {Behlmann}, {Beischer}, {Berdugo}, {Bertucci}, {Bindel}, {Bindi},
  {de Boer}, {Bollweg}, {Bonnivard}, {Borgia}, {Boschini}, {Bourquin}, {Bueno},
  {Burger}, {Burger}, {Cadoux}, {Cai}, {Capell}, {Caroff}, {Casaus},
  {Castellini}, {Cervelli}, {Chae}, {Chang}, {Chen}, {Chen}, {Chen}, {Cheng},
  {Chou}, {Choumilov}, {Choutko}, {Chung}, {Clark}, {Clavero}, {Coignet},
  {Consolandi}, {Contin}, {Corti}, {Creus}, {Crispoltoni}, {Cui}, {Dadzie},
  {Dai}, {Datta}, {Delgado}, {Della Torre}, {Demirk{\"o}z}, {Derome}, {Di
  Falco}, {Dimiccoli}, {D{\'\i}az}, {von Doetinchem}, {Dong}, {Donnini},
  {Duranti}, {D'Urso}, {Egorov}, {Eline}, {Eronen}, {Feng}, {Fiandrini},
  {Fisher}, {Formato}, {Galaktionov}, {Gallucci}, {Garc{\'\i}a-L{\'o}pez},
  {Gargiulo}, {Gast}, {Gebauer}, {Gervasi}, {Ghelfi}, {Giovacchini},
  {G{\'o}mez-Coral}, {Gong}, {Goy}, {Grabski}, {Grandi}, {Graziani}, {Guo},
  {Haino}, {Han}, {He}, {Heil}, {Hsieh}, {Huang}, {Huang}, {Huh}, {Incagli},
  {Ionica}, {Jang}, {Jia}, {Jinchi}, {Kang}, {Kanishev}, {Khiali}, {Kim},
  {Kim}, {Kirn}, {Konak}, {Kounina}, {Kounine}, {Koutsenko}, {Kulemzin}, {La
  Vacca}, {Laudi}, {Laurenti}, {Lazzizzera}, {Lebedev}, {Lee}, {Lee}, {Leluc},
  {Li}, {Li}, {Li}, {Li}, {Li}, {Li}, {Li}, {Lim}, {Lin}, {Lipari}, {Lippert},
  {Liu}, {Liu}, {Lordello}, {Lu}, {Lu}, {Luebelsmeyer}, {Luo}, {Luo}, {Lyu},
  {Machate}, {Ma{\~n}{\'a}}, {Mar{\'\i}n}, {Martin}, {Mart{\'\i}nez}, {Masi},
  {Maurin}, {Menchaca-Rocha}, {Meng}, {Mikuni}, {Mo}, {Mott}, {Nelson}, {Ni},
  {Nikonov}, {Nozzoli}, {Oliva}, {Orcinha}, {Palermo}, {Palmonari},
  {Palomares}, {Paniccia}, {Pauluzzi}, {Pensotti}, {Perrina}, {Phan},
  {Picot-Clemente}, {Pilo}, {Pizzolotto}, {Plyaskin}, {Pohl}, {Poireau},
  {Quadrani}, {Qi}, {Qin}, {Qu}, {R{\"a}ih{\"a}}, {Rancoita}, {Rapin}, {Ricol},
  {Rosier-Lees}, {Rozhkov}, {Rozza}, {Sagdeev}, {Schael}, {Schmidt}, {Schulz
  von Dratzig}, {Schwering}, {Seo}, {Shan}, {Shi}, {Siedenburg}, {Son}, {Song},
  {Tacconi}, {Tang}, {Tang}, {Tescaro}, {Ting}, {Ting}, {Tomassetti}, {Torsti},
  {T{\"u}rko{\v{g}}lu}, {Urban}, {Vagelli}, {Valente}, {Valtonen}, {V{\'a}zquez
  Acosta}, {Vecchi}, {Velasco}, {Vialle}, {Vitale}, {Wang}, {Wang}, {Wang},
  {Wang}, {Wang}, {Wang}, {Wei}, {Weng}, {Whitman}, {Wu}, {Wu}, {Xiong}, {Xu},
  {Yan}, {Yang}, {Yang}, {Yang}, {Yi}, {Yu}, {Yu}, {Zannoni}, {Zeissler},
  {Zhang}, {Zhang}, {Zhang}, {Zhang}, {Zhang}, {Zhang}, {Zheng}, {Zhuang},
  {Zhukov}, {Zichichi}, {Zimmermann}, {Zuccon} \& {AMS
  Collaboration}]{Aguilar+18sec}
{\sc \au{{Aguilar}, M.}, \au{{Ali Cavasonza}, L.}, \au{{Ambrosi}, G.},
  \au{{Arruda}, L.}, \au{{Attig}, N.}, \au{{Aupetit}, S.}, \au{{Azzarello},
  P.}, \au{{Bachlechner}, A.}, \au{{Barao}, F.}, \au{{Barrau}, A.},
  \au{{Barrin}, L.}, \au{{Bartoloni}, A.}, \au{{Basara}, L.},
  \au{{Ba{\textcommabelow s}e{\v{g}}mez-du Pree}, S.}, \au{{Battarbee}, M.},
  \au{{Battiston}, R.}, \au{{Becker}, U.}, \au{{Behlmann}, M.}, \au{{Beischer},
  B.}, \au{{Berdugo}, J.}, \au{{Bertucci}, B.}, \au{{Bindel}, K.~F.},
  \au{{Bindi}, V.}, \au{{de Boer}, W.}, \au{{Bollweg}, K.}, \au{{Bonnivard},
  V.}, \au{{Borgia}, B.}, \au{{Boschini}, M.~J.}, \au{{Bourquin}, M.},
  \au{{Bueno}, E.~F.}, \au{{Burger}, J.}, \au{{Burger}, W.~J.}, \au{{Cadoux},
  F.}, \au{{Cai}, X.~D.}, \au{{Capell}, M.}, \au{{Caroff}, S.}, \au{{Casaus},
  J.}, \au{{Castellini}, G.}, \au{{Cervelli}, F.}, \au{{Chae}, M.~J.},
  \au{{Chang}, Y.~H.}, \au{{Chen}, A.~I.}, \au{{Chen}, G.~M.}, \au{{Chen},
  H.~S.}, \au{{Cheng}, L.}, \au{{Chou}, H.~Y.}, \au{{Choumilov}, E.},
  \au{{Choutko}, V.}, \au{{Chung}, C.~H.}, \au{{Clark}, C.}, \au{{Clavero},
  R.}, \au{{Coignet}, G.}, \au{{Consolandi}, C.}, \au{{Contin}, A.},
  \au{{Corti}, C.}, \au{{Creus}, W.}, \au{{Crispoltoni}, M.}, \au{{Cui}, Z.},
  \au{{Dadzie}, K.}, \au{{Dai}, Y.~M.}, \au{{Datta}, A.}, \au{{Delgado}, C.},
  \au{{Della Torre}, S.}, \au{{Demirk{\"o}z}, M.~B.}, \au{{Derome}, L.},
  \au{{Di Falco}, S.}, \au{{Dimiccoli}, F.}, \au{{D{\'\i}az}, C.}, \au{{von
  Doetinchem}, P.}, \au{{Dong}, F.}, \au{{Donnini}, F.}, \au{{Duranti}, M.},
  \au{{D'Urso}, D.}, \au{{Egorov}, A.}, \au{{Eline}, A.}, \au{{Eronen}, T.},
  \au{{Feng}, J.}, \au{{Fiandrini}, E.}, \au{{Fisher}, P.}, \au{{Formato}, V.},
  \au{{Galaktionov}, Y.}, \au{{Gallucci}, G.}, \au{{Garc{\'\i}a-L{\'o}pez},
  R.~J.}, \au{{Gargiulo}, C.}, \au{{Gast}, H.}, \au{{Gebauer}, I.},
  \au{{Gervasi}, M.}, \au{{Ghelfi}, A.}, \au{{Giovacchini}, F.},
  \au{{G{\'o}mez-Coral}, D.~M.}, \au{{Gong}, J.}, \au{{Goy}, C.},
  \au{{Grabski}, V.}, \au{{Grandi}, D.}, \au{{Graziani}, M.}, \au{{Guo},
  K.~H.}, \au{{Haino}, S.}, \au{{Han}, K.~C.}, \au{{He}, Z.~H.}, \au{{Heil},
  M.}, \au{{Hsieh}, T.~H.}, \au{{Huang}, H.}, \au{{Huang}, Z.~C.}, \au{{Huh},
  C.}, \au{{Incagli}, M.}, \au{{Ionica}, M.}, \au{{Jang}, W.~Y.}, \au{{Jia},
  Yi}, \au{{Jinchi}, H.}, \au{{Kang}, S.~C.}, \au{{Kanishev}, K.},
  \au{{Khiali}, B.}, \au{{Kim}, G.~N.}, \au{{Kim}, K.~S.}, \au{{Kirn}, Th.},
  \au{{Konak}, C.}, \au{{Kounina}, O.}, \au{{Kounine}, A.}, \au{{Koutsenko},
  V.}, \au{{Kulemzin}, A.}, \au{{La Vacca}, G.}, \au{{Laudi}, E.},
  \au{{Laurenti}, G.}, \au{{Lazzizzera}, I.}, \au{{Lebedev}, A.}, \au{{Lee},
  H.~T.}, \au{{Lee}, S.~C.}, \au{{Leluc}, C.}, \au{{Li}, H.~S.}, \au{{Li},
  J.~Q.}, \au{{Li}, Q.}, \au{{Li}, T.~X.}, \au{{Li}, Y.}, \au{{Li}, Z.~H.},
  \au{{Li}, Z.~Y.}, \au{{Lim}, S.}, \au{{Lin}, C.~H.}, \au{{Lipari}, P.},
  \au{{Lippert}, T.}, \au{{Liu}, D.}, \au{{Liu}, Hu}, \au{{Lordello}, V.~D.},
  \au{{Lu}, S.~Q.}, \au{{Lu}, Y.~S.}, \au{{Luebelsmeyer}, K.}, \au{{Luo}, F.},
  \au{{Luo}, J.~Z.}, \au{{Lyu}, S.~S.}, \au{{Machate}, F.}, \au{{Ma{\~n}{\'a}},
  C.}, \au{{Mar{\'\i}n}, J.}, \au{{Martin}, T.}, \au{{Mart{\'\i}nez}, G.},
  \au{{Masi}, N.}, \au{{Maurin}, D.}, \au{{Menchaca-Rocha}, A.}, \au{{Meng},
  Q.}, \au{{Mikuni}, V.~M.}, \au{{Mo}, D.~C.}, \au{{Mott}, P.}, \au{{Nelson},
  T.}, \au{{Ni}, J.~Q.}, \au{{Nikonov}, N.}, \au{{Nozzoli}, F.}, \au{{Oliva},
  A.}, \au{{Orcinha}, M.}, \au{{Palermo}, M.}, \au{{Palmonari}, F.},
  \au{{Palomares}, C.}, \au{{Paniccia}, M.}, \au{{Pauluzzi}, M.},
  \au{{Pensotti}, S.}, \au{{Perrina}, C.}, \au{{Phan}, H.~D.},
  \au{{Picot-Clemente}, N.}, \au{{Pilo}, F.}, \au{{Pizzolotto}, C.},
  \au{{Plyaskin}, V.}, \au{{Pohl}, M.}, \au{{Poireau}, V.}, \au{{Quadrani},
  L.}, \au{{Qi}, X.~M.}, \au{{Qin}, X.}, \au{{Qu}, Z.~Y.}, \au{{R{\"a}ih{\"a}},
  T.}, \au{{Rancoita}, P.~G.}, \au{{Rapin}, D.}, \au{{Ricol}, J.~S.},
  \au{{Rosier-Lees}, S.}, \au{{Rozhkov}, A.}, \au{{Rozza}, D.}, \au{{Sagdeev},
  R.}, \au{{Schael}, S.}, \au{{Schmidt}, S.~M.}, \au{{Schulz von Dratzig}, A.},
  \au{{Schwering}, G.}, \au{{Seo}, E.~S.}, \au{{Shan}, B.~S.}, \au{{Shi},
  J.~Y.}, \au{{Siedenburg}, T.}, \au{{Son}, D.}, \au{{Song}, J.~W.},
  \au{{Tacconi}, M.}, \au{{Tang}, X.~W.}, \au{{Tang}, Z.~C.}, \au{{Tescaro},
  D.}, \au{{Ting}, Samuel C.~C.}, \au{{Ting}, S.~M.}, \au{{Tomassetti}, N.},
  \au{{Torsti}, J.}, \au{{T{\"u}rko{\v{g}}lu}, C.}, \au{{Urban}, T.},
  \au{{Vagelli}, V.}, \au{{Valente}, E.}, \au{{Valtonen}, E.}, \au{{V{\'a}zquez
  Acosta}, M.}, \au{{Vecchi}, M.}, \au{{Velasco}, M.}, \au{{Vialle}, J.~P.},
  \au{{Vitale}, V.}, \au{{Wang}, L.~Q.}, \au{{Wang}, N.~H.}, \au{{Wang},
  Q.~L.}, \au{{Wang}, X.}, \au{{Wang}, X.~Q.}, \au{{Wang}, Z.~X.}, \au{{Wei},
  C.~C.}, \au{{Weng}, Z.~L.}, \au{{Whitman}, K.}, \au{{Wu}, H.}, \au{{Wu}, X.},
  \au{{Xiong}, R.~Q.}, \au{{Xu}, W.}, \au{{Yan}, Q.}, \au{{Yang}, J.},
  \au{{Yang}, M.}, \au{{Yang}, Y.}, \au{{Yi}, H.}, \au{{Yu}, Y.~J.}, \au{{Yu},
  Z.~Q.}, \au{{Zannoni}, M.}, \au{{Zeissler}, S.}, \au{{Zhang}, C.},
  \au{{Zhang}, F.}, \au{{Zhang}, J.}, \au{{Zhang}, J.~H.}, \au{{Zhang}, S.~W.},
  \au{{Zhang}, Z.}, \au{{Zheng}, Z.~M.}, \au{{Zhuang}, H.~L.}, \au{{Zhukov},
  V.}, \au{{Zichichi}, A.}, \au{{Zimmermann}, N.}, \au{{Zuccon}, P.} \&
  \au{{AMS Collaboration}}} \yr{2018{\natexlab{{\em b\/}}}}  \at{{Observation
  of New Properties of Secondary Cosmic Rays Lithium, Beryllium, and Boron by
  the Alpha Magnetic Spectrometer on the International Space Station}}.
  \jt{PRL}  \bvol{120}~(2),  \pg{021101}.

\bibitem[{Aguilar} {\em et~al.\/}(2016{\natexlab{{\em d\/}}}){Aguilar}, {Ali
  Cavasonza}, {Ambrosi}, {Arruda}, {Attig}, {Aupetit}, {Azzarello},
  {Bachlechner}, {Barao}, {Barrau}, {Barrin}, {Bartoloni}, {Basara},
  {Ba{\textcommabelow s}e{\v{g}}mez-du Pree}, {Battarbee}, {Battiston},
  {Becker}, {Behlmann}, {Beischer}, {Berdugo}, {Bertucci}, {Bindel}, {Bindi},
  {Boella}, {de Boer}, {Bollweg}, {Bonnivard}, {Borgia}, {Boschini},
  {Bourquin}, {Bueno}, {Burger}, {Cadoux}, {Cai}, {Capell}, {Caroff}, {Casaus},
  {Castellini}, {Cervelli}, {Chae}, {Chang}, {Chen}, {Chen}, {Chen}, {Cheng},
  {Chou}, {Choumilov}, {Choutko}, {Chung}, {Clark}, {Clavero}, {Coignet},
  {Consolandi}, {Contin}, {Corti}, {Creus}, {Crispoltoni}, {Cui}, {Dai},
  {Delgado}, {Della Torre}, {Demakov}, {Demirk{\"o}z}, {Derome}, {Di Falco},
  {Dimiccoli}, {D{\'\i}az}, {von Doetinchem}, {Dong}, {Donnini}, {Duranti},
  {D'Urso}, {Egorov}, {Eline}, {Eronen}, {Feng}, {Fiand rini}, {Finch},
  {Fisher}, {Formato}, {Galaktionov}, {Gallucci}, {Garc{\'\i}a},
  {Garc{\'\i}a-L{\'o}pez}, {Gargiulo}, {Gast}, {Gebauer}, {Gervasi}, {Ghelfi},
  {Giovacchini}, {Goglov}, {G{\'o}mez-Coral}, {Gong}, {Goy}, {Grabski},
  {Grandi}, {Graziani}, {Guo}, {Haino}, {Han}, {He}, {Heil}, {Hoffman},
  {Hsieh}, {Huang}, {Huang}, {Huh}, {Incagli}, {Ionica}, {Jang}, {Jinchi},
  {Kang}, {Kanishev}, {Kim}, {Kim}, {Kirn}, {Konak}, {Kounina}, {Kounine},
  {Koutsenko}, {Krafczyk}, {La Vacca}, {Laudi}, {Laurenti}, {Lazzizzera},
  {Lebedev}, {Lee}, {Lee}, {Leluc}, {Li}, {Li}, {Li}, {Li}, {Li}, {Li}, {Li},
  {Li}, {Li}, {Lim}, {Lin}, {Lipari}, {Lippert}, {Liu}, {Liu}, {Lordello},
  {Lu}, {Lu}, {Luebelsmeyer}, {Luo}, {Luo}, {Lv}, {Machate}, {Majka},
  {Ma{\~n}{\'a}}, {Mar{\'\i}n}, {Martin}, {Mart{\'\i}nez}, {Masi}, {Maurin},
  {Menchaca-Rocha}, {Meng}, {Mikuni}, {Mo}, {Morescalchi}, {Mott}, {Nelson},
  {Ni}, {Nikonov}, {Nozzoli}, {Oliva}, {Orcinha}, {Palmonari}, {Palomares},
  {Paniccia}, {Pauluzzi}, {Pensotti}, {Pereira}, {Picot-Clemente}, {Pilo},
  {Pizzolotto}, {Plyaskin}, {Pohl}, {Poireau}, {Putze}, {Quadrani}, {Qi},
  {Qin}, {Qu}, {R{\"a}ih{\"a}}, {Rancoita}, {Rapin}, {Ricol}, {Rosier-Lees},
  {Rozhkov}, {Rozza}, {Sagdeev}, {Sandweiss}, {Saouter}, {Schael}, {Schmidt},
  {Schulz von Dratzig}, {Schwering}, {Seo}, {Shan}, {Shi}, {Siedenburg}, {Son},
  {Song}, {Sun}, {Tacconi}, {Tang}, {Tang}, {Tao}, {Tescaro}, {Ting}, {Ting},
  {Tomassetti}, {Torsti}, {T{\"u}rko{\v{g}}lu}, {Urban}, {Vagelli}, {Valente},
  {Vannini}, {Valtonen}, {V{\'a}zquez Acosta}, {Vecchi}, {Velasco}, {Vialle},
  {Vitale}, {Vitillo}, {Wang}, {Wang}, {Wang}, {Wang}, {Wang}, {Wang}, {Wei},
  {Weng}, {Whitman}, {Wienkenh{\"o}ver}, {Wu}, {Wu}, {Xia}, {Xiong}, {Xu},
  {Yan}, {Yang}, {Yang}, {Yang}, {Yi}, {Yu}, {Yu}, {Zeissler}, {Zhang},
  {Zhang}, {Zhang}, {Zhang}, {Zhang}, {Zhang}, {Zheng}, {Zhu}, {Zhuang},
  {Zhukov}, {Zichichi}, {Zimmermann}, {Zuccon} \& {AMS
  Collaboration}]{Aguilar+16BC}
{\sc \au{{Aguilar}, M.}, \au{{Ali Cavasonza}, L.}, \au{{Ambrosi}, G.},
  \au{{Arruda}, L.}, \au{{Attig}, N.}, \au{{Aupetit}, S.}, \au{{Azzarello},
  P.}, \au{{Bachlechner}, A.}, \au{{Barao}, F.}, \au{{Barrau}, A.},
  \au{{Barrin}, L.}, \au{{Bartoloni}, A.}, \au{{Basara}, L.},
  \au{{Ba{\textcommabelow s}e{\v{g}}mez-du Pree}, S.}, \au{{Battarbee}, M.},
  \au{{Battiston}, R.}, \au{{Becker}, U.}, \au{{Behlmann}, M.}, \au{{Beischer},
  B.}, \au{{Berdugo}, J.}, \au{{Bertucci}, B.}, \au{{Bindel}, K.~F.},
  \au{{Bindi}, V.}, \au{{Boella}, G.}, \au{{de Boer}, W.}, \au{{Bollweg}, K.},
  \au{{Bonnivard}, V.}, \au{{Borgia}, B.}, \au{{Boschini}, M.~J.},
  \au{{Bourquin}, M.}, \au{{Bueno}, E.~F.}, \au{{Burger}, J.}, \au{{Cadoux},
  F.}, \au{{Cai}, X.~D.}, \au{{Capell}, M.}, \au{{Caroff}, S.}, \au{{Casaus},
  J.}, \au{{Castellini}, G.}, \au{{Cervelli}, F.}, \au{{Chae}, M.~J.},
  \au{{Chang}, Y.~H.}, \au{{Chen}, A.~I.}, \au{{Chen}, G.~M.}, \au{{Chen},
  H.~S.}, \au{{Cheng}, L.}, \au{{Chou}, H.~Y.}, \au{{Choumilov}, E.},
  \au{{Choutko}, V.}, \au{{Chung}, C.~H.}, \au{{Clark}, C.}, \au{{Clavero},
  R.}, \au{{Coignet}, G.}, \au{{Consolandi}, C.}, \au{{Contin}, A.},
  \au{{Corti}, C.}, \au{{Creus}, W.}, \au{{Crispoltoni}, M.}, \au{{Cui}, Z.},
  \au{{Dai}, Y.~M.}, \au{{Delgado}, C.}, \au{{Della Torre}, S.}, \au{{Demakov},
  O.}, \au{{Demirk{\"o}z}, M.~B.}, \au{{Derome}, L.}, \au{{Di Falco}, S.},
  \au{{Dimiccoli}, F.}, \au{{D{\'\i}az}, C.}, \au{{von Doetinchem}, P.},
  \au{{Dong}, F.}, \au{{Donnini}, F.}, \au{{Duranti}, M.}, \au{{D'Urso}, D.},
  \au{{Egorov}, A.}, \au{{Eline}, A.}, \au{{Eronen}, T.}, \au{{Feng}, J.},
  \au{{Fiand rini}, E.}, \au{{Finch}, E.}, \au{{Fisher}, P.}, \au{{Formato},
  V.}, \au{{Galaktionov}, Y.}, \au{{Gallucci}, G.}, \au{{Garc{\'\i}a}, B.},
  \au{{Garc{\'\i}a-L{\'o}pez}, R.~J.}, \au{{Gargiulo}, C.}, \au{{Gast}, H.},
  \au{{Gebauer}, I.}, \au{{Gervasi}, M.}, \au{{Ghelfi}, A.}, \au{{Giovacchini},
  F.}, \au{{Goglov}, P.}, \au{{G{\'o}mez-Coral}, D.~M.}, \au{{Gong}, J.},
  \au{{Goy}, C.}, \au{{Grabski}, V.}, \au{{Grandi}, D.}, \au{{Graziani}, M.},
  \au{{Guo}, K.~H.}, \au{{Haino}, S.}, \au{{Han}, K.~C.}, \au{{He}, Z.~H.},
  \au{{Heil}, M.}, \au{{Hoffman}, J.}, \au{{Hsieh}, T.~H.}, \au{{Huang}, H.},
  \au{{Huang}, Z.~C.}, \au{{Huh}, C.}, \au{{Incagli}, M.}, \au{{Ionica}, M.},
  \au{{Jang}, W.~Y.}, \au{{Jinchi}, H.}, \au{{Kang}, S.~C.}, \au{{Kanishev},
  K.}, \au{{Kim}, G.~N.}, \au{{Kim}, K.~S.}, \au{{Kirn}, Th.}, \au{{Konak},
  C.}, \au{{Kounina}, O.}, \au{{Kounine}, A.}, \au{{Koutsenko}, V.},
  \au{{Krafczyk}, M.~S.}, \au{{La Vacca}, G.}, \au{{Laudi}, E.},
  \au{{Laurenti}, G.}, \au{{Lazzizzera}, I.}, \au{{Lebedev}, A.}, \au{{Lee},
  H.~T.}, \au{{Lee}, S.~C.}, \au{{Leluc}, C.}, \au{{Li}, H.~S.}, \au{{Li},
  J.~Q.}, \au{{Li}, J.~Q.}, \au{{Li}, Q.}, \au{{Li}, T.~X.}, \au{{Li}, W.},
  \au{{Li}, Y.}, \au{{Li}, Z.~H.}, \au{{Li}, Z.~Y.}, \au{{Lim}, S.}, \au{{Lin},
  C.~H.}, \au{{Lipari}, P.}, \au{{Lippert}, T.}, \au{{Liu}, D.}, \au{{Liu},
  Hu}, \au{{Lordello}, V.~D.}, \au{{Lu}, S.~Q.}, \au{{Lu}, Y.~S.},
  \au{{Luebelsmeyer}, K.}, \au{{Luo}, F.}, \au{{Luo}, J.~Z.}, \au{{Lv}, S.~S.},
  \au{{Machate}, F.}, \au{{Majka}, R.}, \au{{Ma{\~n}{\'a}}, C.},
  \au{{Mar{\'\i}n}, J.}, \au{{Martin}, T.}, \au{{Mart{\'\i}nez}, G.},
  \au{{Masi}, N.}, \au{{Maurin}, D.}, \au{{Menchaca-Rocha}, A.}, \au{{Meng},
  Q.}, \au{{Mikuni}, V.~M.}, \au{{Mo}, D.~C.}, \au{{Morescalchi}, L.},
  \au{{Mott}, P.}, \au{{Nelson}, T.}, \au{{Ni}, J.~Q.}, \au{{Nikonov}, N.},
  \au{{Nozzoli}, F.}, \au{{Oliva}, A.}, \au{{Orcinha}, M.}, \au{{Palmonari},
  F.}, \au{{Palomares}, C.}, \au{{Paniccia}, M.}, \au{{Pauluzzi}, M.},
  \au{{Pensotti}, S.}, \au{{Pereira}, R.}, \au{{Picot-Clemente}, N.},
  \au{{Pilo}, F.}, \au{{Pizzolotto}, C.}, \au{{Plyaskin}, V.}, \au{{Pohl}, M.},
  \au{{Poireau}, V.}, \au{{Putze}, A.}, \au{{Quadrani}, L.}, \au{{Qi}, X.~M.},
  \au{{Qin}, X.}, \au{{Qu}, Z.~Y.}, \au{{R{\"a}ih{\"a}}, T.}, \au{{Rancoita},
  P.~G.}, \au{{Rapin}, D.}, \au{{Ricol}, J.~S.}, \au{{Rosier-Lees}, S.},
  \au{{Rozhkov}, A.}, \au{{Rozza}, D.}, \au{{Sagdeev}, R.}, \au{{Sandweiss},
  J.}, \au{{Saouter}, P.}, \au{{Schael}, S.}, \au{{Schmidt}, S.~M.},
  \au{{Schulz von Dratzig}, A.}, \au{{Schwering}, G.}, \au{{Seo}, E.~S.},
  \au{{Shan}, B.~S.}, \au{{Shi}, J.~Y.}, \au{{Siedenburg}, T.}, \au{{Son}, D.},
  \au{{Song}, J.~W.}, \au{{Sun}, W.~H.}, \au{{Tacconi}, M.}, \au{{Tang},
  X.~W.}, \au{{Tang}, Z.~C.}, \au{{Tao}, L.}, \au{{Tescaro}, D.}, \au{{Ting},
  Samuel C.~C.}, \au{{Ting}, S.~M.}, \au{{Tomassetti}, N.}, \au{{Torsti}, J.},
  \au{{T{\"u}rko{\v{g}}lu}, C.}, \au{{Urban}, T.}, \au{{Vagelli}, V.},
  \au{{Valente}, E.}, \au{{Vannini}, C.}, \au{{Valtonen}, E.}, \au{{V{\'a}zquez
  Acosta}, M.}, \au{{Vecchi}, M.}, \au{{Velasco}, M.}, \au{{Vialle}, J.~P.},
  \au{{Vitale}, V.}, \au{{Vitillo}, S.}, \au{{Wang}, L.~Q.}, \au{{Wang},
  N.~H.}, \au{{Wang}, Q.~L.}, \au{{Wang}, X.}, \au{{Wang}, X.~Q.}, \au{{Wang},
  Z.~X.}, \au{{Wei}, C.~C.}, \au{{Weng}, Z.~L.}, \au{{Whitman}, K.},
  \au{{Wienkenh{\"o}ver}, J.}, \au{{Wu}, H.}, \au{{Wu}, X.}, \au{{Xia}, X.},
  \au{{Xiong}, R.~Q.}, \au{{Xu}, W.}, \au{{Yan}, Q.}, \au{{Yang}, J.},
  \au{{Yang}, M.}, \au{{Yang}, Y.}, \au{{Yi}, H.}, \au{{Yu}, Y.~J.}, \au{{Yu},
  Z.~Q.}, \au{{Zeissler}, S.}, \au{{Zhang}, C.}, \au{{Zhang}, J.}, \au{{Zhang},
  J.~H.}, \au{{Zhang}, S.~D.}, \au{{Zhang}, S.~W.}, \au{{Zhang}, Z.},
  \au{{Zheng}, Z.~M.}, \au{{Zhu}, Z.~Q.}, \au{{Zhuang}, H.~L.}, \au{{Zhukov},
  V.}, \au{{Zichichi}, A.}, \au{{Zimmermann}, N.}, \au{{Zuccon}, P.} \&
  \au{{AMS Collaboration}}} \yr{2016{\natexlab{{\em d\/}}}}  \at{{Precision
  Measurement of the Boron to Carbon Flux Ratio in Cosmic Rays from 1.9 GV to
  2.6 TV with the Alpha Magnetic Spectrometer on the International Space
  Station}}.  \jt{PRL}  \bvol{117}~(23),  \pg{231102}.

\bibitem[Aguilar {\em et~al.\/}(2019{\natexlab{{\em b\/}}})Aguilar,
  Ali~Cavasonza, Ambrosi, Arruda, Attig, Azzarello, Bachlechner, Barao, Barrau,
  Barrin, Bartoloni, Basara, Ba\ifmmode \mbox{\c{s}}\else \c{s}\fi{}e\ifmmode
  \breve{g}\else \u{g}\fi{}mez-du Pree, Battiston, Becker, Behlmann, Beischer,
  Berdugo, Bertucci, Bindi, de~Boer, Bollweg, Borgia, Boschini, Bourquin,
  Bueno, Burger, Burger, Cai, Capell, Caroff, Casaus, Castellini, Cervelli,
  Chang, Chen, Chen, Chen, Cheng, Chou, Choutko, Chung, Clark, Coignet,
  Consolandi, Contin, Corti, Crispoltoni, Cui, Dadzie, Dai, Datta, Delgado,
  Della~Torre, Demirk\"oz, Derome, Di~Falco, Dimiccoli, D\'{\i}az, von
  Doetinchem, Dong, Donnini, Duranti, Egorov, Eline, Eronen, Feng, Fiandrini,
  Fisher, Formato, Galaktionov, Garc\'{\i}a-L\'opez, Gargiulo, Gast, Gebauer,
  Gervasi, Giovacchini, G\'omez-Coral, Gong, Goy, Grabski, Grandi, Graziani,
  Guo, Haino, Han, He, Heil, Hsieh, Huang, Huang, Incagli, Jia, Jinchi,
  Kanishev, Khiali, Kirn, Konak, Kounina, Kounine, Koutsenko, Kulemzin,
  La~Vacca, Laudi, Laurenti, Lazzizzera, Lebedev, Lee, Lee, Leluc, Li, Li, Li,
  Li, Light, Lin, Lippert, Liu, Liu, Liu, Lu, Lu, Luebelsmeyer, Luo, Luo, Luo,
  Lyu, Machate, Ma\~n\'a, Mar\'{\i}n, Martin, Mart\'{\i}nez, Masi, Maurin,
  Menchaca-Rocha, Meng, Mo, Molero, Mott, Mussolin, Nelson, Ni, Nikonov,
  Nozzoli, Oliva, Orcinha, Palermo, Palmonari, Paniccia, Pashnin, Pauluzzi,
  Pensotti, Perrina, Phan, Picot-Clemente, Plyaskin, Pohl, Poireau, Popkow,
  Quadrani, Qi, Qin, Qu, Rancoita, Rapin, Conde, Rosier-Lees, Rozhkov, Rozza,
  Sagdeev, Solano, Schael, Schmidt, Schulz~von Dratzig, Schwering, Seo, Shan,
  Shi, Siedenburg, Song, Sun, Tacconi, Tang, Tang, Tian, Ting, Ting,
  Tomassetti, Torsti, Urban, Vagelli, Valente, Valtonen, V\'azquez~Acosta,
  Vecchi, Velasco, Vialle, Viz\'an, Wang, Wang, Wang, Wang, Wang, Wang, Wei,
  Weng, Wu, Xiong, Xu, Yan, Yang, Yi, Yu, Yu, Zannoni, Zeissler, Zhang, Zhang,
  Zhang, Zhang, Zhao, Zheng, Zhuang, Zhukov, Zichichi, Zimmermann \&
  Zuccon]{Aguilar+19pos}
{\sc \au{Aguilar, M.}, \au{Ali~Cavasonza, L.}, \au{Ambrosi, G.}, \au{Arruda,
  L.}, \au{Attig, N.}, \au{Azzarello, P.}, \au{Bachlechner, A.}, \au{Barao,
  F.}, \au{Barrau, A.}, \au{Barrin, L.}, \au{Bartoloni, A.}, \au{Basara, L.},
  \au{Ba\ifmmode \mbox{\c{s}}\else \c{s}\fi{}e\ifmmode \breve{g}\else
  \u{g}\fi{}mez-du Pree, S.}, \au{Battiston, R.}, \au{Becker, U.},
  \au{Behlmann, M.}, \au{Beischer, B.}, \au{Berdugo, J.}, \au{Bertucci, B.},
  \au{Bindi, V.}, \au{de~Boer, W.}, \au{Bollweg, K.}, \au{Borgia, B.},
  \au{Boschini, M.~J.}, \au{Bourquin, M.}, \au{Bueno, E.~F.}, \au{Burger, J.},
  \au{Burger, W.~J.}, \au{Cai, X.~D.}, \au{Capell, M.}, \au{Caroff, S.},
  \au{Casaus, J.}, \au{Castellini, G.}, \au{Cervelli, F.}, \au{Chang, Y.~H.},
  \au{Chen, G.~M.}, \au{Chen, H.~S.}, \au{Chen, Y.}, \au{Cheng, L.}, \au{Chou,
  H.~Y.}, \au{Choutko, V.}, \au{Chung, C.~H.}, \au{Clark, C.}, \au{Coignet,
  G.}, \au{Consolandi, C.}, \au{Contin, A.}, \au{Corti, C.}, \au{Crispoltoni,
  M.}, \au{Cui, Z.}, \au{Dadzie, K.}, \au{Dai, Y.~M.}, \au{Datta, A.},
  \au{Delgado, C.}, \au{Della~Torre, S.}, \au{Demirk\"oz, M.~B.}, \au{Derome,
  L.}, \au{Di~Falco, S.}, \au{Dimiccoli, F.}, \au{D\'{\i}az, C.}, \au{von
  Doetinchem, P.}, \au{Dong, F.}, \au{Donnini, F.}, \au{Duranti, M.},
  \au{Egorov, A.}, \au{Eline, A.}, \au{Eronen, T.}, \au{Feng, J.},
  \au{Fiandrini, E.}, \au{Fisher, P.}, \au{Formato, V.}, \au{Galaktionov, Y.},
  \au{Garc\'{\i}a-L\'opez, R.~J.}, \au{Gargiulo, C.}, \au{Gast, H.},
  \au{Gebauer, I.}, \au{Gervasi, M.}, \au{Giovacchini, F.}, \au{G\'omez-Coral,
  D.~M.}, \au{Gong, J.}, \au{Goy, C.}, \au{Grabski, V.}, \au{Grandi, D.},
  \au{Graziani, M.}, \au{Guo, K.~H.}, \au{Haino, S.}, \au{Han, K.~C.}, \au{He,
  Z.~H.}, \au{Heil, M.}, \au{Hsieh, T.~H.}, \au{Huang, H.}, \au{Huang, Z.~C.},
  \au{Incagli, M.}, \au{Jia, Yi}, \au{Jinchi, H.}, \au{Kanishev, K.},
  \au{Khiali, B.}, \au{Kirn, Th.}, \au{Konak, C.}, \au{Kounina, O.},
  \au{Kounine, A.}, \au{Koutsenko, V.}, \au{Kulemzin, A.}, \au{La~Vacca, G.},
  \au{Laudi, E.}, \au{Laurenti, G.}, \au{Lazzizzera, I.}, \au{Lebedev, A.},
  \au{Lee, H.~T.}, \au{Lee, S.~C.}, \au{Leluc, C.}, \au{Li, J.~Q.}, \au{Li,
  Q.}, \au{Li, T.~X.}, \au{Li, Z.~H.}, \au{Light, C.}, \au{Lin, C.~H.},
  \au{Lippert, T.}, \au{Liu, F.~Z.}, \au{Liu, Hu}, \au{Liu, Z.}, \au{Lu,
  S.~Q.}, \au{Lu, Y.~S.}, \au{Luebelsmeyer, K.}, \au{Luo, F.}, \au{Luo, J.~Z.},
  \au{Luo, Xi}, \au{Lyu, S.~S.}, \au{Machate, F.}, \au{Ma\~n\'a, C.},
  \au{Mar\'{\i}n, J.}, \au{Martin, T.}, \au{Mart\'{\i}nez, G.}, \au{Masi, N.},
  \au{Maurin, D.}, \au{Menchaca-Rocha, A.}, \au{Meng, Q.}, \au{Mo, D.~C.},
  \au{Molero, M.}, \au{Mott, P.}, \au{Mussolin, L.}, \au{Nelson, T.}, \au{Ni,
  J.~Q.}, \au{Nikonov, N.}, \au{Nozzoli, F.}, \au{Oliva, A.}, \au{Orcinha, M.},
  \au{Palermo, M.}, \au{Palmonari, F.}, \au{Paniccia, M.}, \au{Pashnin, A.},
  \au{Pauluzzi, M.}, \au{Pensotti, S.}, \au{Perrina, C.}, \au{Phan, H.~D.},
  \au{Picot-Clemente, N.}, \au{Plyaskin, V.}, \au{Pohl, M.}, \au{Poireau, V.},
  \au{Popkow, A.}, \au{Quadrani, L.}, \au{Qi, X.~M.}, \au{Qin, X.}, \au{Qu,
  Z.~Y.}, \au{Rancoita, P.~G.}, \au{Rapin, D.}, \au{Conde, A.~Reina},
  \au{Rosier-Lees, S.}, \au{Rozhkov, A.}, \au{Rozza, D.}, \au{Sagdeev, R.},
  \au{Solano, C.}, \au{Schael, S.}, \au{Schmidt, S.~M.}, \au{Schulz~von
  Dratzig, A.}, \au{Schwering, G.}, \au{Seo, E.~S.}, \au{Shan, B.~S.}, \au{Shi,
  J.~Y.}, \au{Siedenburg, T.}, \au{Song, J.~W.}, \au{Sun, Z.~T.}, \au{Tacconi,
  M.}, \au{Tang, X.~W.}, \au{Tang, Z.~C.}, \au{Tian, J.}, \au{Ting, Samuel
  C.~C.}, \au{Ting, S.~M.}, \au{Tomassetti, N.}, \au{Torsti, J.}, \au{Urban,
  T.}, \au{Vagelli, V.}, \au{Valente, E.}, \au{Valtonen, E.},
  \au{V\'azquez~Acosta, M.}, \au{Vecchi, M.}, \au{Velasco, M.}, \au{Vialle,
  J.~P.}, \au{Viz\'an, J.}, \au{Wang, L.~Q.}, \au{Wang, N.~H.}, \au{Wang,
  Q.~L.}, \au{Wang, X.}, \au{Wang, X.~Q.}, \au{Wang, Z.~X.}, \au{Wei, J.},
  \au{Weng, Z.~L.}, \au{Wu, H.}, \au{Xiong, R.~Q.}, \au{Xu, W.}, \au{Yan, Q.},
  \au{Yang, Y.}, \au{Yi, H.}, \au{Yu, Y.~J.}, \au{Yu, Z.~Q.}, \au{Zannoni, M.},
  \au{Zeissler, S.}, \au{Zhang, C.}, \au{Zhang, F.}, \au{Zhang, J.~H.},
  \au{Zhang, Z.}, \au{Zhao, F.}, \au{Zheng, Z.~M.}, \au{Zhuang, H.~L.},
  \au{Zhukov, V.}, \au{Zichichi, A.}, \au{Zimmermann, N.} \& \au{Zuccon, P.}}
  \yr{2019{\natexlab{{\em b\/}}}}  \at{Towards understanding the origin of
  cosmic-ray positrons}.  \jt{Phys. Rev. Lett.}  \bvol{122},  \pg{041102}.

\bibitem[{Aharonian} {\em et~al.\/}(2002){Aharonian}, {Akhperjanian},
  {Beilicke}, {Bernl{\"o}hr}, {B{\"o}rst}, {Bojahr}, {Bolz}, {Coarasa},
  {Contreras}, {Cortina}, {Denninghoff}, {Fonseca}, {Girma}, {G{\"o}tting},
  {Heinzelmann}, {Hermann}, {Heusler}, {Hofmann}, {Horns}, {Jung}, {Kankanyan},
  {Kestel}, {Kettler}, {Kohnle}, {Konopelko}, {Kornmeyer}, {Kranich},
  {Krawczynski}, {Lampeitl}, {Lopez}, {Lorenz}, {Lucarelli}, {Magnussen},
  {Mang}, {Meyer}, {Milite}, {Mirzoyan}, {Moralejo}, {Ona}, {Panter},
  {Plyasheshnikov}, {Prahl}, {P{\"u}hlhofer}, {Rauterberg}, {Reyes}, {Rhode},
  {Ripken}, {R{\"o}hring}, {Rowell}, {Sahakian}, {Samorski}, {Schilling},
  {Schr{\"o}der}, {Siems}, {Sobzynska}, {Stamm}, {Tluczykont}, {V{\"o}lk},
  {Wiedner}, {Wittek}, {Uchiyama}, {Takahashi} \& {HEGRA
  Collaboration}]{2002A&A...393L..37A}
{\sc \au{{Aharonian}, F.}, \au{{Akhperjanian}, A.}, \au{{Beilicke}, M.},
  \au{{Bernl{\"o}hr}, K.}, \au{{B{\"o}rst}, H.}, \au{{Bojahr}, H.}, \au{{Bolz},
  O.}, \au{{Coarasa}, T.}, \au{{Contreras}, J.}, \au{{Cortina}, J.},
  \au{{Denninghoff}, S.}, \au{{Fonseca}, V.}, \au{{Girma}, M.},
  \au{{G{\"o}tting}, N.}, \au{{Heinzelmann}, G.}, \au{{Hermann}, G.},
  \au{{Heusler}, A.}, \au{{Hofmann}, W.}, \au{{Horns}, D.}, \au{{Jung}, I.},
  \au{{Kankanyan}, R.}, \au{{Kestel}, M.}, \au{{Kettler}, J.}, \au{{Kohnle},
  A.}, \au{{Konopelko}, A.}, \au{{Kornmeyer}, H.}, \au{{Kranich}, D.},
  \au{{Krawczynski}, H.}, \au{{Lampeitl}, H.}, \au{{Lopez}, M.}, \au{{Lorenz},
  E.}, \au{{Lucarelli}, F.}, \au{{Magnussen}, N.}, \au{{Mang}, O.},
  \au{{Meyer}, H.}, \au{{Milite}, M.}, \au{{Mirzoyan}, R.}, \au{{Moralejo},
  A.}, \au{{Ona}, E.}, \au{{Panter}, M.}, \au{{Plyasheshnikov}, A.},
  \au{{Prahl}, J.}, \au{{P{\"u}hlhofer}, G.}, \au{{Rauterberg}, G.},
  \au{{Reyes}, R.}, \au{{Rhode}, W.}, \au{{Ripken}, J.}, \au{{R{\"o}hring},
  A.}, \au{{Rowell}, G.~P.}, \au{{Sahakian}, V.}, \au{{Samorski}, M.},
  \au{{Schilling}, M.}, \au{{Schr{\"o}der}, F.}, \au{{Siems}, M.},
  \au{{Sobzynska}, D.}, \au{{Stamm}, W.}, \au{{Tluczykont}, M.},
  \au{{V{\"o}lk}, H.~J.}, \au{{Wiedner}, C.~A.}, \au{{Wittek}, W.},
  \au{{Uchiyama}, Y.}, \au{{Takahashi}, T.} \& \au{{HEGRA Collaboration}}}
  \yr{2002}  \at{{An unidentified TeV source in the vicinity of Cygnus OB2}}.
  \jt{Astronomy and Astrophysics}  \bvol{393},  \pg{L37--L40},  \arxiv{arXiv:
  astro-ph/0207528}.

\bibitem[Aharonian {\em et~al.\/}(2020)Aharonian, Peron, Yang, Casanova \&
  Zanin]{PhysRevD.101.083018}
{\sc \au{Aharonian, Felix}, \au{Peron, Giada}, \au{Yang, Ruizhi}, \au{Casanova,
  Sabrina} \& \au{Zanin, Roberta}} \yr{2020}  \at{Probing the sea of galactic
  cosmic rays with fermi-lat}.  \jt{Phys. Rev. D}  \bvol{101},  \pg{083018}.

\bibitem[{Aharonian} {\em et~al.\/}(2019){Aharonian}, {Yang} \& {de O{\~n}a
  Wilhelmi}]{2019NatAs...3..561A}
{\sc \au{{Aharonian}, Felix}, \au{{Yang}, Ruizhi} \& \au{{de O{\~n}a Wilhelmi},
  Emma}} \yr{2019}  \at{{Massive stars as major factories of Galactic cosmic
  rays}}.  \jt{Nature Astronomy}  \bvol{3},  \pg{561--567},  \arxiv{arXiv:
  1804.02331}.

\bibitem[{Aharonian}(2001)]{2001SSRv...99..187A}
{\sc \au{{Aharonian}, F.~A.}} \yr{2001}  \at{{Gamma Rays From Molecular
  Clouds}}.  \jt{SSR}  \bvol{99},  \pg{187--196},  \arxiv{arXiv:
  astro-ph/0012290}.

\bibitem[{Aharonian}(2004)]{2004vhec.book.....A}
{\sc \au{{Aharonian}, Felix~A.}} \yr{2004} {\em {Very high energy cosmic gamma
  radiation : a crucial window on the extreme Universe}\/}.

\bibitem[{Aharonian} {\em et~al.\/}(2004){Aharonian}, {Akhperjanian}, {Aye},
  {Bazer-Bachi}, {Beilicke}, {Benbow}, {Berge}, {Berghaus}, {Bernl{\"o}hr},
  {Bolz}, {Boisson}, {Borgmeier}, {Breitling}, {Brown}, {Bussons Gordo},
  {Chadwick}, {Chitnis}, {Chounet}, {Cornils}, {Costamante}, {Degrange},
  {Djannati-Ata{\"\i}}, {Drury}, {Ergin}, {Espigat}, {Feinstein}, {Fleury},
  {Fontaine}, {Funk}, {Gallant}, {Giebels}, {Gillessen}, {Goret}, {Guy},
  {Hadjichristidis}, {Hauser}, {Heinzelmann}, {Henri}, {Hermann}, {Hinton},
  {Hofmann}, {Holleran}, {Horns}, {de Jager}, {Jung}, {Kh{\'e}lifi}, {Komin},
  {Konopelko}, {Latham}, {Le Gallou}, {Lemoine}, {Lemi{\`e}re}, {Leroy},
  {Lohse}, {Marcowith}, {Masterson}, {McComb}, {de Naurois}, {Nolan},
  {Noutsos}, {Orford}, {Osborne}, {Ouchrif}, {Panter}, {Pelletier}, {Pita},
  {Pohl}, {P{\"u}hlhofer}, {Punch}, {Raubenheimer}, {Raue}, {Raux}, {Rayner},
  {Redondo}, {Reimer}, {Reimer}, {Ripken}, {Rivoal}, {Rob}, {Rolland},
  {Rowell}, {Sahakian}, {Saug{\'e}}, {Schlenker}, {Schlickeiser}, {Schuster},
  {Schwanke}, {Siewert}, {Sol}, {Steenkamp}, {Stegmann}, {Tavernet},
  {Th{\'e}oret}, {Tluczykont}, {van der Walt}, {Vasileiadis}, {Vincent},
  {Visser}, {V{\"o}lk} \& {Wagner}]{2004Natur.432...75A}
{\sc \au{{Aharonian}, F.~A.}, \au{{Akhperjanian}, A.~G.}, \au{{Aye}, K.~M.},
  \au{{Bazer-Bachi}, A.~R.}, \au{{Beilicke}, M.}, \au{{Benbow}, W.},
  \au{{Berge}, D.}, \au{{Berghaus}, P.}, \au{{Bernl{\"o}hr}, K.}, \au{{Bolz},
  O.}, \au{{Boisson}, C.}, \au{{Borgmeier}, C.}, \au{{Breitling}, F.},
  \au{{Brown}, A.~M.}, \au{{Bussons Gordo}, J.}, \au{{Chadwick}, P.~M.},
  \au{{Chitnis}, V.~R.}, \au{{Chounet}, L.~M.}, \au{{Cornils}, R.},
  \au{{Costamante}, L.}, \au{{Degrange}, B.}, \au{{Djannati-Ata{\"\i}}, A.},
  \au{{Drury}, L.~O'C.}, \au{{Ergin}, T.}, \au{{Espigat}, P.}, \au{{Feinstein},
  F.}, \au{{Fleury}, P.}, \au{{Fontaine}, G.}, \au{{Funk}, S.}, \au{{Gallant},
  Y.~A.}, \au{{Giebels}, B.}, \au{{Gillessen}, S.}, \au{{Goret}, P.},
  \au{{Guy}, J.}, \au{{Hadjichristidis}, C.}, \au{{Hauser}, M.},
  \au{{Heinzelmann}, G.}, \au{{Henri}, G.}, \au{{Hermann}, G.}, \au{{Hinton},
  J.~A.}, \au{{Hofmann}, W.}, \au{{Holleran}, M.}, \au{{Horns}, D.}, \au{{de
  Jager}, O.~C.}, \au{{Jung}, I.}, \au{{Kh{\'e}lifi}, B.}, \au{{Komin}, Nu.},
  \au{{Konopelko}, A.}, \au{{Latham}, I.~J.}, \au{{Le Gallou}, R.},
  \au{{Lemoine}, M.}, \au{{Lemi{\`e}re}, A.}, \au{{Leroy}, N.}, \au{{Lohse},
  T.}, \au{{Marcowith}, A.}, \au{{Masterson}, C.}, \au{{McComb}, T.~J.~L.},
  \au{{de Naurois}, M.}, \au{{Nolan}, S.~J.}, \au{{Noutsos}, A.}, \au{{Orford},
  K.~J.}, \au{{Osborne}, J.~L.}, \au{{Ouchrif}, M.}, \au{{Panter}, M.},
  \au{{Pelletier}, G.}, \au{{Pita}, S.}, \au{{Pohl}, M.}, \au{{P{\"u}hlhofer},
  G.}, \au{{Punch}, M.}, \au{{Raubenheimer}, B.~C.}, \au{{Raue}, M.},
  \au{{Raux}, J.}, \au{{Rayner}, S.~M.}, \au{{Redondo}, I.}, \au{{Reimer}, A.},
  \au{{Reimer}, O.}, \au{{Ripken}, J.}, \au{{Rivoal}, M.}, \au{{Rob}, L.},
  \au{{Rolland}, L.}, \au{{Rowell}, G.}, \au{{Sahakian}, V.}, \au{{Saug{\'e}},
  L.}, \au{{Schlenker}, S.}, \au{{Schlickeiser}, R.}, \au{{Schuster}, C.},
  \au{{Schwanke}, U.}, \au{{Siewert}, M.}, \au{{Sol}, H.}, \au{{Steenkamp},
  R.}, \au{{Stegmann}, C.}, \au{{Tavernet}, J.~P.}, \au{{Th{\'e}oret}, C.~G.},
  \au{{Tluczykont}, M.}, \au{{van der Walt}, D.~J.}, \au{{Vasileiadis}, G.},
  \au{{Vincent}, P.}, \au{{Visser}, B.}, \au{{V{\"o}lk}, H.~J.} \&
  \au{{Wagner}, S.~J.}} \yr{2004}  \at{{High-energy particle acceleration in
  the shell of a supernova remnant}}.  \jt{Nature}  \bvol{432}~(7013),
  \pg{75--77},  \arxiv{arXiv: astro-ph/0411533}.

\bibitem[{Aharonian} \& {Atoyan}(1996)]{1996A&A...309..917A}
{\sc \au{{Aharonian}, F.~A.} \& \au{{Atoyan}, A.~M.}} \yr{1996}  \at{{On the
  emissivity of {\ensuremath{\pi}}\^0\^-decay gamma radiation in the vicinity
  of accelerators of galactic cosmic rays.}}  \jt{A\&A}  \bvol{309},
  \pg{917--928}.

\bibitem[{Aharonian} {\em et~al.\/}(1994){Aharonian}, {Drury} \&
  {Voelk}]{1994A&A...285..645A}
{\sc \au{{Aharonian}, F.~A.}, \au{{Drury}, L.~O'C.} \& \au{{Voelk}, H.~J.}}
  \yr{1994}  \at{{GeV/TeV gamma-ray emission from dense molecular clouds
  overtaken by supernova shells}}.  \jt{A\&A}  \bvol{285},  \pg{645--647}.

\bibitem[{Aharonian} {\em et~al.\/}(2001)]{2001A&A...370..112A}
{\sc \au{{Aharonian}, F.~A.} \& \au{others}} \yr{2001}  \at{{Evidence for TeV
  gamma ray emission from Cassiopeia A}}.  \jt{A\&A}  \bvol{370},
  \pg{112--120},  \arxiv{arXiv: astro-ph/0102391}.

\bibitem[{Ahn} {\em et~al.\/}(2010{\natexlab{{\em a\/}}}){Ahn}, {Allison},
  {Bagliesi}, {Beatty}, {Bigongiari}, {Childers}, {Conklin}, {Coutu},
  {DuVernois}, {Ganel}, {Han}, {Jeon}, {Kim}, {Lee}, {Lutz}, {Maestro},
  {Malinin}, {Marrocchesi}, {Minnick}, {Mognet}, {Nam}, {Nam}, {Nutter},
  {Park}, {Park}, {Seo}, {Sina}, {Wu}, {Yang}, {Yoon}, {Zei} \& {Zinn}]{CREAM}
{\sc \au{{Ahn}, H.~S.}, \au{{Allison}, P.}, \au{{Bagliesi}, M.~G.},
  \au{{Beatty}, J.~J.}, \au{{Bigongiari}, G.}, \au{{Childers}, J.~T.},
  \au{{Conklin}, N.~B.}, \au{{Coutu}, S.}, \au{{DuVernois}, M.~A.},
  \au{{Ganel}, O.}, \au{{Han}, J.~H.}, \au{{Jeon}, J.~A.}, \au{{Kim}, K.~C.},
  \au{{Lee}, M.~H.}, \au{{Lutz}, L.}, \au{{Maestro}, P.}, \au{{Malinin}, A.},
  \au{{Marrocchesi}, P.~S.}, \au{{Minnick}, S.}, \au{{Mognet}, S.~I.},
  \au{{Nam}, J.}, \au{{Nam}, S.}, \au{{Nutter}, S.~L.}, \au{{Park}, I.~H.},
  \au{{Park}, N.~H.}, \au{{Seo}, E.~S.}, \au{{Sina}, R.}, \au{{Wu}, J.},
  \au{{Yang}, J.}, \au{{Yoon}, Y.~S.}, \au{{Zei}, R.} \& \au{{Zinn}, S.~Y.}}
  \yr{2010{\natexlab{{\em a\/}}}}  \at{{Discrepant Hardening Observed in
  Cosmic-ray Elemental Spectra}}.  \jt{ApJ Letters}  \bvol{714}~(1),
  \pg{L89--L93},  \arxiv{arXiv: 1004.1123}.

\bibitem[{Ahn} {\em et~al.\/}(2010{\natexlab{{\em b\/}}}){Ahn}, {Allison},
  {Bagliesi}, {Beatty}, {Bigongiari}, {Childers}, {Conklin}, {Coutu},
  {DuVernois}, {Ganel}, {Han}, {Jeon}, {Kim}, {Lee}, {Lutz}, {Maestro},
  {Malinin}, {Marrocchesi}, {Minnick}, {Mognet}, {Nam}, {Nam}, {Nutter},
  {Park}, {Park}, {Seo}, {Sina}, {Wu}, {Yang}, {Yoon}, {Zei} \& {Zinn}]{Ahn+10}
{\sc \au{{Ahn}, H.~S.}, \au{{Allison}, P.}, \au{{Bagliesi}, M.~G.},
  \au{{Beatty}, J.~J.}, \au{{Bigongiari}, G.}, \au{{Childers}, J.~T.},
  \au{{Conklin}, N.~B.}, \au{{Coutu}, S.}, \au{{DuVernois}, M.~A.},
  \au{{Ganel}, O.}, \au{{Han}, J.~H.}, \au{{Jeon}, J.~A.}, \au{{Kim}, K.~C.},
  \au{{Lee}, M.~H.}, \au{{Lutz}, L.}, \au{{Maestro}, P.}, \au{{Malinin}, A.},
  \au{{Marrocchesi}, P.~S.}, \au{{Minnick}, S.}, \au{{Mognet}, S.~I.},
  \au{{Nam}, J.}, \au{{Nam}, S.}, \au{{Nutter}, S.~L.}, \au{{Park}, I.~H.},
  \au{{Park}, N.~H.}, \au{{Seo}, E.~S.}, \au{{Sina}, R.}, \au{{Wu}, J.},
  \au{{Yang}, J.}, \au{{Yoon}, Y.~S.}, \au{{Zei}, R.} \& \au{{Zinn}, S.~Y.}}
  \yr{2010{\natexlab{{\em b\/}}}}  \at{{Discrepant Hardening Observed in
  Cosmic-ray Elemental Spectra}}.  \jt{ApJL}  \bvol{714}~(1),  \pg{L89--L93},
  \arxiv{arXiv: 1004.1123}.

\bibitem[{Ahnen} {\em et~al.\/}(2017){Ahnen}, {Ansoldi}, {Antonelli}, {Arcaro},
  {Babi{\'c}}, {Banerjee}, {Bangale}, {Barres de Almeida}, {Barrio}, {Becerra
  Gonz{\'a}lez}, {Bednarek}, {Bernardini}, {Berti}, {Bhattacharyya},
  {Biasuzzi}, {Biland }, {Blanch}, {Bonnefoy}, {Bonnoli}, {Carosi}, {Carosi},
  {Chatterjee}, {Colak}, {Colin}, {Colombo}, {Contreras}, {Cortina}, {Covino},
  {Cumani}, {Da Vela}, {Dazzi}, {De Angelis}, {De Lotto}, {de O{\~n}a
  Wilhelmi}, {Di Pierro}, {Doert}, {Dom{\'\i}nguez}, {Dominis Prester},
  {Dorner}, {Doro}, {Einecke}, {Eisenacher Glawion}, {Elsaesser},
  {Engelkemeier}, {Fallah Ramazani}, {Fern{\'a}ndez-Barral}, {Fidalgo},
  {Fonseca}, {Font}, {Fruck}, {Galindo}, {Garc{\'\i}a L{\'o}pez},
  {Garczarczyk}, {Gaug}, {Giammaria}, {Godinovi{\'c}}, {Gora}, {Guberman},
  {Hadasch}, {Hahn}, {Hassan}, {Hayashida}, {Herrera}, {Hose}, {Hrupec},
  {Inada}, {Ishio}, {Konno}, {Kubo}, {Kushida}, {Kuve{\v{z}}di{\'c}}, {Lelas},
  {Lindfors}, {Lombardi}, {Longo}, {L{\'o}pez}, {Maggio}, {Majumdar},
  {Makariev}, {Maneva}, {Manganaro}, {Mannheim}, {Maraschi}, {Mariotti},
  {Mart{\'\i}nez}, {Mazin}, {Menzel}, {Minev}, {Mirzoyan}, {Moralejo},
  {Moreno}, {Moretti}, {Neustroev}, {Niedzwiecki}, {Nievas Rosillo}, {Nilsson},
  {Ninci}, {Nishijima}, {Noda}, {Nogu{\'e}s}, {Paiano}, {Palacio}, {Paneque},
  {Paoletti}, {Paredes}, {Pedaletti}, {Peresano}, {Perri}, {Persic}, {Prada
  Moroni}, {Prand ini}, {Puljak}, {Garcia}, {Reichardt}, {Rhode}, {Rib{\'o}},
  {Rico}, {Righi}, {Saito}, {Satalecka}, {Schroeder}, {Schweizer}, {Shore},
  {Sitarek}, {{\v{S}}nidari{\'c}}, {Sobczynska}, {Stamerra}, {Strzys},
  {Suri{\'c}}, {Takalo}, {Tavecchio}, {Temnikov}, {Terzi{\'c}}, {Tescaro},
  {Teshima}, {Torres-Alb{\`a}}, {Treves}, {Vanzo}, {Vazquez Acosta}, {Vovk},
  {Ward}, {Will} \& {Zari{\'c}}]{2017MNRAS.472.2956A}
{\sc \au{{Ahnen}, M.~L.}, \au{{Ansoldi}, S.}, \au{{Antonelli}, L.~A.},
  \au{{Arcaro}, C.}, \au{{Babi{\'c}}, A.}, \au{{Banerjee}, B.}, \au{{Bangale},
  P.}, \au{{Barres de Almeida}, U.}, \au{{Barrio}, J.~A.}, \au{{Becerra
  Gonz{\'a}lez}, J.}, \au{{Bednarek}, W.}, \au{{Bernardini}, E.}, \au{{Berti},
  A.}, \au{{Bhattacharyya}, W.}, \au{{Biasuzzi}, B.}, \au{{Biland }, A.},
  \au{{Blanch}, O.}, \au{{Bonnefoy}, S.}, \au{{Bonnoli}, G.}, \au{{Carosi},
  R.}, \au{{Carosi}, A.}, \au{{Chatterjee}, A.}, \au{{Colak}, M.}, \au{{Colin},
  P.}, \au{{Colombo}, E.}, \au{{Contreras}, J.~L.}, \au{{Cortina}, J.},
  \au{{Covino}, S.}, \au{{Cumani}, P.}, \au{{Da Vela}, P.}, \au{{Dazzi}, F.},
  \au{{De Angelis}, A.}, \au{{De Lotto}, B.}, \au{{de O{\~n}a Wilhelmi}, E.},
  \au{{Di Pierro}, F.}, \au{{Doert}, M.}, \au{{Dom{\'\i}nguez}, A.},
  \au{{Dominis Prester}, D.}, \au{{Dorner}, D.}, \au{{Doro}, M.},
  \au{{Einecke}, S.}, \au{{Eisenacher Glawion}, D.}, \au{{Elsaesser}, D.},
  \au{{Engelkemeier}, M.}, \au{{Fallah Ramazani}, V.},
  \au{{Fern{\'a}ndez-Barral}, A.}, \au{{Fidalgo}, D.}, \au{{Fonseca}, M.~V.},
  \au{{Font}, L.}, \au{{Fruck}, C.}, \au{{Galindo}, D.}, \au{{Garc{\'\i}a
  L{\'o}pez}, R.~J.}, \au{{Garczarczyk}, M.}, \au{{Gaug}, M.}, \au{{Giammaria},
  P.}, \au{{Godinovi{\'c}}, N.}, \au{{Gora}, D.}, \au{{Guberman}, D.},
  \au{{Hadasch}, D.}, \au{{Hahn}, A.}, \au{{Hassan}, T.}, \au{{Hayashida}, M.},
  \au{{Herrera}, J.}, \au{{Hose}, J.}, \au{{Hrupec}, D.}, \au{{Inada}, T.},
  \au{{Ishio}, K.}, \au{{Konno}, Y.}, \au{{Kubo}, H.}, \au{{Kushida}, J.},
  \au{{Kuve{\v{z}}di{\'c}}, D.}, \au{{Lelas}, D.}, \au{{Lindfors}, E.},
  \au{{Lombardi}, S.}, \au{{Longo}, F.}, \au{{L{\'o}pez}, M.}, \au{{Maggio},
  C.}, \au{{Majumdar}, P.}, \au{{Makariev}, M.}, \au{{Maneva}, G.},
  \au{{Manganaro}, M.}, \au{{Mannheim}, K.}, \au{{Maraschi}, L.},
  \au{{Mariotti}, M.}, \au{{Mart{\'\i}nez}, M.}, \au{{Mazin}, D.},
  \au{{Menzel}, U.}, \au{{Minev}, M.}, \au{{Mirzoyan}, R.}, \au{{Moralejo},
  A.}, \au{{Moreno}, V.}, \au{{Moretti}, E.}, \au{{Neustroev}, V.},
  \au{{Niedzwiecki}, A.}, \au{{Nievas Rosillo}, M.}, \au{{Nilsson}, K.},
  \au{{Ninci}, D.}, \au{{Nishijima}, K.}, \au{{Noda}, K.}, \au{{Nogu{\'e}s},
  L.}, \au{{Paiano}, S.}, \au{{Palacio}, J.}, \au{{Paneque}, D.},
  \au{{Paoletti}, R.}, \au{{Paredes}, J.~M.}, \au{{Pedaletti}, G.},
  \au{{Peresano}, M.}, \au{{Perri}, L.}, \au{{Persic}, M.}, \au{{Prada Moroni},
  P.~G.}, \au{{Prand ini}, E.}, \au{{Puljak}, I.}, \au{{Garcia}, J.~R.},
  \au{{Reichardt}, I.}, \au{{Rhode}, W.}, \au{{Rib{\'o}}, M.}, \au{{Rico}, J.},
  \au{{Righi}, C.}, \au{{Saito}, T.}, \au{{Satalecka}, K.}, \au{{Schroeder},
  S.}, \au{{Schweizer}, T.}, \au{{Shore}, S.~N.}, \au{{Sitarek}, J.},
  \au{{{\v{S}}nidari{\'c}}, I.}, \au{{Sobczynska}, D.}, \au{{Stamerra}, A.},
  \au{{Strzys}, M.}, \au{{Suri{\'c}}, T.}, \au{{Takalo}, L.}, \au{{Tavecchio},
  F.}, \au{{Temnikov}, P.}, \au{{Terzi{\'c}}, T.}, \au{{Tescaro}, D.},
  \au{{Teshima}, M.}, \au{{Torres-Alb{\`a}}, N.}, \au{{Treves}, A.},
  \au{{Vanzo}, G.}, \au{{Vazquez Acosta}, M.}, \au{{Vovk}, I.}, \au{{Ward},
  J.~E.}, \au{{Will}, M.} \& \au{{Zari{\'c}}, D.}} \yr{2017}  \at{{A cut-off in
  the TeV gamma-ray spectrum of the SNR Cassiopeia A}}.  \jt{MNRAS}
  \bvol{472}~(3),  \pg{2956--2962},  \arxiv{arXiv: 1707.01583}.

\bibitem[{Albert} {\em et~al.\/}(2007{\natexlab{{\em a\/}}}){Albert}, {Aliu},
  {Anderhub}, {Antoranz}, {Armada}, {Baixeras}, {Barrio}, {Bartko}, {Bastieri},
  {Becker}, {Bednarek}, {Berger}, {Bigongiari}, {Biland}, {Bock}, {Bordas},
  {Bosch-Ramon}, {Bretz}, {Britvitch}, {Camara}, {Carmona}, {Chilingarian},
  {Coarasa}, {Commichau}, {Contreras}, {Cortina}, {Costado}, {Curtef},
  {Danielyan}, {Dazzi}, {de Angelis}, {Delgado}, {de Los Reyes}, {de Lotto},
  {Domingo-Santamar{\'\i}a}, {Dorner}, {Doro}, {Errando}, {Fagiolini},
  {Ferenc}, {Fern{\'a}ndez}, {Firpo}, {Flix}, {Fonseca}, {Font}, {Fuchs},
  {Galante}, {Garc{\'\i}a-L{\'o}pez}, {Garczarczyk}, {Gaug}, {Giller},
  {Goebel}, {Hakobyan}, {Hayashida}, {Hengstebeck}, {Herrero}, {H{\"o}hne},
  {Hose}, {Hsu}, {Jacon}, {Jogler}, {Kosyra}, {Kranich}, {Kritzer}, {Laille},
  {Lindfors}, {Lombardi}, {Longo}, {L{\'o}pez}, {L{\'o}pez}, {Lorenz},
  {Majumdar}, {Maneva}, {Mannheim}, {Mansutti}, {Mariotti}, {Mart{\'\i}nez},
  {Mazin}, {Merck}, {Meucci}, {Meyer}, {Mirand a}, {Mirzoyan}, {Mizobuchi},
  {Moralejo}, {Nilsson}, {Ninkovic}, {O{\~n}a-Wilhelmi}, {Otte}, {Oya},
  {Paneque}, {Panniello}, {Paoletti}, {Paredes}, {Pasanen}, {Pascoli}, {Pauss},
  {Pegna}, {Persic}, {Peruzzo}, {Piccioli}, {Poller}, {Puchades}, {Prandini},
  {Raymers}, {Rhode}, {Rib{\'o}}, {Rico}, {Rissi}, {Robert}, {R{\"u}gamer},
  {Saggion}, {S{\'a}nchez}, {Sartori}, {Scalzotto}, {Scapin}, {Schmitt},
  {Schweizer}, {Shayduk}, {Shinozaki}, {Shore}, {Sidro}, {Sillanp{\"a}{\"a}},
  {Sobczynska}, {Stamerra}, {Stark}, {Takalo}, {Temnikov}, {Tescaro},
  {Teshima}, {Tonello}, {Torres}, {Turini}, {Vankov}, {Vitale}, {Wagner},
  {Wibig}, {Wittek}, {Zandanel}, {Zanin} \& {Zapatero}]{2007A&A...474..937A}
{\sc \au{{Albert}, J.}, \au{{Aliu}, E.}, \au{{Anderhub}, H.}, \au{{Antoranz},
  P.}, \au{{Armada}, A.}, \au{{Baixeras}, C.}, \au{{Barrio}, J.~A.},
  \au{{Bartko}, H.}, \au{{Bastieri}, D.}, \au{{Becker}, J.~K.}, \au{{Bednarek},
  W.}, \au{{Berger}, K.}, \au{{Bigongiari}, C.}, \au{{Biland}, A.}, \au{{Bock},
  R.~K.}, \au{{Bordas}, P.}, \au{{Bosch-Ramon}, V.}, \au{{Bretz}, T.},
  \au{{Britvitch}, I.}, \au{{Camara}, M.}, \au{{Carmona}, E.},
  \au{{Chilingarian}, A.}, \au{{Coarasa}, J.~A.}, \au{{Commichau}, S.},
  \au{{Contreras}, J.~L.}, \au{{Cortina}, J.}, \au{{Costado}, M.~T.},
  \au{{Curtef}, V.}, \au{{Danielyan}, V.}, \au{{Dazzi}, F.}, \au{{de Angelis},
  A.}, \au{{Delgado}, C.}, \au{{de Los Reyes}, R.}, \au{{de Lotto}, B.},
  \au{{Domingo-Santamar{\'\i}a}, E.}, \au{{Dorner}, D.}, \au{{Doro}, M.},
  \au{{Errando}, M.}, \au{{Fagiolini}, M.}, \au{{Ferenc}, D.},
  \au{{Fern{\'a}ndez}, E.}, \au{{Firpo}, R.}, \au{{Flix}, J.}, \au{{Fonseca},
  M.~V.}, \au{{Font}, L.}, \au{{Fuchs}, M.}, \au{{Galante}, N.},
  \au{{Garc{\'\i}a-L{\'o}pez}, R.}, \au{{Garczarczyk}, M.}, \au{{Gaug}, M.},
  \au{{Giller}, M.}, \au{{Goebel}, F.}, \au{{Hakobyan}, D.}, \au{{Hayashida},
  M.}, \au{{Hengstebeck}, T.}, \au{{Herrero}, A.}, \au{{H{\"o}hne}, D.},
  \au{{Hose}, J.}, \au{{Hsu}, C.~C.}, \au{{Jacon}, P.}, \au{{Jogler}, T.},
  \au{{Kosyra}, R.}, \au{{Kranich}, D.}, \au{{Kritzer}, R.}, \au{{Laille}, A.},
  \au{{Lindfors}, E.}, \au{{Lombardi}, S.}, \au{{Longo}, F.}, \au{{L{\'o}pez},
  J.}, \au{{L{\'o}pez}, M.}, \au{{Lorenz}, E.}, \au{{Majumdar}, P.},
  \au{{Maneva}, G.}, \au{{Mannheim}, K.}, \au{{Mansutti}, O.}, \au{{Mariotti},
  M.}, \au{{Mart{\'\i}nez}, M.}, \au{{Mazin}, D.}, \au{{Merck}, C.},
  \au{{Meucci}, M.}, \au{{Meyer}, M.}, \au{{Mirand a}, J.~M.}, \au{{Mirzoyan},
  R.}, \au{{Mizobuchi}, S.}, \au{{Moralejo}, A.}, \au{{Nilsson}, K.},
  \au{{Ninkovic}, J.}, \au{{O{\~n}a-Wilhelmi}, E.}, \au{{Otte}, N.}, \au{{Oya},
  I.}, \au{{Paneque}, D.}, \au{{Panniello}, M.}, \au{{Paoletti}, R.},
  \au{{Paredes}, J.~M.}, \au{{Pasanen}, M.}, \au{{Pascoli}, D.}, \au{{Pauss},
  F.}, \au{{Pegna}, R.}, \au{{Persic}, M.}, \au{{Peruzzo}, L.}, \au{{Piccioli},
  A.}, \au{{Poller}, M.}, \au{{Puchades}, N.}, \au{{Prandini}, E.},
  \au{{Raymers}, A.}, \au{{Rhode}, W.}, \au{{Rib{\'o}}, M.}, \au{{Rico}, J.},
  \au{{Rissi}, M.}, \au{{Robert}, A.}, \au{{R{\"u}gamer}, S.}, \au{{Saggion},
  A.}, \au{{S{\'a}nchez}, A.}, \au{{Sartori}, P.}, \au{{Scalzotto}, V.},
  \au{{Scapin}, V.}, \au{{Schmitt}, R.}, \au{{Schweizer}, T.}, \au{{Shayduk},
  M.}, \au{{Shinozaki}, K.}, \au{{Shore}, S.~N.}, \au{{Sidro}, N.},
  \au{{Sillanp{\"a}{\"a}}, A.}, \au{{Sobczynska}, D.}, \au{{Stamerra}, A.},
  \au{{Stark}, L.~S.}, \au{{Takalo}, L.}, \au{{Temnikov}, P.}, \au{{Tescaro},
  D.}, \au{{Teshima}, M.}, \au{{Tonello}, N.}, \au{{Torres}, D.~F.},
  \au{{Turini}, N.}, \au{{Vankov}, H.}, \au{{Vitale}, V.}, \au{{Wagner},
  R.~M.}, \au{{Wibig}, T.}, \au{{Wittek}, W.}, \au{{Zandanel}, F.},
  \au{{Zanin}, R.} \& \au{{Zapatero}, J.}} \yr{2007{\natexlab{{\em a\/}}}}
  \at{{Observation of VHE {\ensuremath{\gamma}}-rays from Cassiopeia A with the
  MAGIC telescope}}.  \jt{A\&A}  \bvol{474}~(3),  \pg{937--940},  \arxiv{arXiv:
  0706.4065}.

\bibitem[{Albert} {\em et~al.\/}(2007{\natexlab{{\em b\/}}}){Albert}, {Aliu},
  {Anderhub}, {Antoranz}, {Armada}, {Baixeras}, {Barrio}, {Bartko}, {Bastieri},
  {Becker}, {Bednarek}, {Berger}, {Bigongiari}, {Biland}, {Bock}, {Bordas},
  {Bosch-Ramon}, {Bretz}, {Britvitch}, {Camara}, {Carmona}, {Chilingarian},
  {Coarasa}, {Commichau}, {Contreras}, {Cortina}, {Costado}, {Curtef},
  {Danielyan}, {Dazzi}, {De Angelis}, {Delgado}, {de los Reyes}, {De Lotto},
  {Domingo-Santamar{\'\i}a}, {Dorner}, {Doro}, {Errando}, {Fagiolini},
  {Ferenc}, {Fern{\'a}ndez}, {Firpo}, {Flix}, {Fonseca}, {Font}, {Fuchs},
  {Galante}, {Garc{\'\i}a-L{\'o}pez}, {Garczarczyk}, {Gaug}, {Giller},
  {Goebel}, {Hakobyan}, {Hayashida}, {Hengstebeck}, {Herrero}, {H{\"o}hne},
  {Hose}, {Hsu}, {Jacon}, {Jogler}, {Kosyra}, {Kranich}, {Kritzer}, {Laille},
  {Lindfors}, {Lombardi}, {Longo}, {L{\'o}pez}, {L{\'o}pez}, {Lorenz},
  {Majumdar}, {Maneva}, {Mannheim}, {Mansutti}, {Mariotti}, {Mart{\'\i}nez},
  {Mazin}, {Merck}, {Meucci}, {Meyer}, {Mirand a}, {Mirzoyan}, {Mizobuchi},
  {Moralejo}, {Nieto}, {Nilsson}, {Ninkovic}, {O{\~n}a-Wilhelmi}, {Otte},
  {Oya}, {Paneque}, {Panniello}, {Paoletti}, {Paredes}, {Pasanen}, {Pascoli},
  {Pauss}, {Pegna}, {Persic}, {Peruzzo}, {Piccioli}, {Prandini}, {Puchades},
  {Raymers}, {Rhode}, {Rib{\'o}}, {Rico}, {Rissi}, {Robert}, {R{\"u}gamer},
  {Saggion}, {Saito}, {S{\'a}nchez}, {Sartori}, {Scalzotto}, {Scapin},
  {Schmitt}, {Schweizer}, {Shayduk}, {Shinozaki}, {Shore}, {Sidro},
  {Sillanp{\"a}{\"a}}, {Sobczynska}, {Stamerra}, {Stark}, {Takalo}, {Temnikov},
  {Tescaro}, {Teshima}, {Torres}, {Turini}, {Vankov}, {Vitale}, {Wagner},
  {Wibig}, {Wittek}, {Zandanel}, {Zanin} \& {Zapatero}]{2007ApJ...664L..87A}
{\sc \au{{Albert}, J.}, \au{{Aliu}, E.}, \au{{Anderhub}, H.}, \au{{Antoranz},
  P.}, \au{{Armada}, A.}, \au{{Baixeras}, C.}, \au{{Barrio}, J.~A.},
  \au{{Bartko}, H.}, \au{{Bastieri}, D.}, \au{{Becker}, J.~K.}, \au{{Bednarek},
  W.}, \au{{Berger}, K.}, \au{{Bigongiari}, C.}, \au{{Biland}, A.}, \au{{Bock},
  R.~K.}, \au{{Bordas}, P.}, \au{{Bosch-Ramon}, V.}, \au{{Bretz}, T.},
  \au{{Britvitch}, I.}, \au{{Camara}, M.}, \au{{Carmona}, E.},
  \au{{Chilingarian}, A.}, \au{{Coarasa}, J.~A.}, \au{{Commichau}, S.},
  \au{{Contreras}, J.~L.}, \au{{Cortina}, J.}, \au{{Costado}, M.~T.},
  \au{{Curtef}, V.}, \au{{Danielyan}, V.}, \au{{Dazzi}, F.}, \au{{De Angelis},
  A.}, \au{{Delgado}, C.}, \au{{de los Reyes}, R.}, \au{{De Lotto}, B.},
  \au{{Domingo-Santamar{\'\i}a}, E.}, \au{{Dorner}, D.}, \au{{Doro}, M.},
  \au{{Errando}, M.}, \au{{Fagiolini}, M.}, \au{{Ferenc}, D.},
  \au{{Fern{\'a}ndez}, E.}, \au{{Firpo}, R.}, \au{{Flix}, J.}, \au{{Fonseca},
  M.~V.}, \au{{Font}, L.}, \au{{Fuchs}, M.}, \au{{Galante}, N.},
  \au{{Garc{\'\i}a-L{\'o}pez}, R.~J.}, \au{{Garczarczyk}, M.}, \au{{Gaug}, M.},
  \au{{Giller}, M.}, \au{{Goebel}, F.}, \au{{Hakobyan}, D.}, \au{{Hayashida},
  M.}, \au{{Hengstebeck}, T.}, \au{{Herrero}, A.}, \au{{H{\"o}hne}, D.},
  \au{{Hose}, J.}, \au{{Hsu}, C.~C.}, \au{{Jacon}, P.}, \au{{Jogler}, T.},
  \au{{Kosyra}, R.}, \au{{Kranich}, D.}, \au{{Kritzer}, R.}, \au{{Laille}, A.},
  \au{{Lindfors}, E.}, \au{{Lombardi}, S.}, \au{{Longo}, F.}, \au{{L{\'o}pez},
  J.}, \au{{L{\'o}pez}, M.}, \au{{Lorenz}, E.}, \au{{Majumdar}, P.},
  \au{{Maneva}, G.}, \au{{Mannheim}, K.}, \au{{Mansutti}, O.}, \au{{Mariotti},
  M.}, \au{{Mart{\'\i}nez}, M.}, \au{{Mazin}, D.}, \au{{Merck}, C.},
  \au{{Meucci}, M.}, \au{{Meyer}, M.}, \au{{Mirand a}, J.~M.}, \au{{Mirzoyan},
  R.}, \au{{Mizobuchi}, S.}, \au{{Moralejo}, A.}, \au{{Nieto}, D.},
  \au{{Nilsson}, K.}, \au{{Ninkovic}, J.}, \au{{O{\~n}a-Wilhelmi}, E.},
  \au{{Otte}, N.}, \au{{Oya}, I.}, \au{{Paneque}, D.}, \au{{Panniello}, M.},
  \au{{Paoletti}, R.}, \au{{Paredes}, J.~M.}, \au{{Pasanen}, M.},
  \au{{Pascoli}, D.}, \au{{Pauss}, F.}, \au{{Pegna}, R.}, \au{{Persic}, M.},
  \au{{Peruzzo}, L.}, \au{{Piccioli}, A.}, \au{{Prandini}, E.}, \au{{Puchades},
  N.}, \au{{Raymers}, A.}, \au{{Rhode}, W.}, \au{{Rib{\'o}}, M.}, \au{{Rico},
  J.}, \au{{Rissi}, M.}, \au{{Robert}, A.}, \au{{R{\"u}gamer}, S.},
  \au{{Saggion}, A.}, \au{{Saito}, T.}, \au{{S{\'a}nchez}, A.}, \au{{Sartori},
  P.}, \au{{Scalzotto}, V.}, \au{{Scapin}, V.}, \au{{Schmitt}, R.},
  \au{{Schweizer}, T.}, \au{{Shayduk}, M.}, \au{{Shinozaki}, K.}, \au{{Shore},
  S.~N.}, \au{{Sidro}, N.}, \au{{Sillanp{\"a}{\"a}}, A.}, \au{{Sobczynska},
  D.}, \au{{Stamerra}, A.}, \au{{Stark}, L.~S.}, \au{{Takalo}, L.},
  \au{{Temnikov}, P.}, \au{{Tescaro}, D.}, \au{{Teshima}, M.}, \au{{Torres},
  D.~F.}, \au{{Turini}, N.}, \au{{Vankov}, H.}, \au{{Vitale}, V.},
  \au{{Wagner}, R.~M.}, \au{{Wibig}, T.}, \au{{Wittek}, W.}, \au{{Zandanel},
  F.}, \au{{Zanin}, R.} \& \au{{Zapatero}, J.}} \yr{2007{\natexlab{{\em b\/}}}}
   \at{{Discovery of Very High Energy Gamma Radiation from IC 443 with the
  MAGIC Telescope}}.  \jt{ApJL}  \bvol{664}~(2),  \pg{L87--L90},  \arxiv{arXiv:
  0705.3119}.

\bibitem[{Aloisio} {\em et~al.\/}(2015){Aloisio}, {Blasi} \&
  {Serpico}]{Aloisio+15}
{\sc \au{{Aloisio}, R.}, \au{{Blasi}, P.} \& \au{{Serpico}, P.~D.}} \yr{2015}
  \at{{Nonlinear cosmic ray Galactic transport in the light of AMS-02 and
  Voyager data}}.  \jt{A \& A}  \bvol{583},  \pg{A95},  \arxiv{arXiv:
  1507.00594}.

\bibitem[{Amato}(2014)]{Amato14}
{\sc \au{{Amato}, E.}} \yr{2014}  \at{{The origin of galactic cosmic rays}}.
  \jt{International Journal of Modern Physics D}  \bvol{23}~(7),  \pg{1430013},
   \arxiv{arXiv: 1406.7714}.

\bibitem[{Amato}(2019)]{Amato19}
{\sc \au{{Amato}, E.}} \yr{2019} {The Theory Of Pulsar Wind Nebulae: Recent
  Progress}.  \bt{In {\em High Energy Phenomena in Relativistic Outflows
  VII\/}},  \pg{p.~33}.

\bibitem[{Amato} \& {Blasi}(2006)]{AmatoBlasi06}
{\sc \au{{Amato}, E.} \& \au{{Blasi}, P.}} \yr{2006}  \at{{Non-linear particle
  acceleration at non-relativistic shock waves in the presence of
  self-generated turbulence}}.  \jt{MNRAS}  \bvol{371}~(3),  \pg{1251--1258},
  \arxiv{arXiv: astro-ph/0606592}.

\bibitem[{Amato} \& {Blasi}(2009)]{AmatoBlasi09}
{\sc \au{{Amato}, E.} \& \au{{Blasi}, P.}} \yr{2009}  \at{{A kinetic approach
  to cosmic-ray-induced streaming instability at supernova shocks}}.
  \jt{MNRAS}  \bvol{392}~(4),  \pg{1591--1600},  \arxiv{arXiv: 0806.1223}.

\bibitem[{Amato} \& {Blasi}(2018)]{amatoblasi18}
{\sc \au{{Amato}, Elena} \& \au{{Blasi}, Pasquale}} \yr{2018}  \at{{Cosmic ray
  transport in the Galaxy: A review}}.  \jt{Advances in Space Research}
  \bvol{62}~(10),  \pg{2731--2749},  \arxiv{arXiv: 1704.05696}.

\bibitem[{Ambrogi} {\em et~al.\/}(2018){Ambrogi}, {Celli} \&
  {Aharonian}]{2018APh...100...69A}
{\sc \au{{Ambrogi}, L.}, \au{{Celli}, S.} \& \au{{Aharonian}, F.}} \yr{2018}
  \at{{On the potential of Cherenkov Telescope Arrays and KM3 Neutrino
  Telescopes for the detection of extended sources}}.  \jt{Astroparticle
  Physics}  \bvol{100},  \pg{69--79},  \arxiv{arXiv: 1803.03565}.

\bibitem[{Archambault} {\em et~al.\/}(2017){Archambault}, {Archer}, {Benbow},
  {Bird}, {Bourbeau}, {Buchovecky}, {Buckley}, {Bugaev}, {Cerruti}, {Connolly},
  {Cui}, {Dwarkadas}, {Errando}, {Falcone}, {Feng}, {Finley}, {Fleischhack},
  {Fortson}, {Furniss}, {Griffin}, {H{\"u}tten}, {Hanna}, {Holder}, {Johnson},
  {Kaaret}, {Kar}, {Kelley-Hoskins}, {Kertzman}, {Kieda}, {Krause}, {Kumar},
  {Lang}, {Maier}, {McArthur}, {McCann}, {Moriarty}, {Mukherjee}, {Nieto},
  {O'Brien}, {Ong}, {Otte}, {Park}, {Pohl}, {Popkow}, {Pueschel}, {Quinn},
  {Ragan}, {Reynolds}, {Richards}, {Roache}, {Sadeh}, {Santander}, {Sembroski},
  {Shahinyan}, {Slane}, {Staszak}, {Telezhinsky}, {Trepanier}, {Tyler},
  {Wakely}, {Weinstein}, {Weisgarber}, {Wilcox}, {Wilhelm}, {Williams} \&
  {Zitzer}]{2017ApJ...836...23A}
{\sc \au{{Archambault}, S.}, \au{{Archer}, A.}, \au{{Benbow}, W.}, \au{{Bird},
  R.}, \au{{Bourbeau}, E.}, \au{{Buchovecky}, M.}, \au{{Buckley}, J.~H.},
  \au{{Bugaev}, V.}, \au{{Cerruti}, M.}, \au{{Connolly}, M.~P.}, \au{{Cui},
  W.}, \au{{Dwarkadas}, V.~V.}, \au{{Errando}, M.}, \au{{Falcone}, A.},
  \au{{Feng}, Q.}, \au{{Finley}, J.~P.}, \au{{Fleischhack}, H.}, \au{{Fortson},
  L.}, \au{{Furniss}, A.}, \au{{Griffin}, S.}, \au{{H{\"u}tten}, M.},
  \au{{Hanna}, D.}, \au{{Holder}, J.}, \au{{Johnson}, C.~A.}, \au{{Kaaret},
  P.}, \au{{Kar}, P.}, \au{{Kelley-Hoskins}, N.}, \au{{Kertzman}, M.},
  \au{{Kieda}, D.}, \au{{Krause}, M.}, \au{{Kumar}, S.}, \au{{Lang}, M.~J.},
  \au{{Maier}, G.}, \au{{McArthur}, S.}, \au{{McCann}, A.}, \au{{Moriarty},
  P.}, \au{{Mukherjee}, R.}, \au{{Nieto}, D.}, \au{{O'Brien}, S.}, \au{{Ong},
  R.~A.}, \au{{Otte}, A.~N.}, \au{{Park}, N.}, \au{{Pohl}, M.}, \au{{Popkow},
  A.}, \au{{Pueschel}, E.}, \au{{Quinn}, J.}, \au{{Ragan}, K.}, \au{{Reynolds},
  P.~T.}, \au{{Richards}, G.~T.}, \au{{Roache}, E.}, \au{{Sadeh}, I.},
  \au{{Santander}, M.}, \au{{Sembroski}, G.~H.}, \au{{Shahinyan}, K.},
  \au{{Slane}, P.}, \au{{Staszak}, D.}, \au{{Telezhinsky}, I.},
  \au{{Trepanier}, S.}, \au{{Tyler}, J.}, \au{{Wakely}, S.~P.},
  \au{{Weinstein}, A.}, \au{{Weisgarber}, T.}, \au{{Wilcox}, P.},
  \au{{Wilhelm}, A.}, \au{{Williams}, D.~A.} \& \au{{Zitzer}, B.}} \yr{2017}
  \at{{Gamma-Ray Observations of Tycho{\textquoteright}s Supernova Remnant with
  VERITAS and Fermi}}.  \jt{ApJ}  \bvol{836}~(1),  \pg{23},  \arxiv{arXiv:
  1701.06740}.

\bibitem[{Atoyan} {\em et~al.\/}(2006){Atoyan}, {Buckley} \&
  {Krawczynski}]{2006ApJ...642L.153A}
{\sc \au{{Atoyan}, Armen}, \au{{Buckley}, James} \& \au{{Krawczynski}, Henric}}
  \yr{2006}  \at{{A Gamma-Ray Burst Remnant in Our Galaxy: HESS J1303-631}}.
  \jt{ApJL}  \bvol{642}~(2),  \pg{L153--L156},  \arxiv{arXiv:
  astro-ph/0509615}.

\bibitem[Atoyan {\em et~al.\/}(1995)Atoyan, Aharonian \&
  V\"olk]{PhysRevD.52.3265}
{\sc \au{Atoyan, A.~M.}, \au{Aharonian, F.~A.} \& \au{V\"olk, H.~J.}} \yr{1995}
   \at{Electrons and positrons in the galactic cosmic rays}.  \jt{Phys. Rev. D}
   \bvol{52},  \pg{3265--3275}.

\bibitem[Baade \& Zwicky(1934)]{PhysRev.46.76.2}
{\sc \au{Baade, W.} \& \au{Zwicky, F.}} \yr{1934}  \at{Remarks on super-novae
  and cosmic rays}.  \jt{Phys. Rev.}  \bvol{46},  \pg{76--77}.

\bibitem[Baghmanyan {\em et~al.\/}(2020)Baghmanyan, Peron, Casanova, Aharonian
  \& Zanin]{Baghmanyan_2020}
{\sc \au{Baghmanyan, Vardan}, \au{Peron, Giada}, \au{Casanova, Sabrina},
  \au{Aharonian, Felix} \& \au{Zanin, Roberta}} \yr{2020}  \at{Evidence of
  cosmic-ray excess from local giant molecular clouds}.  \jt{The Astrophysical
  Journal}  \bvol{901}~(1),  \pg{L4}.

\bibitem[{Bartoli} {\em et~al.\/}(2012){Bartoli}, {Bernardini}, {Bi}, {Bleve},
  {Bolognino}, {Branchini}, {Budano}, {Calabrese Melcarne}, {Camarri}, {Cao},
  {Cardarelli}, {Catalanotti}, {Cattaneo}, {Chen}, {Chen}, {Chen}, {Creti},
  {Cui}, {Dai}, {D'Al{\'\i} Staiti}, {Danzengluobu}, {Dattoli}, {De Mitri},
  {D'Ettorre Piazzoli}, {Di Girolamo}, {Ding}, {Di Sciascio}, {Feng}, {Feng},
  {Feng}, {Galeazzi}, {Giroletti}, {Gou}, {Guo}, {He}, {Hu}, {Hu}, {Huang},
  {Iacovacci}, {Iuppa}, {James}, {Jia}, {Labaciren}, {Li}, {Li}, {Li},
  {Liguori}, {Liu}, {Liu}, {Liu}, {Liu}, {Lu}, {Ma}, {Ma}, {Mancarella},
  {Mari}, {Marsella}, {Martello}, {Mastroianni}, {Montini}, {Ning}, {Pagliaro},
  {Panareo}, {Panico}, {Perrone}, {Pistilli}, {Ruggieri}, {Salvini},
  {Santonico}, {Shen}, {Sheng}, {Shi}, {Stanescu}, {Surdo}, {Tan}, {Vallania},
  {Vernetto}, {Vigorito}, {Wang}, {Wang}, {Wu}, {Wu}, {Xu}, {Xue}, {Yang},
  {Yang}, {Yao}, {Yuan}, {Zha}, {Zhang}, {Zhang}, {Zhang}, {Zhang}, {Zhang},
  {Zhang}, {Zhang}, {Zhao}, {Zhaxiciren}, {Zhaxisangzhu}, {Zhou}, {Zhu}, {Zhu},
  {Zizzi} \& {ARGO-YBJ Collaboration}]{2012ApJ...745L..22B}
{\sc \au{{Bartoli}, B.}, \au{{Bernardini}, P.}, \au{{Bi}, X.~J.}, \au{{Bleve},
  C.}, \au{{Bolognino}, I.}, \au{{Branchini}, P.}, \au{{Budano}, A.},
  \au{{Calabrese Melcarne}, A.~K.}, \au{{Camarri}, P.}, \au{{Cao}, Z.},
  \au{{Cardarelli}, R.}, \au{{Catalanotti}, S.}, \au{{Cattaneo}, C.},
  \au{{Chen}, S.~Z.}, \au{{Chen}, T.~L.}, \au{{Chen}, Y.}, \au{{Creti}, P.},
  \au{{Cui}, S.~W.}, \au{{Dai}, B.~Z.}, \au{{D'Al{\'\i} Staiti}, G.},
  \au{{Danzengluobu}}, \au{{Dattoli}, M.}, \au{{De Mitri}, I.}, \au{{D'Ettorre
  Piazzoli}, B.}, \au{{Di Girolamo}, T.}, \au{{Ding}, X.~H.}, \au{{Di
  Sciascio}, G.}, \au{{Feng}, C.~F.}, \au{{Feng}, Zhaoyang}, \au{{Feng},
  Zhenyong}, \au{{Galeazzi}, F.}, \au{{Giroletti}, E.}, \au{{Gou}, Q.~B.},
  \au{{Guo}, Y.~Q.}, \au{{He}, H.~H.}, \au{{Hu}, Haibing}, \au{{Hu}, Hongbo},
  \au{{Huang}, Q.}, \au{{Iacovacci}, M.}, \au{{Iuppa}, R.}, \au{{James}, I.},
  \au{{Jia}, H.~Y.}, \au{{Labaciren}}, \au{{Li}, H.~J.}, \au{{Li}, J.~Y.},
  \au{{Li}, X.~X.}, \au{{Liguori}, G.}, \au{{Liu}, C.}, \au{{Liu}, C.~Q.},
  \au{{Liu}, J.}, \au{{Liu}, M.~Y.}, \au{{Lu}, H.}, \au{{Ma}, L.~L.}, \au{{Ma},
  X.~H.}, \au{{Mancarella}, G.}, \au{{Mari}, S.~M.}, \au{{Marsella}, G.},
  \au{{Martello}, D.}, \au{{Mastroianni}, S.}, \au{{Montini}, P.}, \au{{Ning},
  C.~C.}, \au{{Pagliaro}, A.}, \au{{Panareo}, M.}, \au{{Panico}, B.},
  \au{{Perrone}, L.}, \au{{Pistilli}, P.}, \au{{Ruggieri}, F.}, \au{{Salvini},
  P.}, \au{{Santonico}, R.}, \au{{Shen}, P.~R.}, \au{{Sheng}, X.~D.},
  \au{{Shi}, F.}, \au{{Stanescu}, C.}, \au{{Surdo}, A.}, \au{{Tan}, Y.~H.},
  \au{{Vallania}, P.}, \au{{Vernetto}, S.}, \au{{Vigorito}, C.}, \au{{Wang},
  B.}, \au{{Wang}, H.}, \au{{Wu}, C.~Y.}, \au{{Wu}, H.~R.}, \au{{Xu}, B.},
  \au{{Xue}, L.}, \au{{Yang}, Q.~Y.}, \au{{Yang}, X.~C.}, \au{{Yao}, Z.~G.},
  \au{{Yuan}, A.~F.}, \au{{Zha}, M.}, \au{{Zhang}, H.~M.}, \au{{Zhang},
  Jilong}, \au{{Zhang}, Jianli}, \au{{Zhang}, L.}, \au{{Zhang}, P.},
  \au{{Zhang}, X.~Y.}, \au{{Zhang}, Y.}, \au{{Zhao}, J.}, \au{{Zhaxiciren}},
  \au{{Zhaxisangzhu}}, \au{{Zhou}, X.~X.}, \au{{Zhu}, F.~R.}, \au{{Zhu},
  Q.~Q.}, \au{{Zizzi}, G.} \& \au{{ARGO-YBJ Collaboration}}} \yr{2012}
  \at{{Observation of TeV Gamma Rays from the Cygnus Region with the ARGO-YBJ
  Experiment}}.  \jt{ApJL}  \bvol{745}~(2),  \pg{L22},  \arxiv{arXiv:
  1201.1973}.

\bibitem[{Bartoli} {\em et~al.\/}(2014){Bartoli}, {Bernardini}, {Bi},
  {Branchini}, {Budano}, {Camarri}, {Cao}, {Cardarelli}, {Catalanotti}, {Chen},
  {Chen}, {Creti}, {Cui}, {Dai}, {D'Amone}, {Danzengluobu}, {De Mitri},
  {D'Ettorre Piazzoli}, {Di Girolamo}, {Di Sciascio}, {Feng}, {Feng}, {Feng},
  {Gou}, {Guo}, {He}, {Hu}, {Hu}, {Iacovacci}, {Iuppa}, {Jia}, {Labaciren},
  {Li}, {Liguori}, {Liu}, {Liu}, {Liu}, {Lu}, {Ma}, {Ma}, {Mancarella}, {Mari},
  {Marsella}, {Martello}, {Mastroianni}, {Montini}, {Ning}, {Panareo},
  {Perrone}, {Pistilli}, {Ruggieri}, {Salvini}, {Santonico}, {Shen}, {Sheng},
  {Shi}, {Surdo}, {Tan}, {Vallania}, {Vernetto}, {Vigorito}, {Wang}, {Wu},
  {Wu}, {Xue}, {Yang}, {Yang}, {Yao}, {Yuan}, {Zha}, {Zhang}, {Zhang}, {Zhang},
  {Zhang}, {Zhao}, {Zhaxiciren}, {Zhaxisangzhu}, {Zhou}, {Zhu}, {Zhu}, {Zizzi}
  \& {ARGO-YBJ Collaboration}]{2014ApJ...790..152B}
{\sc \au{{Bartoli}, B.}, \au{{Bernardini}, P.}, \au{{Bi}, X.~J.},
  \au{{Branchini}, P.}, \au{{Budano}, A.}, \au{{Camarri}, P.}, \au{{Cao}, Z.},
  \au{{Cardarelli}, R.}, \au{{Catalanotti}, S.}, \au{{Chen}, S.~Z.},
  \au{{Chen}, T.~L.}, \au{{Creti}, P.}, \au{{Cui}, S.~W.}, \au{{Dai}, B.~Z.},
  \au{{D'Amone}, A.}, \au{{Danzengluobu}}, \au{{De Mitri}, I.}, \au{{D'Ettorre
  Piazzoli}, B.}, \au{{Di Girolamo}, T.}, \au{{Di Sciascio}, G.}, \au{{Feng},
  C.~F.}, \au{{Feng}, Zhaoyang}, \au{{Feng}, Zhenyong}, \au{{Gou}, Q.~B.},
  \au{{Guo}, Y.~Q.}, \au{{He}, H.~H.}, \au{{Hu}, Haibing}, \au{{Hu}, Hongbo},
  \au{{Iacovacci}, M.}, \au{{Iuppa}, R.}, \au{{Jia}, H.~Y.}, \au{{Labaciren}},
  \au{{Li}, H.~J.}, \au{{Liguori}, G.}, \au{{Liu}, C.}, \au{{Liu}, J.},
  \au{{Liu}, M.~Y.}, \au{{Lu}, H.}, \au{{Ma}, L.~L.}, \au{{Ma}, X.~H.},
  \au{{Mancarella}, G.}, \au{{Mari}, S.~M.}, \au{{Marsella}, G.},
  \au{{Martello}, D.}, \au{{Mastroianni}, S.}, \au{{Montini}, P.}, \au{{Ning},
  C.~C.}, \au{{Panareo}, M.}, \au{{Perrone}, L.}, \au{{Pistilli}, P.},
  \au{{Ruggieri}, F.}, \au{{Salvini}, P.}, \au{{Santonico}, R.}, \au{{Shen},
  P.~R.}, \au{{Sheng}, X.~D.}, \au{{Shi}, F.}, \au{{Surdo}, A.}, \au{{Tan},
  Y.~H.}, \au{{Vallania}, P.}, \au{{Vernetto}, S.}, \au{{Vigorito}, C.},
  \au{{Wang}, H.}, \au{{Wu}, C.~Y.}, \au{{Wu}, H.~R.}, \au{{Xue}, L.},
  \au{{Yang}, Q.~Y.}, \au{{Yang}, X.~C.}, \au{{Yao}, Z.~G.}, \au{{Yuan},
  A.~F.}, \au{{Zha}, M.}, \au{{Zhang}, H.~M.}, \au{{Zhang}, L.}, \au{{Zhang},
  X.~Y.}, \au{{Zhang}, Y.}, \au{{Zhao}, J.}, \au{{Zhaxiciren}},
  \au{{Zhaxisangzhu}}, \au{{Zhou}, X.~X.}, \au{{Zhu}, F.~R.}, \au{{Zhu},
  Q.~Q.}, \au{{Zizzi}, G.} \& \au{{ARGO-YBJ Collaboration}}} \yr{2014}
  \at{{Identification of the TeV Gamma-Ray Source ARGO J2031+4157 with the
  Cygnus Cocoon}}.  \jt{ApJ}  \bvol{790}~(2),  \pg{152},  \arxiv{arXiv:
  1406.6436}.

\bibitem[{Beatty} {\em et~al.\/}(2004){Beatty}, {Bhattacharyya}, {Bower},
  {Coutu}, {Duvernois}, {McKee}, {Minnick}, {M{\"u}ller}, {Musser}, {Nutter},
  {Labrador}, {Schubnell}, {Swordy}, {Tarl{\'e}} \&
  {Tomasch}]{2004PhRvL..93x1102B}
{\sc \au{{Beatty}, J.~J.}, \au{{Bhattacharyya}, A.}, \au{{Bower}, C.},
  \au{{Coutu}, S.}, \au{{Duvernois}, M.~A.}, \au{{McKee}, S.}, \au{{Minnick},
  S.~A.}, \au{{M{\"u}ller}, D.}, \au{{Musser}, J.}, \au{{Nutter}, S.},
  \au{{Labrador}, A.~W.}, \au{{Schubnell}, M.}, \au{{Swordy}, S.},
  \au{{Tarl{\'e}}, G.} \& \au{{Tomasch}, A.}} \yr{2004}  \at{{New Measurement
  of the Cosmic-Ray Positron Fraction from 5 to 15GeV}}.  \jt{Physical review
  Letter}  \bvol{93}~(24),  \pg{241102},  \arxiv{arXiv: astro-ph/0412230}.

\bibitem[{Bednarek}(2007)]{2007MNRAS.382..367B}
{\sc \au{{Bednarek}, W.}} \yr{2007}  \at{{{$\gamma$}-ray production in young
  open clusters: Berk 87, Cyg OB2 and Westerlund 2}}.  \jt{MNRAS}  \bvol{382},
  \pg{367--376},  \arxiv{arXiv: 0704.3517}.

\bibitem[{Bednarek} \& {Sitarek}(2007)]{2007MNRAS.377..920B}
{\sc \au{{Bednarek}, W.} \& \au{{Sitarek}, J.}} \yr{2007}  \at{{High-energy
  {$\gamma$}-rays from globular clusters}}.  \jt{MNRAS}  \bvol{377},
  \pg{920--930},  \arxiv{arXiv: astro-ph/0701522}.

\bibitem[{Bell}(1978)]{Bell78}
{\sc \au{{Bell}, A.~R.}} \yr{1978}  \at{{The acceleration of cosmic rays in
  shock fronts - I.}}  \jt{MNRAS}  \bvol{182},  \pg{147--156}.

\bibitem[{Bell}(2004)]{Bell04}
{\sc \au{{Bell}, A.~R.}} \yr{2004}  \at{{Turbulent amplification of magnetic
  field and diffusive shock acceleration of cosmic rays}}.  \jt{MNRAS}
  \bvol{353}~(2),  \pg{550--558}.

\bibitem[{Beresnyak} {\em et~al.\/}(2009){Beresnyak}, {Jones} \&
  {Lazarian}]{Beresnyak+09}
{\sc \au{{Beresnyak}, A.}, \au{{Jones}, T.~W.} \& \au{{Lazarian}, A.}}
  \yr{2009}  \at{{Turbulence-Induced Magnetic Fields and Structure of Cosmic
  Ray Modified Shocks}}.  \jt{ApJ}  \bvol{707}~(2),  \pg{1541--1549},
  \arxiv{arXiv: 0908.2806}.

\bibitem[{Blandford} \& {Ostriker}(1978)]{BlandOstr78}
{\sc \au{{Blandford}, R.~D.} \& \au{{Ostriker}, J.~P.}} \yr{1978}
  \at{{Particle acceleration by astrophysical shocks.}}  \jt{ApJ Letters}
  \bvol{221},  \pg{L29--L32}.

\bibitem[{Blasi}(2013)]{Blasi13}
{\sc \au{{Blasi}, Pasquale}} \yr{2013}  \at{{The origin of galactic cosmic
  rays}}.  \jt{A\&A Rev.}  \bvol{21},  \pg{70},  \arxiv{arXiv: 1311.7346}.

\bibitem[{Blasi} \& {Amato}(2012{\natexlab{{\em a\/}}})]{BlasiAmato12a}
{\sc \au{{Blasi}, Pasquale} \& \au{{Amato}, Elena}} \yr{2012{\natexlab{{\em
  a\/}}}}  \at{{Diffusive propagation of cosmic rays from supernova remnants in
  the Galaxy. I: spectrum and chemical composition}}.  \jt{JCAP}
  \bvol{2012}~(1),  \pg{010},  \arxiv{arXiv: 1105.4521}.

\bibitem[{Blasi} \& {Amato}(2012{\natexlab{{\em b\/}}})]{BlasiAmato12b}
{\sc \au{{Blasi}, Pasquale} \& \au{{Amato}, Elena}} \yr{2012{\natexlab{{\em
  b\/}}}}  \at{{Diffusive propagation of cosmic rays from supernova remnants in
  the Galaxy. II: anisotropy}}.  \jt{JCAP}  \bvol{2012}~(1),  \pg{011},
  \arxiv{arXiv: 1105.4529}.

\bibitem[{Blasi} {\em et~al.\/}(2012){Blasi}, {Amato} \& {Serpico}]{Blasi+12}
{\sc \au{{Blasi}, Pasquale}, \au{{Amato}, Elena} \& \au{{Serpico},
  Pasquale~D.}} \yr{2012}  \at{{Spectral Breaks as a Signature of Cosmic Ray
  Induced Turbulence in the Galaxy}}.  \jt{PRL}  \bvol{109}~(6),  \pg{061101},
  \arxiv{arXiv: 1207.3706}.

\bibitem[{Blum} {\em et~al.\/}(2013){Blum}, {Katz} \& {Waxman}]{Blum+13}
{\sc \au{{Blum}, Kfir}, \au{{Katz}, Boaz} \& \au{{Waxman}, Eli}} \yr{2013}
  \at{{AMS-02 Results Support the Secondary Origin of Cosmic Ray Positrons}}.
  \jt{PRL}  \bvol{111}~(21),  \pg{211101},  \arxiv{arXiv: 1305.1324}.

\bibitem[Blumenthal \& Gould(1970)]{RevModPhys.42.237}
{\sc \au{Blumenthal, George~R.} \& \au{Gould, Robert~J.}} \yr{1970}
  \at{Bremsstrahlung, synchrotron radiation, and compton scattering of
  high-energy electrons traversing dilute gases}.  \jt{Rev. Mod. Phys.}
  \bvol{42},  \pg{237--270}.

\bibitem[{Bresci} {\em et~al.\/}(2019){Bresci}, {Amato}, {Blasi} \&
  {Morlino}]{Bresci+19}
{\sc \au{{Bresci}, V.}, \au{{Amato}, E.}, \au{{Blasi}, P.} \& \au{{Morlino},
  G.}} \yr{2019}  \at{{Effects of re-acceleration and source grammage on
  secondary cosmic rays spectra}}.  \jt{MNRAS}  \bvol{488}~(2),
  \pg{2068--2078},  \arxiv{arXiv: 1904.10282}.

\bibitem[{Buck} {\em et~al.\/}(2020){Buck}, {Pfrommer}, {Pakmor}, {Grand} \&
  {Springel}]{Buck20}
{\sc \au{{Buck}, Tobias}, \au{{Pfrommer}, Christoph}, \au{{Pakmor},
  R{\"u}diger}, \au{{Grand}, Robert J.~J.} \& \au{{Springel}, Volker}}
  \yr{2020}  \at{{The effects of cosmic rays on the formation of Milky Way-mass
  galaxies in a cosmological context}}.  \jt{MNRAS}  \bvol{497}~(2),
  \pg{1712--1737},  \arxiv{arXiv: 1911.00019}.

\bibitem[{Bykov} {\em et~al.\/}(2017){Bykov}, {Amato}, {Petrov},
  {Krassilchtchikov} \& {Levenfish}]{Bykov+17}
{\sc \au{{Bykov}, A.~M.}, \au{{Amato}, E.}, \au{{Petrov}, A.~E.},
  \au{{Krassilchtchikov}, A.~M.} \& \au{{Levenfish}, K.~P.}} \yr{2017}
  \at{{Pulsar Wind Nebulae with Bow Shocks: Non-thermal Radiation and Cosmic
  Ray Leptons}}.  \jt{Space Science Reviews}  \bvol{207}~(1-4),  \pg{235--290},
   \arxiv{arXiv: 1705.00950}.

\bibitem[{Bykov} {\em et~al.\/}(2012){Bykov}, {Ellison} \& {Renaud}]{Bykov+12}
{\sc \au{{Bykov}, Andrei~M.}, \au{{Ellison}, Donald~C.} \& \au{{Renaud},
  Matthieu}} \yr{2012}  \at{{Magnetic Fields in Cosmic Particle Acceleration
  Sources}}.  \jt{Space Science Reviews}  \bvol{166}~(1-4),  \pg{71--95},
  \arxiv{arXiv: 1105.0130}.

\bibitem[{Bykov} {\em et~al.\/}(2020){Bykov}, {Marcowith}, {Amato},
  {Kalyashova}, {Kruijssen} \& {Waxman}]{Bykov+20}
{\sc \au{{Bykov}, Andrei~M.}, \au{{Marcowith}, Alexandre}, \au{{Amato}, Elena},
  \au{{Kalyashova}, Maria~E.}, \au{{Kruijssen}, J.~M.~Diederik} \&
  \au{{Waxman}, Eli}} \yr{2020}  \at{{High-Energy Particles and Radiation in
  Star-Forming Regions}}.  \jt{Space Science Reviews}  \bvol{216}~(3),
  \pg{42},  \arxiv{arXiv: 2003.11534}.

\bibitem[{Bykov} {\em et~al.\/}(2011){Bykov}, {Osipov} \& {Ellison}]{Bykov11}
{\sc \au{{Bykov}, A.~M.}, \au{{Osipov}, S.~M.} \& \au{{Ellison}, D.~C.}}
  \yr{2011}  \at{{Cosmic ray current driven turbulence in shocks with efficient
  particle acceleration: the oblique, long-wavelength mode instability}}.
  \jt{MNRAS}  \bvol{410}~(1),  \pg{39--52},  \arxiv{arXiv: 1010.0408}.

\bibitem[{Bykov} \& {Toptygin}(2001)]{2001AstL...27..625B}
{\sc \au{{Bykov}, A.~M.} \& \au{{Toptygin}, I.~N.}} \yr{2001}  \at{{A Model of
  Particle Acceleration to High Energies by Multiple Supernova Explosions in OB
  Associations}}.  \jt{Astronomy Letters}  \bvol{27}~(10),  \pg{625--633}.

\bibitem[{Caprioli} \& {Spitkovsky}(2014{\natexlab{{\em a\/}}})]{Caprioli14a}
{\sc \au{{Caprioli}, D.} \& \au{{Spitkovsky}, A.}} \yr{2014{\natexlab{{\em
  a\/}}}}  \at{{Simulations of Ion Acceleration at Non-relativistic Shocks. I.
  Acceleration Efficiency}}.  \jt{ApJ}  \bvol{783}~(2),  \pg{91},
  \arxiv{arXiv: 1310.2943}.

\bibitem[{Caprioli} \& {Spitkovsky}(2014{\natexlab{{\em
  b\/}}})]{CaprioliSpitkovsky14B}
{\sc \au{{Caprioli}, D.} \& \au{{Spitkovsky}, A.}} \yr{2014{\natexlab{{\em
  b\/}}}}  \at{{Simulations of Ion Acceleration at Non-relativistic Shocks. II.
  Magnetic Field Amplification}}.  \jt{ApJ}  \bvol{794}~(1),  \pg{46},
  \arxiv{arXiv: 1401.7679}.

\bibitem[{Caprioli} \& {Spitkovsky}(2014{\natexlab{{\em
  c\/}}})]{CaprioliSpitkovsky14C}
{\sc \au{{Caprioli}, D.} \& \au{{Spitkovsky}, A.}} \yr{2014{\natexlab{{\em
  c\/}}}}  \at{{Simulations of Ion Acceleration at Non-relativistic Shocks.
  III. Particle Diffusion}}.  \jt{ApJ}  \bvol{794}~(1),  \pg{47},
  \arxiv{arXiv: 1407.2261}.

\bibitem[{Cardillo} {\em et~al.\/}(2015){Cardillo}, {Amato} \&
  {Blasi}]{Cardillo+15}
{\sc \au{{Cardillo}, Martina}, \au{{Amato}, Elena} \& \au{{Blasi}, Pasquale}}
  \yr{2015}  \at{{On the cosmic ray spectrum from type II supernovae expanding
  in their red giant presupernova wind}}.  \jt{Astroparticle Physics}
  \bvol{69},  \pg{1--10},  \arxiv{arXiv: 1503.03001}.

\bibitem[{Cardillo} {\em et~al.\/}(2016){Cardillo}, {Amato} \&
  {Blasi}]{Cardillo+16}
{\sc \au{{Cardillo}, M.}, \au{{Amato}, E.} \& \au{{Blasi}, P.}} \yr{2016}
  \at{{Supernova remnant W44: a case of cosmic-ray reacceleration}}.  \jt{A\&A}
   \bvol{595},  \pg{A58},  \arxiv{arXiv: 1604.02321}.

\bibitem[{Casanova} {\em et~al.\/}(2010{\natexlab{{\em a\/}}}){Casanova},
  {Aharonian}, {Fukui}, {Gabici}, {Jones}, {Kawamura}, {Onishi}, {Rowell},
  {Sano}, {Torii} \& {Yamamoto}]{2010PASJ...62..769C}
{\sc \au{{Casanova}, S.}, \au{{Aharonian}, F.~A.}, \au{{Fukui}, Y.},
  \au{{Gabici}, S.}, \au{{Jones}, D.~I.}, \au{{Kawamura}, A.}, \au{{Onishi},
  T.}, \au{{Rowell}, G.}, \au{{Sano}, H.}, \au{{Torii}, K.} \& \au{{Yamamoto},
  H.}} \yr{2010{\natexlab{{\em a\/}}}}  \at{{Molecular Clouds as Cosmic-Ray
  Barometers}}.  \jt{PASJ}  \bvol{62},  \pg{769--777},  \arxiv{arXiv:
  0904.2887}.

\bibitem[{Casanova} {\em et~al.\/}(2010{\natexlab{{\em b\/}}}){Casanova},
  {Jones}, {Aharonian}, {Fukui}, {Gabici}, {Kawamura}, {Onishi}, {Rowell},
  {Sano}, {Torii} \& {Yamamoto}]{2010PASJ...62.1127C}
{\sc \au{{Casanova}, S.}, \au{{Jones}, D.~I.}, \au{{Aharonian}, F.~A.},
  \au{{Fukui}, Y.}, \au{{Gabici}, S.}, \au{{Kawamura}, A.}, \au{{Onishi}, T.},
  \au{{Rowell}, G.}, \au{{Sano}, H.}, \au{{Torii}, K.} \& \au{{Yamamoto}, H.}}
  \yr{2010{\natexlab{{\em b\/}}}}  \at{{Modeling the Gamma-Ray Emission
  Produced by Runaway Cosmic Rays in the Environment of RX J1713.7-3946}}.
  \jt{PASJ}  \bvol{62},  \pg{1127--1134},  \arxiv{arXiv: 1003.0379}.

\bibitem[{Cassam-Chena{\"i}} {\em et~al.\/}(2004){Cassam-Chena{\"i}},
  {Decourchelle}, {Ballet}, {Sauvageot}, {Dubner} \&
  {Giacani}]{2004A&A...427..199C}
{\sc \au{{Cassam-Chena{\"i}}, G.}, \au{{Decourchelle}, A.}, \au{{Ballet}, J.},
  \au{{Sauvageot}, J.-L.}, \au{{Dubner}, G.} \& \au{{Giacani}, E.}} \yr{2004}
  \at{{XMM-Newton observations of the supernova remnant RX J1713.7-3946 and its
  central source}}.  \jt{A$\&$A}  \bvol{427},  \pg{199--216}.

\bibitem[{Casse} \& {Paul}(1980)]{1980ApJ...237..236C}
{\sc \au{{Casse}, M.} \& \au{{Paul}, J.~A.}} \yr{1980}  \at{{Local gamma rays
  and cosmic-ray acceleration by supersonic stellar winds}}.  \jt{ApJ}
  \bvol{237},  \pg{236--243}.

\bibitem[{Cesarsky}(1980)]{Cesarsky80}
{\sc \au{{Cesarsky}, C.~J.}} \yr{1980}  \at{{Cosmic-ray confinement in the
  galaxy}}.  \jt{A\&A Rev.}  \bvol{18},  \pg{289--319}.

\bibitem[{Cesarsky} \& {Montmerle}(1983)]{1983SSRv...36..173C}
{\sc \au{{Cesarsky}, C.~J.} \& \au{{Montmerle}, T.}} \yr{1983}  \at{{Gamma-Rays
  from Active Regions in the Galaxy - the Possible Contribution of Stellar
  Winds}}.  \jt{SSR}  \bvol{36}~(2),  \pg{173--193}.

\bibitem[{Cherenkov Telescope Array Consortium} {\em et~al.\/}(2019){Cherenkov
  Telescope Array Consortium}, {Acharya}, {Agudo}, {Al Samarai}, {Alfaro},
  {Alfaro}, {Alispach}, {Alves Batista}, {Amans}, {Amato}, {Ambrosi},
  {Antolini}, {Antonelli}, {Aramo}, {Araya}, {Armstrong}, {Arqueros},
  {Arrabito}, {Asano}, {Ashley}, {Backes}, {Balazs}, {Balbo}, {Ballester},
  {Ballet}, {Bamba}, {Barkov}, {Barres de Almeida}, {Barrio}, {Bastieri},
  {Becherini}, {Belfiore}, {Benbow}, {Berge}, {Bernardini}, {Bernardini},
  {Bernardos}, {Bernl{\"o}hr}, {Bertucci}, {Biasuzzi}, {Bigongiari}, {Biland},
  {Bissaldi}, {Biteau}, {Blanch}, {Blazek}, {Boisson}, {Bolmont}, {Bonanno},
  {Bonardi}, {Bonavolont{\`a}}, {Bonnoli}, {Bosnjak}, {B{\"o}ttcher},
  {Braiding}, {Bregeon}, {Brill}, {Brown}, {Brun}, {Brunetti}, {Buanes},
  {Buckley}, {Bugaev}, {B{\"u}hler}, {Bulgarelli}, {Bulik}, {Burton},
  {Burtovoi}, {Busetto}, {Canestrari}, {Capalbi}, {Capitanio}, {Caproni},
  {Caraveo}, {C{\'a}rdenas}, {Carlile}, {Carosi}, {Carqu{\'\i}n}, {Carr},
  {Casanova}, {Cascone}, {Catalani}, {Catalano}, {Cauz}, {Cerruti}, {Chadwick},
  {Chaty}, {Chaves}, {Chen}, {Chen}, {Chernyakova}, {Chikawa}, {Christov},
  {Chudoba}, {Cie{\'s}lar}, {Coco}, {Colafrancesco}, {Colin}, {Conforti},
  {Connaughton}, {Conrad}, {Contreras}, {Cortina}, {Costa}, {Costantini},
  {Cotter}, {Covino}, {Crocker}, {Cuadra}, {Cuevas}, {Cumani}, {D'A{\`\i}},
  {D'Ammando}, {D'Avanzo}, {D'Urso}, {Daniel}, {Davids}, {Dawson}, {Dazzi}, {De
  Angelis}, {de C{\'a}ssia dos Anjos}, {De Cesare}, {De Franco}, {de Gouveia
  Dal Pino}, {de la Calle}, {de los Reyes Lopez}, {De Lotto}, {De Luca}, {De
  Lucia}, {de Naurois}, {de O{\~n}a Wilhelmi}, {De Palma}, {De Persio}, {de
  Souza}, {Deil}, {Del Santo}, {Delgado}, {della Volpe}, {Di Girolamo}, {Di
  Pierro}, {Di Venere}, {D{\'\i}az}, {Dib}, {Diebold}, {Djannati-Ata{\"\i}},
  {Dom{\'\i}nguez}, {Dominis Prester}, {Dorner}, {Doro}, {Drass}, {Dravins},
  {Dubus}, {Dwarkadas}, {Ebr}, {Eckner}, {Egberts}, {Einecke}, {Ekoume},
  {Els{\"a}sser}, {Ernenwein}, {Espinoza}, {Evoli}, {Fairbairn},
  {Falceta-Goncalves}, {Falcone}, {Farnier}, {Fasola}, {Fedorova}, {Fegan},
  {Fernand ez-Alonso}, {Fern{\'a}ndez-Barral}, {Ferrand}, {Fesquet},
  {Filipovic}, {Fioretti}, {Fontaine}, {Fornasa}, {Fortson}, {Freixas
  Coromina}, {Fruck}, {Fujita}, {Fukazawa}, {Funk}, {F{\"u}{\ss}ling},
  {Gabici}, {Gadola}, {Gallant}, {Garcia}, {Garcia L{\'o}pez}, {Garczarczyk},
  {Gaskins}, {Gasparetto}, {Gaug}, {Gerard}, {Giavitto}, {Giglietto}, {Giommi},
  {Giordano}, {Giro}, {Giroletti}, {Giuliani}, {Glicenstein}, {Gnatyk},
  {Godinovic}, {Goldoni}, {G{\'o}mez-Vargas}, {Gonz{\'a}lez}, {Gonz{\'a}lez},
  {G{\"o}tz}, {Graham}, {Grand i}, {Granot}, {Green}, {Greenshaw}, {Griffiths},
  {Gunji}, {Hadasch}, {Hara}, {Hardcastle}, {Hassan}, {Hayashi}, {Hayashida},
  {Heller}, {Helo}, {Hermann}, {Hinton}, {Hnatyk}, {Hofmann}, {Holder},
  {Horan}, {H{\"o}randel}, {Horns}, {Horvath}, {Hovatta}, {Hrabovsky},
  {Hrupec}, {Humensky}, {H{\"u}tten}, {Iarlori}, {Inada}, {Inome}, {Inoue},
  {Inoue}, {Inoue}, {Iocco}, {Ioka}, {Iori}, {Ishio}, {Iwamura}, {Jamrozy},
  {Janecek}, {Jankowsky}, {Jean}, {Jung-Richardt}, {Jurysek}, {Kaaret},
  {Karkar}, {Katagiri}, {Katz}, {Kawanaka}, {Kazanas}, {Kh{\'e}lifi}, {Kieda},
  {Kimeswenger}, {Kimura}, {Kisaka}, {Knapp}, {Kn{\"o}dlseder}, {Koch},
  {Kohri}, {Komin}, {Kosack}, {Kraus}, {Krause}, {Krau{\ss}}, {Kubo}, {Kukec
  Mezek}, {Kuroda}, {Kushida}, {La Palombara}, {Lamanna}, {Lang}, {Lapington},
  {Le Blanc}, {Leach}, {Lees}, {Lefaucheur}, {Leigui de Oliveira}, {Lenain},
  {Lico}, {Limon}, {Lindfors}, {Lohse}, {Lombardi}, {Longo}, {L{\'o}pez},
  {L{\'o}pez-Coto}, {Lu}, {Lucarelli}, {Luque-Escamilla}, {Lyard}, {Maccarone},
  {Maier}, {Majumdar}, {Malaguti}, {Mandat}, {Maneva}, {Manganaro}, {Mangano},
  {Marcowith}, {Mar{\'\i}n}, {Markoff}, {Mart{\'\i}}, {Martin},
  {Mart{\'\i}nez}, {Mart{\'\i}nez}, {Masetti}, {Masuda}, {Maurin}, {Maxted},
  {Mazin}, {Medina}, {Melandri}, {Mereghetti}, {Meyer}, {Minaya}, {Mirabal},
  {Mirzoyan}, {Mitchell}, {Mizuno}, {Moderski}, {Mohammed}, {Mohrmann},
  {Montaruli}, {Moralejo}, {Morcuende-Parrilla}, {Mori}, {Morlino}, {Morris},
  {Morselli}, {Moulin}, {Mukherjee}, {Mundell}, {Murach}, {Muraishi}, {Murase},
  {Nagai}, {Nagataki}, {Nagayoshi}, {Naito}, {Nakamori}, {Nakamura}, {Niemiec},
  {Nieto}, {Niko{\l}ajuk}, {Nishijima}, {Noda}, {Nosek}, {Novosyadlyj},
  {Nozaki}, {O'Brien}, {Oakes}, {Ohira}, {Ohishi}, {Ohm}, {Okazaki}, {Okumura},
  {Ong}, {Orienti}, {Orito}, {Osborne}, {Ostrowski}, {Otte}, {Oya}, {Padovani},
  {Paizis}, {Palatiello}, {Palatka}, {Paoletti}, {Paredes}, {Pareschi},
  {Parsons}, {Pe'er}, {Pech}, {Pedaletti}, {Perri}, {Persic}, {Petrashyk},
  {Petrucci}, {Petruk}, {Peyaud}, {Pfeifer}, {Piano}, {Pisarski}, {Pita},
  {Pohl}, {Polo}, {Pozo}, {Prandini}, {Prast}, {Principe}, {Prokhorov},
  {Prokoph}, {Prouza}, {P{\"u}hlhofer}, {Punch}, {P{\"u}rckhauer}, {Queiroz},
  {Quirrenbach}, {Rain{\`o}}, {Razzaque}, {Reimer}, {Reimer}, {Reisenegger},
  {Renaud}, {Rezaeian}, {Rhode}, {Ribeiro}, {Rib{\'o}}, {Richtler}, {Rico},
  {Rieger}, {Riquelme}, {Rivoire}, {Rizi}, {Rodriguez}, {Rodriguez Fernandez},
  {Rodr{\'\i}guez V{\'a}zquez}, {Rojas}, {Romano}, {Romeo}, {Rosado}, {Rovero},
  {Rowell}, {Rudak}, {Rugliancich}, {Rulten}, {Sadeh}, {Safi-Harb}, {Saito},
  {Sakaki}, {Sakurai}, {Salina}, {S{\'a}nchez-Conde}, {Sandaker}, {Sandoval},
  {Sangiorgi}, {Sanguillon}, {Sano}, {Santand er}, {Sarkar}, {Satalecka},
  {Saturni}, {Schioppa}, {Schlenstedt}, {Schneider}, {Schoorlemmer},
  {Schovanek}, {Schulz}, {Schussler}, {Schwanke}, {Sciacca}, {Scuderi},
  {Seitenzahl}, {Semikoz}, {Sergijenko}, {Servillat}, {Shalchi}, {Shellard},
  {Sidoli}, {Siejkowski}, {Sillanp{\"a}{\"a}}, {Sironi}, {Sitarek}, {Sliusar},
  {Slowikowska}, {Sol}, {Stamerra}, {Stani{\v{c}}}, {Starling}, {Stawarz},
  {Stefanik}, {Stephan}, {Stolarczyk}, {Stratta}, {Straumann}, {Suomijarvi},
  {Supanitsky}, {Tagliaferri}, {Tajima}, {Tavani}, {Tavecchio}, {Tavernet},
  {Tayabaly}, {Tejedor}, {Temnikov}, {Terada}, {Terrier}, {Terzic}, {Teshima},
  {Testa}, {Thoudam}, {Tian}, {Tibaldo}, {Tluczykont}, {Todero Peixoto},
  {Tokanai}, {Tomastik}, {Tonev}, {Tornikoski}, {Torres}, {Torresi}, {Tosti},
  {Tothill}, {Tovmassian}, {Travnicek}, {Trichard}, {Trifoglio}, {Troyano
  Pujadas}, {Tsujimoto}, {Umana}, {Vagelli}, {Vagnetti}, {Valentino},
  {Vallania}, {Valore}, {van Eldik}, {Vand enbroucke}, {Varner}, {Vasileiadis},
  {Vassiliev}, {V{\'a}zquez Acosta}, {Vecchi}, {Vega}, {Vercellone}, {Veres},
  {Vergani}, {Verzi}, {Vettolani}, {Viana}, {Vigorito}, {Villanueva}, {Voelk},
  {Vollhardt}, {Vorobiov}, {Vrastil}, {Vuillaume}, {Wagner}, {Wagner},
  {Walter}, {Ward}, {Warren}, {Watson}, {Werner}, {White}, {White},
  {Wierzcholska}, {Wilcox}, {Will}, {Williams}, {Wischnewski}, {Wood},
  {Yamamoto}, {Yamazaki}, {Yanagita}, {Yang}, {Yoshida}, {Yoshiike},
  {Yoshikoshi}, {Zacharias}, {Zaharijas}, {Zampieri}, {Zand anel}, {Zanin},
  {Zavrtanik}, {Zavrtanik}, {Zdziarski}, {Zech}, {Zechlin}, {Zhdanov},
  {Ziegler} \& {Zorn}]{2019scta.book.....C}
{\sc \au{{Cherenkov Telescope Array Consortium}}, \au{{Acharya}, B.~S.},
  \au{{Agudo}, I.}, \au{{Al Samarai}, I.}, \au{{Alfaro}, R.}, \au{{Alfaro},
  J.}, \au{{Alispach}, C.}, \au{{Alves Batista}, R.}, \au{{Amans}, J.~P.},
  \au{{Amato}, E.}, \au{{Ambrosi}, G.}, \au{{Antolini}, E.}, \au{{Antonelli},
  L.~A.}, \au{{Aramo}, C.}, \au{{Araya}, M.}, \au{{Armstrong}, T.},
  \au{{Arqueros}, F.}, \au{{Arrabito}, L.}, \au{{Asano}, K.}, \au{{Ashley},
  M.}, \au{{Backes}, M.}, \au{{Balazs}, C.}, \au{{Balbo}, M.}, \au{{Ballester},
  O.}, \au{{Ballet}, J.}, \au{{Bamba}, A.}, \au{{Barkov}, M.}, \au{{Barres de
  Almeida}, U.}, \au{{Barrio}, J.~A.}, \au{{Bastieri}, D.}, \au{{Becherini},
  Y.}, \au{{Belfiore}, A.}, \au{{Benbow}, W.}, \au{{Berge}, D.},
  \au{{Bernardini}, E.}, \au{{Bernardini}, M.~G.}, \au{{Bernardos}, M.},
  \au{{Bernl{\"o}hr}, K.}, \au{{Bertucci}, B.}, \au{{Biasuzzi}, B.},
  \au{{Bigongiari}, C.}, \au{{Biland}, A.}, \au{{Bissaldi}, E.}, \au{{Biteau},
  J.}, \au{{Blanch}, O.}, \au{{Blazek}, J.}, \au{{Boisson}, C.}, \au{{Bolmont},
  J.}, \au{{Bonanno}, G.}, \au{{Bonardi}, A.}, \au{{Bonavolont{\`a}}, C.},
  \au{{Bonnoli}, G.}, \au{{Bosnjak}, Z.}, \au{{B{\"o}ttcher}, M.},
  \au{{Braiding}, C.}, \au{{Bregeon}, J.}, \au{{Brill}, A.}, \au{{Brown},
  A.~M.}, \au{{Brun}, P.}, \au{{Brunetti}, G.}, \au{{Buanes}, T.},
  \au{{Buckley}, J.}, \au{{Bugaev}, V.}, \au{{B{\"u}hler}, R.},
  \au{{Bulgarelli}, A.}, \au{{Bulik}, T.}, \au{{Burton}, M.}, \au{{Burtovoi},
  A.}, \au{{Busetto}, G.}, \au{{Canestrari}, R.}, \au{{Capalbi}, M.},
  \au{{Capitanio}, F.}, \au{{Caproni}, A.}, \au{{Caraveo}, P.},
  \au{{C{\'a}rdenas}, V.}, \au{{Carlile}, C.}, \au{{Carosi}, R.},
  \au{{Carqu{\'\i}n}, E.}, \au{{Carr}, J.}, \au{{Casanova}, S.}, \au{{Cascone},
  E.}, \au{{Catalani}, F.}, \au{{Catalano}, O.}, \au{{Cauz}, D.},
  \au{{Cerruti}, M.}, \au{{Chadwick}, P.}, \au{{Chaty}, S.}, \au{{Chaves},
  R.~C.~G.}, \au{{Chen}, A.}, \au{{Chen}, X.}, \au{{Chernyakova}, M.},
  \au{{Chikawa}, M.}, \au{{Christov}, A.}, \au{{Chudoba}, J.},
  \au{{Cie{\'s}lar}, M.}, \au{{Coco}, V.}, \au{{Colafrancesco}, S.},
  \au{{Colin}, P.}, \au{{Conforti}, V.}, \au{{Connaughton}, V.}, \au{{Conrad},
  J.}, \au{{Contreras}, J.~L.}, \au{{Cortina}, J.}, \au{{Costa}, A.},
  \au{{Costantini}, H.}, \au{{Cotter}, G.}, \au{{Covino}, S.}, \au{{Crocker},
  R.}, \au{{Cuadra}, J.}, \au{{Cuevas}, O.}, \au{{Cumani}, P.},
  \au{{D'A{\`\i}}, A.}, \au{{D'Ammando}, F.}, \au{{D'Avanzo}, P.},
  \au{{D'Urso}, D.}, \au{{Daniel}, M.}, \au{{Davids}, I.}, \au{{Dawson}, B.},
  \au{{Dazzi}, F.}, \au{{De Angelis}, A.}, \au{{de C{\'a}ssia dos Anjos}, R.},
  \au{{De Cesare}, G.}, \au{{De Franco}, A.}, \au{{de Gouveia Dal Pino},
  E.~M.}, \au{{de la Calle}, I.}, \au{{de los Reyes Lopez}, R.}, \au{{De
  Lotto}, B.}, \au{{De Luca}, A.}, \au{{De Lucia}, M.}, \au{{de Naurois}, M.},
  \au{{de O{\~n}a Wilhelmi}, E.}, \au{{De Palma}, F.}, \au{{De Persio}, F.},
  \au{{de Souza}, V.}, \au{{Deil}, C.}, \au{{Del Santo}, M.}, \au{{Delgado},
  C.}, \au{{della Volpe}, D.}, \au{{Di Girolamo}, T.}, \au{{Di Pierro}, F.},
  \au{{Di Venere}, L.}, \au{{D{\'\i}az}, C.}, \au{{Dib}, C.}, \au{{Diebold},
  S.}, \au{{Djannati-Ata{\"\i}}, A.}, \au{{Dom{\'\i}nguez}, A.}, \au{{Dominis
  Prester}, D.}, \au{{Dorner}, D.}, \au{{Doro}, M.}, \au{{Drass}, H.},
  \au{{Dravins}, D.}, \au{{Dubus}, G.}, \au{{Dwarkadas}, V.~V.}, \au{{Ebr},
  J.}, \au{{Eckner}, C.}, \au{{Egberts}, K.}, \au{{Einecke}, S.}, \au{{Ekoume},
  T.~R.~N.}, \au{{Els{\"a}sser}, D.}, \au{{Ernenwein}, J.~P.}, \au{{Espinoza},
  C.}, \au{{Evoli}, C.}, \au{{Fairbairn}, M.}, \au{{Falceta-Goncalves}, D.},
  \au{{Falcone}, A.}, \au{{Farnier}, C.}, \au{{Fasola}, G.}, \au{{Fedorova},
  E.}, \au{{Fegan}, S.}, \au{{Fernand ez-Alonso}, M.},
  \au{{Fern{\'a}ndez-Barral}, A.}, \au{{Ferrand}, G.}, \au{{Fesquet}, M.},
  \au{{Filipovic}, M.}, \au{{Fioretti}, V.}, \au{{Fontaine}, G.},
  \au{{Fornasa}, M.}, \au{{Fortson}, L.}, \au{{Freixas Coromina}, L.},
  \au{{Fruck}, C.}, \au{{Fujita}, Y.}, \au{{Fukazawa}, Y.}, \au{{Funk}, S.},
  \au{{F{\"u}{\ss}ling}, M.}, \au{{Gabici}, S.}, \au{{Gadola}, A.},
  \au{{Gallant}, Y.}, \au{{Garcia}, B.}, \au{{Garcia L{\'o}pez}, R.},
  \au{{Garczarczyk}, M.}, \au{{Gaskins}, J.}, \au{{Gasparetto}, T.},
  \au{{Gaug}, M.}, \au{{Gerard}, L.}, \au{{Giavitto}, G.}, \au{{Giglietto},
  N.}, \au{{Giommi}, P.}, \au{{Giordano}, F.}, \au{{Giro}, E.},
  \au{{Giroletti}, M.}, \au{{Giuliani}, A.}, \au{{Glicenstein}, J.~F.},
  \au{{Gnatyk}, R.}, \au{{Godinovic}, N.}, \au{{Goldoni}, P.},
  \au{{G{\'o}mez-Vargas}, G.}, \au{{Gonz{\'a}lez}, M.~M.}, \au{{Gonz{\'a}lez},
  J.~M.}, \au{{G{\"o}tz}, D.}, \au{{Graham}, J.}, \au{{Grand i}, P.},
  \au{{Granot}, J.}, \au{{Green}, A.~J.}, \au{{Greenshaw}, T.},
  \au{{Griffiths}, S.}, \au{{Gunji}, S.}, \au{{Hadasch}, D.}, \au{{Hara}, S.},
  \au{{Hardcastle}, M.~J.}, \au{{Hassan}, T.}, \au{{Hayashi}, K.},
  \au{{Hayashida}, M.}, \au{{Heller}, M.}, \au{{Helo}, J.~C.}, \au{{Hermann},
  G.}, \au{{Hinton}, J.}, \au{{Hnatyk}, B.}, \au{{Hofmann}, W.}, \au{{Holder},
  J.}, \au{{Horan}, D.}, \au{{H{\"o}randel}, J.}, \au{{Horns}, D.},
  \au{{Horvath}, P.}, \au{{Hovatta}, T.}, \au{{Hrabovsky}, M.}, \au{{Hrupec},
  D.}, \au{{Humensky}, T.~B.}, \au{{H{\"u}tten}, M.}, \au{{Iarlori}, M.},
  \au{{Inada}, T.}, \au{{Inome}, Y.}, \au{{Inoue}, S.}, \au{{Inoue}, T.},
  \au{{Inoue}, Y.}, \au{{Iocco}, F.}, \au{{Ioka}, K.}, \au{{Iori}, M.},
  \au{{Ishio}, K.}, \au{{Iwamura}, Y.}, \au{{Jamrozy}, M.}, \au{{Janecek}, P.},
  \au{{Jankowsky}, D.}, \au{{Jean}, P.}, \au{{Jung-Richardt}, I.},
  \au{{Jurysek}, J.}, \au{{Kaaret}, P.}, \au{{Karkar}, S.}, \au{{Katagiri},
  H.}, \au{{Katz}, U.}, \au{{Kawanaka}, N.}, \au{{Kazanas}, D.},
  \au{{Kh{\'e}lifi}, B.}, \au{{Kieda}, D.~B.}, \au{{Kimeswenger}, S.},
  \au{{Kimura}, S.}, \au{{Kisaka}, S.}, \au{{Knapp}, J.}, \au{{Kn{\"o}dlseder},
  J.}, \au{{Koch}, B.}, \au{{Kohri}, K.}, \au{{Komin}, N.}, \au{{Kosack}, K.},
  \au{{Kraus}, M.}, \au{{Krause}, M.}, \au{{Krau{\ss}}, F.}, \au{{Kubo}, H.},
  \au{{Kukec Mezek}, G.}, \au{{Kuroda}, H.}, \au{{Kushida}, J.}, \au{{La
  Palombara}, N.}, \au{{Lamanna}, G.}, \au{{Lang}, R.~G.}, \au{{Lapington},
  J.}, \au{{Le Blanc}, O.}, \au{{Leach}, S.}, \au{{Lees}, J.~P.},
  \au{{Lefaucheur}, J.}, \au{{Leigui de Oliveira}, M.~A.}, \au{{Lenain},
  J.~P.}, \au{{Lico}, R.}, \au{{Limon}, M.}, \au{{Lindfors}, E.}, \au{{Lohse},
  T.}, \au{{Lombardi}, S.}, \au{{Longo}, F.}, \au{{L{\'o}pez}, M.},
  \au{{L{\'o}pez-Coto}, R.}, \au{{Lu}, C.~C.}, \au{{Lucarelli}, F.},
  \au{{Luque-Escamilla}, P.~L.}, \au{{Lyard}, E.}, \au{{Maccarone}, M.~C.},
  \au{{Maier}, G.}, \au{{Majumdar}, P.}, \au{{Malaguti}, G.}, \au{{Mandat},
  D.}, \au{{Maneva}, G.}, \au{{Manganaro}, M.}, \au{{Mangano}, S.},
  \au{{Marcowith}, A.}, \au{{Mar{\'\i}n}, J.}, \au{{Markoff}, S.},
  \au{{Mart{\'\i}}, J.}, \au{{Martin}, P.}, \au{{Mart{\'\i}nez}, M.},
  \au{{Mart{\'\i}nez}, G.}, \au{{Masetti}, N.}, \au{{Masuda}, S.},
  \au{{Maurin}, G.}, \au{{Maxted}, N.}, \au{{Mazin}, D.}, \au{{Medina}, C.},
  \au{{Melandri}, A.}, \au{{Mereghetti}, S.}, \au{{Meyer}, M.}, \au{{Minaya},
  I.~A.}, \au{{Mirabal}, N.}, \au{{Mirzoyan}, R.}, \au{{Mitchell}, A.},
  \au{{Mizuno}, T.}, \au{{Moderski}, R.}, \au{{Mohammed}, M.}, \au{{Mohrmann},
  L.}, \au{{Montaruli}, T.}, \au{{Moralejo}, A.}, \au{{Morcuende-Parrilla},
  D.}, \au{{Mori}, K.}, \au{{Morlino}, G.}, \au{{Morris}, P.}, \au{{Morselli},
  A.}, \au{{Moulin}, E.}, \au{{Mukherjee}, R.}, \au{{Mundell}, C.},
  \au{{Murach}, T.}, \au{{Muraishi}, H.}, \au{{Murase}, K.}, \au{{Nagai}, A.},
  \au{{Nagataki}, S.}, \au{{Nagayoshi}, T.}, \au{{Naito}, T.}, \au{{Nakamori},
  T.}, \au{{Nakamura}, Y.}, \au{{Niemiec}, J.}, \au{{Nieto}, D.},
  \au{{Niko{\l}ajuk}, M.}, \au{{Nishijima}, K.}, \au{{Noda}, K.}, \au{{Nosek},
  D.}, \au{{Novosyadlyj}, B.}, \au{{Nozaki}, S.}, \au{{O'Brien}, P.},
  \au{{Oakes}, L.}, \au{{Ohira}, Y.}, \au{{Ohishi}, M.}, \au{{Ohm}, S.},
  \au{{Okazaki}, N.}, \au{{Okumura}, A.}, \au{{Ong}, R.~A.}, \au{{Orienti},
  M.}, \au{{Orito}, R.}, \au{{Osborne}, J.~P.}, \au{{Ostrowski}, M.},
  \au{{Otte}, N.}, \au{{Oya}, I.}, \au{{Padovani}, M.}, \au{{Paizis}, A.},
  \au{{Palatiello}, M.}, \au{{Palatka}, M.}, \au{{Paoletti}, R.},
  \au{{Paredes}, J.~M.}, \au{{Pareschi}, G.}, \au{{Parsons}, R.~D.},
  \au{{Pe'er}, A.}, \au{{Pech}, M.}, \au{{Pedaletti}, G.}, \au{{Perri}, M.},
  \au{{Persic}, M.}, \au{{Petrashyk}, A.}, \au{{Petrucci}, P.}, \au{{Petruk},
  O.}, \au{{Peyaud}, B.}, \au{{Pfeifer}, M.}, \au{{Piano}, G.}, \au{{Pisarski},
  A.}, \au{{Pita}, S.}, \au{{Pohl}, M.}, \au{{Polo}, M.}, \au{{Pozo}, D.},
  \au{{Prandini}, E.}, \au{{Prast}, J.}, \au{{Principe}, G.}, \au{{Prokhorov},
  D.}, \au{{Prokoph}, H.}, \au{{Prouza}, M.}, \au{{P{\"u}hlhofer}, G.},
  \au{{Punch}, M.}, \au{{P{\"u}rckhauer}, S.}, \au{{Queiroz}, F.},
  \au{{Quirrenbach}, A.}, \au{{Rain{\`o}}, S.}, \au{{Razzaque}, S.},
  \au{{Reimer}, O.}, \au{{Reimer}, A.}, \au{{Reisenegger}, A.}, \au{{Renaud},
  M.}, \au{{Rezaeian}, A.~H.}, \au{{Rhode}, W.}, \au{{Ribeiro}, D.},
  \au{{Rib{\'o}}, M.}, \au{{Richtler}, T.}, \au{{Rico}, J.}, \au{{Rieger}, F.},
  \au{{Riquelme}, M.}, \au{{Rivoire}, S.}, \au{{Rizi}, V.}, \au{{Rodriguez},
  J.}, \au{{Rodriguez Fernandez}, G.}, \au{{Rodr{\'\i}guez V{\'a}zquez},
  J.~J.}, \au{{Rojas}, G.}, \au{{Romano}, P.}, \au{{Romeo}, G.}, \au{{Rosado},
  J.}, \au{{Rovero}, A.~C.}, \au{{Rowell}, G.}, \au{{Rudak}, B.},
  \au{{Rugliancich}, A.}, \au{{Rulten}, C.}, \au{{Sadeh}, I.}, \au{{Safi-Harb},
  S.}, \au{{Saito}, T.}, \au{{Sakaki}, N.}, \au{{Sakurai}, S.}, \au{{Salina},
  G.}, \au{{S{\'a}nchez-Conde}, M.}, \au{{Sandaker}, H.}, \au{{Sandoval}, A.},
  \au{{Sangiorgi}, P.}, \au{{Sanguillon}, M.}, \au{{Sano}, H.}, \au{{Santand
  er}, M.}, \au{{Sarkar}, S.}, \au{{Satalecka}, K.}, \au{{Saturni}, F.~G.},
  \au{{Schioppa}, E.~J.}, \au{{Schlenstedt}, S.}, \au{{Schneider}, M.},
  \au{{Schoorlemmer}, H.}, \au{{Schovanek}, P.}, \au{{Schulz}, A.},
  \au{{Schussler}, F.}, \au{{Schwanke}, U.}, \au{{Sciacca}, E.}, \au{{Scuderi},
  S.}, \au{{Seitenzahl}, I.}, \au{{Semikoz}, D.}, \au{{Sergijenko}, O.},
  \au{{Servillat}, M.}, \au{{Shalchi}, A.}, \au{{Shellard}, R.~C.},
  \au{{Sidoli}, L.}, \au{{Siejkowski}, H.}, \au{{Sillanp{\"a}{\"a}}, A.},
  \au{{Sironi}, G.}, \au{{Sitarek}, J.}, \au{{Sliusar}, V.}, \au{{Slowikowska},
  A.}, \au{{Sol}, H.}, \au{{Stamerra}, A.}, \au{{Stani{\v{c}}}, S.},
  \au{{Starling}, R.}, \au{{Stawarz}, {\L}.}, \au{{Stefanik}, S.},
  \au{{Stephan}, M.}, \au{{Stolarczyk}, T.}, \au{{Stratta}, G.},
  \au{{Straumann}, U.}, \au{{Suomijarvi}, T.}, \au{{Supanitsky}, A.~D.},
  \au{{Tagliaferri}, G.}, \au{{Tajima}, H.}, \au{{Tavani}, M.},
  \au{{Tavecchio}, F.}, \au{{Tavernet}, J.~P.}, \au{{Tayabaly}, K.},
  \au{{Tejedor}, L.~A.}, \au{{Temnikov}, P.}, \au{{Terada}, Y.}, \au{{Terrier},
  R.}, \au{{Terzic}, T.}, \au{{Teshima}, M.}, \au{{Testa}, V.}, \au{{Thoudam},
  S.}, \au{{Tian}, W.}, \au{{Tibaldo}, L.}, \au{{Tluczykont}, M.}, \au{{Todero
  Peixoto}, C.~J.}, \au{{Tokanai}, F.}, \au{{Tomastik}, J.}, \au{{Tonev}, D.},
  \au{{Tornikoski}, M.}, \au{{Torres}, D.~F.}, \au{{Torresi}, E.}, \au{{Tosti},
  G.}, \au{{Tothill}, N.}, \au{{Tovmassian}, G.}, \au{{Travnicek}, P.},
  \au{{Trichard}, C.}, \au{{Trifoglio}, M.}, \au{{Troyano Pujadas}, I.},
  \au{{Tsujimoto}, S.}, \au{{Umana}, G.}, \au{{Vagelli}, V.}, \au{{Vagnetti},
  F.}, \au{{Valentino}, M.}, \au{{Vallania}, P.}, \au{{Valore}, L.}, \au{{van
  Eldik}, C.}, \au{{Vand enbroucke}, J.}, \au{{Varner}, G.~S.},
  \au{{Vasileiadis}, G.}, \au{{Vassiliev}, V.}, \au{{V{\'a}zquez Acosta}, M.},
  \au{{Vecchi}, M.}, \au{{Vega}, A.}, \au{{Vercellone}, S.}, \au{{Veres}, P.},
  \au{{Vergani}, S.}, \au{{Verzi}, V.}, \au{{Vettolani}, G.~P.}, \au{{Viana},
  A.}, \au{{Vigorito}, C.}, \au{{Villanueva}, J.}, \au{{Voelk}, H.},
  \au{{Vollhardt}, A.}, \au{{Vorobiov}, S.}, \au{{Vrastil}, M.},
  \au{{Vuillaume}, T.}, \au{{Wagner}, S.~J.}, \au{{Wagner}, R.}, \au{{Walter},
  R.}, \au{{Ward}, J.~E.}, \au{{Warren}, D.}, \au{{Watson}, J.~J.},
  \au{{Werner}, F.}, \au{{White}, M.}, \au{{White}, R.}, \au{{Wierzcholska},
  A.}, \au{{Wilcox}, P.}, \au{{Will}, M.}, \au{{Williams}, D.~A.},
  \au{{Wischnewski}, R.}, \au{{Wood}, M.}, \au{{Yamamoto}, T.}, \au{{Yamazaki},
  R.}, \au{{Yanagita}, S.}, \au{{Yang}, L.}, \au{{Yoshida}, T.},
  \au{{Yoshiike}, S.}, \au{{Yoshikoshi}, T.}, \au{{Zacharias}, M.},
  \au{{Zaharijas}, G.}, \au{{Zampieri}, L.}, \au{{Zand anel}, F.}, \au{{Zanin},
  R.}, \au{{Zavrtanik}, M.}, \au{{Zavrtanik}, D.}, \au{{Zdziarski}, A.~A.},
  \au{{Zech}, A.}, \au{{Zechlin}, H.}, \au{{Zhdanov}, V.~I.}, \au{{Ziegler},
  A.} \& \au{{Zorn}, J.}} \yr{2019} {\em {Science with the Cherenkov Telescope
  Array}\/}.

\bibitem[{Cowsik} {\em et~al.\/}(2014){Cowsik}, {Burch} \&
  {Madziwa-Nussinov}]{Cowsik+14}
{\sc \au{{Cowsik}, R.}, \au{{Burch}, B.} \& \au{{Madziwa-Nussinov}, T.}}
  \yr{2014}  \at{{The Origin of the Spectral Intensities of Cosmic-Ray
  Positrons}}.  \jt{APJ}  \bvol{786}~(2),  \pg{124},  \arxiv{arXiv: 1305.1242}.

\bibitem[{Cowsik} \& {Madziwa-Nussinov}(2016)]{Cowsik+16}
{\sc \au{{Cowsik}, R.} \& \au{{Madziwa-Nussinov}, T.}} \yr{2016}  \at{{Spectral
  Intensities of Antiprotons and the Nested Leaky-box Model for Cosmic Rays in
  the Galaxy}}.  \jt{APJ}  \bvol{827}~(2),  \pg{119},  \arxiv{arXiv:
  1505.00305}.

\bibitem[{Cristofari} {\em et~al.\/}(2020){Cristofari}, {Blasi} \&
  {Amato}]{Cristofari20}
{\sc \au{{Cristofari}, Pierre}, \au{{Blasi}, Pasquale} \& \au{{Amato}, Elena}}
  \yr{2020}  \at{{The low rate of Galactic pevatrons}}.  \jt{Astroparticle
  Physics}  \bvol{123},  \pg{102492},  \arxiv{arXiv: 2007.04294}.

\bibitem[{Cristofari} {\em et~al.\/}(2017){Cristofari}, {Gabici}, {Humensky},
  {Santand er}, {Terrier}, {Parizot} \& {Casanova}]{2017MNRAS.471..201C}
{\sc \au{{Cristofari}, P.}, \au{{Gabici}, S.}, \au{{Humensky}, T.~B.},
  \au{{Santand er}, M.}, \au{{Terrier}, R.}, \au{{Parizot}, E.} \&
  \au{{Casanova}, S.}} \yr{2017}  \at{{Supernova remnants in the
  very-high-energy gamma-ray domain: the role of the Cherenkov telescope
  array}}.  \jt{MNRAS}  \bvol{471}~(1),  \pg{201--209},  \arxiv{arXiv:
  1709.01102}.

\bibitem[{Crutcher}(2012)]{Crutcher12}
{\sc \au{{Crutcher}, Richard~M.}} \yr{2012}  \at{{Magnetic Fields in Molecular
  Clouds}}.  \jt{Annual Review of Astronomy and Astrophysics}  \bvol{50},
  \pg{29--63}.

\bibitem[{Cummings} {\em et~al.\/}(2016){Cummings}, {Stone}, {Heikkila}, {Lal},
  {Webber}, {J{\'o}hannesson}, {Moskalenko}, {Orlando} \&
  {Porter}]{Cummings+16}
{\sc \au{{Cummings}, A.~C.}, \au{{Stone}, E.~C.}, \au{{Heikkila}, B.~C.},
  \au{{Lal}, N.}, \au{{Webber}, W.~R.}, \au{{J{\'o}hannesson}, G.},
  \au{{Moskalenko}, I.~V.}, \au{{Orlando}, E.} \& \au{{Porter}, T.~A.}}
  \yr{2016}  \at{{Galactic Cosmic Rays in the Local Interstellar Medium:
  Voyager 1 Observations and Model Results}}.  \jt{ApJ}  \bvol{831}~(1),
  \pg{18}.

\bibitem[{Dame} {\em et~al.\/}(2001){Dame}, {Hartmann} \&
  {Thaddeus}]{2001ApJ...547..792D}
{\sc \au{{Dame}, T.~M.}, \au{{Hartmann}, Dap} \& \au{{Thaddeus}, P.}} \yr{2001}
   \at{{The Milky Way in Molecular Clouds: A New Complete CO Survey}}.
  \jt{APJ}  \bvol{547}~(2),  \pg{792--813},  \arxiv{arXiv: astro-ph/0009217}.

\bibitem[{D'Angelo} {\em et~al.\/}(2016){D'Angelo}, {Blasi} \&
  {Amato}]{DAngelo+16}
{\sc \au{{D'Angelo}, Marta}, \au{{Blasi}, Pasquale} \& \au{{Amato}, Elena}}
  \yr{2016}  \at{{Grammage of cosmic rays around Galactic supernova remnants}}.
   \jt{PRD}  \bvol{94}~(8),  \pg{083003},  \arxiv{arXiv: 1512.05000}.

\bibitem[{Diesing} \& {Caprioli}(2020)]{2020PhRvD.101j3030D}
{\sc \au{{Diesing}, Rebecca} \& \au{{Caprioli}, Damiano}} \yr{2020}
  \at{{Nonsecondary origin of cosmic ray positrons}}.  \jt{Physical Review D}
  \bvol{101}~(10),  \pg{103030},  \arxiv{arXiv: 2001.02240}.

\bibitem[{Donnert} {\em et~al.\/}(2018){Donnert}, {Vazza}, {Br{\"u}ggen} \&
  {ZuHone}]{Donnert18}
{\sc \au{{Donnert}, J.}, \au{{Vazza}, F.}, \au{{Br{\"u}ggen}, M.} \&
  \au{{ZuHone}, J.}} \yr{2018}  \at{{Magnetic Field Amplification in Galaxy
  Clusters and Its Simulation}}.  \jt{Space Science Reviews}  \bvol{214}~(8),
  \pg{122},  \arxiv{arXiv: 1810.09783}.

\bibitem[{Drury} {\em et~al.\/}(1994){Drury}, {Aharonian} \&
  {Voelk}]{1994A&A...287..959D}
{\sc \au{{Drury}, L.~O'C.}, \au{{Aharonian}, F.~A.} \& \au{{Voelk}, H.~J.}}
  \yr{1994}  \at{{The gamma-ray visibility of supernova remnants. A test of
  cosmic ray origin}}.  \jt{A \& A}  \bvol{287},  \pg{959--971},  \arxiv{arXiv:
  astro-ph/9305037}.

\bibitem[{Eichler}(2017)]{Eichler17}
{\sc \au{{Eichler}, David}} \yr{2017}  \at{{An Alternative Explanation of the
  Varying Boron-to-carbon Ratio in Galactic Cosmic Rays}}.  \jt{APJ}
  \bvol{842}~(1),  \pg{50},  \arxiv{arXiv: 1708.05013}.

\bibitem[{Ellison} {\em et~al.\/}(2010){Ellison}, {Patnaude}, {Slane} \&
  {Raymond}]{2010ApJ...712..287E}
{\sc \au{{Ellison}, D.~C.}, \au{{Patnaude}, D.~J.}, \au{{Slane}, P.} \&
  \au{{Raymond}, J.}} \yr{2010}  \at{{Efficient Cosmic Ray Acceleration,
  Hydrodynamics, and Self-Consistent Thermal X-Ray Emission Applied to
  Supernova Remnant RX J1713.7-3946}}.  \jt{ApJ}  \bvol{712},  \pg{287--293},
  \arxiv{arXiv: 1001.1932}.

\bibitem[{Evoli} {\em et~al.\/}(2019){Evoli}, {Aloisio} \& {Blasi}]{Evoli+19}
{\sc \au{{Evoli}, Carmelo}, \au{{Aloisio}, Roberto} \& \au{{Blasi}, Pasquale}}
  \yr{2019}  \at{{Galactic cosmic rays after the AMS-02 observations}}.
  \jt{PRD}  \bvol{99}~(10),  \pg{103023},  \arxiv{arXiv: 1904.10220}.

\bibitem[{Evoli} {\em et~al.\/}(2020{\natexlab{{\em a\/}}}){Evoli}, {Blasi},
  {Amato} \& {Aloisio}]{EvoliPosPRL}
{\sc \au{{Evoli}, Carmelo}, \au{{Blasi}, Pasquale}, \au{{Amato}, Elena} \&
  \au{{Aloisio}, Roberto}} \yr{2020{\natexlab{{\em a\/}}}}  \at{{Signature of
  Energy Losses on the Cosmic Ray Electron Spectrum}}.  \jt{Physical review
  Letters}  \bvol{125}~(5),  \pg{051101},  \arxiv{arXiv: 2007.01302}.

\bibitem[{Evoli} {\em et~al.\/}(2020{\natexlab{{\em b\/}}}){Evoli}, {Morlino},
  {Blasi} \& {Aloisio}]{EvoliBe}
{\sc \au{{Evoli}, Carmelo}, \au{{Morlino}, Giovanni}, \au{{Blasi}, Pasquale} \&
  \au{{Aloisio}, Roberto}} \yr{2020{\natexlab{{\em b\/}}}}  \at{{AMS-02
  beryllium data and its implication for cosmic ray transport}}.  \jt{Physical
  Review D}  \bvol{101}~(2),  \pg{023013},  \arxiv{arXiv: 1910.04113}.

\bibitem[{Ferrand} \& {Marcowith}(2010)]{2010A&A...510A.101F}
{\sc \au{{Ferrand}, G.} \& \au{{Marcowith}, A.}} \yr{2010}  \at{{On the shape
  of the spectrum of cosmic rays accelerated inside superbubbles}}.
  \jt{A$\&$A}  \bvol{510},  \pg{A101},  \arxiv{arXiv: 0911.4457}.

\bibitem[{Fukui} {\em et~al.\/}(2012){Fukui}, {Sano}, {Sato}, {Torii},
  {Horachi}, {Hayakawa}, {McClure-Griffiths}, {Rowell}, {Inoue}, {Inutsuka},
  {Kawamura}, {Yamamoto}, {Okuda}, {Mizuno}, {Onishi}, {Mizuno} \&
  {Ogawa}]{2012ApJ...746...82F}
{\sc \au{{Fukui}, Y.}, \au{{Sano}, H.}, \au{{Sato}, J.}, \au{{Torii}, K.},
  \au{{Horachi}, H.}, \au{{Hayakawa}, T.}, \au{{McClure-Griffiths}, N.~M.},
  \au{{Rowell}, G.}, \au{{Inoue}, T.}, \au{{Inutsuka}, S.}, \au{{Kawamura},
  A.}, \au{{Yamamoto}, H.}, \au{{Okuda}, T.}, \au{{Mizuno}, N.}, \au{{Onishi},
  T.}, \au{{Mizuno}, A.} \& \au{{Ogawa}, H.}} \yr{2012}  \at{{A Detailed Study
  of the Molecular and Atomic Gas toward the {$\gamma$}-Ray Supernova Remnant
  RX J1713.7-3946: Spatial TeV {$\gamma$}-Ray and Interstellar Medium Gas
  Correspondence}}.  \jt{ApJ}  \bvol{746},  \pg{82},  \arxiv{arXiv: 1107.0508}.

\bibitem[{Funk}(2015)]{FunkRev}
{\sc \au{{Funk}, Stefan}} \yr{2015}  \at{{Ground- and Space-Based Gamma-Ray
  Astronomy}}.  \jt{Annual Review of Nuclear and Particle Science}  \bvol{65},
  \pg{245--277},  \arxiv{arXiv: 1508.05190}.

\bibitem[{Gabici} \& {Aharonian}(2014)]{2014MNRAS.445L..70G}
{\sc \au{{Gabici}, S.} \& \au{{Aharonian}, F.~A.}} \yr{2014}  \at{{Hadronic
  gamma-rays from RX J1713.7-3946?}}  \jt{MNRAS}  \bvol{445},  \pg{L70--L73},
  \arxiv{arXiv: 1406.2322}.

\bibitem[{Gabici} {\em et~al.\/}(2007){Gabici}, {Aharonian} \&
  {Blasi}]{2007ApESS.309..365G}
{\sc \au{{Gabici}, S.}, \au{{Aharonian}, F.~A.} \& \au{{Blasi}, P.}} \yr{2007}
  \at{{Gamma rays from molecular clouds}}.  \jt{APSS}  \bvol{309},
  \pg{365--371},  \arxiv{arXiv: astro-ph/0610032}.

\bibitem[{Gabici} {\em et~al.\/}(2009){Gabici}, {Aharonian} \&
  {Casanova}]{2009MNRAS.396.1629G}
{\sc \au{{Gabici}, S.}, \au{{Aharonian}, F.~A.} \& \au{{Casanova}, S.}}
  \yr{2009}  \at{{Broad-band non-thermal emission from molecular clouds
  illuminated by cosmic rays from nearby supernova remnants}}.  \jt{MNRAS}
  \bvol{396}~(3),  \pg{1629--1639},  \arxiv{arXiv: 0901.4549}.

\bibitem[{Gaggero} {\em et~al.\/}(2017){Gaggero}, {Grasso}, {Marinelli},
  {Taoso} \& {Urbano}]{gaggero2017prl}
{\sc \au{{Gaggero}, D.}, \au{{Grasso}, D.}, \au{{Marinelli}, A.}, \au{{Taoso},
  M.} \& \au{{Urbano}, A.}} \yr{2017}  \at{{Diffuse Cosmic Rays Shining in the
  Galactic Center: A Novel Interpretation of H.E.S.S. and Fermi-LAT {$\gamma$}
  -Ray Data}}.  \jt{PRL}  \bvol{119}~(3),  \pg{031101},  \arxiv{arXiv:
  1702.01124}.

\bibitem[Gaggero {\em et~al.\/}(2013)Gaggero, Maccione, Di~Bernardo, Evoli \&
  Grasso]{PhysRevLett.111.021102}
{\sc \au{Gaggero, Daniele}, \au{Maccione, Luca}, \au{Di~Bernardo, Giuseppe},
  \au{Evoli, Carmelo} \& \au{Grasso, Dario}} \yr{2013}  \at{Three-dimensional
  model of cosmic-ray lepton propagation reproduces data from the alpha
  magnetic spectrometer on the international space station}.  \jt{Phys. Rev.
  Lett.}  \bvol{111},  \pg{021102}.

\bibitem[{Ghez} {\em et~al.\/}(2008){Ghez}, {Salim}, {Weinberg}, {Lu}, {Do},
  {Dunn}, {Matthews}, {Morris}, {Yelda}, {Becklin}, {Kremenek}, {Milosavljevic}
  \& {Naiman}]{2008ApJ...689.1044G}
{\sc \au{{Ghez}, A.~M.}, \au{{Salim}, S.}, \au{{Weinberg}, N.~N.}, \au{{Lu},
  J.~R.}, \au{{Do}, T.}, \au{{Dunn}, J.~K.}, \au{{Matthews}, K.}, \au{{Morris},
  M.~R.}, \au{{Yelda}, S.}, \au{{Becklin}, E.~E.}, \au{{Kremenek}, T.},
  \au{{Milosavljevic}, M.} \& \au{{Naiman}, J.}} \yr{2008}  \at{{Measuring
  Distance and Properties of the Milky Way's Central Supermassive Black Hole
  with Stellar Orbits}}.  \jt{ApJ}  \bvol{689}~(2),  \pg{1044--1062},
  \arxiv{arXiv: 0808.2870}.

\bibitem[{Giacalone} \& {Jokipii}(2007)]{GiacaloneJokipii07}
{\sc \au{{Giacalone}, J.} \& \au{{Jokipii}, J.~R.}} \yr{2007}  \at{{Magnetic
  Field Amplification by Shocks in Turbulent Fluids}}.  \jt{ApJL}
  \bvol{663}~(1),  \pg{L41--L44}.

\bibitem[Gillessen {\em et~al.\/}(2009)Gillessen, Eisenhauer, Trippe,
  Alexander, Genzel, Martins \& Ott]{Gillessen_2009}
{\sc \au{Gillessen, S.}, \au{Eisenhauer, F.}, \au{Trippe, S.}, \au{Alexander,
  T.}, \au{Genzel, R.}, \au{Martins, F.} \& \au{Ott, T.}} \yr{2009}
  \at{{MONITORING} {STELLAR} {ORBITS} {AROUND} {THE} {MASSIVE} {BLACK} {HOLE}
  {IN} {THE} {GALACTIC} {CENTER}}.  \jt{The Astrophysical Journal}
  \bvol{692}~(2),  \pg{1075--1109}.

\bibitem[{Ginzburg} \& {Syrovatskii}(1964)]{1964ocr..book.....G}
{\sc \au{{Ginzburg}, V.~L.} \& \au{{Syrovatskii}, S.~I.}} \yr{1964} {\em {The
  Origin of Cosmic Rays}\/}.

\bibitem[{Giuliani} {\em et~al.\/}(2011){Giuliani}, {Cardillo}, {Tavani},
  {Fukui}, {Yoshiike}, {Torii}, {Dubner}, {Castelletti}, {Barbiellini},
  {Bulgarelli}, {Caraveo}, {Costa}, {Cattaneo}, {Chen}, {Contessi}, {Del
  Monte}, {Donnarumma}, {Evangelista}, {Feroci}, {Gianotti}, {Lazzarotto},
  {Lucarelli}, {Longo}, {Marisaldi}, {Mereghetti}, {Pacciani}, {Pellizzoni},
  {Piano}, {Picozza}, {Pittori}, {Pucella}, {Rapisarda}, {Rappoldi},
  {Sabatini}, {Soffitta}, {Striani}, {Trifoglio}, {Trois}, {Vercellone},
  {Verrecchia}, {Vittorini}, {Colafrancesco}, {Giommi} \& {Bignami}]{AgileW44}
{\sc \au{{Giuliani}, A.}, \au{{Cardillo}, M.}, \au{{Tavani}, M.}, \au{{Fukui},
  Y.}, \au{{Yoshiike}, S.}, \au{{Torii}, K.}, \au{{Dubner}, G.},
  \au{{Castelletti}, G.}, \au{{Barbiellini}, G.}, \au{{Bulgarelli}, A.},
  \au{{Caraveo}, P.}, \au{{Costa}, E.}, \au{{Cattaneo}, P.~W.}, \au{{Chen},
  A.}, \au{{Contessi}, T.}, \au{{Del Monte}, E.}, \au{{Donnarumma}, I.},
  \au{{Evangelista}, Y.}, \au{{Feroci}, M.}, \au{{Gianotti}, F.},
  \au{{Lazzarotto}, F.}, \au{{Lucarelli}, F.}, \au{{Longo}, F.},
  \au{{Marisaldi}, M.}, \au{{Mereghetti}, S.}, \au{{Pacciani}, L.},
  \au{{Pellizzoni}, A.}, \au{{Piano}, G.}, \au{{Picozza}, P.}, \au{{Pittori},
  C.}, \au{{Pucella}, G.}, \au{{Rapisarda}, M.}, \au{{Rappoldi}, A.},
  \au{{Sabatini}, S.}, \au{{Soffitta}, P.}, \au{{Striani}, E.},
  \au{{Trifoglio}, M.}, \au{{Trois}, A.}, \au{{Vercellone}, S.},
  \au{{Verrecchia}, F.}, \au{{Vittorini}, V.}, \au{{Colafrancesco}, S.},
  \au{{Giommi}, P.} \& \au{{Bignami}, G.}} \yr{2011}  \at{{Neutral Pion
  Emission from Accelerated Protons in the Supernova Remnant W44}}.  \jt{ApJ
  Letters}  \bvol{742}~(2),  \pg{L30},  \arxiv{arXiv: 1111.4868}.

\bibitem[{Grie{\ss}meier} {\em et~al.\/}(2015){Grie{\ss}meier},
  {Tabataba-Vakili}, {Stadelmann}, {Grenfell} \& {Atri}]{Griessmeier15}
{\sc \au{{Grie{\ss}meier}, J.~M.}, \au{{Tabataba-Vakili}, F.},
  \au{{Stadelmann}, A.}, \au{{Grenfell}, J.~L.} \& \au{{Atri}, D.}} \yr{2015}
  \at{{Galactic cosmic rays on extrasolar Earth-like planets. I. Cosmic ray
  flux}}.  \jt{A\&A}  \bvol{581},  \pg{A44},  \arxiv{arXiv: 1509.00735}.

\bibitem[{H.~E.~S.~S. Collaboration} {\em et~al.\/}(2018){H.~E.~S.~S.
  Collaboration}, {Abdalla}, {Abramowski}, {Aharonian}, {Ait Benkhali},
  {Akhperjanian}, {Andersson}, {Ang{\"u}ner}, {Arrieta}, {Aubert}, {Backes},
  {Balzer}, {Barnard}, {Becherini}, {Becker Tjus}, {Berge}, {Bernhard},
  {Bernl{\"o}hr}, {Blackwell}, {B{\"o}ttcher}, {Boisson}, {Bolmont}, {Bordas},
  {Bregeon}, {Brun}, {Brun}, {Bryan}, {Bulik}, {Capasso}, {Carr}, {Casanova},
  {Cerruti}, {Chakraborty}, {Chalme-Calvet}, {Chaves}, {Chen}, {Chevalier},
  {Chr{\'e}tien}, {Colafrancesco}, {Cologna}, {Condon}, {Conrad}, {Cui},
  {Davids}, {Decock}, {Degrange}, {Deil}, {Devin}, {deWilt}, {Dirson},
  {Djannati-Ata{\"\i}}, {Domainko}, {Donath}, {Drury}, {Dubus}, {Dutson},
  {Dyks}, {Edwards}, {Egberts}, {Eger}, {Ernenwein}, {Eschbach}, {Farnier},
  {Fegan}, {Fernand es}, {Fiasson}, {Fontaine}, {F{\"o}rster}, {Fukuyama},
  {Funk}, {F{\"u}{\ss}ling}, {Gabici}, {Gajdus}, {Gallant}, {Garrigoux},
  {Giavitto}, {Giebels}, {Glicenstein}, {Gottschall}, {Goyal}, {Grondin},
  {Hadasch}, {Hahn}, {Haupt}, {Hawkes}, {Heinzelmann}, {Henri}, {Hermann},
  {Hervet}, {Hinton}, {Hofmann}, {Hoischen}, {Holler}, {Horns}, {Ivascenko},
  {Jacholkowska}, {Jamrozy}, {Janiak}, {Jankowsky}, {Jankowsky}, {Jingo},
  {Jogler}, {Jouvin}, {Jung-Richardt}, {Kastendieck}, {Katarzy{\'n}ski},
  {Katz}, {Kerszberg}, {Kh{\'e}lifi}, {Kieffer}, {King}, {Klepser}, {Klochkov},
  {Klu{\'z}niak}, {Kolitzus}, {Komin}, {Kosack}, {Krakau}, {Kraus}, {Krayzel},
  {Kr{\"u}ger}, {Laffon}, {Lamanna}, {Lau}, {Lees}, {Lefaucheur}, {Lefranc},
  {Lemi{\`e}re}, {Lemoine-Goumard}, {Lenain}, {Leser}, {Lohse}, {Lorentz},
  {Liu}, {L{\'o}pez-Coto}, {Lypova}, {Marandon}, {Marcowith}, {Mariaud},
  {Marx}, {Maurin}, {Maxted}, {Mayer}, {Meintjes}, {Meyer}, {Mitchell},
  {Moderski}, {Mohamed}, {Mohrmann}, {Mor{\r{a}}}, {Moulin}, {Murach}, {de
  Naurois}, {Niederwanger}, {Niemiec}, {Oakes}, {O'Brien}, {Odaka}, {{\"O}ttl},
  {Ohm}, {Ostrowski}, {Oya}, {Padovani}, {Panter}, {Parsons}, {Pekeur},
  {Pelletier}, {Perennes}, {Petrucci}, {Peyaud}, {Piel}, {Pita}, {Poon},
  {Prokhorov}, {Prokoph}, {P{\"u}hlhofer}, {Punch}, {Quirrenbach}, {Raab},
  {Reimer}, {Reimer}, {Renaud}, {de los Reyes}, {Rieger}, {Romoli},
  {Rosier-Lees}, {Rowell}, {Rudak}, {Rulten}, {Sahakian}, {Salek}, {Sanchez},
  {Santangelo}, {Sasaki}, {Schlickeiser}, {Sch{\"u}ssler}, {Schulz},
  {Schwanke}, {Schwemmer}, {Settimo}, {Seyffert}, {Shafi}, {Shilon}, {Simoni},
  {Sol}, {Spanier}, {Spengler}, {Spies}, {Stawarz}, {Steenkamp}, {Stegmann},
  {Stinzing}, {Stycz}, {Sushch}, {Takahashi}, {Tavernet}, {Tavernier},
  {Taylor}, {Terrier}, {Tibaldo}, {Tiziani}, {Tluczykont}, {Trichard}, {Tuffs},
  {Uchiyama}, {van der Walt}, {van Eldik}, {van Rensburg}, {van Soelen},
  {Vasileiadis}, {Veh}, {Venter}, {Viana}, {Vincent}, {Vink}, {Voisin},
  {V{\"o}lk}, {Volpe}, {Vuillaume}, {Wadiasingh}, {Wagner}, {Wagner}, {Wagner},
  {White}, {Wierzcholska}, {Willmann}, {W{\"o}rnlein}, {Wouters}, {Yang},
  {Zabalza}, {Zaborov}, {Zacharias}, {Zdziarski}, {Zech}, {Zefi}, {Ziegler} \&
  {{\.Z}ywucka}]{2018A&A...612A...6H}
{\sc \au{{H.~E.~S.~S. Collaboration}}, \au{{Abdalla}, H.}, \au{{Abramowski},
  A.}, \au{{Aharonian}, F.}, \au{{Ait Benkhali}, F.}, \au{{Akhperjanian},
  A.~G.}, \au{{Andersson}, T.}, \au{{Ang{\"u}ner}, E.~O.}, \au{{Arrieta}, M.},
  \au{{Aubert}, P.}, \au{{Backes}, M.}, \au{{Balzer}, A.}, \au{{Barnard}, M.},
  \au{{Becherini}, Y.}, \au{{Becker Tjus}, J.}, \au{{Berge}, D.},
  \au{{Bernhard}, S.}, \au{{Bernl{\"o}hr}, K.}, \au{{Blackwell}, R.},
  \au{{B{\"o}ttcher}, M.}, \au{{Boisson}, C.}, \au{{Bolmont}, J.},
  \au{{Bordas}, P.}, \au{{Bregeon}, J.}, \au{{Brun}, F.}, \au{{Brun}, P.},
  \au{{Bryan}, M.}, \au{{Bulik}, T.}, \au{{Capasso}, M.}, \au{{Carr}, J.},
  \au{{Casanova}, S.}, \au{{Cerruti}, M.}, \au{{Chakraborty}, N.},
  \au{{Chalme-Calvet}, R.}, \au{{Chaves}, R.~C.~G.}, \au{{Chen}, A.},
  \au{{Chevalier}, J.}, \au{{Chr{\'e}tien}, M.}, \au{{Colafrancesco}, S.},
  \au{{Cologna}, G.}, \au{{Condon}, B.}, \au{{Conrad}, J.}, \au{{Cui}, Y.},
  \au{{Davids}, I.~D.}, \au{{Decock}, J.}, \au{{Degrange}, B.}, \au{{Deil},
  C.}, \au{{Devin}, J.}, \au{{deWilt}, P.}, \au{{Dirson}, L.},
  \au{{Djannati-Ata{\"\i}}, A.}, \au{{Domainko}, W.}, \au{{Donath}, A.},
  \au{{Drury}, L.~O.~'C.}, \au{{Dubus}, G.}, \au{{Dutson}, K.}, \au{{Dyks},
  J.}, \au{{Edwards}, T.}, \au{{Egberts}, K.}, \au{{Eger}, P.},
  \au{{Ernenwein}, J.~P.}, \au{{Eschbach}, S.}, \au{{Farnier}, C.},
  \au{{Fegan}, S.}, \au{{Fernand es}, M.~V.}, \au{{Fiasson}, A.},
  \au{{Fontaine}, G.}, \au{{F{\"o}rster}, A.}, \au{{Fukuyama}, T.}, \au{{Funk},
  S.}, \au{{F{\"u}{\ss}ling}, M.}, \au{{Gabici}, S.}, \au{{Gajdus}, M.},
  \au{{Gallant}, Y.~A.}, \au{{Garrigoux}, T.}, \au{{Giavitto}, G.},
  \au{{Giebels}, B.}, \au{{Glicenstein}, J.~F.}, \au{{Gottschall}, D.},
  \au{{Goyal}, A.}, \au{{Grondin}, M.~H.}, \au{{Hadasch}, D.}, \au{{Hahn}, J.},
  \au{{Haupt}, M.}, \au{{Hawkes}, J.}, \au{{Heinzelmann}, G.}, \au{{Henri},
  G.}, \au{{Hermann}, G.}, \au{{Hervet}, O.}, \au{{Hinton}, J.~A.},
  \au{{Hofmann}, W.}, \au{{Hoischen}, C.}, \au{{Holler}, M.}, \au{{Horns}, D.},
  \au{{Ivascenko}, A.}, \au{{Jacholkowska}, A.}, \au{{Jamrozy}, M.},
  \au{{Janiak}, M.}, \au{{Jankowsky}, D.}, \au{{Jankowsky}, F.}, \au{{Jingo},
  M.}, \au{{Jogler}, T.}, \au{{Jouvin}, L.}, \au{{Jung-Richardt}, I.},
  \au{{Kastendieck}, M.~A.}, \au{{Katarzy{\'n}ski}, K.}, \au{{Katz}, U.},
  \au{{Kerszberg}, D.}, \au{{Kh{\'e}lifi}, B.}, \au{{Kieffer}, M.}, \au{{King},
  J.}, \au{{Klepser}, S.}, \au{{Klochkov}, D.}, \au{{Klu{\'z}niak}, W.},
  \au{{Kolitzus}, D.}, \au{{Komin}, Nu.}, \au{{Kosack}, K.}, \au{{Krakau}, S.},
  \au{{Kraus}, M.}, \au{{Krayzel}, F.}, \au{{Kr{\"u}ger}, P.~P.}, \au{{Laffon},
  H.}, \au{{Lamanna}, G.}, \au{{Lau}, J.}, \au{{Lees}, J.~P.},
  \au{{Lefaucheur}, J.}, \au{{Lefranc}, V.}, \au{{Lemi{\`e}re}, A.},
  \au{{Lemoine-Goumard}, M.}, \au{{Lenain}, J.~P.}, \au{{Leser}, E.},
  \au{{Lohse}, T.}, \au{{Lorentz}, M.}, \au{{Liu}, R.}, \au{{L{\'o}pez-Coto},
  R.}, \au{{Lypova}, I.}, \au{{Marandon}, V.}, \au{{Marcowith}, A.},
  \au{{Mariaud}, C.}, \au{{Marx}, R.}, \au{{Maurin}, G.}, \au{{Maxted}, N.},
  \au{{Mayer}, M.}, \au{{Meintjes}, P.~J.}, \au{{Meyer}, M.}, \au{{Mitchell},
  A.~M.~W.}, \au{{Moderski}, R.}, \au{{Mohamed}, M.}, \au{{Mohrmann}, L.},
  \au{{Mor{\r{a}}}, K.}, \au{{Moulin}, E.}, \au{{Murach}, T.}, \au{{de
  Naurois}, M.}, \au{{Niederwanger}, F.}, \au{{Niemiec}, J.}, \au{{Oakes}, L.},
  \au{{O'Brien}, P.}, \au{{Odaka}, H.}, \au{{{\"O}ttl}, S.}, \au{{Ohm}, S.},
  \au{{Ostrowski}, M.}, \au{{Oya}, I.}, \au{{Padovani}, M.}, \au{{Panter}, M.},
  \au{{Parsons}, R.~D.}, \au{{Pekeur}, N.~W.}, \au{{Pelletier}, G.},
  \au{{Perennes}, C.}, \au{{Petrucci}, P.~O.}, \au{{Peyaud}, B.}, \au{{Piel},
  Q.}, \au{{Pita}, S.}, \au{{Poon}, H.}, \au{{Prokhorov}, D.}, \au{{Prokoph},
  H.}, \au{{P{\"u}hlhofer}, G.}, \au{{Punch}, M.}, \au{{Quirrenbach}, A.},
  \au{{Raab}, S.}, \au{{Reimer}, A.}, \au{{Reimer}, O.}, \au{{Renaud}, M.},
  \au{{de los Reyes}, R.}, \au{{Rieger}, F.}, \au{{Romoli}, C.},
  \au{{Rosier-Lees}, S.}, \au{{Rowell}, G.}, \au{{Rudak}, B.}, \au{{Rulten},
  C.~B.}, \au{{Sahakian}, V.}, \au{{Salek}, D.}, \au{{Sanchez}, D.~A.},
  \au{{Santangelo}, A.}, \au{{Sasaki}, M.}, \au{{Schlickeiser}, R.},
  \au{{Sch{\"u}ssler}, F.}, \au{{Schulz}, A.}, \au{{Schwanke}, U.},
  \au{{Schwemmer}, S.}, \au{{Settimo}, M.}, \au{{Seyffert}, A.~S.},
  \au{{Shafi}, N.}, \au{{Shilon}, I.}, \au{{Simoni}, R.}, \au{{Sol}, H.},
  \au{{Spanier}, F.}, \au{{Spengler}, G.}, \au{{Spies}, F.}, \au{{Stawarz},
  {\L}.}, \au{{Steenkamp}, R.}, \au{{Stegmann}, C.}, \au{{Stinzing}, F.},
  \au{{Stycz}, K.}, \au{{Sushch}, I.}, \au{{Takahashi}, T.}, \au{{Tavernet},
  J.~P.}, \au{{Tavernier}, T.}, \au{{Taylor}, A.~M.}, \au{{Terrier}, R.},
  \au{{Tibaldo}, L.}, \au{{Tiziani}, D.}, \au{{Tluczykont}, M.},
  \au{{Trichard}, C.}, \au{{Tuffs}, R.}, \au{{Uchiyama}, Y.}, \au{{van der
  Walt}, D.~J.}, \au{{van Eldik}, C.}, \au{{van Rensburg}, C.}, \au{{van
  Soelen}, B.}, \au{{Vasileiadis}, G.}, \au{{Veh}, J.}, \au{{Venter}, C.},
  \au{{Viana}, A.}, \au{{Vincent}, P.}, \au{{Vink}, J.}, \au{{Voisin}, F.},
  \au{{V{\"o}lk}, H.~J.}, \au{{Volpe}, F.}, \au{{Vuillaume}, T.},
  \au{{Wadiasingh}, Z.}, \au{{Wagner}, S.~J.}, \au{{Wagner}, P.}, \au{{Wagner},
  R.~M.}, \au{{White}, R.}, \au{{Wierzcholska}, A.}, \au{{Willmann}, P.},
  \au{{W{\"o}rnlein}, A.}, \au{{Wouters}, D.}, \au{{Yang}, R.}, \au{{Zabalza},
  V.}, \au{{Zaborov}, D.}, \au{{Zacharias}, M.}, \au{{Zdziarski}, A.~A.},
  \au{{Zech}, A.}, \au{{Zefi}, F.}, \au{{Ziegler}, A.} \& \au{{{\.Z}ywucka},
  N.}} \yr{2018}  \at{{H.E.S.S. observations of RX J1713.7-3946 with improved
  angular and spectral resolution: Evidence for gamma-ray emission extending
  beyond the X-ray emitting shell}}.  \jt{A\&A}  \bvol{612},  \pg{A6},
  \arxiv{arXiv: 1609.08671}.

\bibitem[{H.~E.~S.~S. Collaboration} {\em et~al.\/}(2015){H.~E.~S.~S.
  Collaboration}, {Abramowski}, {Aharonian}, {Ait Benkhali}, {Akhperjanian},
  {Ang{\"u}ner}, {Backes}, {Balenderan}, {Balzer}, {Barnacka}, {Becherini},
  {Becker-Tjus}, {Berge}, {Bernhard}, {Bernl{\"o}hr}, {Birsin}, {Biteau},
  {B{\"o}ttcher}, {Boisson}, {Bolmont}, {Bordas}, {Bregeon}, {Brun}, {Brun},
  {Bryan}, {Bulik}, {Carrigan}, {Casanova}, {Chadwick}, {Chakraborty},
  {Chalme-Calvet}, {Chaves}, {Chr{\'e}tien}, {Colafrancesco}, {Cologna},
  {Conrad}, {Couturier}, {Cui}, {Dalton}, {Davids}, {Degrange}, {Deil}, {de
  Wilt}, {Djannati-Ata{\"\i}}, {Domainko}, {Donath}, {Drury}, {Dubus},
  {Dutson}, {Dyks}, {Dyrda}, {Edwards}, {Egberts}, {Eger}, {Espigat},
  {Farnier}, {Fegan}, {Feinstein}, {Fernandes}, {Fernandez}, {Fiasson},
  {Fontaine}, {F{\"o}rster}, {F{\"u}{\ss}ling}, {Gabici}, {Gajdus}, {Gallant},
  {Garrigoux}, {Giavitto}, {Giebels}, {Glicenstein}, {Gottschall}, {Grondin},
  {Grudzi{\'n}ska}, {Hadasch}, {H{\"a}ffner}, {Hahn}, {Harris}, {Heinzelmann},
  {Henri}, {Hermann}, {Hervet}, {Hillert}, {Hinton}, {Hofmann}, {Hofverberg},
  {Holler}, {Horns}, {Ivascenko}, {Jacholkowska}, {Jahn}, {Jamrozy}, {Janiak},
  {Jankowsky}, {Jung}, {Kastendieck}, {Katarzy{\'n}ski}, {Katz}, {Kaufmann},
  {Kh{\'e}lifi}, {Kieffer}, {Klepser}, {Klochkov}, {Klu{\'z}niak}, {Kolitzus},
  {Komin}, {Kosack}, {Krakau}, {Krayzel}, {Kr{\"u}ger}, {Laffon}, {Lamanna},
  {Lefaucheur}, {Lefranc}, {Lemi{\`e}re}, {Lemoine-Goumard}, {Lenain}, {Lohse},
  {Lopatin}, {Lu}, {Marandon}, {Marcowith}, {Marx}, {Maurin}, {Maxted},
  {Mayer}, {McComb}, {M{\'e}hault}, {Meintjes}, {Menzler}, {Meyer}, {Mitchell},
  {Moderski}, {Mohamed}, {Mor{\r{a}}}, {Moulin}, {Murach}, {de Naurois},
  {Niemiec}, {Nolan}, {Oakes}, {Odaka}, {Ohm}, {Opitz}, {Ostrowski}, {Oya},
  {Panter}, {Parsons}, {Paz Arribas}, {Pekeur}, {Pelletier}, {Perez},
  {Petrucci}, {Peyaud}, {Pita}, {Poon}, {P{\"u}hlhofer}, {Punch},
  {Quirrenbach}, {Raab}, {Reichardt}, {Reimer}, {Reimer}, {Renaud}, {de los
  Reyes}, {Rieger}, {Rob}, {Romoli}, {Rosier-Lees}, {Rowell}, {Rudak},
  {Rulten}, {Sahakian}, {Salek}, {Sanchez}, {Santangelo}, {Schlickeiser},
  {Sch{\"u}ssler}, {Schulz}, {Schwanke}, {Schwarzburg}, {Schwemmer}, {Sol},
  {Spanier}, {Spengler}, {Spies}, {Stawarz}, {Steenkamp}, {Stegmann},
  {Stinzing}, {Stycz}, {Sushch}, {Tavernet}, {Tavernier}, {Taylor}, {Terrier},
  {Tluczykont}, {Trichard}, {Valerius}, {van Eldik}, {van Soelen},
  {Vasileiadis}, {Veh}, {Venter}, {Viana}, {Vincent}, {Vink}, {V{\"o}lk},
  {Volpe}, {Vorster}, {Vuillaume}, {Wagner}, {Wagner}, {Wagner}, {Ward},
  {Weidinger}, {Weitzel}, {White}, {Wierzcholska}, {Willmann}, {W{\"o}rnlein},
  {Wouters}, {Yang}, {Zabalza}, {Zaborov}, {Zacharias}, {Zdziarski}, {Zech} \&
  {Zechlin}]{2015Sci...347..406H}
{\sc \au{{H.~E.~S.~S. Collaboration}}, \au{{Abramowski}, A.}, \au{{Aharonian},
  F.}, \au{{Ait Benkhali}, F.}, \au{{Akhperjanian}, A.~G.}, \au{{Ang{\"u}ner},
  E.~O.}, \au{{Backes}, M.}, \au{{Balenderan}, S.}, \au{{Balzer}, A.},
  \au{{Barnacka}, A.}, \au{{Becherini}, Y.}, \au{{Becker-Tjus}, J.},
  \au{{Berge}, D.}, \au{{Bernhard}, S.}, \au{{Bernl{\"o}hr}, K.}, \au{{Birsin},
  E.}, \au{{Biteau}, J.}, \au{{B{\"o}ttcher}, M.}, \au{{Boisson}, C.},
  \au{{Bolmont}, J.}, \au{{Bordas}, P.}, \au{{Bregeon}, J.}, \au{{Brun}, F.},
  \au{{Brun}, P.}, \au{{Bryan}, M.}, \au{{Bulik}, T.}, \au{{Carrigan}, S.},
  \au{{Casanova}, S.}, \au{{Chadwick}, P.~M.}, \au{{Chakraborty}, N.},
  \au{{Chalme-Calvet}, R.}, \au{{Chaves}, R.~C.~G.}, \au{{Chr{\'e}tien}, M.},
  \au{{Colafrancesco}, S.}, \au{{Cologna}, G.}, \au{{Conrad}, J.},
  \au{{Couturier}, C.}, \au{{Cui}, Y.}, \au{{Dalton}, M.}, \au{{Davids},
  I.~D.}, \au{{Degrange}, B.}, \au{{Deil}, C.}, \au{{de Wilt}, P.},
  \au{{Djannati-Ata{\"\i}}, A.}, \au{{Domainko}, W.}, \au{{Donath}, A.},
  \au{{Drury}, L.~O'C.}, \au{{Dubus}, G.}, \au{{Dutson}, K.}, \au{{Dyks}, J.},
  \au{{Dyrda}, M.}, \au{{Edwards}, T.}, \au{{Egberts}, K.}, \au{{Eger}, P.},
  \au{{Espigat}, P.}, \au{{Farnier}, C.}, \au{{Fegan}, S.}, \au{{Feinstein},
  F.}, \au{{Fernandes}, M.~V.}, \au{{Fernandez}, D.}, \au{{Fiasson}, A.},
  \au{{Fontaine}, G.}, \au{{F{\"o}rster}, A.}, \au{{F{\"u}{\ss}ling}, M.},
  \au{{Gabici}, S.}, \au{{Gajdus}, M.}, \au{{Gallant}, Y.~A.}, \au{{Garrigoux},
  T.}, \au{{Giavitto}, G.}, \au{{Giebels}, B.}, \au{{Glicenstein}, J.~F.},
  \au{{Gottschall}, D.}, \au{{Grondin}, M.~H.}, \au{{Grudzi{\'n}ska}, M.},
  \au{{Hadasch}, D.}, \au{{H{\"a}ffner}, S.}, \au{{Hahn}, J.}, \au{{Harris},
  J.}, \au{{Heinzelmann}, G.}, \au{{Henri}, G.}, \au{{Hermann}, G.},
  \au{{Hervet}, O.}, \au{{Hillert}, A.}, \au{{Hinton}, J.~A.}, \au{{Hofmann},
  W.}, \au{{Hofverberg}, P.}, \au{{Holler}, M.}, \au{{Horns}, D.},
  \au{{Ivascenko}, A.}, \au{{Jacholkowska}, A.}, \au{{Jahn}, C.},
  \au{{Jamrozy}, M.}, \au{{Janiak}, M.}, \au{{Jankowsky}, F.}, \au{{Jung}, I.},
  \au{{Kastendieck}, M.~A.}, \au{{Katarzy{\'n}ski}, K.}, \au{{Katz}, U.},
  \au{{Kaufmann}, S.}, \au{{Kh{\'e}lifi}, B.}, \au{{Kieffer}, M.},
  \au{{Klepser}, S.}, \au{{Klochkov}, D.}, \au{{Klu{\'z}niak}, W.},
  \au{{Kolitzus}, D.}, \au{{Komin}, Nu.}, \au{{Kosack}, K.}, \au{{Krakau}, S.},
  \au{{Krayzel}, F.}, \au{{Kr{\"u}ger}, P.~P.}, \au{{Laffon}, H.},
  \au{{Lamanna}, G.}, \au{{Lefaucheur}, J.}, \au{{Lefranc}, V.},
  \au{{Lemi{\`e}re}, A.}, \au{{Lemoine-Goumard}, M.}, \au{{Lenain}, J.~P.},
  \au{{Lohse}, T.}, \au{{Lopatin}, A.}, \au{{Lu}, C.~C.}, \au{{Marandon}, V.},
  \au{{Marcowith}, A.}, \au{{Marx}, R.}, \au{{Maurin}, G.}, \au{{Maxted}, N.},
  \au{{Mayer}, M.}, \au{{McComb}, T.~J.~L.}, \au{{M{\'e}hault}, J.},
  \au{{Meintjes}, P.~J.}, \au{{Menzler}, U.}, \au{{Meyer}, M.}, \au{{Mitchell},
  A.~M.~W.}, \au{{Moderski}, R.}, \au{{Mohamed}, M.}, \au{{Mor{\r{a}}}, K.},
  \au{{Moulin}, E.}, \au{{Murach}, T.}, \au{{de Naurois}, M.}, \au{{Niemiec},
  J.}, \au{{Nolan}, S.~J.}, \au{{Oakes}, L.}, \au{{Odaka}, H.}, \au{{Ohm}, S.},
  \au{{Opitz}, B.}, \au{{Ostrowski}, M.}, \au{{Oya}, I.}, \au{{Panter}, M.},
  \au{{Parsons}, R.~D.}, \au{{Paz Arribas}, M.}, \au{{Pekeur}, N.~W.},
  \au{{Pelletier}, G.}, \au{{Perez}, J.}, \au{{Petrucci}, P.~O.}, \au{{Peyaud},
  B.}, \au{{Pita}, S.}, \au{{Poon}, H.}, \au{{P{\"u}hlhofer}, G.}, \au{{Punch},
  M.}, \au{{Quirrenbach}, A.}, \au{{Raab}, S.}, \au{{Reichardt}, I.},
  \au{{Reimer}, A.}, \au{{Reimer}, O.}, \au{{Renaud}, M.}, \au{{de los Reyes},
  R.}, \au{{Rieger}, F.}, \au{{Rob}, L.}, \au{{Romoli}, C.}, \au{{Rosier-Lees},
  S.}, \au{{Rowell}, G.}, \au{{Rudak}, B.}, \au{{Rulten}, C.~B.},
  \au{{Sahakian}, V.}, \au{{Salek}, D.}, \au{{Sanchez}, D.~A.},
  \au{{Santangelo}, A.}, \au{{Schlickeiser}, R.}, \au{{Sch{\"u}ssler}, F.},
  \au{{Schulz}, A.}, \au{{Schwanke}, U.}, \au{{Schwarzburg}, S.},
  \au{{Schwemmer}, S.}, \au{{Sol}, H.}, \au{{Spanier}, F.}, \au{{Spengler},
  G.}, \au{{Spies}, F.}, \au{{Stawarz}, {\L}.}, \au{{Steenkamp}, R.},
  \au{{Stegmann}, C.}, \au{{Stinzing}, F.}, \au{{Stycz}, K.}, \au{{Sushch},
  I.}, \au{{Tavernet}, J.~P.}, \au{{Tavernier}, T.}, \au{{Taylor}, A.~M.},
  \au{{Terrier}, R.}, \au{{Tluczykont}, M.}, \au{{Trichard}, C.},
  \au{{Valerius}, K.}, \au{{van Eldik}, C.}, \au{{van Soelen}, B.},
  \au{{Vasileiadis}, G.}, \au{{Veh}, J.}, \au{{Venter}, C.}, \au{{Viana}, A.},
  \au{{Vincent}, P.}, \au{{Vink}, J.}, \au{{V{\"o}lk}, H.~J.}, \au{{Volpe},
  F.}, \au{{Vorster}, M.}, \au{{Vuillaume}, T.}, \au{{Wagner}, S.~J.},
  \au{{Wagner}, P.}, \au{{Wagner}, R.~M.}, \au{{Ward}, M.}, \au{{Weidinger},
  M.}, \au{{Weitzel}, Q.}, \au{{White}, R.}, \au{{Wierzcholska}, A.},
  \au{{Willmann}, P.}, \au{{W{\"o}rnlein}, A.}, \au{{Wouters}, D.}, \au{{Yang},
  R.}, \au{{Zabalza}, V.}, \au{{Zaborov}, D.}, \au{{Zacharias}, M.},
  \au{{Zdziarski}, A.~A.}, \au{{Zech}, A.} \& \au{{Zechlin}, H.~S.}} \yr{2015}
  \at{{The exceptionally powerful TeV {\ensuremath{\gamma}}-ray emitters in the
  Large Magellanic Cloud}}.  \jt{Science}  \bvol{347}~(6220),  \pg{406--412},
  \arxiv{arXiv: 1501.06578}.

\bibitem[{HESS Collaboration} {\em et~al.\/}(2016){HESS Collaboration},
  {Abramowski}, {Aharonian}, {Benkhali}, {Akhperjanian}, {Ang{\"u}ner},
  {Backes}, {Balzer}, {Becherini}, {Tjus}, {Berge}, {Bernhard}, {Bernl{\"o}hr},
  {Birsin}, {Blackwell}, {B{\"o}ttcher}, {Boisson}, {Bolmont}, {Bordas},
  {Bregeon}, {Brun}, {Brun}, {Bryan}, {Bulik}, {Carr}, {Casanova},
  {Chakraborty}, {Chalme-Calvet}, {Chaves}, {Chen}, {Chr{\'e}tien},
  {Colafrancesco}, {Cologna}, {Conrad}, {Couturier}, {Cui}, {Davids},
  {Degrange}, {Deil}, {Dewilt}, {Djannati-Ata{\"\i}}, {Domainko}, {Donath},
  {Drury}, {Dubus}, {Dutson}, {Dyks}, {Dyrda}, {Edwards}, {Egberts}, {Eger},
  {Ernenwein}, {Espigat}, {Farnier}, {Fegan}, {Feinstein}, {Fernandes},
  {Fernand ez}, {Fiasson}, {Fontaine}, {F{\"o}rster}, {F{\"u}{\ss}ling},
  {Gabici}, {Gajdus}, {Gallant}, {Garrigoux}, {Giavitto}, {Giebels},
  {Glicenstein}, {Gottschall}, {Goyal}, {Grondin}, {Grudzi{\'n}ska}, {Hadasch},
  {H{\"a}ffner}, {Hahn}, {Hawkes}, {Heinzelmann}, {Henri}, {Hermann}, {Hervet},
  {Hillert}, {Hinton}, {Hofmann}, {Hofverberg}, {Hoischen}, {Holler}, {Horns},
  {Ivascenko}, {Jacholkowska}, {Jamrozy}, {Janiak}, {Jankowsky},
  {Jung-Richardt}, {Kastendieck}, {Katarzy{\'n}ski}, {Katz}, {Kerszberg},
  {Kh{\'e}lifi}, {Kieffer}, {Klepser}, {Klochkov}, {Klu{\'z}niak}, {Kolitzus},
  {Komin}, {Kosack}, {Krakau}, {Krayzel}, {Kr{\"u}ger}, {Laffon}, {Lamanna},
  {Lau}, {Lefaucheur}, {Lefranc}, {Lemi{\'e}re}, {Lemoine-Goumard}, {Lenain},
  {Lohse}, {Lopatin}, {Lu}, {Lui}, {Marand on}, {Marcowith}, {Mariaud}, {Marx},
  {Maurin}, {Maxted}, {Mayer}, {Meintjes}, {Menzler}, {Meyer}, {Mitchell},
  {Moderski}, {Mohamed}, {Mor{\r{a}}}, {Moulin}, {Murach}, {de Naurois},
  {Niemiec}, {Oakes}, {Odaka}, {{\"O}ttl}, {Ohm}, {Opitz}, {Ostrowski}, {Oya},
  {Panter}, {Parsons}, {Arribas}, {Pekeur}, {Pelletier}, {Petrucci}, {Peyaud},
  {Pita}, {Poon}, {Prokoph}, {P{\"u}hlhofer}, {Punch}, {Quirrenbach}, {Raab},
  {Reichardt}, {Reimer}, {Reimer}, {Renaud}, {de Los Reyes}, {Rieger},
  {Romoli}, {Rosier-Lees}, {Rowell}, {Rudak}, {Rulten}, {Sahakian}, {Salek},
  {Sanchez}, {Santangelo}, {Sasaki}, {Schlickeiser}, {Sch{\"u}ssler}, {Schulz},
  {Schwanke}, {Schwemmer}, {Seyffert}, {Simoni}, {Sol}, {Spanier}, {Spengler},
  {Spies}, {Stawarz}, {Steenkamp}, {Stegmann}, {Stinzing}, {Stycz}, {Sushch},
  {Tavernet}, {Tavernier}, {Taylor}, {Terrier}, {Tluczykont}, {Trichard},
  {Tuffs}, {Valerius}, {van der Walt}, {van Eldik}, {van Soelen},
  {Vasileiadis}, {Veh}, {Venter}, {Viana}, {Vincent}, {Vink}, {Voisin},
  {V{\"o}lk}, {Vuillaume}, {Wagner}, {Wagner}, {Wagner}, {Weidinger},
  {Weitzel}, {White}, {Wierzcholska}, {Willmann}, {W{\"o}rnlein}, {Wouters},
  {Yang}, {Zabalza}, {Zaborov}, {Zacharias}, {Zdziarski}, {Zech}, {Zefi} \&
  {{\.Z}ywucka}]{2016Natur.531..476H}
{\sc \au{{HESS Collaboration}}, \au{{Abramowski}, A.}, \au{{Aharonian}, F.},
  \au{{Benkhali}, F.~Ait}, \au{{Akhperjanian}, A.~G.}, \au{{Ang{\"u}ner},
  E.~O.}, \au{{Backes}, M.}, \au{{Balzer}, A.}, \au{{Becherini}, Y.},
  \au{{Tjus}, J.~Becker}, \au{{Berge}, D.}, \au{{Bernhard}, S.},
  \au{{Bernl{\"o}hr}, K.}, \au{{Birsin}, E.}, \au{{Blackwell}, R.},
  \au{{B{\"o}ttcher}, M.}, \au{{Boisson}, C.}, \au{{Bolmont}, J.},
  \au{{Bordas}, P.}, \au{{Bregeon}, J.}, \au{{Brun}, F.}, \au{{Brun}, P.},
  \au{{Bryan}, M.}, \au{{Bulik}, T.}, \au{{Carr}, J.}, \au{{Casanova}, S.},
  \au{{Chakraborty}, N.}, \au{{Chalme-Calvet}, R.}, \au{{Chaves}, R.~C.~G.},
  \au{{Chen}, A.}, \au{{Chr{\'e}tien}, M.}, \au{{Colafrancesco}, S.},
  \au{{Cologna}, G.}, \au{{Conrad}, J.}, \au{{Couturier}, C.}, \au{{Cui}, Y.},
  \au{{Davids}, I.~D.}, \au{{Degrange}, B.}, \au{{Deil}, C.}, \au{{Dewilt},
  P.}, \au{{Djannati-Ata{\"\i}}, A.}, \au{{Domainko}, W.}, \au{{Donath}, A.},
  \au{{Drury}, L.~O'C.}, \au{{Dubus}, G.}, \au{{Dutson}, K.}, \au{{Dyks}, J.},
  \au{{Dyrda}, M.}, \au{{Edwards}, T.}, \au{{Egberts}, K.}, \au{{Eger}, P.},
  \au{{Ernenwein}, J.~P.}, \au{{Espigat}, P.}, \au{{Farnier}, C.}, \au{{Fegan},
  S.}, \au{{Feinstein}, F.}, \au{{Fernandes}, M.~V.}, \au{{Fernand ez}, D.},
  \au{{Fiasson}, A.}, \au{{Fontaine}, G.}, \au{{F{\"o}rster}, A.},
  \au{{F{\"u}{\ss}ling}, M.}, \au{{Gabici}, S.}, \au{{Gajdus}, M.},
  \au{{Gallant}, Y.~A.}, \au{{Garrigoux}, T.}, \au{{Giavitto}, G.},
  \au{{Giebels}, B.}, \au{{Glicenstein}, J.~F.}, \au{{Gottschall}, D.},
  \au{{Goyal}, A.}, \au{{Grondin}, M.~H.}, \au{{Grudzi{\'n}ska}, M.},
  \au{{Hadasch}, D.}, \au{{H{\"a}ffner}, S.}, \au{{Hahn}, J.}, \au{{Hawkes},
  J.}, \au{{Heinzelmann}, G.}, \au{{Henri}, G.}, \au{{Hermann}, G.},
  \au{{Hervet}, O.}, \au{{Hillert}, A.}, \au{{Hinton}, J.~A.}, \au{{Hofmann},
  W.}, \au{{Hofverberg}, P.}, \au{{Hoischen}, C.}, \au{{Holler}, M.},
  \au{{Horns}, D.}, \au{{Ivascenko}, A.}, \au{{Jacholkowska}, A.},
  \au{{Jamrozy}, M.}, \au{{Janiak}, M.}, \au{{Jankowsky}, F.},
  \au{{Jung-Richardt}, I.}, \au{{Kastendieck}, M.~A.}, \au{{Katarzy{\'n}ski},
  K.}, \au{{Katz}, U.}, \au{{Kerszberg}, D.}, \au{{Kh{\'e}lifi}, B.},
  \au{{Kieffer}, M.}, \au{{Klepser}, S.}, \au{{Klochkov}, D.},
  \au{{Klu{\'z}niak}, W.}, \au{{Kolitzus}, D.}, \au{{Komin}, Nu.},
  \au{{Kosack}, K.}, \au{{Krakau}, S.}, \au{{Krayzel}, F.}, \au{{Kr{\"u}ger},
  P.~P.}, \au{{Laffon}, H.}, \au{{Lamanna}, G.}, \au{{Lau}, J.},
  \au{{Lefaucheur}, J.}, \au{{Lefranc}, V.}, \au{{Lemi{\'e}re}, A.},
  \au{{Lemoine-Goumard}, M.}, \au{{Lenain}, J.~P.}, \au{{Lohse}, T.},
  \au{{Lopatin}, A.}, \au{{Lu}, C.~C.}, \au{{Lui}, R.}, \au{{Marand on}, V.},
  \au{{Marcowith}, A.}, \au{{Mariaud}, C.}, \au{{Marx}, R.}, \au{{Maurin}, G.},
  \au{{Maxted}, N.}, \au{{Mayer}, M.}, \au{{Meintjes}, P.~J.}, \au{{Menzler},
  U.}, \au{{Meyer}, M.}, \au{{Mitchell}, A.~M.~W.}, \au{{Moderski}, R.},
  \au{{Mohamed}, M.}, \au{{Mor{\r{a}}}, K.}, \au{{Moulin}, E.}, \au{{Murach},
  T.}, \au{{de Naurois}, M.}, \au{{Niemiec}, J.}, \au{{Oakes}, L.},
  \au{{Odaka}, H.}, \au{{{\"O}ttl}, S.}, \au{{Ohm}, S.}, \au{{Opitz}, B.},
  \au{{Ostrowski}, M.}, \au{{Oya}, I.}, \au{{Panter}, M.}, \au{{Parsons},
  R.~D.}, \au{{Arribas}, M.~Paz}, \au{{Pekeur}, N.~W.}, \au{{Pelletier}, G.},
  \au{{Petrucci}, P.~O.}, \au{{Peyaud}, B.}, \au{{Pita}, S.}, \au{{Poon}, H.},
  \au{{Prokoph}, H.}, \au{{P{\"u}hlhofer}, G.}, \au{{Punch}, M.},
  \au{{Quirrenbach}, A.}, \au{{Raab}, S.}, \au{{Reichardt}, I.}, \au{{Reimer},
  A.}, \au{{Reimer}, O.}, \au{{Renaud}, M.}, \au{{de Los Reyes}, R.},
  \au{{Rieger}, F.}, \au{{Romoli}, C.}, \au{{Rosier-Lees}, S.}, \au{{Rowell},
  G.}, \au{{Rudak}, B.}, \au{{Rulten}, C.~B.}, \au{{Sahakian}, V.},
  \au{{Salek}, D.}, \au{{Sanchez}, D.~A.}, \au{{Santangelo}, A.}, \au{{Sasaki},
  M.}, \au{{Schlickeiser}, R.}, \au{{Sch{\"u}ssler}, F.}, \au{{Schulz}, A.},
  \au{{Schwanke}, U.}, \au{{Schwemmer}, S.}, \au{{Seyffert}, A.~S.},
  \au{{Simoni}, R.}, \au{{Sol}, H.}, \au{{Spanier}, F.}, \au{{Spengler}, G.},
  \au{{Spies}, F.}, \au{{Stawarz}, {\L}.}, \au{{Steenkamp}, R.},
  \au{{Stegmann}, C.}, \au{{Stinzing}, F.}, \au{{Stycz}, K.}, \au{{Sushch},
  I.}, \au{{Tavernet}, J.~P.}, \au{{Tavernier}, T.}, \au{{Taylor}, A.~M.},
  \au{{Terrier}, R.}, \au{{Tluczykont}, M.}, \au{{Trichard}, C.}, \au{{Tuffs},
  R.}, \au{{Valerius}, K.}, \au{{van der Walt}, J.}, \au{{van Eldik}, C.},
  \au{{van Soelen}, B.}, \au{{Vasileiadis}, G.}, \au{{Veh}, J.}, \au{{Venter},
  C.}, \au{{Viana}, A.}, \au{{Vincent}, P.}, \au{{Vink}, J.}, \au{{Voisin},
  F.}, \au{{V{\"o}lk}, H.~J.}, \au{{Vuillaume}, T.}, \au{{Wagner}, S.~J.},
  \au{{Wagner}, P.}, \au{{Wagner}, R.~M.}, \au{{Weidinger}, M.}, \au{{Weitzel},
  Q.}, \au{{White}, R.}, \au{{Wierzcholska}, A.}, \au{{Willmann}, P.},
  \au{{W{\"o}rnlein}, A.}, \au{{Wouters}, D.}, \au{{Yang}, R.}, \au{{Zabalza},
  V.}, \au{{Zaborov}, D.}, \au{{Zacharias}, M.}, \au{{Zdziarski}, A.~A.},
  \au{{Zech}, A.}, \au{{Zefi}, F.} \& \au{{{\.Z}ywucka}, N.}} \yr{2016}
  \at{{Acceleration of petaelectronvolt protons in the Galactic Centre}}.
  \jt{Nature}  \bvol{531}~(7595),  \pg{476--479},  \arxiv{arXiv: 1603.07730}.

\bibitem[{Heyer} \& {Dame}(2015)]{2015ARA&A..53..583H}
{\sc \au{{Heyer}, Mark} \& \au{{Dame}, T.~M.}} \yr{2015}  \at{{Molecular Clouds
  in the Milky Way}}.  \jt{Annual Review of Astronomy and Astrophysics}
  \bvol{53},  \pg{583--629}.

\bibitem[{Higdon} \& {Lingenfelter}(2005)]{2005ApJ...628..738H}
{\sc \au{{Higdon}, J.~C.} \& \au{{Lingenfelter}, R.~E.}} \yr{2005}  \at{{OB
  Associations, Supernova-generated Superbubbles, and the Source of Cosmic
  Rays}}.  \jt{ApJ}  \bvol{628}~(2),  \pg{738--749}.

\bibitem[{Holcomb} \& {Spitkovsky}(2019)]{Holcomb19}
{\sc \au{{Holcomb}, Cole} \& \au{{Spitkovsky}, Anatoly}} \yr{2019}  \at{{On the
  Growth and Saturation of the Gyroresonant Streaming Instabilities}}.
  \jt{ApJ}  \bvol{882}~(1),  \pg{3},  \arxiv{arXiv: 1811.01951}.

\bibitem[Hona {\em et~al.\/}(2019)Hona, Fleischhack \&
  Huentemeyer]{Hona:2019g7}
{\sc \au{Hona, Binita}, \au{Fleischhack, Henrike} \& \au{Huentemeyer, Petra}}
  \yr{2019}  \at{{Testing the Limits of Particle Acceleration in Cygnus OB2
  with HAWC}}.  \jt{PoS}  \bvol{ICRC2019},  \pg{699}.

\bibitem[{Ibarra} {\em et~al.\/}(2013){Ibarra}, {Tran} \&
  {Weniger}]{2013IJMPA..2830040I}
{\sc \au{{Ibarra}, Alejandro}, \au{{Tran}, David} \& \au{{Weniger}, Christoph}}
  \yr{2013}  \at{{Indirect Searches for Decaying Dark Matter}}.
  \jt{International Journal of Modern Physics A}  \bvol{28}~(27),
  \pg{1330040},  \arxiv{arXiv: 1307.6434}.

\bibitem[{Inoue} {\em et~al.\/}(2012){Inoue}, {Yamazaki}, {Inutsuka} \&
  {Fukui}]{2012ApJ...744...71I}
{\sc \au{{Inoue}, T.}, \au{{Yamazaki}, R.}, \au{{Inutsuka}, S.-i.} \&
  \au{{Fukui}, Y.}} \yr{2012}  \at{{Toward Understanding the Cosmic-Ray
  Acceleration at Young Supernova Remnants Interacting with Interstellar
  Clouds: Possible Applications to RX J1713.7-3946}}.  \jt{ApJ}  \bvol{744},
  \pg{71},  \arxiv{arXiv: 1106.3080}.

\bibitem[{Issa} \& {Wolfendale}(1981)]{1981Natur.292..430I}
{\sc \au{{Issa}, M.~R.} \& \au{{Wolfendale}, A.~W.}} \yr{1981}  \at{{Gamma rays
  from the cosmic ray irradiation of local molecular clouds}}.  \jt{Nature}
  \bvol{292},  \pg{430--433}.

\bibitem[Kafexhiu {\em et~al.\/}(2014)Kafexhiu, Aharonian, Taylor \&
  Vila]{PhysRevD.90.123014}
{\sc \au{Kafexhiu, Ervin}, \au{Aharonian, Felix}, \au{Taylor, Andrew~M.} \&
  \au{Vila, Gabriela~S.}} \yr{2014}  \at{Parametrization of gamma-ray
  production cross sections for $pp$ interactions in a broad proton energy
  range from the kinematic threshold to pev energies}.  \jt{Phys. Rev. D}
  \bvol{90},  \pg{123014}.

\bibitem[Kelner {\em et~al.\/}(2006)Kelner, Aharonian \&
  Bugayov]{PhysRevD.74.034018}
{\sc \au{Kelner, S.~R.}, \au{Aharonian, F.~A.} \& \au{Bugayov, V.~V.}}
  \yr{2006} Energy spectra of gamma rays, electrons, and neutrinos produced at
  proton-proton interactions in the very high energy regime.

\bibitem[{Klepach} {\em et~al.\/}(2000){Klepach}, {Ptuskin} \&
  {Zirakashvili}]{2000APh....13..161K}
{\sc \au{{Klepach}, E.~G.}, \au{{Ptuskin}, V.~S.} \& \au{{Zirakashvili},
  V.~N.}} \yr{2000}  \at{{Cosmic ray acceleration by multiple spherical
  shocks}}.  \jt{Astroparticle Physics}  \bvol{13},  \pg{161--172}.

\bibitem[{Konopelko} {\em et~al.\/}(2007)]{2007ApJ...658.1062K}
{\sc \au{{Konopelko}, A.} \& \au{others}} \yr{2007}  \at{{Observations of the
  Unidentified TeV {$\gamma$}-Ray Source TeV J2032+4130 with the Whipple
  Observatory 10 m Telescope}}.  \jt{ApJ}  \bvol{658},  \pg{1062--1068},
  \arxiv{arXiv: astro-ph/0611730}.

\bibitem[{Korsmeier} {\em et~al.\/}(2018){Korsmeier}, {Donato} \& {Di
  Mauro}]{Korsmeier+18}
{\sc \au{{Korsmeier}, Michael}, \au{{Donato}, Fiorenza} \& \au{{Di Mauro},
  Mattia}} \yr{2018}  \at{{Production cross sections of cosmic antiprotons in
  the light of new data from the NA61 and LHCb experiments}}.  \jt{PRD}
  \bvol{97}~(10),  \pg{103019},  \arxiv{arXiv: 1802.03030}.

\bibitem[{Lagage} \& {Cesarsky}(1983)]{LagageCesarsky83}
{\sc \au{{Lagage}, P.~O.} \& \au{{Cesarsky}, C.~J.}} \yr{1983}  \at{{The
  maximum energy of cosmic rays accelerated by supernova shocks.}}  \jt{A\&A}
  \bvol{125},  \pg{249--257}.

\bibitem[{Lazarian} {\em et~al.\/}(2020){Lazarian}, {Eyink}, {Jafari}, {Kowal},
  {Li}, {Xu} \& {Vishniac}]{Lazarian20}
{\sc \au{{Lazarian}, Alex}, \au{{Eyink}, Gregory~L.}, \au{{Jafari}, Amir},
  \au{{Kowal}, Grzegorz}, \au{{Li}, Hui}, \au{{Xu}, Siyao} \& \au{{Vishniac},
  Ethan~T.}} \yr{2020}  \at{{3D turbulent reconnection: Theory, tests, and
  astrophysical implications}}.  \jt{Physics of Plasmas}  \bvol{27}~(1),
  \pg{012305},  \arxiv{arXiv: 2001.00868}.

\bibitem[{Lipari}(2017)]{Lipari17}
{\sc \au{{Lipari}, Paolo}} \yr{2017}  \at{{Interpretation of the cosmic ray
  positron and antiproton fluxes}}.  \jt{PRD}  \bvol{95}~(6),  \pg{063009},
  \arxiv{arXiv: 1608.02018}.

\bibitem[{Malone} \& {HAWC Collaboration}(2017)]{2017APS..APR.X4008M}
{\sc \au{{Malone}, K.} \& \au{{HAWC Collaboration}}} \yr{2017} {The gamma-ray
  sky above 50 TeV with the HAWC Observatory}.  \bt{In {\em APS April Meeting
  Abstracts\/}},  \pg{p. X4.008}.

\bibitem[{Montmerle} \& {Cesarsky}(1979)]{1979ICRC....1..191M}
{\sc \au{{Montmerle}, T.} \& \au{{Cesarsky}, C.}} \yr{1979} {Gamma-Ray Emission
  from Snobs}.  \bt{In {\em International Cosmic Ray Conference\/}},
  \st{International Cosmic Ray Conference},  \vol{vol.~1},  \pg{p. 191}.

\bibitem[{Morlino} \& {Amato}(2020)]{MorlinoFe20}
{\sc \au{{Morlino}, Giovanni} \& \au{{Amato}, Elena}} \yr{2020}  \at{{Impact of
  transport modeling on the $^{60}$Fe abundance inside Galactic cosmic ray
  sources}}.  \jt{Phys. Rev. D}  \bvol{101}~(8),  \pg{083017},  \arxiv{arXiv:
  2003.04700}.

\bibitem[{Morlino} {\em et~al.\/}(2009){Morlino}, {Amato} \&
  {Blasi}]{MorlinoRXJ}
{\sc \au{{Morlino}, G.}, \au{{Amato}, E.} \& \au{{Blasi}, P.}} \yr{2009}
  \at{{Gamma-ray emission from SNR RX J1713.7-3946 and the origin of galactic
  cosmic rays}}.  \jt{MNRAS}  \bvol{392}~(1),  \pg{240--250},  \arxiv{arXiv:
  0810.0094}.

\bibitem[{Morrison}(1957)]{1957RvMP...29..235M}
{\sc \au{{Morrison}, P.}} \yr{1957}  \at{{On the Origins of Cosmic Rays}}.
  \jt{Reviews of Modern Physics}  \bvol{29},  \pg{235--243}.

\bibitem[{Moskalenko} \& {Strong}(1998)]{1998ApJ...493..694M}
{\sc \au{{Moskalenko}, I.~V.} \& \au{{Strong}, A.~W.}} \yr{1998}
  \at{{Production and Propagation of Cosmic-Ray Positrons and Electrons}}.
  \jt{ApJ}  \bvol{493}~(2),  \pg{694--707},  \arxiv{arXiv: astro-ph/9710124}.

\bibitem[Murray(2011)]{Murray_2011}
{\sc \au{Murray, Norman}} \yr{2011}  \at{{STAR} {FORMATION} {EFFICIENCIES}
  {AND} {LIFETIMES} {OF} {GIANT} {MOLECULAR} {CLOUDS} {IN} {THE} {MILKY}
  {WAY}}.  \jt{The Astrophysical Journal}  \bvol{729}~(2),  \pg{133}.

\bibitem[{Nava} {\em et~al.\/}(2016){Nava}, {Gabici}, {Marcowith}, {Morlino} \&
  {Ptuskin}]{Nava+16}
{\sc \au{{Nava}, L.}, \au{{Gabici}, S.}, \au{{Marcowith}, A.}, \au{{Morlino},
  G.} \& \au{{Ptuskin}, V.~S.}} \yr{2016}  \at{{Non-linear diffusion of cosmic
  rays escaping from supernova remnants - I. The effect of neutrals}}.
  \jt{MNRAS}  \bvol{461}~(4),  \pg{3552--3562},  \arxiv{arXiv: 1606.06902}.

\bibitem[{Ohira}(2016)]{Ohira16}
{\sc \au{{Ohira}, Yutaka}} \yr{2016}  \at{{Magnetic Field Amplification by
  Collisionless Shocks in Partially Ionized Plasmas}}.  \jt{ApJ}
  \bvol{817}~(2),  \pg{137},  \arxiv{arXiv: 1512.04167}.

\bibitem[{Oka} {\em et~al.\/}(1998){Oka}, {Hasegawa}, {Sato}, {Tsuboi} \&
  {Miyazaki}]{1998ApJS..118..455O}
{\sc \au{{Oka}, Tomoharu}, \au{{Hasegawa}, Tetsuo}, \au{{Sato}, Fumio},
  \au{{Tsuboi}, Masato} \& \au{{Miyazaki}, Atsushi}} \yr{1998}  \at{{A
  Large-Scale CO Survey of the Galactic Center}}.  \jt{ApJS}  \bvol{118}~(2),
  \pg{455--515}.

\bibitem[{Padovani} {\em et~al.\/}(2020){Padovani}, {Ivlev}, {Galli}, {Offner},
  {Indriolo}, {Rodgers-Lee}, {Marcowith}, {Girichidis}, {Bykov} \&
  {Kruijssen}]{Padovani20}
{\sc \au{{Padovani}, Marco}, \au{{Ivlev}, Alexei~V.}, \au{{Galli}, Daniele},
  \au{{Offner}, Stella S.~R.}, \au{{Indriolo}, Nick}, \au{{Rodgers-Lee},
  Donna}, \au{{Marcowith}, Alexandre}, \au{{Girichidis}, Philipp}, \au{{Bykov},
  Andrei~M.} \& \au{{Kruijssen}, J.~M.~Diederik}} \yr{2020}  \at{{Impact of
  Low-Energy Cosmic Rays on Star Formation}}.  \jt{Space Science Reviews}
  \bvol{216}~(2),  \pg{29},  \arxiv{arXiv: 2002.10282}.

\bibitem[{Parizot} {\em et~al.\/}(2004){Parizot}, {Marcowith}, {van der
  Swaluw}, {Bykov} \& {Tatischeff}]{2004A&A...424..747P}
{\sc \au{{Parizot}, E.}, \au{{Marcowith}, A.}, \au{{van der Swaluw}, E.},
  \au{{Bykov}, A.~M.} \& \au{{Tatischeff}, V.}} \yr{2004}  \at{{Superbubbles
  and energetic particles in the Galaxy. I. Collective effects of particle
  acceleration}}.  \jt{Astronomy and Astrophysics}  \bvol{424},  \pg{747--760},
   \arxiv{arXiv: astro-ph/0405531}.

\bibitem[{Particle Data Group} {\em et~al.\/}(2020){Particle Data Group},
  {Zyla}, {Barnett}, {Beringer}, {Dahl}, {Dwyer}, {Groom}, {Lin}, {Lugovsky},
  {Pianori}, {Robinson}, {Wohl}, {Yao}, {Agashe}, {Aielli}, {Allanach},
  {Amsler}, {Antonelli}, {Aschenauer}, {Asner}, {Baer}, {Banerjee}, {Baudis},
  {Bauer}, {Beatty}, {Belousov}, {Bethke}, {Bettini}, {Biebel}, {Black},
  {Blucher}, {Buchmuller}, {Burkert}, {Bychkov}, {Cahn}, {Carena}, {Ceccucci},
  {Cerri}, {Chakraborty}, {Chivukula}, {Cowan}, {D'Ambrosio}, {Damour}, {de
  Florian}, {de Gouv{\^e}a}, {DeGrand}, {de Jong}, {Dissertori}, {Dobrescu},
  {D'Onofrio}, {Doser}, {Drees}, {Dreiner}, {Eerola}, {Egede}, {Eidelman},
  {Ellis}, {Erler}, {Ezhela}, {Fetscher}, {Fields}, {Foster}, {Freitas},
  {Gallagher}, {Garren}, {Gerber}, {Gerbier}, {Gershon}, {Gershtein},
  {Gherghetta}, {Godizov}, {Gonzalez-Garcia}, {Goodman}, {Grab}, {Gritsan},
  {Grojean}, {Gr{\"u}newald}, {Gurtu}, {Gutsche}, {Haber}, {Hanhart},
  {Hashimoto}, {Hayato}, {Hebecker}, {Heinemeyer}, {Heltsley},
  {Hern{\'a}ndez-Rey}, {Hikasa}, {Hisano}, {H{\"o}cker}, {Holder}, {Holtkamp},
  {Huston}, {Hyodo}, {Johnson}, {Kado}, {Karliner}, {Katz}, {Kenzie}, {Khoze},
  {Klein}, {Klempt}, {Kowalewski}, {Krauss}, {Kreps}, {Krusche}, {Kwon},
  {Lahav}, {Laiho}, {Lellouch}, {Lesgourgues}, {Liddle}, {Ligeti}, {Lippmann},
  {Liss}, {Littenberg}, {Lourengo}, {Lugovsky}, {Lusiani}, {Makida}, {Maltoni},
  {Mannel}, {Manohar}, {Marciano}, {Masoni}, {Matthews}, {Mei{\ss}ner},
  {Mikhasenko}, {Miller}, {Milstead}, {Mitchell}, {M{\"o}nig}, {Molaro},
  {Moortgat}, {Moskovic}, {Nakamura}, {Narain}, {Nason}, {Navas}, {Neubert},
  {Nevski}, {Nir}, {Olive}, {Patrignani}, {Peacock}, {Petcov}, {Petrov},
  {Pich}, {Piepke}, {Pomarol}, {Profumo}, {Quadt}, {Rabbertz}, {Rademacker},
  {Raffelt}, {Ramani}, {Ramsey-Musolf}, {Ratcliff}, {Richardson}, {Ringwald},
  {Roesler}, {Rolli}, {Romaniouk}, {Rosenberg}, {Rosner}, {Rybka}, {Ryskin},
  {Ryutin}, {Sakai}, {Salam}, {Sarkar}, {Sauli}, {Schneider}, {Scholberg},
  {Schwartz}, {Schwiening}, {Scott}, {Sharma}, {Sharpe}, {Shutt}, {Silari},
  {Sj{\"o}strand}, {Skands}, {Skwarnicki}, {Smoot}, {Soffer}, {Sozzi},
  {Spanier}, {Spiering}, {Stahl}, {Stone}, {Sumino}, {Sumiyoshi}, {Syphers},
  {Takahashi}, {Tanabashi}, {Tanaka}, {Ta{\v{s}}evsk{\'y}}, {Terashi},
  {Terning}, {Thoma}, {Thorne}, {Tiator}, {Titov}, {Tkachenko}, {Tovey},
  {Trabelsi}, {Urquijo}, {Valencia}, {Van de Water}, {Varelas}, {Venanzoni},
  {Verde}, {Vincter}, {Vogel}, {Vogelsang}, {Vogt}, {Vorobyev}, {Wakely},
  {Walkowiak}, {Walter}, {Wands}, {Wascko}, {Weinberg}, {Weinberg}, {White},
  {Wiencke}, {Willocq}, {Woody}, {Workman}, {Yokoyama}, {Yoshida}, {Zand
  erighi}, {Zeller}, {Zenin}, {Zhu}, {Zhu}, {Zimmermann}, {Anderson},
  {Basaglia}, {Lugovsky}, {Schaffner} \& {Zheng}]{2020PTEP.2020h3C01P}
{\sc \au{{Particle Data Group}}, \au{{Zyla}, P.~A.}, \au{{Barnett}, R.~M.},
  \au{{Beringer}, J.}, \au{{Dahl}, O.}, \au{{Dwyer}, D.~A.}, \au{{Groom},
  D.~E.}, \au{{Lin}, C.~J.}, \au{{Lugovsky}, K.~S.}, \au{{Pianori}, E.},
  \au{{Robinson}, D.~J.}, \au{{Wohl}, C.~G.}, \au{{Yao}, W.~M.}, \au{{Agashe},
  K.}, \au{{Aielli}, G.}, \au{{Allanach}, B.~C.}, \au{{Amsler}, C.},
  \au{{Antonelli}, M.}, \au{{Aschenauer}, E.~C.}, \au{{Asner}, D.~M.},
  \au{{Baer}, H.}, \au{{Banerjee}, Sw}, \au{{Baudis}, L.}, \au{{Bauer}, C.~W.},
  \au{{Beatty}, J.~J.}, \au{{Belousov}, V.~I.}, \au{{Bethke}, S.},
  \au{{Bettini}, A.}, \au{{Biebel}, O.}, \au{{Black}, K.~M.}, \au{{Blucher},
  E.}, \au{{Buchmuller}, O.}, \au{{Burkert}, V.}, \au{{Bychkov}, M.~A.},
  \au{{Cahn}, R.~N.}, \au{{Carena}, M.}, \au{{Ceccucci}, A.}, \au{{Cerri}, A.},
  \au{{Chakraborty}, D.}, \au{{Chivukula}, R.~Sekhar}, \au{{Cowan}, G.},
  \au{{D'Ambrosio}, G.}, \au{{Damour}, T.}, \au{{de Florian}, D.}, \au{{de
  Gouv{\^e}a}, A.}, \au{{DeGrand}, T.}, \au{{de Jong}, P.}, \au{{Dissertori},
  G.}, \au{{Dobrescu}, B.~A.}, \au{{D'Onofrio}, M.}, \au{{Doser}, M.},
  \au{{Drees}, M.}, \au{{Dreiner}, H.~K.}, \au{{Eerola}, P.}, \au{{Egede}, U.},
  \au{{Eidelman}, S.}, \au{{Ellis}, J.}, \au{{Erler}, J.}, \au{{Ezhela},
  V.~V.}, \au{{Fetscher}, W.}, \au{{Fields}, B.~D.}, \au{{Foster}, B.},
  \au{{Freitas}, A.}, \au{{Gallagher}, H.}, \au{{Garren}, L.}, \au{{Gerber},
  H.~J.}, \au{{Gerbier}, G.}, \au{{Gershon}, T.}, \au{{Gershtein}, Y.},
  \au{{Gherghetta}, T.}, \au{{Godizov}, A.~A.}, \au{{Gonzalez-Garcia}, M.~C.},
  \au{{Goodman}, M.}, \au{{Grab}, C.}, \au{{Gritsan}, A.~V.}, \au{{Grojean},
  C.}, \au{{Gr{\"u}newald}, M.}, \au{{Gurtu}, A.}, \au{{Gutsche}, T.},
  \au{{Haber}, H.~E.}, \au{{Hanhart}, C.}, \au{{Hashimoto}, S.}, \au{{Hayato},
  Y.}, \au{{Hebecker}, A.}, \au{{Heinemeyer}, S.}, \au{{Heltsley}, B.},
  \au{{Hern{\'a}ndez-Rey}, J.~J.}, \au{{Hikasa}, K.}, \au{{Hisano}, J.},
  \au{{H{\"o}cker}, A.}, \au{{Holder}, J.}, \au{{Holtkamp}, A.}, \au{{Huston},
  J.}, \au{{Hyodo}, T.}, \au{{Johnson}, K.~F.}, \au{{Kado}, M.},
  \au{{Karliner}, M.}, \au{{Katz}, U.~F.}, \au{{Kenzie}, M.}, \au{{Khoze},
  V.~A.}, \au{{Klein}, S.~R.}, \au{{Klempt}, E.}, \au{{Kowalewski}, R.~V.},
  \au{{Krauss}, F.}, \au{{Kreps}, M.}, \au{{Krusche}, B.}, \au{{Kwon}, Y.},
  \au{{Lahav}, O.}, \au{{Laiho}, J.}, \au{{Lellouch}, L.~P.},
  \au{{Lesgourgues}, J.}, \au{{Liddle}, A.~R.}, \au{{Ligeti}, Z.},
  \au{{Lippmann}, C.}, \au{{Liss}, T.~M.}, \au{{Littenberg}, L.},
  \au{{Lourengo}, C.}, \au{{Lugovsky}, S.~B.}, \au{{Lusiani}, A.},
  \au{{Makida}, Y.}, \au{{Maltoni}, F.}, \au{{Mannel}, T.}, \au{{Manohar},
  A.~V.}, \au{{Marciano}, W.~J.}, \au{{Masoni}, A.}, \au{{Matthews}, J.},
  \au{{Mei{\ss}ner}, U.~G.}, \au{{Mikhasenko}, M.}, \au{{Miller}, D.~J.},
  \au{{Milstead}, D.}, \au{{Mitchell}, R.~E.}, \au{{M{\"o}nig}, K.},
  \au{{Molaro}, P.}, \au{{Moortgat}, F.}, \au{{Moskovic}, M.}, \au{{Nakamura},
  K.}, \au{{Narain}, M.}, \au{{Nason}, P.}, \au{{Navas}, S.}, \au{{Neubert},
  M.}, \au{{Nevski}, P.}, \au{{Nir}, Y.}, \au{{Olive}, K.~A.},
  \au{{Patrignani}, C.}, \au{{Peacock}, J.~A.}, \au{{Petcov}, S.~T.},
  \au{{Petrov}, V.~A.}, \au{{Pich}, A.}, \au{{Piepke}, A.}, \au{{Pomarol}, A.},
  \au{{Profumo}, S.}, \au{{Quadt}, A.}, \au{{Rabbertz}, K.}, \au{{Rademacker},
  J.}, \au{{Raffelt}, G.}, \au{{Ramani}, H.}, \au{{Ramsey-Musolf}, M.},
  \au{{Ratcliff}, B.~N.}, \au{{Richardson}, P.}, \au{{Ringwald}, A.},
  \au{{Roesler}, S.}, \au{{Rolli}, S.}, \au{{Romaniouk}, A.}, \au{{Rosenberg},
  L.~J.}, \au{{Rosner}, J.~L.}, \au{{Rybka}, G.}, \au{{Ryskin}, M.},
  \au{{Ryutin}, R.~A.}, \au{{Sakai}, Y.}, \au{{Salam}, G.~P.}, \au{{Sarkar},
  S.}, \au{{Sauli}, F.}, \au{{Schneider}, O.}, \au{{Scholberg}, K.},
  \au{{Schwartz}, A.~J.}, \au{{Schwiening}, J.}, \au{{Scott}, D.},
  \au{{Sharma}, V.}, \au{{Sharpe}, S.~R.}, \au{{Shutt}, T.}, \au{{Silari}, M.},
  \au{{Sj{\"o}strand}, T.}, \au{{Skands}, P.}, \au{{Skwarnicki}, T.},
  \au{{Smoot}, G.~F.}, \au{{Soffer}, A.}, \au{{Sozzi}, M.~S.}, \au{{Spanier},
  S.}, \au{{Spiering}, C.}, \au{{Stahl}, A.}, \au{{Stone}, S.~L.},
  \au{{Sumino}, Y.}, \au{{Sumiyoshi}, T.}, \au{{Syphers}, M.~J.},
  \au{{Takahashi}, F.}, \au{{Tanabashi}, M.}, \au{{Tanaka}, J.},
  \au{{Ta{\v{s}}evsk{\'y}}, M.}, \au{{Terashi}, K.}, \au{{Terning}, J.},
  \au{{Thoma}, U.}, \au{{Thorne}, R.~S.}, \au{{Tiator}, L.}, \au{{Titov}, M.},
  \au{{Tkachenko}, N.~P.}, \au{{Tovey}, D.~R.}, \au{{Trabelsi}, K.},
  \au{{Urquijo}, P.}, \au{{Valencia}, G.}, \au{{Van de Water}, R.},
  \au{{Varelas}, N.}, \au{{Venanzoni}, G.}, \au{{Verde}, L.}, \au{{Vincter},
  M.~G.}, \au{{Vogel}, P.}, \au{{Vogelsang}, W.}, \au{{Vogt}, A.},
  \au{{Vorobyev}, V.}, \au{{Wakely}, S.~P.}, \au{{Walkowiak}, W.},
  \au{{Walter}, C.~W.}, \au{{Wands}, D.}, \au{{Wascko}, M.~O.}, \au{{Weinberg},
  D.~H.}, \au{{Weinberg}, E.~J.}, \au{{White}, M.}, \au{{Wiencke}, L.~R.},
  \au{{Willocq}, S.}, \au{{Woody}, C.~L.}, \au{{Workman}, R.~L.},
  \au{{Yokoyama}, M.}, \au{{Yoshida}, R.}, \au{{Zand erighi}, G.},
  \au{{Zeller}, G.~P.}, \au{{Zenin}, O.~V.}, \au{{Zhu}, R.~Y.}, \au{{Zhu},
  S.~L.}, \au{{Zimmermann}, F.}, \au{{Anderson}, J.}, \au{{Basaglia}, T.},
  \au{{Lugovsky}, V.~S.}, \au{{Schaffner}, P.} \& \au{{Zheng}, W.}} \yr{2020}
  \at{{Review of Particle Physics}}.  \jt{Progress of Theoretical and
  Experimental Physics}  \bvol{2020}~(8),  \pg{083C01}.

\bibitem[Peron {\em et~al.\/}(2020)Peron, Aharonian, Casanova, Zanin \&
  Romoli]{Peron_2020}
{\sc \au{Peron, Giada}, \au{Aharonian, Felix}, \au{Casanova, Sabrina},
  \au{Zanin, Roberta} \& \au{Romoli, Carlo}} \yr{2020}  \at{On the gamma-ray
  emission of w44 and its surroundings}.  \jt{The Astrophysical Journal}
  \bvol{896}~(2),  \pg{L23}.

\bibitem[Pfeffermann \& Aschenbach(1996)]{article1996}
{\sc \au{Pfeffermann, Elmar} \& \au{Aschenbach, B.}} \yr{1996}  \at{Rosat
  observation of a new supernova remnant in the constellation scorpius.}
  \bvol{-1},  \pg{267--268}.

\bibitem[{Pohl} {\em et~al.\/}(2005){Pohl}, {Yan} \& {Lazarian}]{Pohl+05}
{\sc \au{{Pohl}, M.}, \au{{Yan}, H.} \& \au{{Lazarian}, A.}} \yr{2005}
  \at{{Magnetically Limited X-Ray Filaments in Young Supernova Remnants}}.
  \jt{ApJL}  \bvol{626}~(2),  \pg{L101--L104}.

\bibitem[{Recchia} {\em et~al.\/}(2019){Recchia}, {Gabici}, {Aharonian} \&
  {Vink}]{Recchia19}
{\sc \au{{Recchia}, S.}, \au{{Gabici}, S.}, \au{{Aharonian}, F.~A.} \&
  \au{{Vink}, J.}} \yr{2019}  \at{{Local fading accelerator and the origin of
  TeV cosmic ray electrons}}.  \jt{Physical Review D}  \bvol{99}~(10),
  \pg{103022},  \arxiv{arXiv: 1811.07551}.

\bibitem[{Reid} {\em et~al.\/}(2014){Reid}, {Menten}, {Brunthaler}, {Zheng},
  {Dame}, {Xu}, {Wu}, {Zhang}, {Sanna}, {Sato}, {Hachisuka}, {Choi}, {Immer},
  {Moscadelli}, {Rygl} \& {Bartkiewicz}]{2014ApJ...783..130R}
{\sc \au{{Reid}, M.~J.}, \au{{Menten}, K.~M.}, \au{{Brunthaler}, A.},
  \au{{Zheng}, X.~W.}, \au{{Dame}, T.~M.}, \au{{Xu}, Y.}, \au{{Wu}, Y.},
  \au{{Zhang}, B.}, \au{{Sanna}, A.}, \au{{Sato}, M.}, \au{{Hachisuka}, K.},
  \au{{Choi}, Y.~K.}, \au{{Immer}, K.}, \au{{Moscadelli}, L.}, \au{{Rygl},
  K.~L.~J.} \& \au{{Bartkiewicz}, A.}} \yr{2014}  \at{{Trigonometric Parallaxes
  of High Mass Star Forming Regions: The Structure and Kinematics of the Milky
  Way}}.  \jt{ApJ}  \bvol{783}~(2),  \pg{130},  \arxiv{arXiv: 1401.5377}.

\bibitem[{Reimer} {\em et~al.\/}(2006){Reimer}, {Pohl} \&
  {Reimer}]{2006ApJ...644.1118R}
{\sc \au{{Reimer}, A.}, \au{{Pohl}, M.} \& \au{{Reimer}, O.}} \yr{2006}
  \at{{Nonthermal High-Energy Emission from Colliding Winds of Massive Stars}}.
   \jt{ApJ}  \bvol{644}~(2),  \pg{1118--1144},  \arxiv{arXiv:
  astro-ph/0510701}.

\bibitem[{Rettig} \& {Pohl}(2012)]{Rettig2012}
{\sc \au{{Rettig}, R.} \& \au{{Pohl}, M.}} \yr{2012}  \at{{The properties of
  non-thermal X-ray filaments in young supernova remnants}}.  \jt{A\&A}
  \bvol{545},  \pg{A47},  \arxiv{arXiv: 1208.5322}.

\bibitem[{Reville} {\em et~al.\/}(2008){Reville}, {O'Sullivan}, {Duffy} \&
  {Kirk}]{Reville+08}
{\sc \au{{Reville}, B.}, \au{{O'Sullivan}, S.}, \au{{Duffy}, P.} \& \au{{Kirk},
  J.~G.}} \yr{2008}  \at{{The transport of cosmic rays in self-excited magnetic
  turbulence}}.  \jt{MNRAS}  \bvol{386}~(1),  \pg{509--515},  \arxiv{arXiv:
  0802.0109}.

\bibitem[{Riquelme} \& {Spitkovsky}(2009)]{RiquelmeSpitkovsky09}
{\sc \au{{Riquelme}, Mario~A.} \& \au{{Spitkovsky}, Anatoly}} \yr{2009}
  \at{{Nonlinear Study of Bell's Cosmic Ray Current-Driven Instability}}.
  \jt{ApJ}  \bvol{694}~(1),  \pg{626--642},  \arxiv{arXiv: 0810.4565}.

\bibitem[{Rogachevskii} {\em et~al.\/}(2012){Rogachevskii}, {Kleeorin},
  {Brandenburg} \& {Eichler}]{Rogachevskii12}
{\sc \au{{Rogachevskii}, Igor}, \au{{Kleeorin}, Nathan}, \au{{Brandenburg},
  Axel} \& \au{{Eichler}, David}} \yr{2012}  \at{{Cosmic-Ray Current-driven
  Turbulence and Mean-field Dynamo Effect}}.  \jt{ApJ}  \bvol{753}~(1),
  \pg{6},  \arxiv{arXiv: 1204.4246}.

\bibitem[{Roman-Duval} {\em et~al.\/}(2009){Roman-Duval}, {Jackson}, {Heyer},
  {Johnson}, {Rathborne}, {Shah} \& {Simon}]{2009ApJ...699.1153R}
{\sc \au{{Roman-Duval}, Julia}, \au{{Jackson}, James~M.}, \au{{Heyer}, Mark},
  \au{{Johnson}, Alexis}, \au{{Rathborne}, Jill}, \au{{Shah}, Ronak} \&
  \au{{Simon}, Robert}} \yr{2009}  \at{{Kinematic Distances to Molecular Clouds
  Identified in the Galactic Ring Survey}}.  \jt{ApJ}  \bvol{699}~(2),
  \pg{1153--1170},  \arxiv{arXiv: 0905.0723}.

\bibitem[{Schure} \& {Bell}(2014)]{SchureBell14}
{\sc \au{{Schure}, K.~M.} \& \au{{Bell}, A.~R.}} \yr{2014}  \at{{From cosmic
  ray source to the Galactic pool}}.  \jt{MNRAS}  \bvol{437}~(3),
  \pg{2802--2805},  \arxiv{arXiv: 1310.7027}.

\bibitem[{Schure} {\em et~al.\/}(2012){Schure}, {Bell}, {O'C Drury} \&
  {Bykov}]{Schure+12}
{\sc \au{{Schure}, K.~M.}, \au{{Bell}, A.~R.}, \au{{O'C Drury}, L.} \&
  \au{{Bykov}, A.~M.}} \yr{2012}  \at{{Diffusive Shock Acceleration and
  Magnetic Field Amplification}}.  \jt{Space Science Review}  \bvol{173}~(1-4),
   \pg{491--519},  \arxiv{arXiv: 1203.1637}.

\bibitem[{Shikaze} {\em et~al.\/}(2007){Shikaze}, {Haino}, {Abe}, {Fuke},
  {Hams}, {Kim}, {Makida}, {Matsuda}, {Mitchell}, {Moiseev}, {Nishimura},
  {Nozaki}, {Orito}, {Ormes}, {Sanuki}, {Sasaki}, {Seo}, {Streitmatter},
  {Suzuki}, {Tanaka}, {Yamagami}, {Yamamoto}, {Yoshida} \&
  {Yoshimura}]{2007APh....28..154S}
{\sc \au{{Shikaze}, Y.}, \au{{Haino}, S.}, \au{{Abe}, K.}, \au{{Fuke}, H.},
  \au{{Hams}, T.}, \au{{Kim}, K.~C.}, \au{{Makida}, Y.}, \au{{Matsuda}, S.},
  \au{{Mitchell}, J.~W.}, \au{{Moiseev}, A.~A.}, \au{{Nishimura}, J.},
  \au{{Nozaki}, M.}, \au{{Orito}, S.}, \au{{Ormes}, J.~F.}, \au{{Sanuki}, T.},
  \au{{Sasaki}, M.}, \au{{Seo}, E.~S.}, \au{{Streitmatter}, R.~E.},
  \au{{Suzuki}, J.}, \au{{Tanaka}, K.}, \au{{Yamagami}, T.}, \au{{Yamamoto},
  A.}, \au{{Yoshida}, T.} \& \au{{Yoshimura}, K.}} \yr{2007}  \at{{Measurements
  of 0.2 20 GeV/n cosmic-ray proton and helium spectra from 1997 through 2002
  with the BESS spectrometer}}.  \jt{Astroparticle Physics}  \bvol{28}~(1),
  \pg{154--167},  \arxiv{arXiv: astro-ph/0611388}.

\bibitem[{Skilling}(1975)]{Skilling75}
{\sc \au{{Skilling}, J.}} \yr{1975}  \at{{Cosmic ray streaming - II. Effect of
  particles on Alfv{\'e}n waves.}}  \jt{MNRAS}  \bvol{173},  \pg{245--254}.

\bibitem[{Stone} {\em et~al.\/}(2013){Stone}, {Cummings}, {McDonald},
  {Heikkila}, {Lal} \& {Webber}]{Stone+13}
{\sc \au{{Stone}, E.~C.}, \au{{Cummings}, A.~C.}, \au{{McDonald}, F.~B.},
  \au{{Heikkila}, B.~C.}, \au{{Lal}, N.} \& \au{{Webber}, W.~R.}} \yr{2013}
  \at{{Voyager 1 Observes Low-Energy Galactic Cosmic Rays in a Region Depleted
  of Heliospheric Ions}}.  \jt{Science}  \bvol{341}~(6142),  \pg{150--153}.

\bibitem[{Sveshnikova} {\em et~al.\/}(2013){Sveshnikova}, {Strelnikova} \&
  {Ptuskin}]{Sveshnikova+13}
{\sc \au{{Sveshnikova}, L.~G.}, \au{{Strelnikova}, O.~N.} \& \au{{Ptuskin},
  V.~S.}} \yr{2013}  \at{{Spectrum and anisotropy of cosmic rays at
  TeV-PeV-energies and contribution of nearby sources}}.  \jt{Astroparticle
  Physics}  \bvol{50},  \pg{33--46},  \arxiv{arXiv: 1301.2028}.

\bibitem[{Tanaka} {\em et~al.\/}(2008){Tanaka}, {Uchiyama}, {Aharonian},
  {Takahashi}, {Bamba}, {Hiraga}, {Kataoka}, {Kishishita}, {Kokubun}, {Mori},
  {Nakazawa}, {Petre}, {Tajima} \& {Watanabe}]{2008ApJ...685..988T}
{\sc \au{{Tanaka}, T.}, \au{{Uchiyama}, Y.}, \au{{Aharonian}, F.~A.},
  \au{{Takahashi}, T.}, \au{{Bamba}, A.}, \au{{Hiraga}, J.~S.}, \au{{Kataoka},
  J.}, \au{{Kishishita}, T.}, \au{{Kokubun}, M.}, \au{{Mori}, K.},
  \au{{Nakazawa}, K.}, \au{{Petre}, R.}, \au{{Tajima}, H.} \& \au{{Watanabe},
  S.}} \yr{2008}  \at{{Study of Nonthermal Emission from SNR RX J1713.7-3946
  with Suzaku}}.  \jt{ApJ}  \bvol{685},  \pg{988--1004},  \arxiv{arXiv:
  0806.1490}.

\bibitem[{Tavani} {\em et~al.\/}(2010){Tavani}, {Giuliani}, {Chen}, {Argan},
  {Barbiellini}, {Bulgarelli}, {Caraveo}, {Cattaneo}, {Cocco}, {Contessi},
  {D'Ammand o}, {Costa}, {De Paris}, {Del Monte}, {Di Cocco}, {Donnarumma},
  {Evangelista}, {Ferrari}, {Feroci}, {Fuschino}, {Galli}, {Gianotti},
  {Labanti}, {Lapshov}, {Lazzarotto}, {Lipari}, {Longo}, {Marisaldi},
  {Mastropietro}, {Mereghetti}, {Morelli}, {Moretti}, {Morselli}, {Pacciani},
  {Pellizzoni}, {Perotti}, {Piano}, {Picozza}, {Pilia}, {Pucella}, {Prest},
  {Rapisarda}, {Rappoldi}, {Scalise}, {Rubini}, {Sabatini}, {Striani},
  {Soffitta}, {Trifoglio}, {Trois}, {Vallazza}, {Vercellone}, {Vittorini},
  {Zambra}, {Zanello}, {Pittori}, {Verrecchia}, {Santolamazza}, {Giommi},
  {Colafrancesco}, {Antonelli} \& {Salotti}]{AgileIC443}
{\sc \au{{Tavani}, M.}, \au{{Giuliani}, A.}, \au{{Chen}, A.~W.}, \au{{Argan},
  A.}, \au{{Barbiellini}, G.}, \au{{Bulgarelli}, A.}, \au{{Caraveo}, P.},
  \au{{Cattaneo}, P.~W.}, \au{{Cocco}, V.}, \au{{Contessi}, T.}, \au{{D'Ammand
  o}, F.}, \au{{Costa}, E.}, \au{{De Paris}, G.}, \au{{Del Monte}, E.}, \au{{Di
  Cocco}, G.}, \au{{Donnarumma}, I.}, \au{{Evangelista}, Y.}, \au{{Ferrari},
  A.}, \au{{Feroci}, M.}, \au{{Fuschino}, F.}, \au{{Galli}, M.},
  \au{{Gianotti}, F.}, \au{{Labanti}, C.}, \au{{Lapshov}, I.},
  \au{{Lazzarotto}, F.}, \au{{Lipari}, P.}, \au{{Longo}, F.}, \au{{Marisaldi},
  M.}, \au{{Mastropietro}, M.}, \au{{Mereghetti}, S.}, \au{{Morelli}, E.},
  \au{{Moretti}, E.}, \au{{Morselli}, A.}, \au{{Pacciani}, L.},
  \au{{Pellizzoni}, A.}, \au{{Perotti}, F.}, \au{{Piano}, G.}, \au{{Picozza},
  P.}, \au{{Pilia}, M.}, \au{{Pucella}, G.}, \au{{Prest}, M.}, \au{{Rapisarda},
  M.}, \au{{Rappoldi}, A.}, \au{{Scalise}, E.}, \au{{Rubini}, A.},
  \au{{Sabatini}, S.}, \au{{Striani}, E.}, \au{{Soffitta}, P.},
  \au{{Trifoglio}, M.}, \au{{Trois}, A.}, \au{{Vallazza}, E.},
  \au{{Vercellone}, S.}, \au{{Vittorini}, V.}, \au{{Zambra}, A.},
  \au{{Zanello}, D.}, \au{{Pittori}, C.}, \au{{Verrecchia}, F.},
  \au{{Santolamazza}, P.}, \au{{Giommi}, P.}, \au{{Colafrancesco}, S.},
  \au{{Antonelli}, L.~A.} \& \au{{Salotti}, L.}} \yr{2010}  \at{{Direct
  Evidence for Hadronic Cosmic-Ray Acceleration in the Supernova Remnant IC
  443}}.  \jt{ApJ Letters}  \bvol{710}~(2),  \pg{L151--L155},  \arxiv{arXiv:
  1001.5150}.

\bibitem[{Thoudam} \& {H{\"o}randel}(2012)]{HorandelHardening}
{\sc \au{{Thoudam}, Satyendra} \& \au{{H{\"o}randel}, J{\"o}rg~R.}} \yr{2012}
  \at{{Nearby supernova remnants and the cosmic ray spectral hardening at high
  energies}}.  \jt{MNRAS}  \bvol{421}~(2),  \pg{1209--1214},  \arxiv{arXiv:
  1112.3020}.

\bibitem[{Tomassetti}(2012)]{TomassettiHardening}
{\sc \au{{Tomassetti}, Nicola}} \yr{2012}  \at{{Origin of the Cosmic-Ray
  Spectral Hardening}}.  \jt{ApJ Letters}  \bvol{752}~(1),  \pg{L13},
  \arxiv{arXiv: 1204.4492}.

\bibitem[{Torres} {\em et~al.\/}(2004){Torres}, {Domingo-Santamar{\'\i}a} \&
  {Romero}]{2004ApJ...601L..75T}
{\sc \au{{Torres}, Diego~F.}, \au{{Domingo-Santamar{\'\i}a}, Eva} \&
  \au{{Romero}, Gustavo~E.}} \yr{2004}  \at{{High-Energy Gamma Rays from
  Stellar Associations}}.  \jt{ApJL}  \bvol{601}~(1),  \pg{L75--L78},
  \arxiv{arXiv: astro-ph/0312128}.

\bibitem[{Tsuboi} {\em et~al.\/}(1999){Tsuboi}, {Handa} \&
  {Ukita}]{1999ApJS..120....1T}
{\sc \au{{Tsuboi}, M.}, \au{{Handa}, T.} \& \au{{Ukita}, N.}} \yr{1999}
  \at{{Dense Molecular Clouds in the Galactic Center Region. I. Observations
  and Data}}.  \jt{APJS}  \bvol{120},  \pg{1--39}.

\bibitem[{Uchiyama} {\em et~al.\/}(2007){Uchiyama}, {Aharonian}, {Tanaka},
  {Takahashi} \& {Maeda}]{2007Natur.449..576U}
{\sc \au{{Uchiyama}, Yasunobu}, \au{{Aharonian}, Felix~A.}, \au{{Tanaka},
  Takaaki}, \au{{Takahashi}, Tadayuki} \& \au{{Maeda}, Yoshitomo}} \yr{2007}
  \at{{Extremely fast acceleration of cosmic rays in a supernova remnant}}.
  \jt{Nature}  \bvol{449}~(7162),  \pg{576--578}.

\bibitem[{Uchiyama} {\em et~al.\/}(2010){Uchiyama}, {Blandford}, {Funk},
  {Tajima} \& {Tanaka}]{Uchiyama+10}
{\sc \au{{Uchiyama}, Yasunobu}, \au{{Blandford}, Roger~D.}, \au{{Funk},
  Stefan}, \au{{Tajima}, Hiroyasu} \& \au{{Tanaka}, Takaaki}} \yr{2010}
  \at{{Gamma-ray Emission from Crushed Clouds in Supernova Remnants}}.  \jt{ApJ
  Letters}  \bvol{723}~(1),  \pg{L122--L126},  \arxiv{arXiv: 1008.1840}.

\bibitem[{Uchiyama} {\em et~al.\/}(2012){Uchiyama}, {Funk}, {Katagiri},
  {Katsuta}, {Lemoine-Goumard}, {Tajima}, {Tanaka} \&
  {Torres}]{2012ApJ...749L..35U}
{\sc \au{{Uchiyama}, Yasunobu}, \au{{Funk}, Stefan}, \au{{Katagiri}, Hideaki},
  \au{{Katsuta}, Junichiro}, \au{{Lemoine-Goumard}, Marianne}, \au{{Tajima},
  Hiroyasu}, \au{{Tanaka}, Takaaki} \& \au{{Torres}, Diego~F.}} \yr{2012}
  \at{{Fermi Large Area Telescope Discovery of GeV Gamma-Ray Emission from the
  Vicinity of SNR W44}}.  \jt{ApJ Letters}  \bvol{749}~(2),  \pg{L35},
  \arxiv{arXiv: 1203.3234}.

\bibitem[{Vink}(2012)]{Vink12}
{\sc \au{{Vink}, Jacco}} \yr{2012}  \at{{Supernova remnants: the X-ray
  perspective}}.  \jt{A\&A Rev.}  \bvol{20},  \pg{49},  \arxiv{arXiv:
  1112.0576}.

\bibitem[{Voelk} \& {Forman}(1982)]{1982ApJ...253..188V}
{\sc \au{{Voelk}, H.~J.} \& \au{{Forman}, M.}} \yr{1982}  \at{{Cosmic rays and
  gamma-rays from OB stars}}.  \jt{ApJ}  \bvol{253},  \pg{188--198}.

\bibitem[Wang {\em et~al.\/}(1997)Wang, Qu \& Chen]{article1997}
{\sc \au{Wang, Zong-Ren}, \au{Qu, Q.-Y} \& \au{Chen, Yang}} \yr{1997} Is rx
  j1713.7-3946 the remnant of the ad393 guest star?

\bibitem[{Wentzel}(1974)]{Wentzel74}
{\sc \au{{Wentzel}, D.~G.}} \yr{1974}  \at{{Cosmic-ray propagation in the
  Galaxy: collective effects.}}  \jt{A\&A Rev.}  \bvol{12},  \pg{71--96}.

\bibitem[{Yanasak} {\em et~al.\/}(2001){Yanasak}, {Wiedenbeck}, {Mewaldt},
  {Davis}, {Cummings}, {George}, {Leske}, {Stone}, {Christian}, {von
  Rosenvinge}, {Binns}, {Hink} \& {Israel}]{Yanasak2001}
{\sc \au{{Yanasak}, N.~E.}, \au{{Wiedenbeck}, M.~E.}, \au{{Mewaldt}, R.~A.},
  \au{{Davis}, A.~J.}, \au{{Cummings}, A.~C.}, \au{{George}, J.~S.},
  \au{{Leske}, R.~A.}, \au{{Stone}, E.~C.}, \au{{Christian}, E.~R.}, \au{{von
  Rosenvinge}, T.~T.}, \au{{Binns}, W.~R.}, \au{{Hink}, P.~L.} \& \au{{Israel},
  M.~H.}} \yr{2001}  \at{{Measurement of the Secondary Radionuclides $^{10}$Be,
  $^{26}$Al, $^{36}$Cl, $^{54}$Mn, and $^{14}$C and Implications for the
  Galactic Cosmic-Ray Age}}.  \jt{ApJ}  \bvol{563}~(2),  \pg{768--792}.

\bibitem[Yin {\em et~al.\/}(2013)Yin, Yu, Yuan \& Bi]{PhysRevD.88.023001}
{\sc \au{Yin, Peng-Fei}, \au{Yu, Zhao-Huan}, \au{Yuan, Qiang} \& \au{Bi,
  Xiao-Jun}} \yr{2013}  \at{Pulsar interpretation for the ams-02 result}.
  \jt{Phys. Rev. D}  \bvol{88},  \pg{023001}.

\bibitem[{Zabalza}(2015)]{naima}
{\sc \au{{Zabalza}, V.}} \yr{2015}  \at{naima: a python package for inference
  of relativistic particle energy distributions from observed nonthermal
  spectra}.  \jt{Proc.~of International Cosmic Ray Conference 2015}  \pg{p.
  922},  \arxiv{arXiv: 1509.03319}.

\bibitem[{Zirakashvili} \& {Aharonian}(2010)]{2010ApJ...708..965Z}
{\sc \au{{Zirakashvili}, V.~N.} \& \au{{Aharonian}, F.~A.}} \yr{2010}
  \at{{Nonthermal Radiation of Young Supernova Remnants: The Case of RX
  J1713.7-3946}}.  \jt{ApJ}  \bvol{708},  \pg{965--980},  \arxiv{arXiv:
  0909.2285}.

\bibitem[{Zweibel}(2013)]{Zweibel13}
{\sc \au{{Zweibel}, Ellen~G.}} \yr{2013}  \at{{The microphysics and
  macrophysics of cosmic rays}}.  \jt{Physics of Plasmas}  \bvol{20}~(5),
  \pg{055501}.

\end{thebibliography}

\end{document}